\documentclass[12pt,a4paper]{report}
\usepackage{xifthen}
\usepackage[algochapter]{algorithm2e}
\usepackage[]{algorithmic}
\usepackage{rotating}
\usepackage{pdflscape}
\usepackage{packages/scienceThesis}
\usepackage{verbatim}
\usepackage{amsfonts}
\usepackage{subfig}
\usepackage{courier}
\usepackage{pdfpages}
\usepackage{lipsum}
\usepackage[acronym,toc,nonumberlist]{glossaries}
\usepackage{breqn}
\usepackage{xcolor,colortbl}
\usepackage{amsthm}

\theoremstyle{definition}

\theoremstyle{definition}

\theoremstyle{definition}

\theoremstyle{definition}

\theoremstyle{definition}

\theoremstyle{definition}

\usepackage{fancyhdr}
\makeatletter
\renewcommand{\chaptermark}[1]{\markboth{\small\textsc{\@chapapp}\ \thechapter:\ \sc{#1}}{}}
\makeatother
\usepackage{sectsty}
\chapterfont{\large\sc\centering}
\chaptertitlefont{\sc\centering}
\subsubsectionfont{\centering}

\usepackage[Bjornstrup]{fncychap}
\newsavebox{\ChpNumBox}

\makeatletter
\ChNameVar{\Large\rm} 
\ChNumVar{\Huge} 
\ChTitleVar{\Huge\rm} 
\ChRuleWidth{4pt} 
\ChNameUpperCase 

\renewcommand{\DOTI}[1]{%
\vskip -100pt%
\begin{minipage}[c][][b]{0.82\textwidth}\linespread{0}
\centering\sc\large\color{white}{Chapter}\color{black}\\
\begin{tabular}{@{}m{0cm}@{}@{}c@{}}
\rule{0cm}{2.8cm} & \begin{minipage}{1\textwidth}\linespread{0.9}\centering\Huge\sc{#1}\end{minipage}%
\end{tabular}
\end{minipage}\hfill%
             \begin{minipage}[c][][b]{0.17\textwidth}%
 \centering\sc\large Chapter\\
 \feline@chm[\textwidth]%
\end{minipage}%
\par\nobreak
\vskip 35\p@}

\renewcommand{\DOTIS}[1]{%
\vskip -100pt%
\begin{minipage}[c][][b]{0.82\textwidth}\linespread{0}
\centering\sc\large\color{white}{Chapter}\color{black}\\
\begin{tabular}{@{}m{0cm}@{}@{}c@{}}
\rule{0cm}{2.8cm} & \begin{minipage}{1\textwidth}\linespread{0.9}\centering\Huge\sc{#1}\end{minipage}%
\end{tabular}
\end{minipage}\hfill%
             \begin{minipage}[c][][b]{0.17\textwidth}%
 \centering\sc\large\color{white}\sc Chapter\\
 \feline@chmn[\textwidth]%
\end{minipage}%
\par\nobreak
\vskip 35\p@}
\makeatother

\renewcommand{\abstractcom}[1]{%
\ifthenelse{\equal{#1}{}}{}{%
\clearpage
\thispagestyle{plain}
\singlespacing%
\addcontentsline{toc}{chapter}{Abstract}%
\DOTIS{Abstract}
#1
\par
}}
\renewcommand{\acknowledgementscom}[1]{%
\ifthenelse{\equal{#1}{}}{}{%
    \clearpage
    \thispagestyle{plain}
    \singlespacing
    \addcontentsline{toc}{chapter}{Acknowledgements}
	\DOTIS{Acknowledgements}
	#1
	\par
	}
}

\usepackage[colorlinks]{hyperref}
\hypersetup{
        bookmarksnumbered,
        pdfstartview={FitH},
        citecolor={black},
        linkcolor={black},
        urlcolor={black},
        pdfpagemode={UseOutlines}
}
\usepackage{subfiles}

\usepackage{lmodern}
\usepackage[T2A,T1]{fontenc}
\usepackage[english]{babel}
\usepackage{amsmath}
\usepackage{tcolorbox}
\usepackage{amsfonts}
\usepackage{amssymb}
\usepackage{graphicx}
\usepackage{tabularx}
\usepackage{listings}   

\usepackage{booktabs}
\usepackage{multirow}
\usepackage{float}
\usepackage{placeins}
\usepackage[makeroom]{cancel}
\usepackage[utf8]{inputenc}
\usepackage{booktabs}
\usepackage{tabularx}
\usepackage{amsmath}
\usepackage{bm}
\usepackage{amsfonts}
\usepackage[makeroom]{cancel}
\usepackage[algochapter]{algorithm2e}
\usepackage{setspace}
\usepackage{algorithm2e}
\SetKwInput{As}{Assumptions}
\SetKwInput{KwPar}{Parameters}

\SetKwFor{ForAllP}{for all}{do in parallel}{end}
\SetKwFor{ForAll}{for all}{do}{end}

\SetKwInput{KwTunParam}{Tunable Parameters}
\SetKwInput{KwConfParam}{Configuration Parameters}
\SetKwInput{KwThreads}{Threads Allocated}
\SetKwInput{KwInit}{Initilise}
\SetKwInput{KwOptParam}{Optimisation Parameters}
\usepackage[none]{hyphenat}
\usepackage[numbers,sort&compress]{natbib}

\usepackage{IEEEtrantools}
\usepackage{enumerate}
\newcolumntype{L}{>{\raggedright\arraybackslash}X}
\newcolumntype{P}[1]{>{\raggedright\arraybackslash}p{#1}}

\newcommand{\etal}{\textit{et al. }}

\newcommand{\optarg}[1][]{%
  \ifthenelse{\isempty{#1}}%
    {}
    {(((#1)))}
}
\newcommand{\shah}[1]{\text{III}_{#1}}
\newcommand{\gshah}[1]{\overline{\text{III}}_{#1}}

\newcommand{\rect}[1]{\text{rect}_{#1}}
\newcommand{\sinc}[1]{\text{sinc}_{#1}}
\newcommand{\freq}[0]{\xi}
\newcommand{\freqel}[0]{\xi}
\newcommand{\freqset}[0]{\Xi}

\newcommand{\myvec}[1]{\boldsymbol{#1}}
\newcommand{\myfreq}[0]{\myvec{\freq}}
\newcommand{\myfreqel}[0]{\myvec{\freqel}}
\newcommand{\x}[0]{x}
\newcommand{\myx}[0]{\myvec{\x}}
\newcommand{\mybeta}[0]{\beta} 
\newcommand{\myN}[0]{N} 

\newcommand{\myDeltax}[0]{\Delta\myx} 

\newcommand{\appendgpumetric}[0]{P100:}
\newcommand{\metric}[1]{\texttt{#1} }
\newcommand{\gpumetric}[1]{\metric{\appendgpumetric#1}}
\newcommand{\commitrate}[0]{\metric{Commit Rate}}
\newcommand{\commitratestop}[0]{\metric{Commit Rate.}}
\newcommand{\commitratecomma}[0]{\metric{Commit Rate,}}
\newcommand{\globalhitrate}[0]{\gpumetric{global\_hit\_rate}}
\newcommand{\globalhitratecomma}[0]{\gpumetric{global\_hit\_rate,}}

\newcommand{\gridrate}[0]{\metric{Gridding Rate}}
\newcommand{\gridratestop}[0]{\metric{Gridding Rate.}}
\newcommand{\gridratecomma}[0]{\metric{Gridding Rate,}}

\newcommand{\bestgridrate}[0]{\metric{Best Gridding Rate}}
\newcommand{\bestgridratestop}[0]{\metric{Best Gridding Rate.}}
\newcommand{\bestgridratecomma}[0]{\metric{Best Gridding Rate,}}
\newcommand{\maxbestgridrate}[0]{\metric {Max Best Gridding Rate}}

\newcommand{\ltexrate}[0]{\gpumetric{l2\_tex\_hit\_rate}}
\newcommand{\ltexratestop}[0]{\gpumetric{l2\_tex\_hit\_rate.}}

\newcommand{\polgain}[0]{\metric{Polarisation Gain}}
\newcommand{\polgaincomma}[0]{\metric{Polarisation Gain,}}
\newcommand{\polgainstop}[0]{\metric{Polarisation Gain.}}

\newcommand{\dramreadtransactionsrate}[0]{\metric{DRAM Read Rate}}

\newcommand{\efficiency}[0]{\metric{Efficiency}}

\newcommand{\optimalityfactor}[0]{\metric{Optimality Factor}}
\newcommand{\wsplitoptimalityfactor}[1]{\metric{W\textsubscript{split}=#1 Optimality Factor}}

\newcommand{\gridderadvantage}[0]{\metric{Gridder Advantage}}
\newcommand{\gridderadvantageonhybrid}[0]{\metric{NN Gridder Advantage over Hybrid }}
\newcommand{\rowpruninggain}[0]{\metric{Row Pruning Gain}}
\newcommand{\columnpruninggain}[0]{\metric{Column Pruning Gain}}
\newcommand{\columnpruninggainstop}[0]{\metric{Column Pruning Gain.}}
\newcommand{\columnpruninggaincomma}[0]{\metric{Column Pruning Gain,}}
\newcommand{\accumulatedpruninggain}[0]{\metric{Accumulated Pruning Gain}}
\newcommand{\accumulatedpruninggainstop}[0]{\metric{Accumulated Pruning Gain.}}
\newcommand{\columnprunerfootprint}[0]{\metric{Column Pruner Footprint}}

\newcommand{\pruninggain}[0]{\metric{Pruning Gain}}
\newcommand{\computeutilisation}[0]{\metric{Compute Utilisation}}
\newcommand{\computerate}[0]{\metric{Compute Rate}}

\newcommand{\computeratestop}[0]{\metric{Compute Rate.}}

\newcommand{\texutilisation}[0]{\gpumetric{tex\_utilization}}
\newcommand{\ltwoutilisation}[0]{\gpumetric{l2\_utilization}}
\newcommand{\sharedutilisation}[0]{\gpumetric{shared\_utilization}}
\newcommand{\dramutilisation}[0]{\gpumetric{dram\_utilization}}
\newcommand{\dramutilisationstop}[0]{\gpumetric{dram\_utilization.}}
\newcommand{\dramutilisationcomma}[0]{\gpumetric{dram\_utilization,}}
\newcommand{\executiontime}[0]{\metric{CUDA Kernel execution time}}
\newcommand{\executiontimestop}[0]{\metric{CUDA Kernel execution time.}}
\newcommand{\executiontimecomma}[0]{\metric{CUDA Kernel execution time,}}

\newcommand{\griddedrecords}[0]{\metric{Gridded Records}}
\newcommand{\griddedrecordsstop}[0]{\metric{Gridded Records.}}

\definecolor{mygrey}{rgb}{0.7,0.7,0.7}
\definecolor{mylightgrey}{rgb}{0.95,0.95,0.95}
\usepackage{accents}
\usepackage{amsmath}
\usepackage{amssymb}
\usepackage{amsthm}



\usepackage{mdframed}
\mdfsetup{%
linecolor=black,
linewidth=1pt,
backgroundcolor=mylightgrey}
\newcommand{\myalgoline}[0]{\vskip -9pt \rule{\linewidth}{1pt}
\vskip 6pt}
\newcommand{\di}[1]{\text{d}#1}

\begin{document}
\title{High-Performance Gridding For Radio Interferometric Image Synthesis}
\author{Daniel Muscat}
\date{January 2021}
\supervisor{Prof. Kristian Zarb Adami}

\department{Physics}
\degree{Ph.D}
\acknowledgements{As in most ambitious and successful projects, success is only possible with the help of others. \\[8pt]
I wish to thank my supervisor, Kristian Zarb Adami, for his support, dedication and patience throughout these years. His guidance and feedback helped me to get this doctorate done. \\[8pt]
During my PhD tenure, I met many great scientists who somehow inspired me and helped me. From the bottom of my heart I wish to thank William Cotton, Tony Willis, Andre Offringa, Steve Gull, Haoyang Ye and Bram Veenboer.\\[8pt]
I also had great support from my workmates, in particular, my direct manager Dave Mifsud who was ready to adapt my work as to be able to finish this PhD.\\[8pt]
I wish to thank my late father, Joseph Muscat, for his support. He proofread the transfer report. As a remembrance I have replicated the drawings in Figures 1.1 and 1.2 from my Masters, which he drew for me.\\[8pt]
A big thank you and a heartfelt hug go to my wife Nicola Muscat, who had the patience to support me while doing this PhD and proofread this thesis. It was a tough time, especially with two kids.\\[8pt]
To my children, Timothy and Amy, thank you for always putting a smile on my face even on days where things were challenging.\\[8pt]
This research made use of APLpy, an open-source plotting package for Python (Robitaille and Bressert \cite{2012ascl.soft08017R})
}

\abstract{Convolutional Gridding is a technique (algorithm) extensively used in Radio Interferometric Image Synthesis for fast inversion of functions sampled with irregular intervals on the Fourier plane. In this thesis, we propose some modifications to the technique to execute faster on a GPU. These modifications give rise to \textit{Hybrid Gridding} and \textit{Pruned NN Interpolation}, which take advantage of the oversampling of the Gridding Convolutional Function in Convolutional Gridding to try to make gridding faster with no reduction in the quality of the output. Our experiments showed that given the right conditions, Hybrid Gridding executes up to $6.8\times$ faster than Convolutional Gridding, and Pruned NN Interpolation is generally slower than Hybrid Gridding. \\  
\\
The two new techniques feature the downsampling of an oversampled grid through convolution to accelerate the Fourier inversion.  It is a well-known approximate technique which suffers from aliasing. In this thesis, we are re-proposing the technique as a \textit{Convolution-Based FFT Pruning} algorithm able to suppress aliasing below arithmetic noise. The algorithm uses the recently discovered least-misfit gridding functions, which through our experiments gave promising results, although not as good as expected from the related published work on the stated gridding functions.  Nevertheless, our experiments showed that, given the right conditions, Convolutional-Based Pruning reduces the Fourier inversion execution time on a GPU by approximately a factor of $8\times$.  }


\include{}

\chapter{Introduction}
\label{chap:introduction}

In Radio Interferometry, a technique (algorithm) know as Convolutional Gridding is widely used for image synthesis. In this thesis, we seek to modify Convolutional Gridding so that it executes faster on the NVIDIA\textsuperscript{\textregistered} Tesla\textsuperscript{\textregistered} P100 GPU Accelerator \cite{Corporation2016}, with no degradation in the ability of the technique to suppress aliasing.

This introduction's main scope is to review Radio Interferometric Image Synthesis, where we build up the use case of Convolutional Gridding and the proposed modifications. At the end of the chapter, we discuss publications and how this thesis is laid out. 

In the first sections until Section \ref{sec:introduction:deconvolution}, we discuss various topics related to Radio Interferometry and Image Synthesis and build the use case for Convolutional Gridding. Then in Section \ref{sec:introduction:vistoimagetransform} we review the many Visibility-to-image algorithms that use Convolutional Gridding. The said section leads to Section \ref{sec:introduction:convgriddingdef}. After giving a short description of Convolutional Gridding and our proposed modifications, we identify which Visibility-to-image algorithms are compatible with the proposed modifications.

Before moving on to the first section, we need to clarify some terminology used in this thesis. In general, we shall define notations, acronyms and terms as we progress through the thesis, but hereunder are definitions of some important terms used.

\begin{itemize}
    \item \textbf{Aliasing:} The term aliasing exclusively refers to the unwanted distortion in an output image induced while regularly sampling the image on its spatial frequency plane with a rate lower than the Nyquist Rate.
    \item \textbf{Visibility record:} A Visibility record (or record) is a sample of Visibility data, including the Visibility value and the sample's coordinates.  When used in contexts with multi-polarisation, the record contains the Visibility values for each polarisation.  
    \item \textbf{$uv$-profile:} This is the set of all Visibility records' coordinates (including the $w$ coordinate) in a given telescope observation to be imaged.   
    \item \textbf{Performance:}  Any mention of Performance refers to how fast something executes in terms of the workload, where such workload is generally the number of records to process. The term is intentionally capitalised, but its verb or adjective forms are not capitalised while retaining the same meaning.
    \item \textbf{Precision:} The term Precision (with a capital P) always refers to the arithmetic precision used for floating-point numbers to either compute with or digitally represent such floating-point numbers. The reader should always assume that any compute done with a given Precision will retain the same Precision throughout the whole process, with inputs and outputs also represented with the same Precision. In this thesis we will consider two types of Precision which are \textit{Single-Precision} and  \textit{Double-Precision} defined in  ISO/IEC 60559:2020 \cite{standard:floating1} as \textit{binary32} and \textit{binary64} respectively.
    \item \textbf{Arithmetic noise:} Arithmetic noise is the rounding error introduced in a given calculation because we are using finite Precision.
    \item \textbf{P100:} The term P100 refers to the  NVIDIA\textsuperscript{\textregistered} Tesla\textsuperscript{\textregistered} P100 GPU Accelerator.
    \item \textbf{Compute:} The term compute is used as a noun and a verb. In this thesis, the noun refers to executed instructions on the P100 that are not memory-related, while the verb refers to executing such instructions on the P100.
     \item \textbf{Logic:} The term Logic refers to all compute instructions, excluding those related to floating-point arithmetic.
    \item \textbf{Algorithm:} The term algorithm retains its normal definition, with the understanding that it does not refer to an implementation (defined below).
    \item \textbf{Implementation: } The term means the implementation of an algorithm or part of, to execute on computing hardware, which, unless otherwise stated, is the P100.
    \item \textbf{Convolutional Gridding:} An algorithm studied in this thesis and defined in detail through Algorithm \ref{algo:maths:convgriddingoversampling}.
    \item \textbf{Hybrid Gridding:}  An algorithm studied in this thesis and defined in detail through Algorithm \ref{algo:maths:hybrid}.
    \item \textbf{Pruned NN Interpolation:}  An algorithm studied in this thesis and defined in detail through Algorithm \ref{algo:maths:purenninterpolation}.
     
    \item \textbf{Modified gridding:} The term modified gridding is used in this thesis as an adjective to refer to Hybrid Gridding and Pruned NN Interpolation collectively.
    \item \textbf{Studied:} The term studied is an adjective used in this thesis to refer to the Convolutional Gridding, Hybrid Gridding and Pruned NN Interpolation collectively.
    
\end{itemize}
\section{The Radio Interferometer}

Albert Abraham Michelson and Francis G. Pease were the first to make use of Interferometry in astronomy. In 1921 they made the first diameter measurements of the star Betelgeuse \cite{Michelson1921} using the interference properties of light. Discussions by Michelson on the use of Interferometers are known to have taken place at least 30 years before \cite{CI_Michelson_stellar,Michelson1920}. The first use of a Radio Interferometer in astronomy is attributed to Ryle and Vonberg \cite{RYLE1948AnSun}, whereby the Ryle-Vonberg receiver (Ryle and Vonberg \cite{RyleVonberg1946})  was used (Sullivan \cite{Sullivan1991h}). From that time onwards, the theory and technology of interferometers have made enormous advances, so much so that today humanity is building the Square Kilometre Array (SKA) \cite{SKAURL2015,skakeydocs}. 
The SKA is a multi-nation project based in South Africa and Australia with very ambitious goals (Braun \etal \cite{Braun2014}). Other examples of Radio Interferometers are the Low-Frequency Array known as the LOFAR (van Haarlem \etal \cite{Haarlem2013}), the Karl G. Jansky Very Large Array known as the VLA (Napier \etal \cite{Napier1983}) and the Giant Metrewave Radio Telescope known as the GMRT (Gupta \cite{Gupta2011}) all of which are considered as pathfinders to the SKA \cite{pathfindersite}.

An Interferometer is composed of an array of two or more antennas. It can achieve a high angular resolution, which is limited by the distance between the two antennas that are furthest apart in an inversely proportional way. In contrast, the maximum angular resolution of a single dish is limited by its diffraction limit,  inversely proportional to the diameter of the said dish. For a single-dish to reach a sub-arc-second resolution, it needs diameters that are prohibitive. However, for an Interferometer, such sub-arc-second resolution is just a question of having antennas as far away as necessary from each other. A good example is the \textit{Very Long Baseline Interferometry} (VLBI) \textit{Space Observatory Programme} (VSOP) (Hirabayashi \etal \cite{Hirabayashi1998}) were part of the array orbited in space. 

The basic Interferometer device is composed of two antennas, each receiving sky radiation. The radiation received is correlated using a complex correlator, and the measured quantity of the Interferometer as outputted by the correlator is called Visibility.  Visibility records are given against the baseline, which is the vector distance between the two dishes.

In general, Radio Interferometers measure four polarisations for each Visibility. Radiation reaching any given antenna is read in two orthogonal polarisations, say R and L, which are cross-correlated with polarised readings from another antenna to generate a quadruple, say RR, RL, LR, and LL, which we shall still refer to as \textit{polarisations}. In the context of this thesis work, each  polarisation is imaged in the same way and therefore, we will continue this review without mentioning polarisation. 

The Interferometer can measure many Visibility records in a short time. Each combination of antenna pair in the array is a basic Interferometer that simultaneously measures Visibility against a different baseline. Due to the rotation of the Earth, each 2-element Interferometer baseline changes with time, implying that subsequent Visibility measurements are against a different baseline and therefore considered different readings.  The bandwidth of the correlated input flux needs to be as narrow as possible, or otherwise, accuracy in the measurement of Visibility will be degraded, especially for wide-field imaging. The entire bandwidth is split into smaller channels, and each channel is correlated independently. The Visibility for each channel is measured against a different unique baseline since baselines are measured in wavelengths.

\section{The Measurement Equation}
\label{sec:introduction:measurmentequation}
The main goal of a Radio Interferferometric Imager is to transform the irregularly distributed multi-polarised Visibility records in a regularly sampled multi-polarised intensity image equal to the flux distribution of the real sky being observed. 

The relationship between Visibility and intensity in a Radio Interferometer is well known (Smirnov \cite{Smirnov2011}, Hamaker \etal \cite{Hamaker1996}), and here we will derive this relationship known as the Measurement Equation. The derivation is based on Thompson \etal \cite{thompson2008interferometry}.

\begin{figure}[h]
\centerline{\includegraphics[width=\linewidth,scale=0.5, trim=0cm 0cm 0cm 0cm, clip=true]{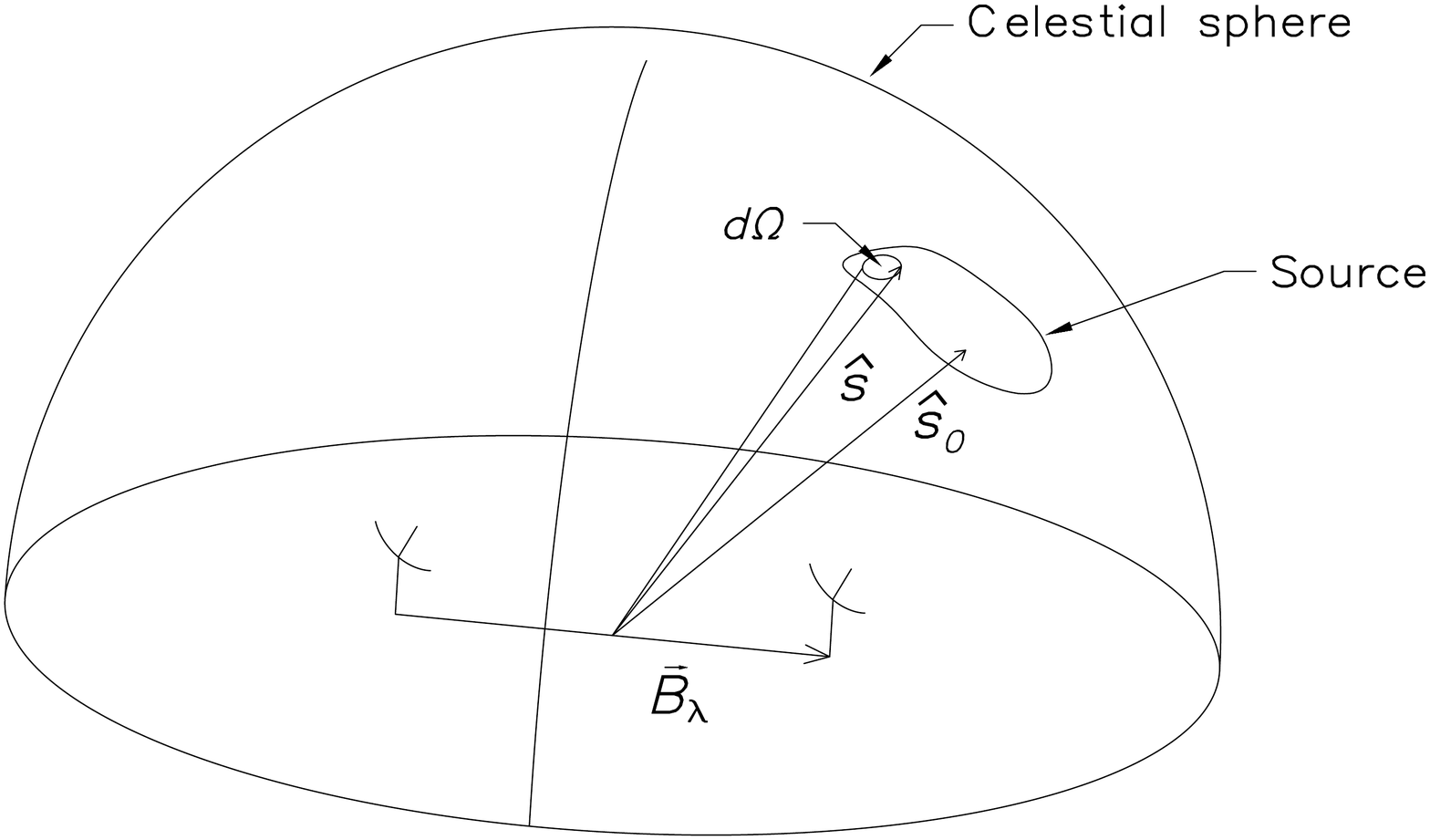}}
\caption[Defined quantities for the Measurement Equation]{Diagram illustrating the various quantities defined in Section \ref{sec:introduction:measurmentequation} and used in Equation \ref{equ:introduction:measurmentvector}\\
\newline
The diagram is a replica from Muscat \cite{Muscat2014}.
}
\label{fig:introduction:interferometer_sphere}
\end{figure}
The reader is referred to Figure \ref{fig:introduction:interferometer_sphere} for an illustration of the various vector quantities we now define. 

Let $\hat{s}$ be a unit vector that points to an arbitrary position on the celestial sphere such that $I(\hat{s})$ denotes intensity on the surface of the celestial sphere identified by the solid angle $\text{d}\Omega$. $I(\hat{s})$ is only defined on the surface of the celestial sphere. 

Let $\hat{s}_0$ be another unit vector acting as a reference direction, fixed to the sky.  Such a reference direction is generally referred to as the \textit{phase centre}.

Let $\vec{B}$ denote the vectorial distance in meters between two antennas known as the \textit{baseline}. Additionally, let $\lambda$ denote the wavelength of the radio signal such that $\vec{B}_\lambda=\vec{B}/\lambda$ is the baseline measured in wavelengths. 

The Radio Interferometer measures  Visibility denoted by $\mathcal{V}(\vec{B}_\lambda)$ through Equation \ref{equ:introduction:measurmentvector}.
\begin{equation}
\label{equ:introduction:measurmentvector}
\mathcal{V}(\vec{B}_\lambda)=\int_{4\pi}A(\hat{s})I(\hat{s})e^{-j2\pi\vec{B}_\lambda\cdot(\hat{s}-\hat{s_0})}\di{\Omega}
\end{equation}

The definition made through Equation \ref{equ:introduction:measurmentvector} is consistent with Thompson \etal \cite{thompson2008interferometry} and follows the sign convention of the exponent used in Born \etal \cite{born1999principles} and Bracewell \cite{Bracewell1958}.

$A(\hat{s})$ is known as the A-term. It encapsulates the effective collective area of the antennas and distortions in the sky intensity while the radio waves reach the antennas, including those distortions caused by the ionosphere. The A-term is quite complicated and can vary with time and other factors. In this thesis, we have little interest in this term, and for convenience and simplicity, we subsume the term in what we call the \textit{measured intensity} $I_{\text{m}}(\hat{s})$, defined by Equation \ref{equ:introduction:measuredintensitys}.

\begin{equation}
\label{equ:introduction:measuredintensitys}
   I_{\text{m}}(\hat{s})=A(\hat{s})I(\hat{s})
\end{equation}

We note to the reader that we will slightly modify the definition of measured intensity in the following few paragraphs. 

By means of Equation \ref{equ:introduction:measuredintensitys}, the Measurement Equation \ref{equ:introduction:measurmentvector} reduces to Equation \ref{equ:introduction:measurmentsimpler}.

\begin{equation}
\label{equ:introduction:measurmentsimpler}
    \mathcal{V}(\vec{B}_\lambda)=\int_{4\pi}I_{\text{m}}(\hat{s})e^{-j2\pi\vec{B}_\lambda\cdot(\hat{s}-\hat{s}_0)}\text{d}\Omega
\end{equation}

\begin{figure}
\centerline{\includegraphics[scale=0.5]{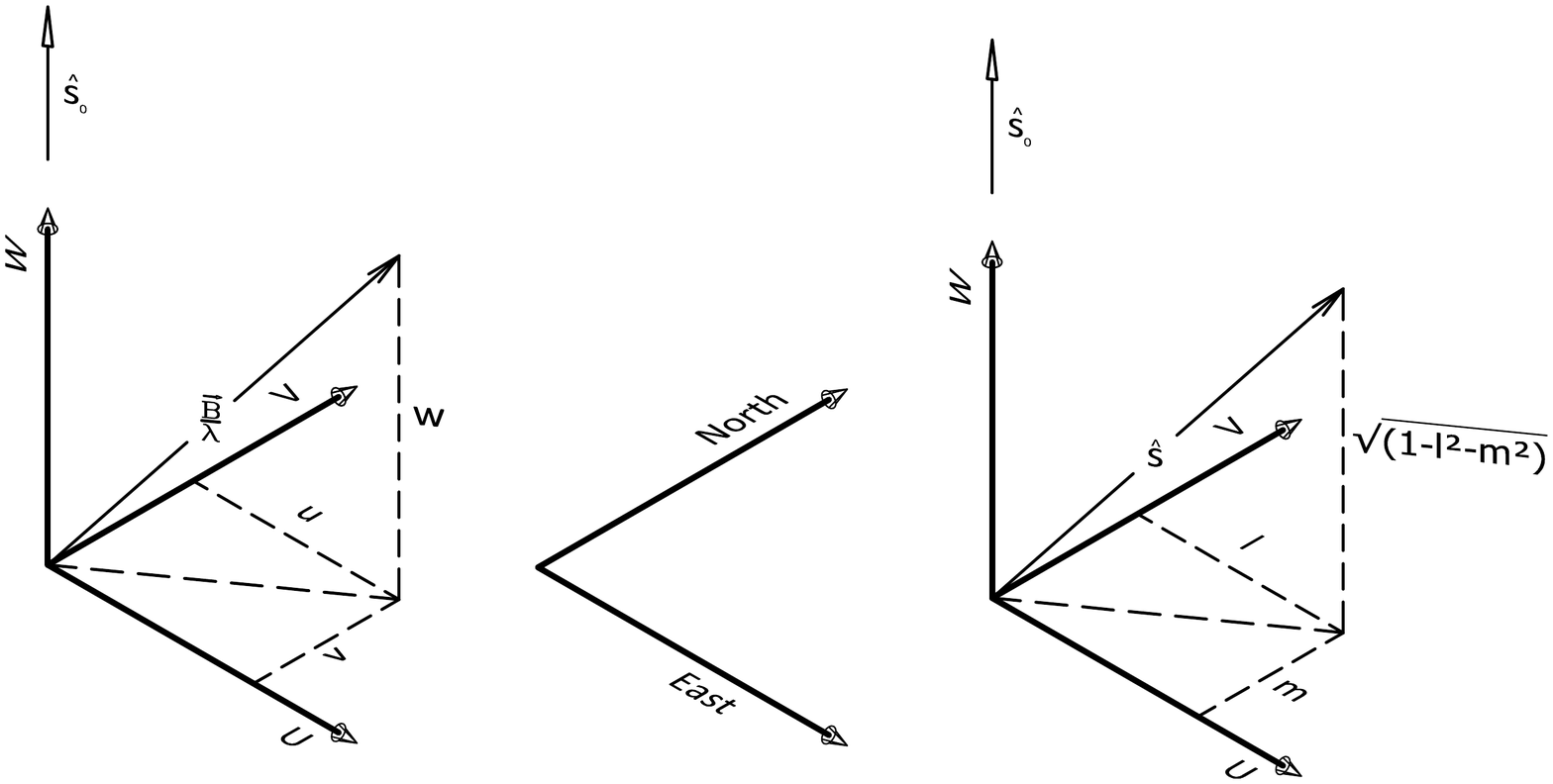}}

\caption[The UVW coordinate system]{The UVW coordinate system. The left diagram  shows how a baseline ($\vec{B}/\lambda$) is expressed in its $(u,v,w)$ components. The right diagram shows how the direction cosines $l$ and $m$ express a position on the celestial sky pointed at by the unit vector $\hat{s}=(l,m,\sqrt{1-l^2-m^2})$. $\hat{s}_0$ is a unit vector representing the \textit{phase centre}, which is a reference direction fixed to the sky, usually pointing to the observed part of the sky. 
\vspace{4pt}\newline
The UVW coordinate system is a right-handed Cartesian system. The U and V axes are normal to the phase centre. The U-axis is in the East-West direction, while the V-axis is in the North-South direction. The W-axis is in the direction of the \textit{phase centre}.
\vspace{4pt}\newline
Image is a replica from Muscat \cite{Muscat2014}.
}
\label{fig:introduction:uvw}
\end{figure}

We now introduce the UVW coordinate system, defined and illustrated in Figure \ref{fig:introduction:uvw}. In this coordinate system $\hat{s}=(l,m,\sqrt{1-l^2-m^2})$, while $\vec{B}_\lambda=(u,v,w)$. Note that $s_0=(0,0,1)$.

We denote Visibility expressed in terms of $(u,v,w)$ as $\mathcal{V}_w(u,v)$, while we denote measured intensity distribution in the $(l,m,\sqrt{1-l^2-m^2})$ direction as $I_{\text{m}}(l,m)$, for which we slightly change the definition as per Equation \ref{equ:introduction:measuredintensitylm}.

\begin{equation}
\label{equ:introduction:measuredintensitylm}
    I_{\text{m}}(l,m)=\frac{A(\hat{s})I(\hat{s})}{\sqrt{1-l^2-m^2}}
\end{equation}

The new quotient is included in $I_{\text{m}}(l,m)$ to simplify the forthcoming equation.

Through a derivation given in Thompson \etal \cite{thompson2008interferometry} the Measurement Equation \ref{equ:introduction:measurmentsimpler} is expressed in terms of the UVW coordinate system by Equation \ref{equ:introduction:measurmentequation}.

\begin{equation}
\label{equ:introduction:measurmentequation}
\mathcal{V}_w(u,v)=\int_{\infty}{I_{\text{m}}(l,m)e^{-j2\pi(ul+vm+w[\sqrt{1-l^2-m^2} -1])}} \text{d}l\text{d}m
\end{equation}

where the integral $\int_{\infty}$ denotes a definite integral on all integration variables, taken over the $-\infty$ to $\infty$ range. 

The term $w\left(\sqrt{1-l^2-m^2}-1\right)$ is known as the W-term, and we remind the reader that in this section, we also introduced the A-term, which is subsumed in $I_{\text{m}}(l,m)$.

For convenience, we define $\mathcal{V}'_w(u,v)$ defined by Equation \ref{equ:introduction:visphasor}.

\begin{equation}
\label{equ:introduction:visphasor}
\mathcal{V}'_w(u,v)=\mathcal{V}_w(u,v)e^{-j2\pi w}
\end{equation}

On multiplying the two sides of the Measurement Equation, \ref{equ:introduction:measurmentequation} with the phasor $e^{-j2\pi w}$, the phase-tracker in the W-term cancels out such that the Measurement Equation can take the form of Equation \ref{equ:introduction:measurmentequationwithV}.

\begin{equation}
\label{equ:introduction:measurmentequationwithV}
\mathcal{V}'_w(u,v)=\int_{\infty}{I_{\text{m}}(l,m)e^{-j2\pi\left(ul+vm+w\sqrt{1-l^2-m^2}\right)}} \text{d}l\text{d}m
\end{equation}

\section{Properties of the Measurement Equation}
\label{sec:introduction:properties}

In the forthcoming Section \ref{sec:introduction:wtermhandling}, we  review the many available Visibility-to-image algorithms that compute an intensity image from the measured Visibility data. They take advantage of the various properties the Measurement Equation holds, which we now list and discuss.

\subsubsection{Three-Dimensional Fourier Transform}

The Measurement Equation \ref{equ:introduction:measurmentequationwithV} can be expressed as a Three-Dimensional Fast Fourier Transform (Clark \cite{Clark1973}, Cornwell \etal  \cite{Cornwell2008}, Perley \cite{Perley1999}) as per Set of Equations \ref{equ:introduction:measurmentequation3d}.

\begin{subequations}
\label{equ:introduction:measurmentequation3d}
\begin{equation}
\mathcal{V}'_w(u,v)=\int_\infty I'_{\text{m}}(l,m,n)e^{-j2\pi(ul+vm+wn)} \di{l}\di{m}\di{n}
\end{equation}
\begin{equation}
I'_{\text{m}}(l,m,n)=I_{\text{m}}(l,m)\delta\left(n-\sqrt{1-l^2-m^2}\right)
\end{equation}
\end{subequations}
where $\delta\left(n-\sqrt{1-l^2-m^2}\right)$ is introduced to project $I_{\text{m}}(l,m)$ to the surface of the unit celestial sphere in a three-dimensional space with independent coordinates $l,m$ and $n$. Such a setup is identical to that of Equation \ref{equ:introduction:measuredintensitys} and as illustrated in Figure \ref{fig:introduction:interferometer_sphere}.  Note that excluding the unit sphere's surface where  $I_{\text{m}}(l,m)$  resides, the whole space is void from any information.

\subsubsection{The Visibility Plane at $w=0$}

On setting $w=0$ in Measurement Equation \ref{equ:introduction:measurmentequation}, we get Equation \ref{equ:introduction:measurmentw0}.

\begin{equation}
\label{equ:introduction:measurmentw0}
\mathcal{V}_0(u,v)=\int_\infty I_{\text{m}}(l,m)e^{-j2\pi(ul+vm)} \di{l}\di{m}
\end{equation}

Equation \ref{equ:introduction:measurmentw0} is a Two-Dimensional Fourier Transform as per Equation \ref{equ:introduction:measurment2DP}.

\begin{equation}
\label{equ:introduction:measurment2DP}
\mathcal{V}_0(u,v)=\mathcal{F}I_{\text{m}}(l,m)
\end{equation}

where $\mathcal{F}$ denotes a Two-Dimensional Fourier Transform. 

Therefore, the Visibility plane at $w=0$ is related to $I_{\text{m}}(l,m)$ by a simple Fourier Transform.  

\subsubsection{A Visibility plane perpendicular to the W-axis at $w=c$}
Let $c$ be a constant and let $g_w(l,m)$ contain the W-term as per Equation \ref{equ:introduction:gw}
\begin{equation}
\label{equ:introduction:gw}
g_w(l,m)=e^{-j2\pi w\left(\sqrt{1-l^2-m^2}-1\right)}
\end{equation}

$1/g_w(l,m)$ clearly exists and we denote it with $\overline{g}_w(l,m)$, that is $\overline{g}_w(l,m)=1/g_w(l,m)$ 

On considering the Visibilities on plane $w=c$, that is $\mathcal{V}_c(u,v)$, and substituting with Equation \ref{equ:introduction:gw}, the Measurement Equation \ref{equ:introduction:measuredintensitylm} transforms into:

\begin{equation}
\mathcal{V}_c(u,v)=\int_{\infty}I_{\text{m}}(l,m)g_w(l,m)e^{-j2\pi(ul+vm)} \di{l}\di{m}
\end{equation}

which is a Two-Dimensional Fourier Transform as better shown by Equation \ref{equ:introduction:measurmentequationfourier}.
\begin{equation}
\label{equ:introduction:measurmentequationfourier}
\mathcal{V}_c(u,v)=\mathcal{F}\left[I_{\text{m}}(l,m)g_c(l,m)\right] 
\end{equation}

On applying an inverse Fourier Transform $\left( \mathcal{F}^{-1} \right)$ on both sides of the above equation and put  $I_{\text{m}}(l,m)$ subject of the formula, we get:

\begin{equation}
\label{equ:introduction:measurmentequationfourierIsubject}
I_{\text{m}}(l,m)=\overline{g}_c(l,m)\mathcal{F}^{-1}\mathcal{V}_c(u,v)
\end{equation}
Therefore, Visibility on a plane perpendicular to the W-axis is related to $I_{\text{m}}(l,m)$ via a Fourier Transform of the plane and a simple point-wise multiplication with a phase corrector $\overline{g}_c(l,m)$.

Let us now define $G_w(u,v)$ and $\overline{G}_w(u,v)$ as the Fourier Transform of $g_w(l,m)$ and $\overline{g}_w(l,m)$ respectively, that is 
\begin{subequations}
\begin{equation}
    G_w(u,v)=\mathcal{F}g_w(l,m)
\end{equation}    
\begin{equation}
\overline{G}_w(u,v)=\mathcal{F}\overline{g}_w(l,m)
\end{equation}
\end{subequations}
Using the Convolution Theorem, we can express Equation \ref{equ:introduction:measurmentequationfourier} as a convolution (Frater and Docherty \cite{Frater1980}).

\begin{equation}
\label{equ:introduction:vcdef}
    \mathcal{V}_c=[\mathcal{F}I_{\text{m}}(l,m)]*G_c(u,v)
\end{equation}
where $*$ is a two-dimensional convolution.
On substituting Equation \ref{equ:introduction:measurment2DP} in Equation \ref{equ:introduction:vcdef} we get 
\begin{equation}
    \mathcal{V}_c(u,v)=\mathcal{V}_0(u,v)*G_c(u,v)
\end{equation}
and since the inverse of $g_w(l,m)$, that is $\overline{g}_w(l,m)$ exists, it follows that:
\begin{equation}
\label{equ:introduction:v00}
    \mathcal{V}_0(u,v)=\mathcal{V}_c(u,v)*\overline{G}_c(u,v)
\end{equation}
Therefore, a Visibility plane perpendicular to the W-axis is related to the Visibility plane at $w=0$ by a simple convolution.

\subsubsection{General Visibility Plane defined by $w=au+bv+c$}

We now investigate how a general Visibility Plane defined with $w=au+bv+c$ is related to $I_{\text{m}}(l,m)$ where $a$, $b$ and $c$ are constants.

For convenience, we shall use the three-dimensional form of the Measurement Equation, as expressed in Equation \ref{equ:introduction:measurmentequation3d} and reproduced as Equation \ref{equ:introduction:measurmentequation3dreproduced}.

\begin{equation}
\label{equ:introduction:measurmentequation3dreproduced}
    \mathcal{V}'_w(u,v)=\int_{\infty}I_{\text{m}}(l,m)\delta(n-\sqrt{1-l^2-m^2})e^{-j2\pi(ul+vm+wn)} \text{d}l\text{d}m\text{d}n
\end{equation}

Let's substitute $w=au+bv+c$. For better readability $V'_{au+bv+c}(u,v)$ is denoted with $\mathcal{V}'_{\text{plane}}(u,v)$.

\begin{equation}
\label{equ:introduction:vplane}
    \mathcal{V}'_{\text{plane}}(u,v)=\int_{\infty}I_{\text{m}}(l,m)\delta(n-\sqrt{1-l^2-m^2})e^{-j2\pi(ul+vm+[au+bv+c]n)}\di l\di m\di{n}
\end{equation}

We now use the Reverse Chain Rule to simplify Equation \ref{equ:introduction:vplane}.

Let $l'=l+an$ and $m'=m+bn$ and $n'=n$

On applying the Reverse Chain Rule to Equation \ref{equ:introduction:vplane} while re-arranging and cancelling terms, Equation \ref{equ:introduction:chainruleapplied} is derived. Note that the Jacobian of the transformation is equal to 1, and integral limits remain the same. 

\begin{equation}
\label{equ:introduction:chainruleapplied}
    \mathcal{V}'_{\text{plane}}(u,v)=\int_{\infty}I_{\text{m}}(l,m)\delta(n'-\sqrt{1-l^2-m^2})e^{-j2\pi(ul'+vm'+cn')}\di l'\di m'\di n'
\end{equation}

Equation \ref{equ:introduction:chainruleapplied} shows that a Three-Dimensional Fourier relationship exists between Visibility on a plane $w=au+bv+c$ and $I_{\text{m}}(l,m)$ but coordinates are distorted. The intensity distribution is on the surface of a spheroid for the distorted system with $(l',m',n')$ coordinates.

Let's now split Equation \ref{equ:introduction:chainruleapplied} in the Set of Equations \ref{equ:introduction:splitequation} as to express the equation as a Two-Dimensional Transform.
\begin{subequations}
\label{equ:introduction:splitequation}
\begin{equation}
\label{equ:introduction:splitequationno1}
    \mathcal{V}'_{\text{plane}}(u,v)=\int_{\infty} I'_{\text{m}}(l',m')e^{-j2\pi(ul'+vm')}\di l'\di m'
\end{equation}
\begin{equation}
\label{equ:introduction:integralin}
    I'_{\text{m}}(l',m')=\int_{\infty}I_{\text{m}}(l,m,n)\delta(n'-\sqrt{1-l^2-m^2})e^{-j2\pi cn'} \di n'
\end{equation}
\end{subequations}

The integral in Equation \ref{equ:introduction:integralin} can be worked out by noting that  $\delta(n'-\sqrt{1-l^2-m^2})$ is not equal to zero only at $n'=\sqrt{1-l^2-m^2}$ and therefore:
\begin{equation}
    I'_{\text{m}}(l',m')=I_{\text{m}}(l'',m'')e^{-j2\pi c\sqrt{1-l^2-m^2}} 
\end{equation}

where $l''=l'-a\sqrt{1-l^2-m^2}$ and $m''=m'-b\sqrt{1-l^2-m^2}$.

Substituting in Equation \ref{equ:introduction:splitequationno1} we get 

\begin{equation}
\label{equ:introduction:measurmentdistorted}
    \mathcal{V}'_{\text{plane}}(u,v)=\int_{\infty} I_{\text{m}}(l'',m'')e^{-j2\pi(ul'+vm'+c\sqrt{1-l^2-m^2})}\di l'\di m'
\end{equation}

On comparing Equation \ref{equ:introduction:measurmentdistorted} with the Measurement Equation \ref{equ:introduction:measurmentequationwithV}, we realise that they are similar with the exception that the coordinates are distorted. Furthermore, if we consider the subset of planes where $c=0$, that is, Visibility planes defined by $w=au+bv$, a two-dimensional Fourier relationship with distorted coordinates results. 

\begin{equation}
\label{equ:introduction:snapshotequation}
    \mathcal{V}'_{\text{plane}}(u,v)=\int_{\infty} I_{\text{m}}(l'',m'')e^{-j2\pi(ul'+vm')}\di l'\di m'
\end{equation}

Using the same rationale used to derive Equation \ref{equ:introduction:v00}, we get 
\begin{equation}
    \mathcal{V}'_{au+bv}(u,v)=\mathcal{V}'_{au+bv+c}(u,v)*\overline{G}'_c(u,v)
\end{equation}

where $\overline{G}'_c(u,v)=\mathcal{F}\left[\overline{g}_c(l,m)e^{j2\pi c}\right]$

Therefore, a Visibility plane defined by $w=au+bv+c$ is related to the parallel plane defined by $w=au+bv$ by a simple convolution.

\section{Irregular sampling of Visibility}
\begin{figure}[]
\includegraphics[width=\linewidth]{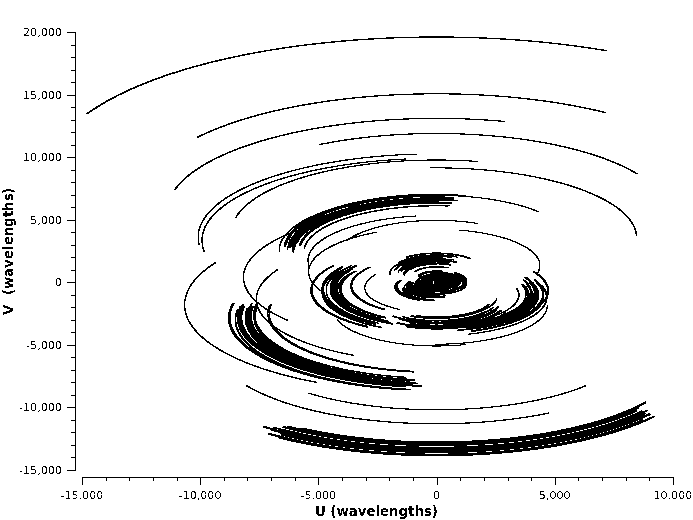}
\caption[$uv$-profile of a real LOFAR observation]{$uv$-profile of a real LOFAR observation. Every baseline forms a trajectory as it samples the true sky, controlled by Earth's movement, relative to a fixed point in the sky. This observation is used in most of our experiments described in Chapters \ref{chap:2dgridding}, \ref{chap:hybrid}, \ref{chap:purenn} and \ref{chap:comparative}.}
\label{fig:introduction:uvcoverage}
\end{figure}

The Radio Interferometer relies on its movement relative to a fixed point in the sky to sample $V_w(u,v)$ at different $(u,v,w)$ coordinates. $V_w(u,v)$ is irregularly sampled on all three axes as illustrated by the LOFAR observation $uv$-profile in Figure \ref{fig:introduction:uvcoverage}.

Let us mathematically explain the effects of irregular sampling. Using Equation \ref{equ:introduction:measurmentequationfourierIsubject} and the various arguments given in Section \ref{sec:introduction:properties}, we can express the Measurement Equation as shown in Equation \ref{equ:introduction:forirregularsampling}. 
\begin{equation}
\label{equ:introduction:forirregularsampling}
\mathcal{F}^{-1}\mathcal{V}_w(u,v)=I_{\text{m}}(l,m)g_w(l,m)
\end{equation}

Now let $S_w(u,v)$ be the sampling function of the Radio Interferometer, and let $s_w(l,m)=\mathcal{F}S_w(u,v)$ many times referred to as the Point Spread Function (PSF).

The sampling of the sky by the Radio Interferometer can be modelled with a point-wise multiplication of $V_w(u,v)$ with $S_w(u,v)$ which by the Convolution Theorem will result in a convolution of $s_w(l,m)$ with $I_{\text{m}}(l,m)g_w(l,m)$ as per Equation \ref{equ:introduction:psfconvolution}.

\begin{equation}
\label{equ:introduction:psfconvolution}
\mathcal{F}^{-1}\left[\mathcal{V}_w(u,v)S_w(u,v)\right]=\left[I_{\text{m}}(l,m)g_w(l,m)\right]*s_w(l,m)
\end{equation}

In general, the convolution with the PSF is irreversible, making it impossible to recover $I_{\text{m}}(l,m)$ by a linear process. Therefore irregular sampling distorts $I_{\text{m}}(l,m)$ in a way that in general $I_{\text{m}}(l,m)$ can only be recovered through a non-linear process known as Deconvolution.

\section{Deconvolution}
\label{sec:introduction:deconvolution}

Deconvolution is not the scope of this thesis. Nevertheless, it is the requirements of modern Deconvolution algorithms that set the stage for increasing the Performance of Convolutional Gridding.

There are a vast number of Deconvolution algorithms, and  a non-exhaustive list follows: 

\begin{itemize}
    \item CLEAN (Hogbom \cite{Hogbom1974})  and derivatives such as Clark \cite{Clark1980}, Cotton-Schwab (described in a paragraph of Schwab \cite{Schwab1984}), Multi-Resolution CLEAN (Wakker and Schwarz \cite{Wakker1988}), Double Deconvolution (DD) for Multi-Frequency Synthesis(MFS) (Conway \etal  \cite{Conway1990}), MS-CLEAN (Cornwell \cite{Cornwell2008c}), MF-CLEAN (Salt and Wieringa \cite{Sault1994}), MS-MFS (Rau and Cornwell \cite{Rau2011}), ASP-CLEAN (Bhatnagar and Cornwell \cite{Bhatnagar2004}), and Joined Channel(JC) CLEAN (Offringa and Smirnov \cite{Offringa2017}).
    \item Compressed sensing based algorithms\footnote{It is worthwhile to note that there are some local contributions to the application of Compressed Sensing to Radio Interferometry, in particular, Gauci \etal\cite{Gauci2015}.} (Wiaux \etal \cite{Wiaux2009}) including Li \etal \cite{Li2011}, SARA (Carrillo \etal \cite{Carrillo2012}),  HyperSARA (Abdulaziz \etal \cite{Abdulaziz2019}), Purify (Carrillo \etal \cite{Carrillo2014}), MORESANE (Dabbach \etal \cite{Dabbech2015a}), and SASIR (Garsden \etal  \cite{Garsden2015} and Girard \etal \cite{Girard2015}).
    \item Bayesian-based Deconvolution such as RESOLVE (Junklewitz \etal \cite{Junklewitz2016}).
\end{itemize}

\begin{figure}
    \centering
    \includegraphics[page=1,width=\linewidth,trim=1cm 1cm 1cm 1cm, clip]{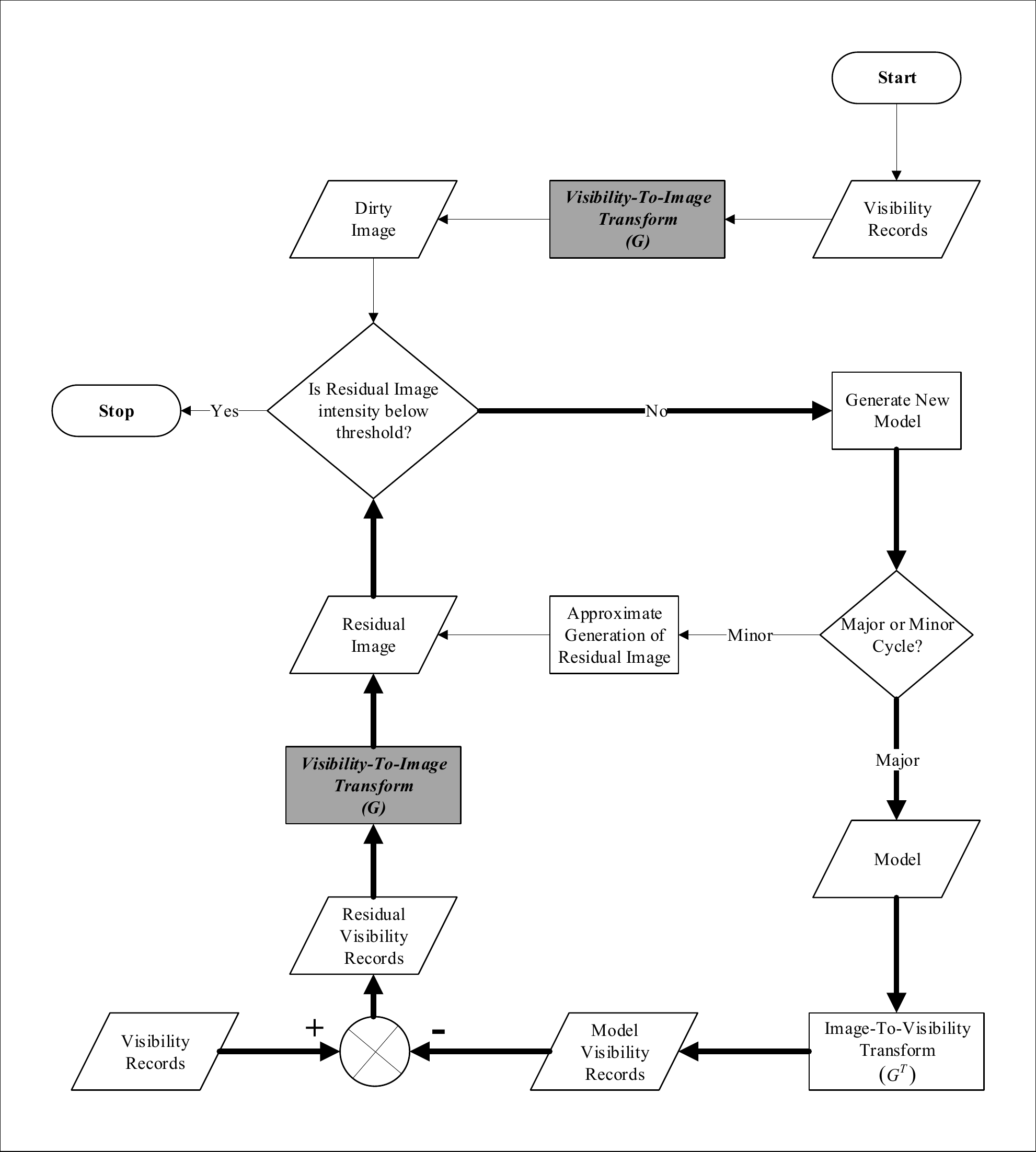}
    \\[20pt]
    \caption[Imaging Pipeline Flowchart]{Flowchart of a general imaging pipeline. The pipeline first generates a dirty image and then applies Deconvolution. The process coloured in grey is the process that applies Convolutional Gridding, which this thesis focuses on. The fat arrows track the major cycle. The output of the pipeline is a model of the observed sky.}
    \label{fig:introduction:deconvolution}
\end{figure}

A flowchart of an imaging pipeline meant to serve as an aid for this section is drawn in Figure \ref{fig:introduction:deconvolution}. 

In general, three images are involved in Deconvolution. These are called the model, dirty and residual images. The model is the predicted image of the real sky, while the dirty image is that image generated from the Interferometer's measured Visibilities. The residual is the image that arises when the intensity predicted in the model gets subtracted from the dirty image. One notes that the residual image is equal to the dirty image if the model has zero intensity.

Deconvolution works via an iterative method using a feedback loop. In every iteration (more known as a cycle), the predicted model is updated based on the residual image and a priori knowledge, generally in the form of an assumption.\footnote{For example, in CLEAN (Hogbom \cite{Hogbom1974}) it is assumed that a source exists at the position where the residual image contains a peak.}  Once the model is updated, the residual is re-calculated for use in the next cycle. A successful Deconvolution process ends once the intensity present in the residual is below a given threshold.

There are two methods of how a residual image is calculated in any cycle. In a  minor cycle, updates made to the model within the cycle are convolved with a truncated PSF (Clark \cite{Clark1980}). The result is subtracted from the residual image generated by the previous cycle. The minor cycle is designed to be fast in execution but is an approximate approach, prone to a build-up of errors. On the other hand, in a major cycle, a more accurate approach but much slower to execute is taken. As pioneered by a paragraph in Schwab \cite{Schwab1984}, the subtraction to calculate the residual image happen in the Visibility space, and the residual image is generated from scratch.

Many Deconvolution algorithms, which we gave a non-exhaustive list at the beginning of this section, use minor and major cycles together. Many iterations will be minor cycles, and major cycles are executed intermittently to reset the build-up of errors. Some Deconvolution algorithms such as Purify and RESOLVE only use major cycles and do not execute minor cycles.

\section{Convolution Gridding use case}
\label{sec:introduction:residualcalculation}

Let us do a simple formulation to explain how the residual is calculated in a major cycle to understand where Convolutional Gridding is used. 

Let $\mathcal{G}$ be a backward Visibility-to-image operator whereby irregularly sampled Visibility data (records) is transformed into an intensity image (the dirty or the residual image). Let $\mathcal{G}^T$ be a forward image-to-visibility operator that transforms an intensity image (the model) to Visibility data irregularly sampled by the $uv$-profile of $V_w(u,v)S_w(u,v)$. Denoting the model image as $I_{\text{model}}(l,m)$ and the  residual image as $I_{\text{residual}}(l,m)$, then a major cycle calculates the residual image by applying the equation:
\begin{equation}
\label{equ:introduction:residualcalc}
    I_{\text{residual}}(l,m)=\mathcal{G}\left[\mathcal{V}_w(u,v)S_w(u,v)-\mathcal{G}^T I_{\text{model}}(l,m)\right]
\end{equation}

The Direct use of the Measurement Equation to execute $\mathcal{G}$ or $\mathcal{G}^T$ is in most cases evaded as it is computationally too expensive. In general, algorithms to compute  $\mathcal{G}$ or $\mathcal{G^T}$ are based on a convolution-based process called \textit{gridding} (\textit{Convolution Gridding}) for $\mathcal{G}$ or \textit{degridding} (\textit{Convolutional Degridding}) for $G^T$.  Convolution Gridding and Degridding rely on a two-dimensional convolution with a Gridding Convolution Function (GCF) of finite width. Convolutional Gridding and Degridding cannot handle the A-term and W-term, and algorithms computing $G$ or $G^T$ adjust as to handle the two terms. Adjustments come in two flavours, either modify the GCF by including in it information on the terms or correct for the terms outside Convolutional Gridding such as not to modify the GCF.

In this thesis, Degridding is out of scope, and we put all our focus on Convolutional Gridding using a non-modified GCF. We are proposing modifications to increase the Performance of Convolutional Gridding when using an unmodified (that is, no handling of the A-term or W-term within Convolutional Gridding). 

Convolutional Gridding together with Convolutional Degridding is likely to dominate the execution time of the whole imaging process. Noting that sometimes there can be as many as $10^5$ major cycles (Arras \etal \cite{Arras2021}), one can easily infer that increasing the Performance of Convolutional Gridding benefits a whole class of Deconvolution algorithms. Similar benefits can also be obtained by increasing the Performance of Convolutional Degridding, and it is unfortunate that we did not do any work related to degridding.   However, we are tagging such an endeavour for future work and note that the modifications that we are proposing are easily adaptable to the simple form of Convolutional Degridding.

We have little interest in cases where Convolutional Gridding is used with a modified GCF to correct for the A-term and W-term. The reader should always assume that unless otherwise stated, any reference to Convolutional Gridding in this thesis implicitly implies the use of a non-modified GCF.   

\section{Visibility-to-image algorithms}
\label{sec:introduction:vistoimagetransform}
This section will review the various algorithms available for the backward Visibility-to-Image Transform. The intent is to understand how these algorithms cater for the W-term and A-term and how Convolutional Gridding is integrated. A reader naive on Convolutional Gridding might find difficulties following up some of the content in this section, and in such circumstance, we recommend that the reader read Section \ref{sec:introduction:convgriddingdef}.

For clarity, let us formalise the Visibility-to-image operator ($\mathcal{G}$): There is a finite set of Visibility records that we would like to turn into a dirty  or residual image, now denoted by $I_{\text{D}}(l,m)$ using the  Measurement Equation \ref{equ:introduction:measurmentequation}. That is, given a set of $P$ Visibility records $\{\mathcal{V}_{w_p}(u_p,v_p); 0\le p < P \text{ and } w_p,u_p,v_p \text{ are coordinates}  \}$, we would like to calculate in the fastest way possible a sampled and bounded (that is, of a finite number of samples) version of $I_{\text{D}}(l,m)$ as expressed by Equation \ref{equ:introduction:ID}.
\begin{equation}
\label{equ:introduction:ID}
    I_{\text{D}}(l,m)=\sum\limits_{p=0}^{P-1}\mathcal{V}_{w_p}(u_p,v_p)e^{j2\pi(u_p l+v_p m)}A_p(l,m)\overline{g}_{w_p}(l,m)
\end{equation}
where $\overline{g}_{w}(l,m)$ was defined in Section \ref{sec:introduction:properties} and $A_p(l,m)$ is $A(l,m)/\sqrt{1-l^2-m^2}$  for the $(p+1)^{\text{th}}$ Visibility record. We are including the quotient $\sqrt{1-l^2-m^2}$ in the A-term ($A(l,m)$) just for convenience and we are binding the value of the A-term with a given Visibility record to clarify that the A-term is a function of time and other factors. 

As pointed out before, a direct application of Equation \ref{equ:introduction:ID} to compute $\mathcal{G}$ is computationally too expensive and generally avoided. Therefore, other algorithms are used to compute $\mathcal{G}$ generally based on Convolutional Gridding. Convolutional Gridding does not implicitly cater to the A-term or W-term. What most Visibility-to-image algorithms do is to take advantage of various properties of the Measurement Equation, discussed in Section \ref{sec:introduction:properties}, to handle the W-term or A-term. 

Let us now review various Visibility-to-image algorithms, categorised by the term they cater to and how they cater to the term.

\subsection{Catering for the W-term}
\label{sec:introduction:wtermhandling}

We can categorise four different approaches taken in Radio Interferometry to handle the W-term, which we shall now discuss.

\subsubsection{Ignore the W-term} 

Thompson \cite{Thompson1999} states that given a short field of observation such that
\begin{equation}
    \left(\sqrt{1-l^2-m^2}-1\right)w \approx -0.5(l^2+m^2)w \approx 0
\end{equation}
it is acceptable to ignore the W-term and proceed with Convolutional Gridding. 

For a wider field of view, Faceting techniques (Kogan and Greisen \cite{Kogan2009}, Cornwell and Perley \cite{Cornwell1992a}) are used to partition the image into smaller images (facets) whose field of observation is small enough to ignore the W-term. Each facet can be generated independently using Convolutional Gridding, and at the end, the facets are combined to generate the needed image. 

\subsubsection{Include correction for the W-term in the convolution}

Through Equation \ref{equ:introduction:v00} we showed that:

\begin{equation}
\label{equ:maths:v0subject}
    \mathcal{V}_0(u,v)=\mathcal{V}_w(u,v)\overline{G}_w(u,v)
\end{equation}

Based on Equation \ref{equ:maths:v0subject}, Cornwell \etal \cite{Cornwell2008} designed the W-projection algorithm whereby any Visibility record is projected on the $w=0$ plane utilising a convolution with $\overline{G}_w(u,v)$ integrated with the GCF of Convolutional Gridding. Therefore, the GCF is modified and made dependent on the coordinates of the Visibility records.

\subsubsection{Correction of the W-term in the $(l,m)$-plane}

We showed in Equation \ref{equ:introduction:measurmentequationfourierIsubject} that:
\begin{equation}
\label{equ:introduction:lmwcorr}
    I_{m}(l,m)=\left[\mathcal{F}^{-1}\mathcal{V}_w(u,v)\right]\overline{g}_w(l,m)
\end{equation}
In W-stacking (Humphreys and Cornwell \cite{Humphreys2011}, Offringa \etal \cite{Offringa2014}) Visibility records with a similar value of $w$ are gridded using Convolutional Gridding on separate grids. By virtue of Equation \ref{equ:introduction:lmwcorr}, the W-term is corrected by applying $\overline{g}_w(l,m)$ on the output of Convolutional Gridding.  All the resultant images are summed together after correction is applied. 

Another approach to correct the W-term on the $(l,m)$-plane is known as  Snapshots (Cornwell and Perley \cite{Cornwell1992a}, Cornwell \etal \cite{Cornwell2012}, Ord \etal \cite{Ord2010}, Brouw \cite{brouw1971data}). In a given short period of observation time or just one \textit{snapshot}, the Radio Interferometer array approximates a coplanar array whereby $w$ is a linear combination of $u$ and $v$ for the records in that given a short period of observation time. As revealed by Equation \ref{equ:introduction:snapshotequation}, if such records are gridded alone using Convolutional Gridding, the image results with warped coordinates correctable with a simple $(l,m)$-correction. Therefore in Snapshots, one generates such images using  Convolutional Gridding and adds them together after correction.

\subsubsection{Hybrid correction}

All the approaches to handle the W-term mentioned above can and are integrated together. For example, in CASA (McMullin \etal \cite{McMullin2007}), a popular imaging tool, W-projection, can be used on Facets. In DDFacet (Tasse \etal \cite{Tasse2018}), a modified version of W-projection is integrated with Faceting. In W-snapshots (Cornwell \etal \cite{Cornwell2012}), Snapshots and W-projection are used together. Lately, Pratley \etal \cite{Pratley2018} proposed integrating W-projection with W-stacking in a way that is accurate and scalable. 

\subsubsection{3D FFT}

We showed through the Set of Equations \ref{equ:introduction:measurmentequation3d} that the Measurement Equation could be expressed as a Three-Dimensional Fast Fourier Transform. Therefore, the three-dimensional version of Convolutional Gridding can be applied to create a three-dimensional image. The problem with using Three-Dimensional FFTs is that the generated cube is mostly empty of true emission (Cornwell \etal \cite{Cornwell2008}), but lately, Smith \etal \cite{Smith2017} proposed a reconsideration of the Three-Dimensional FFT and Offringa \cite{offringaprivate2020} implemented the Three-Dimensional FFT in WSClean (Offringa \etal \cite{Offringa2014}).  

\subsection{Handling the A-term}
Handling the A-term is a tougher challenge than handling the W-term since the A-term is a function of time, polarisation and the particular telescope measuring the sky. In literature, there are fewer techniques for handling the A-term in comparison with handling the W-term.

A-Projection (Bhatnagar \etal \cite{Bhatnagar2008}) and Wide-Band (WB) A-Projection (Bhatnagar \etal \cite{Bhatnagar2013a}) handle the A-term by injection data in the GCF in a similar way to W-projection.  Tasse \etal \cite{Tasse2013} integrated A-projection with W-projection to create what is known as AW-projection for the use in an imager for LOFAR known as the AWImager. Also, in ASKAPSoft (Cornwell \etal \cite{Cornwell2016}), A-projection is integrated with W-stacking. 

Facets can handle the A-term. In the already mentioned DDFacet, the primary use of Facets is to handle the A-term. Van Weeren \etal \cite{VanWeeren2016} describes a Facet based approach for calibration in LOFAR, used by Williams \etal \cite{Williams2016}. 

\subsection{Image Domain Gridding}
Image Domain Gridding (van der Tol \etal \cite{VanderTol2019}, Veenboer \etal \cite{Veenboer2017})  is a new technique that re-defines the way a convolution is calculated. Low-Resolution Images are generated from neighbouring Visibility records via the direct use of the Measurement Equation. The convolution is executed by multiplying with the Low-Resolution Image. The result is subsequently transformed back to the $uv$-space using an FFT and added back to the $uv$-grid in the appropriate region. Once all Visibility records are processed, the output image is generated via the usual FFT and correction.  

IDG requires more computation (van der Tol \etal \cite{VanderTol2019}, Veenboer \etal \cite{Veenboer2017}, Offringa \etal \cite{Offringa2019}) than Convolutional Gridding. However, it adapts to parallelism and is proven to work efficiently on GPUs (Veenboer \etal \cite{Veenboer2017}). Its main advantage is that it does not use convolution functions in the UV-domain, eliminating oversampling and removing the need to pre-calculate convolution functions. In AW-projection for wide-field SKA images, such convolution functions contain W-term and A-term data, which, if computed, may dominate the imaging process in terms of execution time (van der Tol \etal \cite{VanderTol2019}).

\section{Introduction to Convolutional Gridding and proposed modifications}
\label{sec:introduction:convgriddingdef}

\begin{sidewaysfigure}[]
\centerline{\includegraphics[width=\linewidth, trim=0.5cm 0.5cm 0.5cm 0.5cm, clip=true]{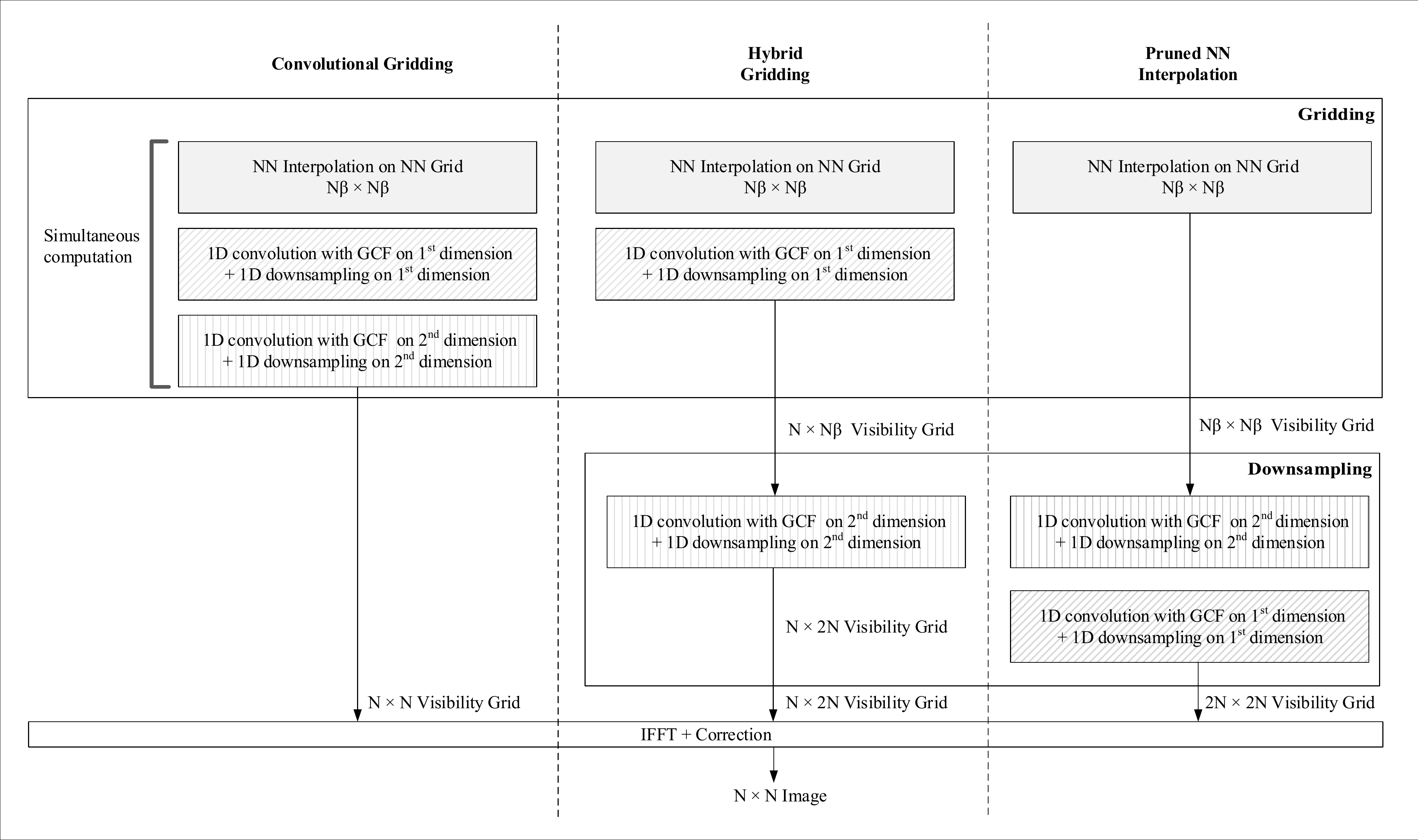}}
\caption[The studied algorithms]{Diagram illustrating how Convolutional Gridding morphs into Hybrid Gridding and Pruned NN Interpolation 
}
\label{fig:introduction:algointro}
\end{sidewaysfigure}
We have already stated that the direct application of Equation \ref{equ:introduction:ID} to calculate $I_{\text{D}}(l,m)$ is not feasible in Radio Interferometry, and Convolutional Gridding is a popular technique used as part of the solution to the problem. A detailed review of Convolutional Gridding and proposed modifications are provided in Chapter \ref{chap:theory}. However, here we give a preliminary description of Convolutional Gridding and the modifications we propose for better Performance.

When re-stating the problem, we argued that we are interested in a sampled and bounded version of $I_{\text{D}}(l,m)$, and this is because we are using a digital computer. Convolutional Gridding tries to take advantage of the Fourier Transform revealed in Equation \ref{equ:introduction:measurmentequationfourier}, using Fast Fourier Transforms (FFTs) (Cooley and Tukey \cite{Cooley1965}). FFTs operate only on a regularly sampled grid of finite width (a bounded grid). Convolutional Gridding maps Visibility records to such a grid by convolving the Visibility records with a real-valued multiplicatively separable Gridding Convolutional Function (GCF) of finite width and at the same time sample the result to the said grid. Once all records are gridded, an Inverse Fast Fourier Transform (IFFT) is applied over the grid and afterwards corrected to generate an output image. 

The output image will not be equal to the sampled and bounded version of $I_{\text{D}}(l,m)$ for two main reasons. First and foremost, as stated before, the simple form of  Convolutional Gridding we are describing does not take care of the W-term denoted by $\overline{g}_{w_p}(l,m)$, and the A-term denoted by $A_p(l,m)$. Another reason is that Convolutional Gridding suffers from aliasing, since $I_{\text{D}}(l,m)$ is expected to be aperiodic and infinitely wide in all dimensions, implying no sampling satisfies the requirements of the Nyquist Rate. 

Let us now introduce in brief our proposed modifications. A detailed analysis with necessary proofs is given in Chapter \ref{chap:theory} while Figure \ref{fig:introduction:algointro} gives a simple illustration. We take advantage of the fact that in real implementations, the GCF is oversampled, that is, sampled at finer intervals than the prior mentioned grid, by a factor of $\beta$, and stored in a lookup table. The Convolutional Gridder will then consult with the lookup table using a Nearest Neighbour (NN) Interpolation scheme to calculate the GCF for each record during gridding. We realise and prove in Chapter \ref{chap:theory} that such an approach is identical to gridding records using NN Interpolation on a virtual grid oversampled by a factor of $\beta$, which we call the NN Grid. Afterwards, the NN Grid is downsampled by convolving with the GCF. Convolutional Gridding does these two steps concurrently. In our modifications, we split these two steps to morph the Convolutional Gridding into two modified gridding algorithms, collectively referred to as the \textit{modified gridding} algorithms and individually named \textit{Hybrid Gridding} and \textit{Pruned NN Interpolation}. In Hybrid Gridding, we split the two steps for one dimension, while in Pruned NN Interpolation, we split the two steps for the two dimensions. 

Finally, we modify the downsampling step of the modified gridding algorithms to try to suppress aliasing generated by the downsampling step below arithmetic noise. This latter modification pushed us to re-review the use of convolution to downsample an oversampled regular grid to accelerate its Fourier inversion. We refer to this process as \textit{Convolution-Based FFT Pruning}, which is analysed in Chapter \ref{chap:theory}.  

One can quickly realise that the modified gridding algorithms can replace Convolutional Gridding in a Visibility-to-image algorithm if and only if the  GCF in Convolutional Gridding is multiplicative separable and invariant for all records gridded, that is, the GCF is not dependent on any Visibility record to be gridded. For this reason, the modified gridding algorithms cannot be directly applied for those Visibility-to-image algorithms that include W-term and A-term information in the GCF of Convolutional Gridding, since the GCF loses its multiplicative separable and invariant properties.    

Table \ref{tab:introduction:vtoialgos} lists all the Visibility-to-image algorithms previously reviewed and state if Convolutional Gridding whit-in each Visibility-to-image algorithm can be replaced with the modified gridding algorithms. The rules are straightforward and based on what is explained in the previous paragraph. The modified gridding algorithms can be applied to all Visibility-to-image algorithms that do not modify the GCF in their Convolutional Gridding step by replacing Convolutional Gridding with the modified gridding algorithm. Suppose a Visibility-to-image algorithm modifies the GCF. In that case, the modified gridding algorithms can only be applied if the Visibility-to-image algorithm is first modified to re-establish the use of an unmodified, multiplicative separable and invariant GCF. Suppose it is not possible, or there is no sense in modifying the given Visibility-to-image algorithm. In that case, the modified gridding algorithms are incompatible with the stated Visibility-to-image algorithm.

\begin{table}[]
\centering
\begin{tabular}{@{\vspace{5pt}}l@{\hspace{2cm}}l@{}}
\toprule
Visibility-to-image algorithm                    & \begin{minipage}{6cm}{Compatible with the modified gridding algorithms?}\end{minipage}  \\ \midrule
Ignore w-term                                    & Yes                                                           \\ 
Facets  & Yes                                                     \\
W-projection                                   & No                                                    \\
W-stacking                                       & Yes                                                        \\
Snapshots & Yes                                                        \\
Facets with W-projection                         & Maybe$^\dagger$        \\
DDFacet                                         & Maybe$^\dagger$   \\
W-snapshots                                     & Maybe$^\dagger$\\
W-projection + W-stacking                     & No     
\\
3D FFT  & Maybe$^\star$ \\
A-projection    & No \\
Wide-Band A-Projection & No  \\
AW-Projection & No \\
Facets handling the A-term & Maybe$^\ddagger$ \\
Image Domain Gridding & Not Applicable \\
\bottomrule
\end{tabular}
\caption[List of Visibilty-to-Image algorithms and their compatibility with the modified gridding algorithms]{Table listing all mentioned Visibility-to-image algorithms mentioned in Section \ref{sec:introduction:vistoimagetransform}, and stating if they are compatible with the modified gridding algorithms.\\[4pt]
$^\dagger$ The use of W-projection in the given Visibility-to-image algorithm needs to be changed to W-stacking for the modified gridding algorithms to be compatible.\\[4pt]
$^\star$ The modified gridding algorithms need to be applied for three dimensions.\\[4pt]
$^\ddagger$ Any Visibility-to-image algorithm that use Facets handling the A-term is compatible with the modified gridding algorithms given the  A-term is corrected in the image plane and W-stacking is used instead of W-projection.}
\label{tab:introduction:vtoialgos}
\end{table}

\section{Thesis layout}
\label{sec:introduction:thesislayout}
This thesis is laid out as follows:
 
At the start of this chapter, we stated our thesis' primary goal. Afterwards, we gave context to our goal by reviewing Radio Interferometry and how image synthesis works. Such a review led us to briefly introduce Convolutional Gridding and our modified gridding algorithms called Hybrid Gridding and Pruned NN Interpolation. It was also identified in which algorithms that handle the W-term and A-term the modified gridding algorithms could be integrated.

Chapter \ref{chap:theory} is an in-depth theoretical and mathematical review of Convolutional Gridding building towards a proper formulation of the modified gridding algorithms, including Convolution-Based FFT Pruning. From theory, we derive expectations (claims) for these modified gridding algorithms, where such expectations will be experimentally tested at the end of the thesis.  

From Chapter \ref{chap:methodology} onwards, we discuss the implementations made to be able to verify our expectations of the modified gridding algorithms experimentally.

Chapter \ref{chap:methodology} introduces various concepts of CUDA programming which we will use in subsequent chapters, and delivers essential notes on our experimental setup and the general structure of the studied implementations. At the end of the chapter we define the many Performance metrics that we are to measure in experiments and use for an in-depth analysis of Performance.     

The three subsequent chapters specialise in discussing in full our implemented Gridders\footnote{A Gridder is an implementation of a Gridding Step, defined in Chapter \ref{chap:theory}.} of Convolution Gridding (Chapter \ref{chap:2dgridding}), Hybrid Gridding  (Chapter \ref{chap:hybrid}) and Pruned NN Interpolation (Chapter \ref{chap:purenn}). Each chapter includes an in-depth analysis on the Performance of the Gridder in question, based on experimental results.

Chapter \ref{chap:pruning} discusses in detail our implementation of Convolution-Based FFT Pruning as needed by Hybrid Gridding and Pruned NN Interpolation. The discussion progresses in a similar way to how discussions in previous chapters progress. At the end of the chapter, some experimental analyses are included on the suppression of aliasing in Convolution-Based FFT Pruning.

In all chapters we mentioned so far, we report on the studied implementations of Gridders and Pruners, and in Chapter \ref{chap:comparative} we put everything together to provide a comparative experimental analysis on Performance and aliasing between our studied implementations. The ultimate goal is to show if any of the modified gridding algorithms can perform better than Convolutional Gridding with no decrease in the suppression of aliasing. 

In the previous chapters, we consciously avoid discussing the validation of the gridders in terms of image quality (particularly alias distortions) and Performance. We left such validation to the currently discussed Chapter \ref{chap:comparative}, since all validations are done through comparisons. In particular, at the beginning of the chapter, we experimentally validate our implementation of Convolutional Gridding with other well-known gridders since the Convolutional Gridder is used as a reference for validating and comparing the other studied implementations.

Finally, we conclude in Chapter \ref{chap:conclusion} by  summarising what we discussed in this thesis and citing the main results. A mandatory future work section then follows.

\section{Publications}
At the writing of this thesis, we did not yet publish any content in peer-reviewed journals. However, we refer to Muscat and Bernardi \cite{Muscat2014b}, which is a Square Kilometre Array Science Data Processor (SDP) Memo authored by ourselves, as part of an international collaboration done during the tenure of the PhD with the said SDP.

We intend to publish most of the content in this thesis. In particular, we are planning to publish Chapter \ref{chap:2dgridding} (The Convolutional Gridder), Chapter \ref{chap:hybrid} (The Hybrid Gridder),  Chapter \ref{chap:purenn} (The NN Gridder), and Chapter \ref{chap:pruning} (The Pruners) as separate papers. We envisage integrating content from the other chapters in the said papers.

\section{Conclusion}
In this chapter, we briefly introduced Image Synthesis in Radio Interferometry and the main thesis' goal with the modifications in Convolutional Gridding that we are to examine to reach our goal. In Section \ref{sec:introduction:thesislayout}, we explained the thesis layout, and afterwards, we discussed content from this thesis that we intend to publish in peer-reviewed journals.

\chapter{Theory and Mathematical Review}
\label{chap:theory}
This chapter gives a theoretical and mathematical review of Convolutional Gridding that builds towards and reviews the proposed modifications that morph Convolutional Gridding to Hybrid Gridding and Pruned Nearest Neighbour (NN) Interpolation, collectively referred to as the modified gridding algorithms.   

The modified gridding algorithms downsample a grid using convolution, which is a well-known technique affected by aliasing. At the end of the chapter, we will re-consider this technique as Convolution-Based FFT Pruning, whereby with the use of the least-misfit gridding functions, we try to suppress aliasing below arithmetic noise.

We have organised this chapter as follows: In Section \ref{sec:maths:notationandterminology} we define notations, terminology, function and assumptions that we use throughout the chapter. In Section \ref{sec:maths:problemformulation} we reformulate the problem Convolutional Gridding tries to solve, and in the subsequent section, we review Convolutional Gridding in detail. In Section \ref{sec:maths:prunednninterpolation}, we review simple NN Interpolation and progress to define and review the modified gridding algorithms. At the end we dedicate Section \ref{sec:maths:fftpruning} to discuss Convolution-Based FFT Pruning.

\section{Notations, terminology and assumptions}
\label{sec:maths:notationandterminology}
This section defines the notation we are using in this chapter, together with some terminology and assumptions. Each sub-section discusses some particular topic indicated by its title. 

\subsection{Spaces, and multiplicative separability}

For our analyses, we define two Euclidean spaces of two dimensions, which are the $lm$-space and its Fourier equivalent $uv$-space.  Vectors $\myx=(l,m)$ and $\myfreq=(u,v)$ are Cartesian co-ordinates for the respective spaces.

All functions presented in this chapter are multiplicative separable, implying that they can be re-adapted to cater for any number of dimensions.  A function $D(\myx)$ is multiplicative separable, if there exists functions $D_1(l)$ and $D_2(m)$ such that $D(\myx)=D_1(l)D_2(m)$. We refer to the work of Tan \cite{TanSeptemebr1986} for proof that Convolutional Gridding is multiplicative separable. 

\subsection{Grids and normalisation}
\label{sec:maths:normalisation}
Directly related to the $lm$-space and $uv$-space are the $lm$-grid and $uv$-grid. Since computation is on a digital machine, many of the functions we will define are regularly sampled with a given interval and laid out on the said grids which are bounded. We will use the term pixel to refer to a point on a grid and set the origin, $\myx=(0,0)$ and $\myfreq=(0,0)$, on a pixel at the centre.

In order to simplify our workings, we will always assume that the final output intensity image of any algorithm laid out on the $lm$-grid is of dimensions $N \times N$ pixels with a two-dimensional sampling interval $\myDeltax=(\Delta l,\Delta m)$. We normalise all our work such that $N\Delta l=N\Delta m=1$, implying $\Delta l=\Delta m$. We do stress out that the output image is square in size just for convenience and all the work we present here can be re-adapted to cater for rectangular images.

\subsection{Functions and notations}
We group functions into two, the so-called \textit{helper functions} and the other functions. We define helper functions in Section \ref{sec:maths:helperfunctions} and discuss notation of the other functions below.

Most of the other functions, hereafter referred to just as functions, are defined with a domain of two or more variables, with the last variable being the oversampling factor $\beta$, provided it has relevance in the context of the discussion being pursued. 

We denote all functions that exist in the $lm$-space with a small letter, and the Fourier equivalents in the $uv$-space with the same letter but capital. 

We affix to the right, a superscript with the letter $g$ to any function that is sampled over a grid. The helper functions $\gshah{\Psi}(\myx)$, and $\gshah{\Psi}(\myfreq)$  are purposely defined in Section \ref{sec:maths:helperfunctions} to sample a function on a bounded grid.  

We denote functions describing the same things in different algorithms with the same symbol and affix a two-letter suffix to the subscript to distinguish for which algorithm the function is defined. The letters used for the said suffix are an abbreviation of the algorithm in question and given in Table \ref{tab:maths:abbreviations}. For example, the output of Hybrid Gridding is $f^{\text{g}}_{\text{hy}}(\myx,\beta)$, while the output of Pruned NN Interpolation is $f^{\text{g}}_{\text{pn}}(\myx,\beta)$.

\subsection{Helper functions}
\label{sec:maths:helperfunctions}
We now define all helper functions. One notes that these functions operate on the $uv$-space differently than on the $lm$-space. Notation of the helper functions is different than that of the other functions and most of them have a suffix $\Psi$ on which the helper function depends. In many cases, the suffix $\Psi$  has some relation with the oversampling factor $\beta$, and in various equations, $\Psi$ will be set to $\beta$. 

For simplification, we define all helper functions in one dimension, which, given that they are multiplicative separable, means that a defined helper function $D_{\Psi}(l)$ or $D_{\Psi}(u)$ is to be considered defined in two dimensions using the relationship $D_{\Psi}(\myx)=D_{\Psi}(l)D_{\Psi}(m)$ or $D_{\Psi}(\myfreq)=D_{\Psi}(u)D_{\Psi}(v)$ respectively.

\subsubsection{Dirac Delta Function $\boldsymbol{\delta(l)}$ and $\boldsymbol{\delta(u)}$}

The Dirac Delta Function is the unit impulse function defined in the Set Of Equations \ref{equ:maths:diracdelta}.
\begin{subequations}
\label{equ:maths:diracdelta}
\begin{equation}
\delta(l)=\delta(u)=\begin{cases} 
   +\infty & \text{if } l=u=0 \\
   0 & \text{otherwise} \\
  \end{cases} 
\end{equation}
\begin{equation}
    \int_\infty \delta(l)=\int_\infty \delta(u)=1
\end{equation}
\end{subequations}

\subsubsection{Shah function $\boldsymbol{\shah{\Psi}(l)}$ and $\boldsymbol{\shah{\Psi}(u)}$}

The Shah function is an infinite impulse train which we define for the $lm$-space and $uv$-space as Fourier Transforms of each other as shown in Equations \ref{equ:maths:shahl} and \ref{equ:maths:shahu}. 

\begin{equation}
\label{equ:maths:shahl}
    \shah{\Psi}(l)=\sum\limits_{b\in\mathbb{Z}}\delta(l-\Psi b)
\end{equation}

\begin{equation}
\label{equ:maths:shahu}
    \shah{\Psi}(u)=\sum\limits_{b\in\mathbb{Z}}\delta(u-\Psi^{-1} b)
\end{equation}

where $\mathbb{Z}$ denotes the set of integers.

\subsubsection{$\boldsymbol{\rect{\Psi}(l)}$ and $\boldsymbol{\rect{\Psi}(u)}$}

We define the one-dimensional $\rect{\Psi}(l)$ and $\rect{\Psi}(u)$ through Equations  \ref{equ:maths:rectl} and \ref{equ:maths:rectu}.
\begin{equation}
\label{equ:maths:rectl}
\rect{\Psi}(l)=\begin{cases} 
   1 & \text{if } -0.5\Psi<l\le 0.5\Psi \\
   0 & \text{otherwise}
  \end{cases}  
\end{equation}

\begin{equation}
\label{equ:maths:rectu}
\rect{\Psi}(u)=\begin{cases} 
   1 & \text{if } -(2\Psi)^{-1}<u\le (2\Psi)^{-1} \\ 
   0 & \text{otherwise}
  \end{cases}
\end{equation}

\subsubsection{$\boldsymbol{\sinc{\Psi}(l)}$ and $\boldsymbol{\sinc{\Psi}(u)}$}
We define the one-dimensional $\sinc{\Psi}(l)$ and $\sinc{\Psi}(u)$ through Equations  \ref{equ:maths:sincl} and \ref{equ:maths:sincu}

\begin{equation}
\label{equ:maths:sincl}
  \sinc{\Psi}(l)=\frac{\text{sin}(\pi \Psi^{-1}l)}{\pi \Psi^{-1}l}  
\end{equation}
\begin{equation}
\label{equ:maths:sincu}
  \sinc{\Psi}(u)=\frac{\text{sin}(\pi \Psi u)}{\pi \Psi u}  
\end{equation}

We note that $\sinc{\Psi}(u)$ is the Fourier transform of $\rect{\Psi}(l)$ while $\rect{\Psi}(u)$ is the Fourier transform of $\sinc{\Psi}(l)$.  

\subsubsection{$\boldsymbol{\gshah{\Psi} (l)}$ and $\boldsymbol{\gshah{\Psi} (u)}$}
We will use $\gshah{\Psi} (l)$ and $\gshah{\Psi} (u)$ to sample a given function on a regular bounded grid. 

\begin{equation}
\gshah{\Psi}(l)=\sum\limits_{b\in\mathbb{Z}} \delta(l-b\Delta l)\rect{\Psi}(l)
\end{equation}
\begin{equation}
\gshah{\Psi}(u)=\shah{\Psi}(u)\rect{1}(u\Delta l)
\end{equation}

We note that the value of $\Psi$ in $\gshah{\Psi}(l)$ affects only the boundaries of the $lm$-grid while in $\gshah{\Psi}(u)$ the value of $\Psi$ controls the sampling interval of the $uv$-grid.

\begin{table}
\centering
\begin{tabular}{lll}
\toprule
Algorithm                   & Suffix \\
\midrule                                    
Convolutional Gridding & $\text{co}$  \\
Nearest Neighbour Interpolation & $\text{nn}$ \\
Pruned Nearest Neighbour Interpolation & $\text{pn}$ \\
Hybrid Gridding & $\text{hy}$ \\
Convolution-Based FFT Pruning & $\text{pr}$ \\
\bottomrule
\end{tabular}
\caption[Algorithm abbreviations]{Algorithm abbreviations used  in suffixes.}
\label{tab:maths:abbreviations}
\end{table}

\section{Problem formulation}
\label{sec:maths:problemformulation}
We already formulated the problem through Equation \ref{equ:introduction:ID}, but that equation contains various terms which, as discussed in the previous chapter, our modified gridding algorithms do not handle directly. Therefore, we here re-formulate the problem that Convolutional Gridding and the modified gridding algorithms try to solve in a more convenient form.

Let $F(\myfreq)$ be a function on the $uv$-space. $F(\myfreq)$ is zero-valued everywhere, except for $P$ values of $\myfreq$ exclusive members of the set $\freqset$, that is $F(\myfreq)=0 \text{ if } \myfreq\notin\freqset$. Note that $F(\myfreq)$ is equivalent to the irregularly sampled $V_w(u,v)$ with the exception that we are dropping $w$. We still use the term record or Visibility record to refer to elements in $\freqset$ together with the value of $F(\myfreq)$ for the given element in $\freqset$.  

Let $f(x)$ be the Inverse Fourier Transform of $F(\myfreq)$. Equation \ref{equ:maths:dftanalogue} defines the relationship between $f(\myx)$ and $F(\myfreq)$.
\begin{equation}
   f(\myx)=\sum\limits_{\myfreq \in \freqset}F(\myfreq)e^{-j2\pi(\myx\cdot\myfreq)}
\label{equ:maths:dftanalogue}   
\end{equation}
where '$\cdot$'  is the dot product, that is $\myx\cdot\myfreq=ul+vm$.
$f(\myx)$ is similar to $I_D(l,m)$ but without the A-term and W-term. 

The problem is to calculate the regularly sampled and bounded version of $f(\myx)$ which we denote with $f^{\text{g}}(\myx)$ and define by the DFT Equation \ref{equ:maths:dft}. 

\begin{equation}
\label{equ:maths:dft}
   f^{\text{g}}(\myx)=f(\myx)\gshah{1}(\myx)=\sum\limits_{\myfreq \in \freqset}F(\myfreq)e^{-j2\pi(\myx\cdot\myfreq)}\gshah{1}(\myx)
\end{equation}

We remind the reader that for convenience we are assuming that $f^{\text{g}}(\myx)$ is  of size $N\times N$ pixels. 

\section{Convolutional Gridding}
\label{sec:maths:convolutioinalgridding}
In this section, we will define and analyse Convolutional Gridding. We will only introduce and discuss oversampling in Section \ref{sec:maths:gcfoversampling} and therefore, from this point onwards until the stated section, the reader should not assume any oversampling of the GCF.

\begin{figure}
\centering
\includegraphics[page=1,width=0.9\linewidth,trim=0.1cm 0.1cm 0.1cm 0.1cm,clip]{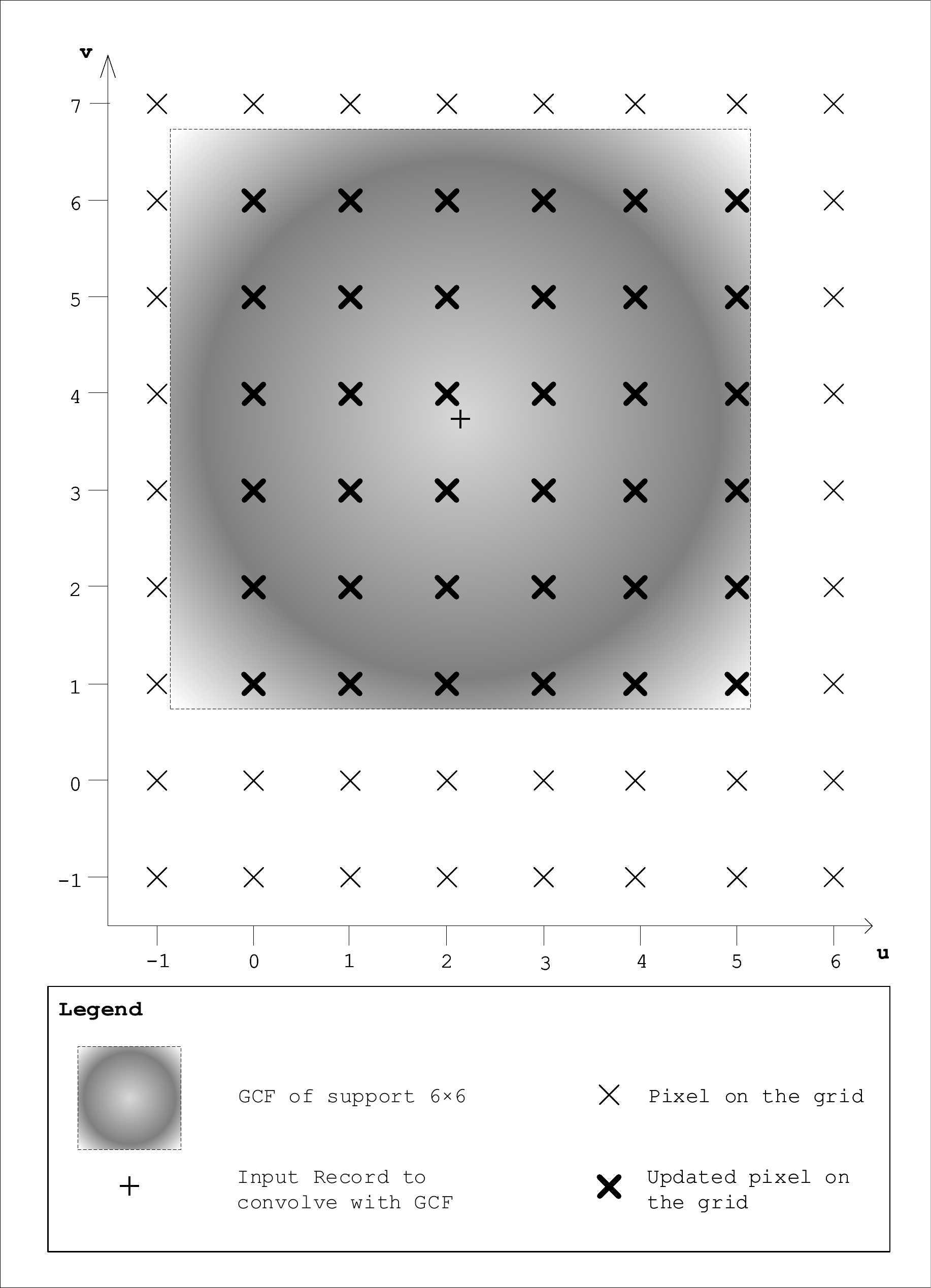}
\caption[Convolutional Gridding]{Diagram illustrating the Gridding Step (Equation \ref{equ:maths:griddingstep1}) in Convolutional Gridding. A record in $F(\myfreq)$  is being convolved with $C_{\text{co}}(\myfreq)$ and sampled over the grid. Only a small region of the grid is drawn in the diagram.}
\label{fig:maths:convgridding}
\end{figure}

The general algorithm for Convolutional Gridding is laid out in Algorithm \ref{algo:maths:convgridding}. The GCF is denoted by $C_{\text{co}}(\myfreq)$ and the operator * denotes a convolution. An illustration of the Gridding Step (Equation \ref{equ:maths:griddingstep1}) is given in Figure \ref{fig:maths:convgridding}.

\setlength{\algomargin}{0.1em}
\begin{algorithm}
\begin{mdframed}
\setstretch{1.5}
\KwIn{$F(\myfreq)$, $C_{\text{co}}(\myfreq)$} 
\SetKwInput{KwHelp}{Helper Input}
\KwHelp{GCF: $C_{\text{co}}(\myfreq)$, Corrector: $h_{\text{co}}(\myx)$}
\KwOut{$f^{\text{g}}_{\text{co}}(\myx)$} 
\SetKwInput{KwAs}{Assumptions}
\KwAs{The Bounded Records Assumption}
\myalgoline
\SetKwBlock{step}{Gridding Step:}{}
\step{Each non-zero sample of $F(\myfreq)$ is convolved with  $C_{\text{co}}(\myfreq)$ and simultaneously sampled on the $uv$-grid:
\begin{equation}
W^{\text{g}}_{\text{co}}(\myfreq)=\left[F(\myfreq)*C_{\text{co}}(\myfreq)\right]\gshah{1}\left(\myfreq\right)
\label{equ:maths:griddingstep1}
\end{equation}
}
\SetKwBlock{step}{IFFT Step:}{}
\step{
An Inverse Fast Fourier Transform (IFFT) is performed over $W^{\text{g}}_{\text{co}}(\myfreq)$ to get $w^{\text{g}}_{\text{co}}(\myx)$:
\begin{equation}
\label{equ:maths:wgcoconv}
w^{\text{g}}_{\text{co}}(\myx)=\left[f(\myx)c_{\text{co}}(\myx)*\shah{1}(\myx)\right]\gshah{1}(\myx)
\end{equation}
}

\SetKwBlock{step}{Correction Step:}{}
\step{A corrector $h_{\text{co}}(\myx)$ is applied to obtain $f^{\text{g}}_{\text{co}}(\myx)$ which approximates $f_g(\myx)$:
\begin{equation}
f^{\text{g}}_{\text{co}}(\myx)=\left[f(\myx)c_{\text{co}}(\myx)*\shah{1}\left(\myx\right)
\right]\gshah{1}(\myx)h_{\text{co}}(\myx)
\label{equ:maths:griddingoutput}
\end{equation}

}

\end{mdframed}
\caption[General Convolutional Gridding]{The general Convolutional Gridding Algorithm. We discuss the Bounded Records Assumption in Section \ref{sec:maths:outofbounds}.}
\label{algo:maths:convgridding} 
\end{algorithm}

\subsection{The Bounded Record Assumption}
\label{sec:maths:outofbounds}

In formulating all the gridding algorithms in this thesis, it is assumed that $F(\myfreq)*C_{\text{co}}(\myfreq)$ is zero-valued outside the $uv$-grid boundaries. We refer to this assumption as the Bounded Records Assumption, which is in line with how other publications describe Convolutional Gridding such as Jackson \etal\cite{Jackson1991} and Greisen \cite{Greisen1979}.  The said assumption is also implemented in various Radio Interferometric Imagers such as CASA (McMullin \etal \cite{McMullin2007}), lwimager and WSClean (Offringa \etal \cite{Offringa2014}), whereby any record not abiding to this assumption is ignored.

The rationale behind the stated assumption is related to the fact that $f^{\text{g}}_{\text{co}}(\myx)$ is sampled at an interval $\myDeltax$ and for reasons rooted in the Sampling Theorem the IFFT will only use the pixels in the $uv$-grid that are within the boundaries of the $uv$-grid.

We point out that it is possible to override the Bounded Records Assumption  in  Convolutional Gridding by modifying the Gridding Step to implement Equation \ref{equ:maths:step1newformula} instead of Equation \ref{equ:maths:griddingstep1}.

\begin{equation}
\label{equ:maths:step1newformula}
    W^{\text{g}}_{\text{co}}(\myfreq)=[F(\myfreq)*C_{\text{co}}(\myfreq)*\shah{1}(\myfreq\odot\myvec{\Delta}\myx)]\gshah{1}(\myfreq)
\end{equation}
where $\odot$ is the Hadamard product, that is $\myfreq\odot\myvec{\Delta}\myx=(u \Delta l, v\Delta m)$. The convolution with $\shah{1}(\myfreq\odot\myvec{\Delta}\myx)$ in Equation \ref{equ:maths:step1newformula} aliases back any records that are outside the boundaries mimicking the aliasing effects in the $uv$-space induced by sampling  $f(\myx)$ to $f^{\text{g}}(\myx)$.

\subsection{Component analyses}
\label{sec:maths:componentanalysis}
We find it useful to formulate how every record in $F(\myfreq)$ gets distorted by  Convolutional Gridding based on the value of $\myfreq$. We measure distortion per component using the function  $l_{\text{co}}(\delta\myfreq,\myx)$ as defined  by Equation \ref{equ:maths:ldef}. 
\begin{equation}
\label{equ:maths:ldef}
    f^{\text{g}}_{\text{co}}(\myx)=\sum\limits_{\myfreq \in \freqset}\left[F(\myfreq)e^{-j2\pi\myfreq\cdot\myx}l_{\text{co}}(\delta\myfreq,\myx)\right]\gshah{1}(\myx)h_{\text{co}}(\myx)
\end{equation}
The vector $\delta\myfreq$ is equal to the vectorial distance of a record at $\myfreq$ from the nearest pixel with $uv$-coordinates $\myvec{k} \in \mathbb{Z}^2$, where $\mathbb{Z}^2$ is the set of all two-element integer vectors $\left[a\in\mathbb{Z},b \in \mathbb{Z}\right]$. Therefore $\forall\delta\freq_k \in \myfreq$, $0.5\le\delta\freq_k<0.5$ and $\myfreq=\myvec{k}+\delta\myfreq$.

Substituting the Equation \ref{equ:maths:dftanalogue} in Equation \ref{equ:maths:griddingoutput} we get

\begin{equation}
f^{\text{g}}_{\text{co}}(\myvec{x})=\left[\sum\limits_{\myfreqel \in \freqset} F(\myvec{\freqel})e^{-j2\pi\myvec{\freqel}\cdot\myx}c_{\text{co}}(\myx)*\shah{1}\left(\myx\right)
\right]\gshah{1}(\myx)h_{\text{co}}(\myx)
\end{equation}
Substituting $\myfreq=\myvec{k}+\myvec{\delta\freq}$ and re-expressing convolution with $\shah{1}\left(\myx\right)$ as a summation (refer to Equation \ref{equ:maths:shahl}) we get:
\begin{equation}
f^{\text{g}}_{\text{co}}(\myvec{x})=\sum\limits_{\myfreqel \in \freqset}\left[F(\myvec{\freqel})\sum\limits_{\myvec{b}\in \mathbb{Z}^2}e^{-j2\pi(\myvec{k}+\myvec{\delta\freqel})\cdot(\myx-\myvec{b})}c(\myx-\myvec{b})\right]\gshah{1}(\myx) h_{\text{co}}(\myx)
\end{equation}
Which leads to:
\begin{equation}
\label{equ:maths:companalysisout}
f^{\text{g}}_{\text{co}}(\myvec{x})=\sum\limits_{\myfreqel \in \freqset}\left[F(\myvec{\freqel})e^{-j2\pi\myvec{\delta\freq}\cdot\myx}\sum\limits_{\myvec{b}\in \mathbb{Z}^2}\cancelto{1}{e^{-j2\pi\myvec{k}\cdot\myvec{ b}}}e^{-j2\pi\myvec{\delta\freqel}\cdot(\myx-\myvec{b})}c(\myx-\myvec{b})\right]\gshah{1}(\myx)h_{\text{co}}(\myx)
\end{equation}

Comparing Equation  \ref{equ:maths:companalysisout} with Equation \ref{equ:maths:ldef} we get:
\begin{equation}
    l_{\text{co}}(\myvec{\delta\freq},\myx)=\sum\limits_{\myvec{b} \in \mathbb{Z}^2}e^{-j2\pi\myvec{\delta\freq}\cdot(\myx-\myvec{b})}c(\myx-\myvec{b})
\end{equation}

One notes that $l_{\text{co}}(\myvec{\delta\freq},\myx)$ is neither dependent on $F(\myfreq)$ nor $\myvec{k}$ which implies that records with equal $\myvec{\delta\freq}$ are distorted with a constant, only dependent on the $lm$-coordinate. Therefore, if all records have equal $\myvec{\delta\freq}$ there exists a corrector $h_{\text{co}}(\myx)$ that can cancel all distortions. Conversely, if some of the records have a different $\myvec{\delta\freq}$, a corrector cancelling all distortions is unlikely to exist. What we just stated is a way to explain the existence of aliasing in Convolutional Gridding and in the next sub-section we give a second approach to explain and measure the distortion caused by aliasing.

\subsection{Aliasing}
\label{sec:maths:aliasing}
As pointed out many times, Convolutional Gridding suffers from aliasing. Therefore, excluding exceptional circumstances, $f^{\text{g}}_{\text{co}}(\myx)$ can only approximate $f^{\text{g}}(\myx)$. The Gridding Step induces aliasing when $F(\myfreq)*C_{\text{co}}(\myfreq)$ is sampled to the grid. $f(\myx)c_{\text{co}}(\myx)=\mathcal{F}^{-1}[F(\myfreq)*C_{\text{co}}(\myfreq)]$  is generally infinitely wide and therefore, there exits no sampling rate that qualifies as the Nyquist Rate for the sampling of $F(\myfreq)*C_{\text{co}}(\myfreq)$. 

It is worth to analyse the effects of aliasing by converting the convolution in Equation \ref{equ:maths:griddingoutput} into a sum and rewriting it into the following set of equations:
\begin{subequations}
\begin{equation}
    f^{\text{g}}_{\text{co}}(\myx)=\left[o^{\text{g}}_{\text{co}}(\myx)+a^{\text{g}}_{\text{co}}(\myx)\right]\gshah{1}(\myx)h_{\text{co}}(\myx)
\end{equation}
\begin{equation}
    o^{\text{g}}_{\text{co}}(\myx)=f(\myx)c_{\text{co}}(\myx)
\end{equation}
\begin{equation}
    a^{\text{g}}_{\text{co}}(\myx)=\sum\limits_{\myvec{b} \in \myvec{\mathbb{Z}^2},\myvec{b} \ne \myvec{0}  } f(\myvec{x}-\myvec{b})c_{\text{co}}(\myvec{x}-\myvec{b})
\end{equation}
\label{equ:maths:aliasing}
\end{subequations}

$o^{\text{g}}_{\text{co}}(\myx)$ is the term containing the data from which we can recover $f^{\text{g}}(\myx)$ while $a^{\text{g}}_{\text{co}}(\myx)$ defines all the irreversible distortions caused by aliasing. In the unlikely case that $a^{\text{g}}_{\text{co}}(\myx)=0$ and provided  $h_{\text{co}}(\myx)=1/c_{\text{co}}(\myx)$ then $o^{\text{g}}_{\text{co}}(\myx)h_{\text{co}}(\myx)=f^{\text{g}}(\myx)$.

The above set of equations show that aliasing causes $f(\myx)c_{\text{co}}(\myx)$ to wrap around, which leads to some crucial statements: 
\begin{enumerate}
    \item Structures in $f(\myx)$ outside the grid boundaries will appear in $f^{\text{g}}_{\text{co}}(\myx)$. Therefore, aliasing is a function of $f(\myx)$ and the level of distortion caused by aliasing is dependent on the values of $f(\myx)$ outside the boundaries of the $lm$-grid.
    \item A good choice of $c_{\text{co}}(\myx)$ tries to reduce aliasing by attenuating or zeroing $f(\myx)$  outside the boundaries of the grid prior sampling occurring in the $uv$-space. Therefore, $C_{\text{co}}(\myfreq)$ is an anti-aliasing $lm$-space filter.
    \item For best anti-aliasing results, the ratio $a^{\text{g}}_{\text{co}}(\myx)/o^{\text{g}}_{\text{co}}(\myx)$ should approach zero. Therefore, $c_{\text{co}}(\myx)$ should have values as high as possible within the boundaries and approach zero elsewhere. Due to the needed properties of $C_{\text{co}}(\myfreq)$ (refer to Section \ref{sec:maths:gcfform} ), $c_{\text{co}}(\myx)$ is expected to be continuous, infinitely wide, and unable to abruptly change from a high value to zero at the boundaries of the $lm$-grid. Instead, there is a region near the boundaries where $c_{\text{co}}(\myx)$ rolls-off from high values to values approaching zero as the  $lm$-coordinates move away from the origin. $f^{\text{g}}_{\text{co}}(\myx)$ is generally ignored in this region, by means of padding, since it suffers from a severe level of aliasing distortion (Jackson \etal \cite{Jackson1991}, O'Sullivan \cite{OSullivan1985}).
\end{enumerate}

\subsubsection{Exceptions}
At the beginning of this section, we have hinted that in exceptional circumstances, Convolutional Gridding can give accurate results with no aliasing effects. We now list some of these exceptions. 

A trivial case is when $F(\myfreq)=0$, which beyond any doubt will lead to $f^{\text{g}}_{\text{co}}(\myx)=f^{\text{g}}(\myx)=0$

Another exception is, when all records have coordinates with equal values of $\myvec{\delta\freq}$, say $\myvec{y}$. We already discussed this scenario  in Section \ref{sec:maths:componentanalysis} and if the corrector is set to $h_{\text{co}}(\myx)=1/l_{\text{co}}(\myvec{y},\myx)$ Convolutional Gridding will output $f^{\text{g}}_{\text{co}}(\myx)=f^{\text{g}}(\myx)$. Our claim can be verified by making the necessary substitutions in Equation \ref{equ:maths:ldef}.
\begin{equation}
f^{\text{g}}_{\text{co}}(\myvec{x})=\sum\limits_{\myfreqel \in \freqset}\left[F(\myfreqel)e^{-j2\pi\myfreqel\cdot\myx}l_{\text{co}}(\myvec{y},\myx)\right]/l_{\text{co}}(\myvec{y},\myx)
\end{equation}
\begin{equation}
f^{\text{g}}_{\text{co}}(\myvec{x})=\sum\limits_{\myfreqel \in \freqset}\left[F(\myvec{\freqel})e^{-j2\pi\myvec{\freqel}\cdot\myx}\right]\cancelto{1}{l_{\text{co}}(\myvec{y},\myx)/l_{\text{co}}(\myvec{y},\myx)}=f^{\text{g}}(\myx)
\end{equation}

The same claim can also be verified through Equation \ref{equ:maths:griddingoutput} by considering $f(\myx)$ as periodic on all dimensions, since $\myvec{\delta\freq}$ is equal for all records, implying $f(\myx)=f(\myx-\myvec{b})$ where $\myvec{b}\in \mathbb{Z}^2$. 

Note that the just stated exception is trivial since we are applying Convolutional Gridding over an input already regularly sampled  with an interval equal to that of the $uv$-grid. Therefore, a simple Nearest Neighbour Interpolation Scheme with phase correction is enough to generate $f^{\text{g}}(\myx)$ accurately.  

A more interesting exception is when $c_{\text{co}}(\myx)$ is zero outside the grid boundaries, which will lead to $a^{\text{g}}_{\text{co}}(\myx)=0$, and therefore no aliasing. O'Sullivan\cite{OSullivan1985} shows that $c_{\text{co}}(\myx)=\rect{1}(\myx)$, implying $C_{\text{co}}(\myfreq)=
\sinc{1}(\myfreq)$ is the ideal solution, but realise that $\sinc{1}(\myfreq)$ is infinitely wide, making it impossible to implement (O'Sullivan\cite{OSullivan1985}, Jackson \etal\cite{Jackson1991}). However, we point out that the modification in Convolutional Gridding presented in Section \ref{sec:maths:outofbounds} makes the use of such a function possible. The Gridding Step can be implemented as a circular convolution on the $uv$-grid with $J(\myfreq)$, defined as follows:

\begin{subequations}
\label{equ:maths:circsinc}
\begin{equation}
    J(\myfreq)=\left[\sinc{1}(\myfreq)*T(\myfreq)\right]R(\myfreq)
\end{equation}
where
\begin{equation}
    T(\myfreq)=\sum\limits_{\myvec{b} \in \mathbb{Z}^2} \delta(\myfreq+N\myvec{b})
\end{equation}
and
\begin{equation}
    R(\myfreq)=\begin{cases} 
   1 & \text{if } -N/2\le u,v < N/2 \\ 
   0 & \text{otherwise}
   \end{cases}
\end{equation}
\end{subequations}
\subsection{The form of the GCF}
\label{sec:maths:gcfform}
Let us now discuss the form of the GCF for Convolutional Gridding in Radio Interferometry.

Greisen \cite{Greisen1979} argues that $C_{\text{co}}(\myfreq)$ should be real, symmetric, multiplicative separable (Jackson \etal\cite{Jackson1991}) of  small finite support (Jackson \etal\cite{Jackson1991}, O'Sullivan\cite{OSullivan1985}) and stored in a lookup-table. We note that the assumption that $C_{\text{co}}(\myfreq)$ is multiplicative separable is fundamental in this thesis as already stated and in general, the functions that  $C_{\text{co}}(\myfreq)$  separates to are equivalent. Storing in a look-up table leads to $C_{\text{co}}(\myfreq)$ getting oversampled, which we discuss in Section \ref{sec:maths:gcfoversampling}. 

In this thesis, we define \textit{support}  of a GCF as the maximum amount of neighbour pixels on the $uv$-grid a record maps to when convolved with the said GCF. Support is given per dimension, and in Convolutional Gridding we set the support of the GCF at $S\times S$, $S \in \mathbb{N}$, where $\mathbb{N}$ is the set of natural numbers.

The need for $C_{\text{co}}(\myfreq)$ to have small finite support, independent from the $uv$-grid size arises from the desire to have the Gridding Step scale with $S^2$ rather than $N^2$. The use of the infinite sinc as modified in Section \ref{sec:maths:aliasing} has to be ruled out since $S$ is equal to $N$. Note that we deliver an analysis of time complexity in Section \ref{sec:maths:comparisons}.

There are various studies proposing different GCFs for use in Radio Interferometry. Ye \etal \cite{Ye2019} claims that Elizabeth Waldram was using a Gaussian, the Truncated sinc function and Gaussian times sinc function for gridding since 1961. She is possibly the first to use Convolutional Gridding, provided we do not consider Nearest Neighbour Interpolation as Convolutional Gridding (refer to Section \ref{sec:maths:nnint}). 

At the time of writing this thesis, the most commonly used function is the Prolate Spheroidal Function \cite{Stratton1935} of order one  which has a one dimension support of six. It was singled out by Schwab \cite{SchwabOct1980,Schwab1984a}, after assessing its anti-aliasing properties through a weighted ratio between the concentration of $c_{\text{co}}(\myx)$ in the region of interest versus all $c_{\text{co}}(\myx)$. The metric is a modified version of a simpler metric proposed by Brouw \cite{BROUW1975131}. Before the Prolate Spheroidal Function of order one became the main-stream choice for a GCF, the Prolate Spheroidal Wave function of order zero based on works by Slepian and Pollak \cite{Slepian1961} and Landau and Pollak \cite{Landau1961} was regarded as the best function (Schwab \cite{SchwabOct1980}). Nevertheless, Kaiser-Bessel functions (Kuo and Keiser \cite{kuo1966system}) were also used (and are still used, such as in WSClean), in view that such functions are a good approximation of the Prolate Spheroidal Wave function of order zero. Kaiser-Bessel functions are also easier to implement than the Prolate Spheroidal Wave function (Greisen \cite{Greisen1979} and Jackson \etal \cite{Jackson1991}). The Gaussian function was also previously used by Brouw \cite{brouw1971data}. 

Lately Ye \etal \cite{Ye2019}, in extending the work of Tan \cite{TanSeptemebr1986}, discovered and proposed the  least-misfit gridding functions for use as GCFs. Ye \etal \cite{Ye2019} derived these functions by minimising the difference between the DFT, and the corrected output of Convolutional Gridding, within a defined central region described by a parameter $x_0$. Ye \etal \cite{Ye2019} claims  that some of these least-misfit gridding functions suppress aliasing below arithmetic noise in the previously mentioned central region. However, we warn the reader that as per Ye \etal \cite{Ye2019}, to attain such suppression, the oversampling factor $\beta$, needs to be at least $10^6$.

\subsection{The form of the corrector}

Traditionally the corrector $h_{\text{co}}(\myx)$ is set to $1/c_{\text{co}}(\myx)$ (Greisen \cite{Greisen1979}). Such a corrector choice will amplify $o^{\text{g}}_{\text{co}}(\myx)$ to the desired output $f^{\text{g}}(\myx)$ but will inevitably amplify the undesired  $a^{\text{g}}_{\text{co}}(\myx)$ term (Jackson \etal \cite{Jackson1991}). In the inner region where $a^{\text{g}}_{\text{co}}(\myx)/o^{\text{g}}_{\text{co}}(\myx)$ approaches 0, such a setting is fine, but in the outer region with elevated values of $a^{\text{g}}_{\text{co}}(\myx)/o^{\text{g}}_{\text{co}}(\myx)$, setting  $h_{\text{co}}(\myx)<1/c_{\text{co}}(\myx)$ can be beneficial. Tan \cite{TanSeptemebr1986} and subsequently Ye \etal  \cite{Ye2019} argue that based on a least misfit calculation a better choice of corrector is:
\begin{equation}
h_{\text{co}}(\myx)=\frac{c_{\text{co}}(\myx)}{\sum\limits_{\myvec{b}\in\mathbb{Z}^2}c_{\text{co}}(\myx-\myvec{b})^2}
\end{equation}

Tan \cite{TanSeptemebr1986} and Ye \etal \cite{Ye2019} also clarify that this corrector is approximately equal to $1/c_{\text{co}}(\myx)$ in the central region. 

\subsection{GCF oversampling}
\label{sec:maths:gcfoversampling}
As previously noted, Greisen \cite{Greisen1979} states that $C_{\text{co}}(\myfreq)$ should be stored in a lookup table, as to reduce computation. This practice, which we refer to as \textit{GCF oversampling} is until today a standard rule in the implementation of Convolutional Gridding (for example CASA), since the convolution function is usually computationally intensive to calculate for each record during the gridding process (Merry \cite{Merry2016}).

The lookup table is filled with GCF values sampled at a rate higher than the sampling rate of the $uv$-grid. The ratio between the sampling rate of the GCF and that of the $uv$-grid is known as the oversampling factor $\beta \in \mathbb{N}$. The gridder will consult the lookup table to retrieve the value of $C_{\text{co}}(\myvec{\freq})$ using Nearest Neighbour Interpolation. Though it is possible to have a different oversampling factor for each dimension, we will assume that $\beta$ is constant over all dimensions, since it is generally the case in main stream imagers as there is no reason to treat separate directions differently. When  $C_{\text{co}}(\myfreq)$ is multiplicatively separable into identical functions, the lookup table can be populated with the one-dimensional version of the GCF as argued by Ye \etal \cite{Ye2019} and implemented in lwimager and CASA for Interferometric Gridding with no w-projection.

It is obvious that GCF oversampling affects the output of Convolutional Gridding and we here investigate how. Greisen \cite{Greisen1979} argues that we are changing the convolution function from $C_{\text{co}}(\myfreq)$ to $C_{\text{co}}(\myfreq,\beta)$  as follows (Muscat \cite{Muscat2014}):
\begin{equation}
C_{\text{co}}(\myfreq,\beta)=\left[C_{\text{co}}(\myfreq)\shah{\beta}(\myfreq)\right]*\rect{\beta}(\myfreq)
\end{equation}
$c_{\text{co}}(\myx,\beta)$ is given by Equation \ref{equ:maths:cco}.

\begin{equation}
c_{\text{co}}(\myx,\beta)=\left\{c_{\text{co}}(\myx)*\shah{\beta}(\myx)\right\}\sinc{\beta}(\myx)
\label{equ:maths:cco}
\end{equation}
Substituting in Equation \ref{equ:maths:wgcoconv} the output of the IFFT Step in Convolutional Gridding using GCF oversampling $w^{\text{g}}_{\text{co}}(\myx,\beta)$ is:

\begin{equation}
w^{\text{g}}_{\text{co}}(\myx,\beta)=\left[f(\myx)\left\{c_{\text{co}}(\myx)*\shah{\beta}(\myx)\right\}\sinc{\beta}(\myx)*\shah{1}\left(\myx\right)
\right]
\label{equ:maths:wgcfoversmapling}
\end{equation}
and the output of Convolution Gridding using GCF oversampling $f^{\text{g}}_{\text{co}}(\myx,\beta)$ is:
\begin{equation}
f^{\text{g}}_{\text{co}}(\myx,\beta)=\left[f(\myx)\left\{c_{\text{co}}(\myx)*\shah{\beta}(\myx)\right\}\sinc{\beta}(\myx)*\shah{1}\left(\myx\right)
\right]\shah{1}(\myx)h_{\text{co}}(\myx,\beta)
\end{equation}

In gridding implementations such as in CASA and lwimager the corrector $h_{\text{co}}(\myx,\beta)$ is set as follows when w-projection is enabled:
\begin{equation}
    h_{\text{co}}(\myx,\beta)=\frac{1}{c_{\text{co}}(\myx)\sinc{\beta}(\myx)}
\end{equation}
 The corrector is clearly not equal to the inverse of $c_{\text{co}}(\myx,\beta)$ as suggested by Greisen \cite{Greisen1979} and Jackson \etal \cite{Jackson1991} but given that the chosen GCF has good anti-aliasing properties, it should approximate it well. Obviously Tan's \cite{TanSeptemebr1986} proposition for the corrector can also be considered.

\section{Nearest Neighbour (NN) Interpolation and the modified gridding  algorithms}
\label{sec:maths:prunednninterpolation}
In this section we derive and describe in detail our proposed modified gridding algorithms, known as Hybrid Gridding and Pruned NN Interpolation. We first get to the fundamentals and analyse NN Interpolation over an oversampled grid and its relation with Convolutional Gridding using an oversampled GCF. Once all necessary analyses are made, we define in detail the modified gridding algorithms and do more analyses.  

\subsection{NN Interpolation on an oversampled \textit{uv}-grid}
\label{sec:maths:nnint}
In Radio Interferometry, Nearest Neighbour Interpolation is a precursor technique to Convolutional Gridding. Hogg \etal \cite{Hogg1969} makes a simple implementation of the technique.  Flavours of this technique in Radio Interferometry are known as Cell Averaging (Thompson and Bracewell \cite{Thompson1974}) and Cell Summing (Mathur \cite{Mathur1969}). Pan and Kak \cite{Pan1983} and others also published work on NN Interpolation. 

\begin{figure}
    \centering
\includegraphics[page=1,width=0.9\linewidth,trim=0.1cm 0.5cm 0.1cm 0.5cm,clip]{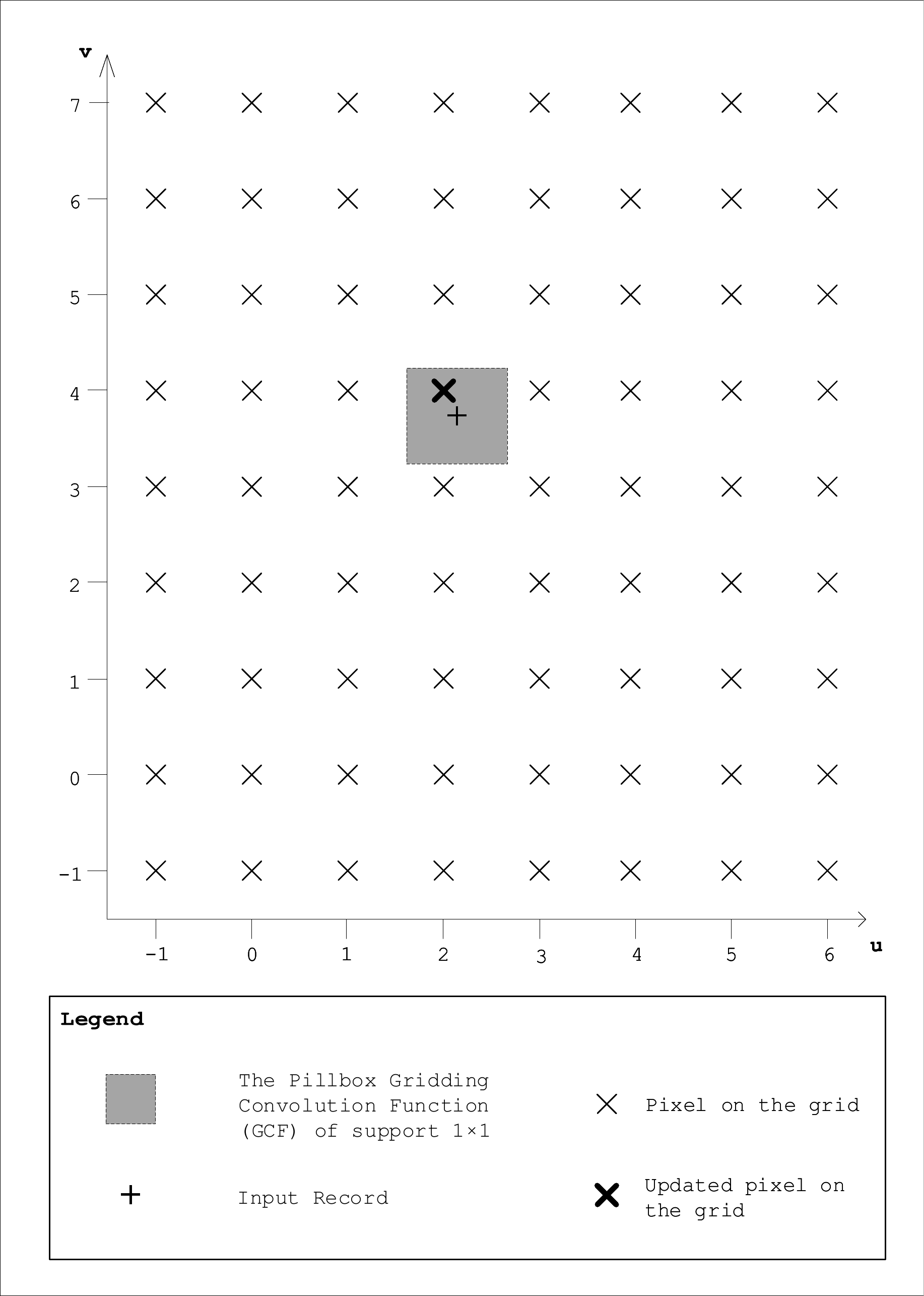}
\caption[NN Interpolation]{Diagram illustrating NN Interpolation. A record in $F(\myfreq)$  is mapped to the nearest pixel. One can model NN Interpolation as Convolutional Gridding with the GCF set to a pillbox function. Only a small region of the grid is drawn in the diagram.}
\label{fig:maths:nngridding}
\end{figure}

In NN Interpolation, each record is mapped to the nearest pixel on the $uv$-grid with no modification to the Visibility value of the record. It can be modelled as Convolutional Gridding with the GCF set to a pillbox function as illustrated in Figure \ref{fig:maths:nngridding}.

We are interested in NN Interpolation applied over a $uv$-grid oversampled by a factor of $\beta$. We refer to the oversampled $uv$-grid as the \textit{NN Grid}, whose normalised form has a sampling interval of $\beta^{-1}$ on all dimensions and is of size $N\beta \times N\beta$ pixels. 

Algorithm \ref{algo:maths:nninterpolation} describes NN Interpolation over an oversampled $uv$-grid. The Gridding Step of NN Interpolation is equivalent to that of Convolutional Gridding with a GCF equal to $C_{\text{nn}}(\myfreq,\beta)=\rect{\beta}(\myfreq)$. The reader can refer to Figure \ref{fig:maths:prunednngridding} fro an illustration of the Gridding Step. If we stick to Greisen's \cite{Greisen1979} proposal, the corrector $h_{\text{nn}}(\myx,\beta)$ should be set to:

\begin{equation}
    h_{\text{nn}}(\myx,\beta)=\frac{1}{\sinc{\beta}(\myx)}
\end{equation}

\setlength{\algomargin}{0.1em}
\begin{algorithm}
\begin{mdframed}
\setstretch{1.5}

\KwIn{$F(\myfreq)$} 
\KwPar{Oversampling factor: $\beta \in \mathbb{N}$}
\SetKwInput{KwHelp}{Helper Input}
\KwHelp{Corrector: $h_{\text{nn}}(\myx,\beta)$}
\KwOut{$f^{\text{g}}_{\text{nn}}(\myx,\beta)$} 
\SetKwInput{KwAs}{Assumptions}
\KwAs{The Bounded Records Assumption.}
\myalgoline
\SetKwBlock{step}{Gridding Step:}{}
\step{  Each record is mapped to the nearest pixel on the oversampled $uv$-grid. We model this as a convolution with  $C_{\text{nn}}(\myfreq,\beta)=\rect{\beta}(\myx)$ and simultaneous sampling using $\gshah{\beta}(\myfreq)$.  

\begin{equation}
\label{equ:maths:griddingnninterpolation}
W^{\text{g}}_{\text{nn}}(\myfreq,\beta)=\left[F(\myvec{\freq})*\rect{\beta}(\myfreq)\right]\gshah{\beta}(\myfreq)
\end{equation}

}
\SetKwBlock{step}{IFFT Step:}{}
\step{
An IFFT is performed over $W^{\text{g}}_{\text{nn}}(\myfreq,\beta)$ to get $w^{\text{g}}_{\text{nn}}(\myx,\beta)$. Note that the latter is over an $lm$-grid $\beta$ times larger than needed on each dimension, but the grid interval is still $\myvec{\Delta}\myx$.

\begin{equation}
w^{\text{g}}_{\text{nn}}(\myx,\beta)=\left[f(\myx)\sinc{\beta}(\myx)*\shah{\beta}(\myx)
\right]\gshah{\beta}(\myx)
\label{equ:nnstep2}
\end{equation}

}
\SetKwBlock{step}{Correction Step:}{}
\step{Apply the corrector $h_{\text{nn}}(\myx,\beta)$ and cut out the un-needed part of $w^{\text{g}}_{\text{nn}}(\myx,\beta)$ (represented here with a multiplication with $\gshah{1}(\myx)$).
\begin{equation}
f^{\textrm{g}}_{\text{nn}}(\myx,\beta)=w^{\text{g}}_{\text{nn}}(\myx,\beta)\gshah{1}(\myx)h_{\text{nn}}(\myx,\beta)
\label{equ:nnoutput}
\end{equation}
}
\end{mdframed}
\caption[General Nearest Neighbour Interpolation]{General Algorithm for Nearest Neighbour Interpolation.}
\label{algo:maths:nninterpolation}
\end{algorithm}

\subsection{Nearest Neighbour Interpolation and GCF oversampling}
\label{sec:maths:nnintgcfoversampling}

In Section \ref{sec:maths:gcfoversampling} we modelled Convolutional Gridding with an oversampled GCF by changing the form of the GCF. We shall now establish that Convolutional Gridding with a GCF oversampled by a factor of $\beta$ can also be modelled by splitting the Gridding Step into two. In the first step, records of $F(\myfreq)$ are NN interpolated to a $uv$-grid oversampled by a factor of $\beta$. In the second step, the resulting oversampled $uv$-grid is downsampled through a convolution with the GCF. Details are laid out in Algorithm \ref{algo:maths:equivalentalgo}.

The output of Algorithm \ref{algo:maths:equivalentalgo} is $\overline{f}^{\text{g}}_{\text{co}}(\myx,\beta)$, which we now prove to be equivalent to the output of Convolutional Gridding with a GCF oversampled by a factor of $\beta$ as modelled in Section \ref{sec:maths:gcfoversampling}. Since the Correction Step is identical in the two algorithms, it suffices to prove that $\overline{w}^{\text{g}}_{\text{co}}(\myx,\beta)=w^{\text{g}}_{\text{co}}(\myx,\beta)$.

\setlength{\algomargin}{0.1em}
\begin{algorithm}
\begin{mdframed}
\setstretch{1.5}
\KwIn{$F(\myfreq)$} 
\SetKwInput{KwHelp}{Helper Input}
\KwHelp{GCF: $C_{\text{co}}(\myfreq,\beta)$, Corrector: $h_{\text{co}}(\myx,\beta)$}
\KwOut{$\overline{f}^{\text{g}}_{\text{co}}(\myx,\beta)$} 
\SetKwInput{KwAs}{Assumptions}
\KwAs{The Bounded Records Assumption}
\myalgoline
\SetKwBlock{step}{NN Interpolation Step:}{}
\step{ Records are Interpolated over the Oversampled Grid. This step is identical to the Gridding Step of Algorithm \ref{algo:maths:nninterpolation}. 

\begin{equation}
\label{equ:maths:nninterpolationconvgridding}
W^{\text{g}}_{\text{nn}}(\myfreq,\beta)=\left[F(\myvec{\freq})*\rect{\beta}(\myfreq)\right]\gshah{\beta}(\myfreq)
\end{equation}
}
\SetKwBlock{step}{Convolution Step:}{}
\step{ $W^{\text{g}}_{\text{nn}}(\myfreq,\beta)$ is convolved with $C_{\text{co}}(\myfreq)$ and the output is re-sampled. The output of this step is a function overlaid on a grid with no oversampling.

\begin{equation}
\label{equ:maths:interpolationstepout}
    \overline{W}^{\text{g}}_{\text{co}}(\myfreq,\beta)=\left[W^{\text{g}}_{\text{nn}}(\myfreq,\beta)*C_{\text{co}}(\myfreq)\right]\gshah{1}(\myfreq)
\end{equation}
}

\SetKwBlock{step}{IFFT Step:}{}
\step{
Apply an IFFT to obtain $\overline{w}^{\text{g}}_{\text{co}}(\myx,\beta)$
\begin{equation}
\label{equ:maths:ifftstepout}
    \overline{w}^{\text{g}}_{\text{co}}(\myx,\beta)=\left[\left\{\mathcal{F}^{-1}W^{\text{g}}_{\text{nn}}(\myfreq,\beta)\right\}c_{\text{co}}(\myx)*\shah{1}(\myx)\right]\gshah{1}(\myx)
\end{equation}
\begin{equation}
\label{equ:maths:overlinew}
    \overline{w}^{\text{g}}_{\text{co}}(\myx,\beta)=\left[\left\{f(\myx)\sinc{\beta}(\myx)*\shah{\beta}(\myx)\right\}c_{\text{co}}(\myx)*\shah{1}(\myx)\right]\gshah{1}(\myx)
\end{equation}

}

\SetKwBlock{step}{Correction Step:}{}
\step{Apply correction $h_{\text{co}}(\myx,\beta)$. Output is $\overline{f}^{\text{g}}_{\text{co}}(\myx,\beta)$
\begin{equation}
    \overline{f}^{\text{g}}_{\text{co}}(\myx,\beta)=\overline{w}^{\text{g}}_{\text{co}}(\myx,\beta)h_{\text{co}}(\myx,\beta)    
\end{equation}
}
\end{mdframed}
\caption[Equivalent algorithm for Convolutional Gridding using an oversampled GCF]{Equivalent Algorithm for Convolutional Gridding with GCF Oversampling. The output of this algorithm is set to $\overline{f}^{\text{g}}_{\text{co}}(\myx,\beta)$, which we prove in Section \ref{sec:maths:nnintgcfoversampling} that it is equal to $f^{\text{g}}_{\text{co}}(\myx,\beta)$.}
\label{algo:maths:equivalentalgo} 
\end{algorithm}

$w^{\text{g}}_{\text{co}}(\myx,\beta)$ is derived in Equation \ref{equ:maths:wgcfoversmapling} and reproduced here-under:
\begin{equation}
w^{\text{g}}_{\text{co}}(\myx,\beta)=\left[f(\myx)\left\{c_{\text{co}}(\myx)*\shah{\beta}(\myx)\right\}\sinc{\beta}(\myx)*\shah{1}(\myx)\right]\gshah{1}(\myx)
\end{equation}

Expressing the two convolutions with the Shah function as summations we get:
\begin{equation}
w^{\text{g}}_{\text{co}}(\myx,\beta)=\left[\,{\sum\limits_{\myvec{p_{1}} \in \myvec{\mathbb{Z}^2}}} f(\myx-\myvec{p_{1}})\left\{\sum\limits_{\myvec{p_{2}}\in\myvec{\mathbb{Z}^2}} c(\myx+\myvec{p_{2}}\beta-\myvec{p_{1}})\right\}\sinc{\beta}\left(\myx-\myvec{p_{1}}\right)\right]\,\gshah{1}(\myx)
\label{equ:maths:equivalencestep1}
\end{equation}

Let $\myvec{p_{3}}=\myvec{p_{1}}-\myvec{p_{2}}\beta$

Implying $\myvec{p_{1}}=\myvec{p_{3}}+\myvec{p_{2}}\beta$

Substituting in Equation \ref{equ:maths:equivalencestep1} we get
\begin{multline}
w^{\text{g}}_{\text{co}}(\myx,\beta)=\\
\left[\sum\limits_{\myvec{p_{3}} \in \myvec{\mathbb{Z}^2}} \sum\limits_{\myvec{p_{2}}\in\myvec{\mathbb{Z}^2}}f(\myx-\myvec{p_{3}}-\myvec{p_{2}}\beta) c(\myx+\cancel{\myvec{p_{2}}\beta}-\cancel{\myvec{p_{2}}\beta}-\myvec{p_{3}})\sinc{\beta}\left(\myx-\myvec{p_{3}}-\myvec{p_{2}\beta}\right)\right]\gshah{1}(\myx)
\end{multline}
\begin{equation}
w^{\text{g}}_{\text{co}}(\myx,\beta)=\left[\sum\limits_{\myvec{p_{3}} \in \myvec{\mathbb{Z}^2}} \sum\limits_{\myvec{p_{2}}\in\myvec{\mathbb{Z}^2}}f(\myx-\myvec{p_{3}}-\myvec{p_{2}}\beta) c(\myx-\myvec{p_{3}})\sinc{\beta}\left(\myx-\myvec{p_{3}}-\myvec{p_{2}\beta}\right)\right]\gshah{1}(\myx)
\end{equation}

Re-grouping

\begin{equation}
w^{\text{g}}_{co,\beta}=\left[\sum\limits_{\myvec{p_{3}} \in \myvec{\mathbb{Z}^2}} \left\{\sum\limits_{\myvec{p_{2}}\in\myvec{\mathbb{Z}^2}}f(\myx-\myvec{p_{3}}-\myvec{p_{2}}\beta) \sinc{\beta}\left(\myx-\myvec{p_{3}}-\myvec{p_{2}}\right)\right\}c(\myx-\myvec{p_{3}})\right]\gshah{1}(\myx)
\end{equation}

\begin{equation}
w^{\text{g}}_{co,\beta}=\left[\sum\limits_{\myvec{p_{3}} \in \myvec{\mathbb{Z}^2}} \left\{f(\myx-\myvec{p_{3}}) \sinc{\beta}\left(\myx-\myvec{p_{3}}\right)*\shah{\beta}(\myx)\right\}c(\myx-\myvec{p_{3}})\right]\gshah{1}(\myx)
\end{equation}

\begin{equation}
w^{\text{g}}_{co,\beta}=\left[\left\{f(\myx) \sinc{\beta}(\myx)*\shah{\beta}(\myx)\right\}c_{\text{co}}(\myx)*\shah{1}(\myx)\right]\gshah{1}(\myx)
\label{equ:maths:derivedwg}
\end{equation}

Comparing Equation \ref{equ:maths:overlinew} with Equation \ref{equ:maths:derivedwg}, we find they are equivalent and therefore $w^{\text{g}}_{\text{co}}(\myx,\beta)=\overline{w}^{\text{g}}_{\text{co}}(\myx,\beta)$, proving our claim.

\subsection{Contributions of the convolution step}
\label{sec:maths:convstepcont}
We now review the contributions of the Convolution Step to Algorithm \ref{algo:maths:equivalentalgo}. We note that without the Convolution Step, Algorithm \ref{algo:maths:equivalentalgo} is identical to Algorithm \ref{algo:maths:nninterpolation}. Until otherwise stated, we will refer to Algorithm \ref{algo:maths:equivalentalgo} as Convolutional Gridding and refer to Algorithm \ref{algo:maths:nninterpolation} as NN Interpolation. The reader must always assume that the $uv$-grid output of the NN Interpolation Step in Convolutional Gridding (Algorithm \ref{algo:maths:equivalentalgo}) and the $uv$-grid in NN Interpolation (Algorithm \ref{algo:maths:nninterpolation}) are both oversampled with a factor of $\beta$. It is also important for the reader to realise that the Gridding Step in NN Interpolation is equivalent to the NN Interpolation Step of Convolutional Gridding since Equations \ref{equ:maths:griddingnninterpolation} and \ref{equ:maths:nninterpolationconvgridding} are equivalent.

\subsubsection{Aliasing}

Let us analyse Equations \ref{equ:maths:interpolationstepout} and \ref{equ:maths:ifftstepout}, reproduced here:

\begin{equation}
\label{equ:maths:repinterpolationstepout}
    \overline{W}^{\text{g}}_{\text{co}}(\myfreq,\beta)=\left[W^{\text{g}}_{\text{nn}}(\myfreq,\beta)*C_{\text{co}}(\myfreq)\right]\gshah{1}(\myfreq)
\end{equation}

\begin{equation}
\label{equ:maths:repifftstepout}
    \overline{w}^{\text{g}}_{\text{co}}(\myx,\beta)=\left[\left\{\mathcal{F}^{-1}W^{\text{g}}_{\text{nn}}(\myfreq,\beta)\right\}c_{\text{co}}(\myx)*\shah{1}(\myx)\right]\gshah{1}(\myx)
\end{equation}

These two equations are Fourier equivalent to each other and indicate that in Convolutional Gridding, the Convolution Step generates aliasing over and above that of the NN Interpolation Step. The main indication comes from the convolution in Equation \ref{equ:maths:repifftstepout}, whereby we point out that $\mathcal{F}^{-1}W^{\text{g}}_{\text{nn}}(\myfreq,\beta)$ is the inverse Fourier equivalent of the NN Interpolation Step.

Therefore, we conclude that NN Interpolation suffers less from aliasing than Convolutional Gridding. The output of NN Interpolation is expected to be a better approximation to $f^{\text{g}}(\myx)$ than the output of Convolutional Gridding. 

\subsubsection{Memory Footprint}
The Convolution Step is reducing the memory footprint. The output of NN Interpolation is a grid $\beta^2$ times larger than that of the Convolution Step. Therefore, the IFFT Step of NN Interpolation needs to handle a larger grid than the same step in Convolutional Gridding,  directly implying more memory use. 

\subsubsection{Computation Footprint}
The Convolution Step is increasing the computation footprint of the Gridding Step. As a matter of fact, the NN Interpolation Step requires only one complex addition to a complex-valued pixel while Convolution Gridding requires $S^2$ multiplications of real GCF values with a complex-valued record and a further $S^2$ complex additions to lay the convolution on the $uv$-grid. On the other hand, the Convolution Step is reducing the IFFT Step computation footprint since IFFT needs to handle smaller grids.

\subsubsection{Conclusion}
It is clear that  Convolution Gridding, when compared to NN Interpolation, is trading for less intensive IFFTs and memory footprint at the expense of a higher level of aliasing and a more computationally taxing gridding. We extend the statement by stating that the Convolution Step is pruning the IFFT, but carefully clarify that \textit{a prima facie} while pruning, the Convolution Step is injecting aliasing.

\subsection{The modified gridding algorithms}
\label{sec:maths:newgriddingalgorithms}
Based on the above deliberation, we define the new proposed modified gridding algorithms named \textit{Hybrid Gridding} and \textit{Pruned NN Interpolation}. These modified gridding algorithms are based on what we proved about Convolutional Gridding, that is, the Gridding Step is equivalent to NN Interpolation with simultaneous downsampling of the $uv$-grid using a convolution. In the modified gridding algorithms, we modify Convolutional Gridding by moving downsampling of the $uv$-grid from the Gridding Step to what we call the Pruning Step. The Pruning Step is executed after all records are gridded and therefore is part of finalisation. In Hybrid Gridding, downsampling by convolution is moved to the Pruning Step in one dimension. In contrast, in Pruned NN Interpolation, downsampling by convolution is moved to the Pruning Step in all dimensions. 

We  give a bit of spice to the Pruning Step, whereby we propose the use of the least-misfit gridding functions in a bid to have the Pruning Step virtually alias-free by having distortions caused by aliasing below arithmetic noise. In order to reach the desired effect, the $uv$-grid is downsampled by an integer factor $\alpha$ divisor of $\beta$.  

We are very cautious in the use of the term \textit{pruning} in our algorithms. The definition given by Markel \cite{Markel1971a} on FFT Pruning is about eliminating operations where zeros are involved or eliminating other operations that calculate output pixels that are not required. Using a convolution to downsample changes operations rather than eliminating them and might not qualify as pruning, especially if one considers the fact that downsampling using convolution injects aliasing. However, since our proposal is intended to be virtually alias-free, we feel that we can categorise our downsampling idea as \textit{Convolution-Based FFT Pruning}, which we discuss in detail in Section \ref{sec:maths:fftpruning}.

\subsection{Details on the studied gridding algorithms}
\label{sec:maths:studiedalgorithms}

Let us now get to the details on how we implemented Convolutional Gridding, Hybrid Gridding and Pruned NN Interpolation, such that we can experimentally study their Performance and aliasing behaviour. 

We shall collectively refer to Convolutional Gridding, Hybrid Gridding and Pruned NN Interpolation in the form studied in this thesis as the \textit{studied algorithms}. Furthermore, we will refer to their implementation over the P100 as the \textit{studied implementations}. We also define the terms \textit{Gridder} and \textit{Pruner} to mean an implementation of the Gridding Step and Pruning Step, respectively.   

Algorithms \ref{algo:maths:convgriddingoversampling}, \ref{algo:maths:hybrid} and \ref{algo:maths:purenninterpolation} define the studied algorithms as implemented, and experimented upon in this thesis. Further details follow in the next paragraphs.

In all the studied algorithms, the Bounded Record Assumption still applies to the extent that a linear or a circular convolution in the Pruning Step will output the same result. We assume $\beta$ to be an even positive number greater than 2, such that the Pruning Step can downsample the grid to an oversampling factor of two, which implies a  downsampling factor of $\alpha=\beta/2$. Note that it is still acceptable to use $\beta=1$ or $\beta=2$ for the studied algorithms, but the  Pruning Step becomes useless and should be removed. For $\beta=1$, there is no sense in using any of the algorithms, and NN Interpolation with no grid oversampling is enough.   

We now clarify about the form of the GCF used in the Gridding Step of Hybrid Gridding (Algorithm \ref{algo:maths:hybrid}) denoted by  $C_{\text{hy}}(\myfreq,\beta)$, and set to $C_{\text{co}}(u,\beta)\rect{\beta}(v)$. In all the studied algorithms, the Gridding Step is always modelled as a two-dimensional convolution. Therefore the form of $C_{\text{hy}}(\myfreq,\beta)$ is set in such a way that it  causes a convolution with $C_{\text{co}}(u,\beta)$ in the first dimension and a Nearest Neighbour Interpolation in the second dimension modelled as a convolution with $\rect{\beta}(v)$. We use the GCF of Convolutional Gridding as to make it clear that we will use the same one-dimensional form of the GCF, in the Gridding Step of the Hybrid and Convolutional Gridding. Finally, we state that the support of $C_{\text{hy}}(\myfreq,\beta)$ is equal to  $S \times 1$, since the function is in two dimensions.

Some clarification on the Pruning Step and the GCF used are now in order. We denote the GCF used in pruning with $Z(\myfreq)$  or $Z(v)$. One dimensional support is denoted with $S_z$. There is no need to consider the fact that the GCF will be sampled and put in a lookup-table since the input to the Pruning Step is a regular grid. We note that $Z(\myfreq)$ and $Z(u)$ have the form used if we were going to downsample the grid to an oversampling factor of 1. We want to downsample the grid to an oversampling factor of 2, and therefore we need to convolve with a shortened convolution function $Z(2\myfreq)$ or $Z(2v)$. 

Finally, Equations \ref{equ:maths:hco}, \ref{equ:maths:hpr} and \ref{equ:maths:hhy} give our choice for the algorithms' corrector which are according to the suggestion given by Greisen \cite{Greisen1979}  and Jackson \etal \cite{Jackson1991}.
\begin{equation}
    h_{\text{co}}(\myx,\beta)=\frac{1}{\left[c_{\text{co}}(\myx)*\shah{\beta}(\myx)\right]\sinc{\beta}(\myx)}
\label{equ:maths:hco}
\end{equation}
\begin{equation}
    h_{\text{pn}}(\myx,\beta)=\frac{1}{\left[z(0.5\myx)*\shah{\beta/2}(\myx)\right]\sinc{\beta}(\myx)}
\label{equ:maths:hpr}    
\end{equation}

\begin{equation}
    h_{\text{hy}}(\myx,\beta)=h_{\text{co}}(l,\beta)h_{\text{pr}}(m,\beta)
\label{equ:maths:hhy}
\end{equation}

\setlength{\algomargin}{0.1em}
\begin{algorithm}
\begin{mdframed}
\setstretch{1.5}
\KwIn{Records in $F(\myfreq)$} 
\SetKwInput{KwPar}{Parameters}
\KwPar{Oversampling factor: $\beta=2(n+1)$ $n \in \mathbb{N}$}
\SetKwInput{KwHelp}{Helper Input}
\KwHelp{Oversampled GCF: $C_{\text{co}}(\myfreq,\beta)$, Corrector: $h_{\text{co}}(\myx,\beta)$}
\KwOut{$f^{\text{g}}_{\text{co}}(\myx,\beta)$} 
\SetKwInput{KwAs}{Assumptions}
\KwAs{As per Section \ref{sec:maths:studiedalgorithms}}
\myalgoline
\SetKwBlock{step}{Gridding Step:}{}
\step{ Each non-zero sample of $F(\myfreq)$ is convolved with the oversampled  $C_{\text{co}}(\myfreq,\beta)$ and simultaneously sampled on the $uv$-grid:
\begin{equation}
W^{\text{g}}_{\text{co}}(\myfreq,\beta)=\left[F(\myfreq)*C_{\text{co}}(\myfreq,\beta)\right]\shah{1}\left(\myfreq\right)
\end{equation}
}
\SetKwBlock{step}{IFFT Step:}{}
\step{
An IFFT is performed over $W^{\text{g}}_{\text{co}}(\myfreq,\beta)$ to get $w^{\text{g}}_{\text{co}}(\myx,\beta)$:
\begin{equation}
w^{\text{g}}_{\text{co}}(\myx,\beta)=\left[f(\myx)\left\{c_{\text{co}}(\myx)*\shah{\beta}(\myx)\right\}\sinc{\beta}(\myx)*\shah{1}\left(\myx\right)
\right]\gshah{1}(\myx)
\end{equation}
}
\SetKwBlock{step}{Correction Step:}{}
\step{The corrector $h_{\text{co}}(\myx,\beta)$ is applied to obtain the output $f^{\text{g}}_{\text{co}}(\myx,\beta)$
\begin{equation}
f^{\text{g}}_{\text{co}}(\myx,\beta)=\left[f(\myx)\left\{c_{\text{co}}(\myx)*\shah{\beta}(\myx)\right\}\sinc{\beta}(\myx)*\shah{1}\left(\myx\right)
\right]\gshah{1}(\myx)h_{\text{co}}(\myx,\beta)
\end{equation}
}
\end{mdframed}
\caption[The studied Convolutional Gridding Algorithm.]{The studied Convolutional Gridding Algorithm using an oversampled GCF.}

\label{algo:maths:convgriddingoversampling} 
\end{algorithm}

\setlength{\algomargin}{0.1em}
\begin{algorithm}
\begin{mdframed}
\setstretch{1.2}
\KwIn{Records in $F(\myfreq)$} 
\KwPar{Oversampling Factor: $\beta=2(n+1)$ $n \in \mathbb{N}$}
\SetKwInput{KwHelp}{Helper Input}
\KwHelp{Oversampled GCF: $C_{\text{hy}}(\myfreq,\beta)=C_{\text{co}}(u,\beta)\rect{\beta}(v)$, Pruning GCF: $\delta(u)Z(v)$,  Corrector: $h_{\text{hy}}(\myx,\beta)$}
\KwOut{$f^{\text{g}}_{\text{hy}}(\myx,\beta)$} 
\SetKwInput{KwAs}{Assumptions}
\KwAs{As per Section \ref{sec:maths:studiedalgorithms}}
\myalgoline
\SetKwBlock{step}{Gridding Step:}{}
\step{Each record is mapped to the nearest pixel on the oversampled $uv$-grid with oversampling factor $\beta$. 

\begin{equation}
\label{equ:maths:hybridgriddingstep}
W^{\text{g}}_{\text{hy}}(\myfreq,\beta)=\left[F(\myvec{\freq})*C_{\text{hy}}(\myfreq,\beta)\right]\gshah{1}(u)\gshah{\beta}(v)
\end{equation}

}
\SetKwBlock{step}{Pruning Step:}{}
\step{$W^{\text{g}}_{\text{hy}}(\myfreq,\beta)$ is sampled down on the second dimension by a factor of $\beta/2$ through a convolution.

\begin{equation}
\label{equ:maths:hybridpruner}
V^{\text{g}}_{\text{hy}}(\myfreq,\beta)=\left[W^{\text{g}}_{\text{hy}}(\myfreq,\beta)*\left\{\delta(u)Z(2v)\right\}\right]\gshah{1}(u)\gshah{2}(v)
\end{equation}

}
\SetKwBlock{step}{IFFT Step:}{}
\step{
An IFFT is performed over $V^{\text{g}}_{\text{hy}}(\myfreq,\beta)$ to get $v^{\text{g}}_{\text{hy}}(\myx,\beta)$. 

\begin{subequations}
\begin{equation}
v^{\text{g}}_{\text{hy}}(\myx,\beta)=\left[\left\{w^{\text{g}}_{\text{hy}}(\myx,\mybeta)z(0.5 l)\right\}*\left\{\shah{1}(l)\shah{2}(m)\right\}\right]\gshah{1}(l)\gshah{2}(m)
\end{equation}
\begin{equation}
    w^{\text{g}}_{\text{hy}}(\myx,\beta)=\left[f(\myx)c_{\text{hy}}(\myx,\beta)\right]*\left\{\shah{1}(l)\shah{\beta}(m)\right\}\gshah{1}(l)\gshah{\beta}(m)
\end{equation}
\begin{equation}
    c_{\text{hy}}(\myx,\beta)=\left[\left\{c_{\text{co}}(l)\delta(m)\right\}*\left\{\shah{\beta}(l)\delta(m)\right\}\right] \sinc{\beta}(\myx)
\end{equation}
\end{subequations}
}
\SetKwBlock{step}{Correction Step:}{}
\step{Apply the corrector $h_{\text{hy}}(\myx,\beta)$ and cut out the un-needed part of $v^{\text{g}}_{\text{hy}}(\myx,\beta)$ (represented here with a multiplication with $\gshah{1}(\myx)$).
\begin{equation}
f^{\text{g}}_{\text{hy}}(\myx,\beta)=v^{\text{g}}_{\text{hy}}(\myx,\beta)h_{\text{hy}}(\myx,\beta)\gshah{1}(\myx)
\end{equation}
}
\end{mdframed}
\caption[The studied Hybrid Gridding Algorithm]{The studied Hybrid Gridding Algorithm.}
\label{algo:maths:hybrid}
\end{algorithm}

\setlength{\algomargin}{0.1em}
\begin{algorithm}
\begin{mdframed}
\setstretch{1.5}
\KwIn{Records in $F(\myfreq)$} 
\KwPar{Oversampling Factor: $\beta=2(n+1)$ $n \in \mathbb{N}$}
\SetKwInput{KwHelp}{Helper Input}
\KwHelp{Pruning GCF: $Z(\myfreq)$, Corrector: $h_{\text{nn}}(\myx,\beta)$}
\KwOut{$f^{\text{g}}_{\text{pn}}(\myx,\beta)$} 
\SetKwInput{KwAs}{Assumptions}
\KwAs{As per Section \ref{sec:maths:studiedalgorithms}}
\myalgoline
\SetKwBlock{step}{Gridding Step:}{}
\step{Each record is mapped to the nearest pixel on the oversampled $uv$-grid with oversampling factor $\beta$. 

\begin{equation}
\label{equ:maths:prunednninterpolationgriddingstep}
W^{\text{g}}_{\text{pn}}(\myfreq,\beta)=\left[F(\myvec{\freq})*\rect{\beta}(\myfreq)\right]\gshah{\beta}(\myfreq)
\end{equation}

}
\SetKwBlock{step}{Pruning Step:}{}
\step{$W^{\text{g}}_{\text{pn}}(\myfreq,\beta)$ is sampled down by a factor of $\beta/2$ through a convolution. 
\begin{equation}
\label{equ:maths:nnpruner}
V^{\text{g}}_{\text{pn}}(\myfreq,\beta)=\left[W^{\text{g}}_{\text{pn}}(\myfreq,\beta)*Z(2\myfreq)\right]\gshah{2}(\myfreq)
\end{equation}

}
\SetKwBlock{step}{IFFT Step:}{}
\step{
An IFFT is performed over $V^{\text{g}}_{\text{pn}}(\myfreq,\beta)$ to get $v^{\text{g}}_{\text{pn}}(\myx,\beta)$. 

\begin{subequations}
\begin{equation}
    v^{\text{g}}_{\text{pn}}(\myx,\beta)=\left[\left\{w^{\text{g}}_{\text{pn}}(\myx,\beta)z(0.5\myx)\right\}*\shah{2}(\myx)\right]\gshah{2}(\myx)
\end{equation}
\begin{equation}
    w^{\text{g}}_{\text{pn}}(\myfreq,\beta)=\left\{\left[f(\myx)\sinc{\beta}(\myx)\right]*\shah{\beta}(\myx)\right\}\gshah{\beta}(\myx)
\end{equation}
\end{subequations}
}
\SetKwBlock{step}{Correction Step:}{}
\step{Apply the corrector $h_{\text{pn}}(\myx,\beta)$  and cut out the un-needed part of $v^{\text{g}}_{\text{pn}}(\myx,\beta)$ (represented here with a multiplication with $\gshah{1}(\myx)$).
\begin{equation}
f^{\text{g}}_{\text{pn}}(\myx,\beta)=\left\{\left[w^{\text{g}}_{\text{pn}}(\myfreq,\beta)z(0.5\myx)\right]*\shah{2}(\myx)\right\}h_{\text{pn}}(\myx,\beta)\gshah{1}(\myx)
\end{equation}
}
\end{mdframed}
\caption[The studied Pruned NN Interpolation Algorithm]{The studied Pruned NN Interpolation Algorithm.}
\label{algo:maths:purenninterpolation}
\end{algorithm}
\begin{figure}
\centering
\includegraphics[page=1,width=0.9\linewidth,trim=0.1cm 0.5cm 0.1cm 0.5cm,clip]{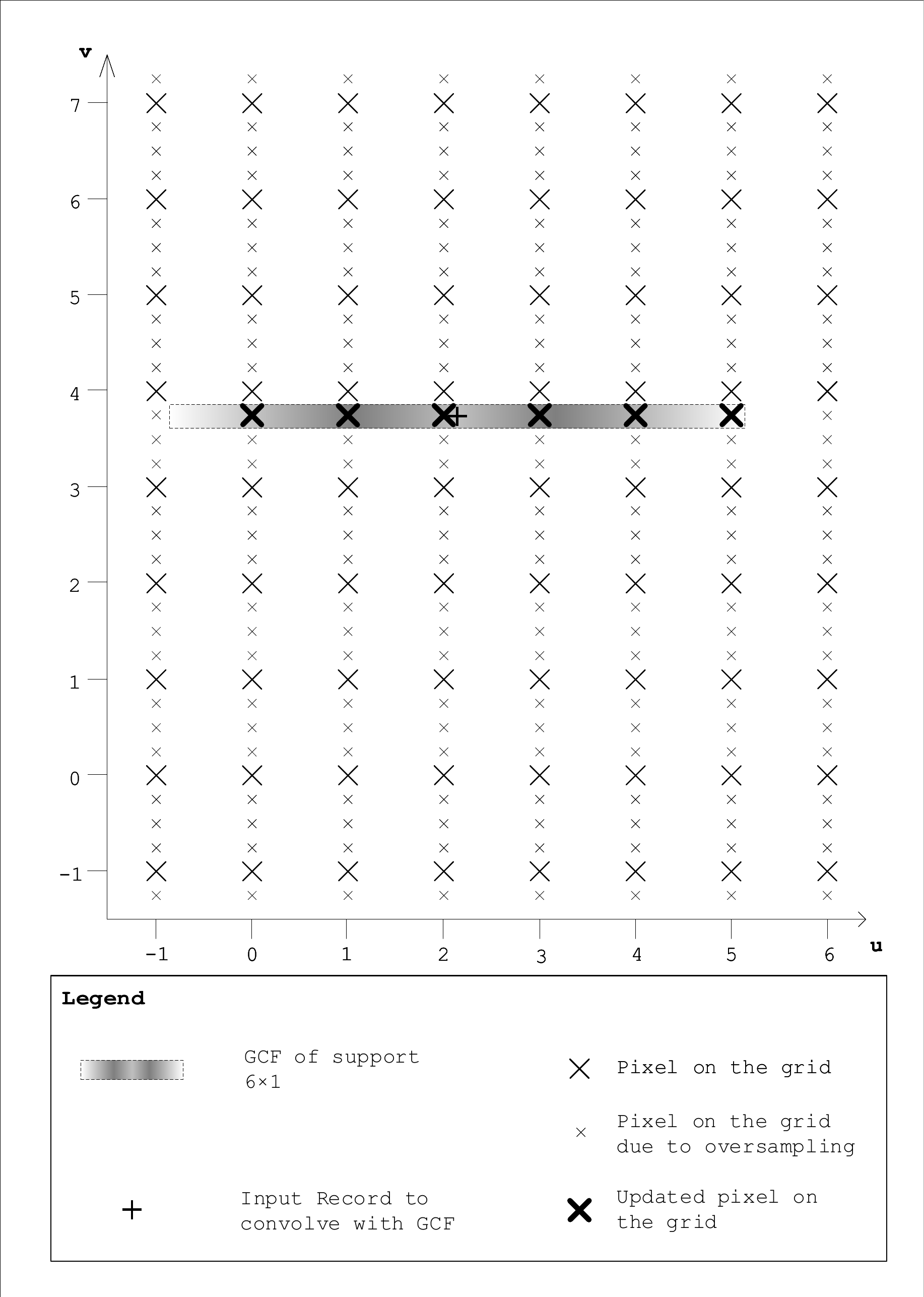}
\caption[Hybrid Gridding]{Diagram illustrating the Gridding Step (Equation \ref{equ:maths:hybridgriddingstep}) of Hybrid Gridding. A record in $F(\myfreq)$  is being convolved with $C_{\text{hy}}(\myfreq)$ of support $6 \times 1$ and sampled over the grid. The grid is oversampled on one of its dimensions. Only a small region of the grid is drawn in the diagram.}
\label{fig:maths:hybridgridding}
\end{figure}
\begin{figure}
\centering
\includegraphics[page=1,width=0.9\linewidth,trim=0.1cm 0.5cm 0.1cm 0.5cm,clip]{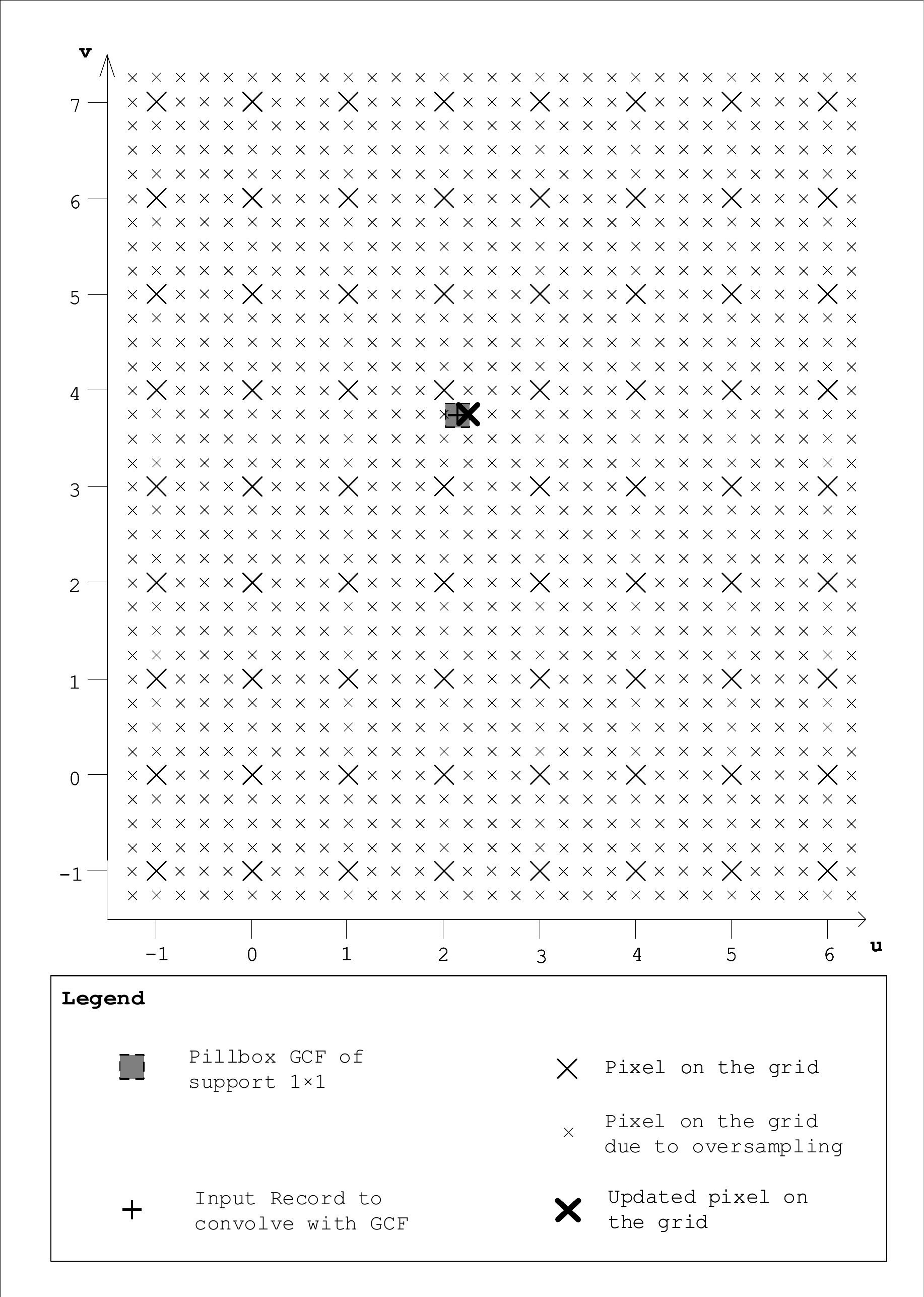}
\caption[Pruned NN Interpolation]{Diagram illustrating the Gridding Step (Equations \ref{equ:maths:prunednninterpolationgriddingstep} or \ref{equ:maths:griddingnninterpolation}) of Pruned NN Interpolation. A record in $F(\myfreq)$  is moved to the nearest pixel on the oversampled grid, modelled as a convolution with $\text{rect}_\beta(\myfreq)$ and a subsequent sampling to the grid.  Only a small region of the grid is drawn in the diagram.}
\label{fig:maths:prunednngridding}
\end{figure}

\subsection{Comparison of aliasing, and computational complexity}
\label{sec:maths:comparisons}

We finalise our discussion and analyses on the studied algorithms by theoretically comparing their level of aliasing, time complexity and memory consumption. 

\subsubsection{Aliasing}

Referring to what we already discussed in Section \ref{sec:maths:convstepcont}, we expect Pruned NN Interpolation to be the least to suffer from aliasing effects. Hybrid Gridding should suffer from a higher level of aliasing while Convolutional Gridding should suffer the most.

\subsubsection{Time Complexity}
\label{sec:maths:timecomplexitygrid}
\begin{table}[]
\centering
\begin{tabular}{@{}llll@{}}
\toprule
\begin{minipage}[]{4cm}Step\end{minipage}    & \begin{minipage}[]{3.2cm}Convolutional Gridding\end{minipage}       & \begin{minipage}[]{3.2cm}Hybrid \\Gridding\end{minipage}                              & \begin{minipage}[]{3.2cm} \begin{flushleft}Pruned NN\\Interpolation\end{flushleft}\end{minipage}                                               \\ \midrule
Gridding           & $\mathcal{O}(PS^2)$ & $\mathcal{O}(PS)$                         & $\mathcal{O}(P)$                                               \\
Pruning                   &                      & $\mathcal{O}(N^2\beta S_z)$             & $\mathcal{O}(N^2\beta^2{S_z}^2)$            \\
IFFT                       & $\mathcal{O}(N^2\text{log}N^2)$    & $\mathcal{O}(2N^2\text{log}[2N^2])$                   & $\mathcal{O}(4N^2\text{log}[4N^2])$                                 \\
Correction                & $\mathcal{O}(N^2)$             & $\mathcal{O}(N^2)$                            & $\mathcal{O}(N^2)$                                          \\ \bottomrule
\end{tabular}
\caption[Time complexity of the studied algorithms]{Time complexity of  the studied algorithms' steps using the Big-$\mathcal{O}$ notation.}
\label{tab:maths:computation}
\end{table}

Table \ref{tab:maths:computation} gives the time complexity in Big-$\mathcal{O}$ notation for each step in the studied algorithms. We already implied in Section \ref{sec:maths:convstepcont} that the removal of convolution in the Gridding Step reduces the computation footprint of the said step. Such a statement is well reflected with the time complexities given for the Gridding Steps of the studied algorithms.

Hybrid Gridding and Pruned NN  Interpolation feature a Pruning Step which contributes to more computation for the stated algorithms, but at the same time reduces the time complexity of the IFFT Step. If the Pruning Step is not applied in the said algorithms, then the IFFT  Step time complexity of Hybrid Gridding would increase to $\mathcal{O}(N^2\beta\text{log}[N^2\beta])$ and of Pruned NN Interpolation would increase to $\mathcal{O}(N^2\beta^2\text{log}[N^2\beta^2])$. We note that even with the Pruning Step enabled time complexities of the IFFT Step for Hybrid Gridding and Pruned NN Interpolation are still higher than that of Convolutional Gridding, because the Pruning Step downsamples to an oversampling factor of two instead of one. 

We note that the time complexity of the Gridding Steps of all algorithms is dependent on $P$ (the number of input records) and independent from $\myN$ and $\beta$. In contrast, the Pruning Step and IFFT step are dependent on $\myN$ and $\beta$ but independent of $P$, which in conjunction with our other observations  leads us to make the following claims which we try to support via experimentation in Section \ref{sec:comparative:performancestudied}:   
\begin{enumerate}
    \item Convolutional Gridding is favourable for a small set of records (low value of $P$).
    \item Hybrid Gridding is favourable for a larger set of records (increased value of $P$) that can overcome the inclusion of the Pruning Step and the extra time taken by the IFFT step to transform a grid double in size to that of Convolutional Gridding.
    \item Pruned NN Interpolation should only be favourable for a much larger set of records.
 \end{enumerate}

\subsubsection{Memory Consumption}

It is easy to verify that the output $uv$-grid of the studied algorithms' Gridding Step increases by a factor of $\beta$ from one algorithm to another in the order of Convolutional Gridding, Hybrid Gridding and finally Pruned NN Interpolation. The Pruning Step will then reduce such grids such that the IFFT is presented with grids that only increase for every algorithm by a factor of two in the same order previously given.

\section{Convolution based FFT Pruning}
\label{sec:maths:fftpruning}

In Section \ref{sec:maths:newgriddingalgorithms}, we discussed the possibility of downsampling an oversampled grid using convolution as a means to prune an IFFT of the said grid. Downsampling a grid using convolution is a well-known technique, and one can view this idea as a re-application of Convolutional Gridding using as input the regularly oversampled grid. Such a technique is inherently susceptible to aliasing. However, with the use of the least-misfit gridding functions  (Ye \etal  \cite{Ye2019}), we hope that aliasing gets suppressed below arithmetic noise in order to re-propose downsampling using convolution as Convolution-Based FFT Pruning.

In this section, we discuss such Convolution-based FFT Pruning, whereby in the first sub-section, we discuss FFT Pruning and give a review of published work on FFT Pruning. In the subsequent sub-section, we lay out a general algorithm for Convolution-Based Pruning in one-dimension, which in Chapter \ref{chap:pruning} we adapt, implement and test for the use in Hybrid Gridding and Pruned NN Interpolation. We then finalise our discussion in the last two sub-sections, by describing aliasing and time complexity of the algorithm.

We clarify that all the work we are proposing applies for FFT and IFFT pruning. Convolutional Gridding can also be applied inversely. We tend to speak about pruning an IFFT because it is of primary interest in this thesis, but we call it FFT Pruning as it is the general way such algorithms are called irrespective if they are applied for FFT or IFFT.

\subsection{Definition and review of FFT Pruning}
\label{sec:maths:pruningreview}
An inherent property of FFT algorithms such as the original radix-2 (Cooley and Tukey \cite{Cooley1965}), higher radix (Bergland \cite{Bergland1969}), mixed-radix (Singleton \cite{Singleton1969}), prime factor (Kolba and Parks \cite{Kolba1977}), Winograd \cite{Winograd1978}, split-radix (Sorensen \etal \cite{Sorensen1986}) and others is that the output has the same number of pixels as the input. While this is a reasonable assumption for most applications using FFTs,  some applications can benefit from better computational efficiency resulting from ignoring zero-valued pixels in the input or output pixels that the application does not need. FFT Pruning is a group of algorithms that try to increase the computational efficiency of an FFT by eliminating operations that do not give any contribution to the final required result. Operations on zeros, caused by the input containing zeros do not contribute to the final results and therefore can be eliminated. Operations that contribute solely to output pixels that are not required by an application can also be eliminated.  

There are many different published algorithms for FFT Pruning, and we here mention a few. Merkel \cite{Markel1971} proposed the first FFT Pruning Algorithm whereby zero-valued pixels in the input grid are ignored by pruning using a Decimation in Frequency (DIF) method. Subsequently, Skinner \cite{Skinner1976} proposed a pruned Decimation in Time (DIT) FFT algorithm to operations contributing to  non-required output. Sreenivas and Rao \cite{Sreenivas1979, Sreenivas1980}  enhanced Skinner’s algorithm to handle zero-valued input and the non-required output while Nagi \cite{Nagi1986} proposed an alternative DIT based FFT pruning algorithm using frequency shift. All the above mentioned FFT Pruning algorithms work only for when the non-zero input or required output is in a continuous range within their respective grids, and the range has a size equal to $2^m$ $m \in \mathbb{N}$ pixels. Alves \etal \cite{Alves2000} proposed a general FFT pruning algorithm that overcomes this limitation.

Sorensen and Burrus \cite{Sorensen1993} proposed another pruning method known as Transform Decomposition since it decomposes the FFT into several smaller FFTs. Medina-Melendrez \etal \cite{Medina-Melendrez2009} further enhanced it to handle ignorable input.  Yuan \etal \cite{Yuan2011} discusses pruning for split-radix FFT, and lately, Qin \etal  \cite{Qin2018} proposed a pruned algorithm targeted for use in 5G mobile technology. 

\setlength{\algomargin}{0.1em}
\begin{algorithm}
\begin{mdframed}
\setstretch{1.5}
\KwIn{$G(u,\beta)$} 
\KwPar{$\zeta=\beta/\alpha$ where $\zeta, \alpha, \beta \in \mathbb{N}$}
\SetKwInput{KwHelp}{Helper Input}
\KwHelp{Pruning GCF: $Z(u)$, Corrector: $h_{\text{pr}}(l,\beta,\zeta)$ }
\KwOut{$g^{\text{g}}_{\text{pr}}(l,\beta,\zeta)$}
\myalgoline
\SetKwBlock{step}{Downsampling Step:}{}
\step{ Apply a circular convolution between $G(u,\beta)$ and $Z(\zeta u)$ and down-sample instantaneously to an oversampling grid of oversampling factor of $\zeta$. The output of this step is $W^{\text{g}}_{\text{pr}}(u,\zeta)$

\begin{equation}
W^{\text{g}}_{\text{pr}}(u,\beta,\zeta)=\left[G(u,\beta)*Z(\zeta u)*\shah{(N\zeta)^{-1}}(u)\right]\gshah{\zeta}(u)
\label{equ:maths:prungrid}
\end{equation}

The extra convolution with $\shah{(N\zeta)^{-1}}(u)$ in Equation \ref{equ:maths:prungrid} is to cater for circular convolution. 
}
\SetKwBlock{step}{IFFT Step:}{}
\step{
Apply an IFFT over $W^{\text{g}}_{\text{pr}}(u,\beta,\zeta)$ to get $w^{\text{g}}_{\text{pr}}(l,\beta,\zeta)$:
\begin{equation}
w^{\text{g}}_{\text{pr}}(l,\beta,\zeta)=\left[\sum_{p_2\in \mathbb{N} } g(l-p_2\zeta,\beta)\left\{z([l-p_2\zeta]\zeta^{-1})*\shah{\beta}(l)\right\}\right]\gshah{\zeta}(l)
\label{equ:maths:wgpr}
\end{equation}
where $g(l,\beta)=\mathcal{F}^{-1}G(u,\beta)$ and the derivation of $w^{\text{g}}_{\text{pr}}(l,\zeta)$ is given in Section \ref{sec:maths:pruningalgo}.
}
\SetKwBlock{step}{Correction Step:}{}
\step{Apply corrector $h_{\text{pr}}(l)$ and cut out the un-needed outer region to obtain $g^{\text{g}}_{\text{pr}}(l,\beta,\zeta)$ which approximates $g^{\text{g}}(l,\beta)=g(l,\beta)\gshah{1}(l)$:
\begin{equation}
g^{\text{g}}_{\text{pr}}(l,\beta,\zeta)=\left[\sum_{p_2\in \mathbb{N} } g(l-{p}_2\zeta)\left\{z([l-p_2\zeta]\zeta^{-1})*\shah{\beta}(l)\right\}\right]h_{\text{pr}}(l,\beta)\gshah{1}(l)
\label{equ:maths:prunningoutput}
\end{equation}
}
\end{mdframed}
\caption[Convolutional-Based FFT Pruning]{A General Algorithm for Convolution-based FFT Pruning in one dimension.}
\label{algo:maths:convprunning} 
\end{algorithm}

\subsection{The Algorithm}
\label{sec:maths:pruningalgo}

We describe a general form of  one-dimensional Convolution-Based FFT Pruning in Algorithm \ref{algo:maths:convprunning}. It prunes by ignoring an outer non-required region of the output since the input grid is oversampled. We remind the reader that we are assuming that the origin of any grid is at the centre. In our description, we are still assuming the normalisation set in Section \ref{sec:maths:normalisation} but applied for one-dimension.

The input to Algorithm \ref{algo:maths:convprunning} is $G(u,\beta)$, a one-dimensional function oversampled on a $u$-grid by a factor of $\beta$. Its size is $N\beta$ pixels, and by definition, it obeys the relationship $G(u,\beta)=G(u,\beta)\gshah{\beta}(u)$. The Inverse Fourier Transform is $g(l,\beta)$ and due to sampling in $G(u,\beta)$ the relationship $g(l,\beta)=g(l-p\beta,\beta)$ $\forall p \in \mathbb{Z}$ applies. Our interest is to try to calculate $g^{\text{g}}(l,\beta)=g(l,\beta)\gshah{1}(l)$.

$G(u,\beta)$ is down-sampled by a factor of $\alpha \in \mathbb{N}$ to an oversampling factor of $\zeta \in \mathbb{N}$ using a convolution with a GCF $Z(u)$. $Z(u)$ is of the form needed if $\zeta=1$ and therefore in Equation \ref{equ:maths:prungrid} $Z(u)$ is contracted according to the value of $\zeta$. 

A good choice of $\zeta \in \mathbb{N}$ depends on the aliasing behaviour of the GCF and in general $\zeta$ must be set to a value greater or equal to the inverse of the central interval length of $z(l)$ where the GCF is expected to suppress aliasing below arithmetic noise. For example, in the Pruning Step of the studied algorithms, we set $\zeta=2$ since Ye \etal \cite{Ye2019} claimed that the least-misfit gridding functions we intend to use, suppress aliasing below arithmetic noise in the half central region. 

The output of Algorithm \ref{algo:maths:convprunning} is $g^{\text{g}}_{\text{pr}}(l,\beta,\zeta)$ and lies on an $l$-grid of size $N$.

Finally, in Equation \ref{equ:maths:wgpr} of Algorithm \ref{algo:maths:convprunning}  we left out the derivation for $w^{\text{g}}_{\text{pr}}(l,\beta,\zeta)$ which we give hereunder. 

On applying the IFFT to $W^{\text{g}}_{\text{pr}}(u,\beta,\zeta)$ we get $w^{\text{g}}_{\text{pr}}(l,\beta,\zeta)$
\begin{equation}
w^{\text{g}}_{\text{pr}}(l,\beta,\zeta)=\left[\mathcal{F}^{-1}W^{\text{g}}_{\text{pr}}(u,\beta,\zeta)\right]\gshah{\zeta}(l)
\label{equ:maths:wgpr2}
\end{equation}
Now substituting Equation \ref{equ:maths:prungrid} in Equation \ref{equ:maths:wgpr2} and working out the Inverse Fourier Transform, we get:
\begin{equation}
    w^{\text{g}}_{\text{pr}}(l,\beta,\zeta)=\left\{\left[g(l,\beta)z(\zeta^{-1}l)\shah{(N\zeta)^{-1}}(l)\right]*\shah{\zeta}(l)\right\}\gshah{\zeta}(l)
\end{equation}
and expressing convolution with $\shah{\zeta}(l)$ as a summation, we get:
\begin{equation}
w^{\text{g}}_{\text{pr}}(l,\beta,\zeta)=\left[\sum\limits_{p_1 \in \mathbb{Z}} g(l- p_1 \zeta,\beta)z([l-p_1\zeta]\zeta^{-1})\shah{(N\zeta)^{-1}}(l-p_1\zeta)\right]\gshah{\zeta}(l)
\end{equation}
One notes that $\shah{(N\zeta)^{-1}}(l-p_1\zeta)\gshah{\zeta}(l)=\gshah{\zeta}(l)$ and therefore:
\begin{equation}
w^{\text{g}}_{\text{pr}}(l,\beta,\zeta)=\left[\sum\limits_{p_1 \in \mathbb{Z}} g(l- p_1 \zeta)z([l-p_1\zeta]\zeta^{-1})\right]\gshah{\zeta}(l)
\end{equation}
We also note that since $G(u,\beta)=G(u,\beta)\gshah{\beta}(u)$ and $g(l,\beta)=g(l+\beta p,\beta)$ $\forall p \in \mathbb{Z}$.

\begin{equation}
w^{\text{g}}_{\text{pr}}(l,\beta,\zeta)=\left[\sum_{p_2\in \mathbb{Z}}\left\{\sum_{p_1\in \mathbb{Z}} g(l-p_1\zeta)z\left(\left[l-p_2 \beta-p_1\zeta\right]\zeta^{-1}\right)\right\}\right]\gshah{\zeta}(l)
\end{equation}
\begin{equation}
w^{\text{g}}_{\text{pr}}(l,\beta,\zeta)=\left[\sum_{p_1\in \mathbb{Z}} g(l-p_1\zeta)\left\{\sum_{p_2\in \mathbb{Z}}z\left(\left[l-p_2 \beta-p_1\zeta\right]\zeta^{-1}\right)\right\}\right]\gshah{\zeta}(l)
\end{equation}

\subsection{Aliasing and the corrector}
\label{sec:maths:prunealias}
With no reason for doubt, the effects of aliasing discussed in previous sections still apply for the proposed Convolution-Based FFT Pruning Algorithm \ref{algo:maths:convprunning}. However, since we know that the input lies on a regularly sampled grid, extra considerations are in order.

Let us again analyse the effects of aliasing like we did in Section \ref{sec:maths:aliasing} and rewrite Equation \ref{equ:maths:prunningoutput} into the following set of equations:
\begin{subequations}
    \begin{equation}
        g^{\text{g}}_{\text{pr}}(l,\beta,\zeta)=o^{\text{g}}_{\text{pr}}(l,\beta,\zeta) + a^{\text{g}}_{\text{pr}}(l,\beta,\zeta)
    \end{equation}
    \begin{equation}
        o^{\text{g}}_{\text{pr}}(l,\beta,\zeta)=g(l,\beta)\left\{z(l\zeta^{-1})*\shah{\beta}(l)\right\}h_{\text{pr}}(l,\beta,\zeta)
\label{equ:maths:pr-og}
    \end{equation}
    \begin{equation}
       a^{\text{g}}_{\text{pr}}(l,\beta,\zeta)=\left[\sum_{p_1\in \mathbb{Z}, p_1\ne 0} g(l-p_1\zeta)\left\{z([l-p_1\zeta]\zeta^{-1})*\shah{\beta}(l)\right\}\right]h_{\text{pr}}(l,\beta,\zeta)
    \end{equation}
\label{equ:maths:aliasprun}
\end{subequations}

On comparing the set of Equations \ref{equ:maths:aliasprun} with  the set of Equations \ref{equ:maths:aliasing}, it is clear that $a^{\text{g}}_{\text{pr}}(l,\beta,\zeta)$ and $o^{\text{g}}_{\text{pr}}(l,\beta,\zeta)$ are different from the respective $a^{\text{g}}_{\text{co}}(l)$ and $o^{\text{g}}_{\text{co}}$. $g(l,\beta)$  is periodic and some summation terms in Equation \ref{equ:maths:pr-og} embedded in the convolution with $\shah{\beta}(l)$ that are considered as part of irreversible aliasing distortion in Convolutional Gridding are now reversible, for any GCF choice.

We note that if the GCF has the required anti-aliasing properties that suppress aliasing below arithmetic noise, then $\{z([l-p_1\zeta]\zeta^{-1})*\shah{\beta}(l)\}$ should approach 0 $\forall p_1\in \mathbb{Z}$, $p_1 \ne 0$ such that $a^{\text{g}}_{\text{pr}}(l,\beta,\zeta) \approx 0$.

As a corrector choice, we want to have $g^{\text{g}}_{\text{pr}}(l,\beta,\zeta)=o^{\text{g}}_{\text{pr}}(l,\beta,\zeta)$, and therefore we choose:

\begin{equation}
h_{\text{pr}}(l,\beta,\zeta)=\frac{1}{z\left(l\zeta^{-1}\right)*\shah{\beta}(l)}
\end{equation}

\subsection{Time complexity}

\begin{table}[]
\centering
\begin{tabular}{@{}l@{\rule{1.5cm}{0cm}}l@{}}
\toprule
\begin{minipage}[]{4cm}Step\end{minipage} & \begin{minipage}[]{4cm}Time complexity\end{minipage} \\ \midrule
Downsampling       .
& $\mathcal{O}(N\beta S_z)$ \\
IFFT                       & $\mathcal{O}(N\zeta\text{log}[N\zeta])$\\
Correction & $\mathcal{O}(N)$ \\
Direct IFFT of Input & $\mathcal{O}(N\alpha\zeta\text{log}[N\alpha\zeta])$ = $\mathcal{O}(N\beta\text{log}[N\beta])$ \\
\bottomrule
\end{tabular}
\caption[Time complexity of Convolution-Based FFT Pruning]{Time complexity of the steps in Convolution-Based FFT Pruning (Algorithm \ref{algo:maths:convprunning}) using the Big-$\mathcal{O}$ notation. The last row gives the time complexity for an IFFT without pruning. Note that $\beta=\alpha\zeta$.}
\label{tab:maths:pruningtimecomplexity}
\end{table}

Algorithm \ref{algo:maths:convprunning} is geared towards accelerating the IFFT, and therefore it is worth to make a comparison of time complexity between the said algorithm and a direct IFFT without any pruning. All the time complexities are stated in Table \ref{tab:maths:pruningtimecomplexity}.

For the trivial case of $\zeta=\beta$, the IFFT Step has equal time complexity with the direct FFT, and therefore since there is still the Downsampling and the Correction Steps to add, Algorithm \ref{algo:maths:convprunning} should be slower. Increasing $\beta$ or decreasing $\zeta$ causes the IFFT step to decrease in time complexity when compared with the direct IFFT. Therefore, there is an increase in the possibility that the computation required by the Downsampling and Correction Steps is less than what is saved in the IFFT and therefore  Algorithm \ref{algo:maths:convprunning} delivers more Performance than a direct IFFT. Hence we conclude that our Convolution-Based FFT Pruning is mostly suited for a high value of $\beta$ with a low value of $\zeta$. 

\section{Conclusion}
This chapter gave a theoretical and mathematical review of Convolutional Gridding building towards reviewing the new modified gridding algorithms, named Hybrid Gridding and Pruned NN Interpolation. At the end of this chapter Convolution-Based FFT Pruning is reviewed.

\chapter{Methodology}
\label{chap:methodology}

In this chapter and those that follow, we shall be discussing our novel implementations of Convolutional Gridding (Algorithm \ref{algo:maths:convgriddingoversampling}), Hybrid Gridding (Algorithm \ref{algo:maths:hybrid}) and Pruned NN Interpolation (Algorithm \ref{algo:maths:purenninterpolation}) collectively referred to as the studied algorithms. This chapter is meant to serve as an introduction to all the work we are to present,  define terminology and introduce various concepts that are used in subsequent chapters.

The layout of this chapter is as follows: 

In Section \ref{sec:methodology:cuda}, we discuss various topics related to CUDA and programming on the P100. The main goal of the section is to define various terms and review useful concepts in GPU programming that we will refer to in subsequent chapters. 

In Section \ref{sec:methodology:implimentation}, we give notes on the implementation of the studied algorithms that are common in all three algorithms. The section includes how the Correction Step and the IFFT Step were implemented. We will not be discussing the implementation of these two steps elsewhere.

In Section \ref{sec:methodology:tuning}, we give details on the Brute Force Search, a manual process we used to tune our Gridders and Pruners.

In Section \ref{sec:methodology:experimentalsetup}, we describe our Experimental Setup, including details of how we conducted experiments and presented results in plots. 

We conclude with Section \ref{sec:methodology:performance}, whereby we list and define what we call Performance Metrics, through which we will measure and analyse Performance.

\section{Notes on CUDA and GPUs}
\label{sec:methodology:cuda}

CUDA (Nickolls \etal \cite{Nickolls2008}) is an acronym standing for Compute Unified Device Architecture and is a platform developed by NVIDIA. It provides a toolkit and programming interface for NVIDIA GPUs (Graphics Processing Units). We used CUDA version 10.1 to develop all the Gridders and Pruners to execute on the P100.

Originally, GPUs were specialised processors capable of accelerating graphics rendering (Das and Deka \cite{das2016history}). They evolved, and today GPUs are also used as general processors to achieve a high level of Performance. They are parallel devices capable of impressive computational power through their ability to execute thousand of threads concurrently. The P100 \cite{Corporation2016} is a GPU specialised for high-Performance computing, based on the Pascal microarchitecture developed by NVIDIA and equipped with one  GP100 chip and HBM2s (High Bandwidth Memory 2) \cite{GP100GPU}. As of the writing of this thesis, the P100 is not the latest generation GPU. NVIDIA released GPUs based on the  Volta and Turing microarchitectures that succeed the Pascal microarchitecture. More information on the P100, including specifications, is given while progressing through this chapter, mainly in Table \ref{tab:methodology:p100specs}.

\begin{figure}
    \centering
    \includegraphics[page=1,width=\linewidth]{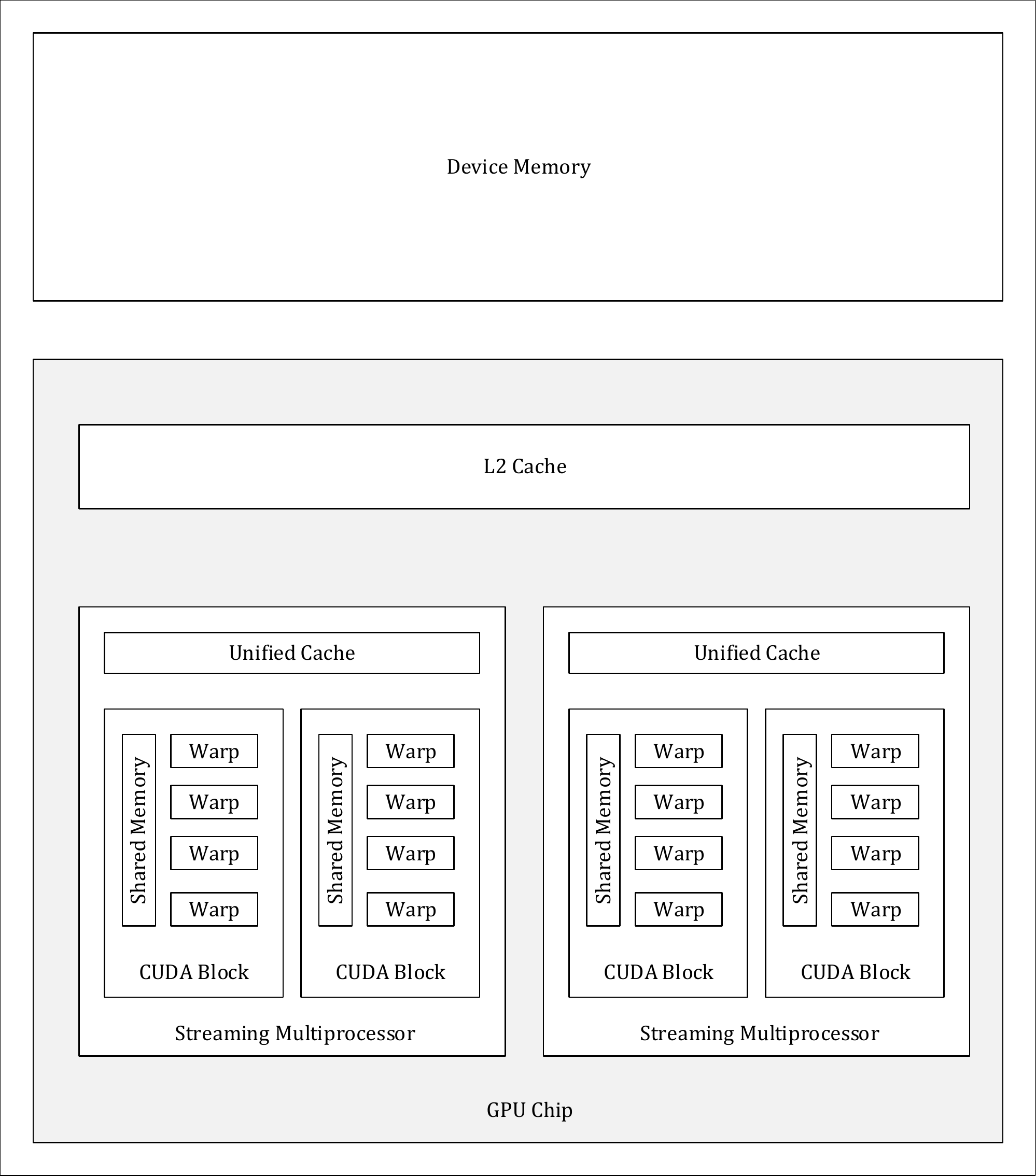}
    \caption[GPU components]{Diagram depicting the various GPU components that are mention in Section \ref{sec:methodology:cuda}. Note that different GPUs have a different number of components. Therefore, the number of the same components appearing in this diagram is not equal to the number of the same components a given GPU has, including the P100.}
    \label{fig:methodology:gpudiagram}
\end{figure}

Most of the information given in this section is based on documentation provided by NVIDIA on CUDA, mainly the Programming Guide \cite{NvidiaProgGuide}, the Best Practices Guide \cite{NVIDIAbestpractice}, the Pascal Tuning Guide \cite{NVIDIATuningPascal}, the Profiler User's Guide \cite{NvidiaProfiler} and the CUDA Runtime API Reference Manual \cite{cudareference}. A simple diagram depicting the various GPU components that will be mentioned through out this section is available in Figure \ref{fig:methodology:gpudiagram}.

Our main goal in this section is to introduce concepts in CUDA and discuss some related terms that are used in this thesis. The information is given in various sub-sections, with each sub-section discussing a particular topic.

\subsection{The CUDA grid layout, warps and SMs} 

The programming model of CUDA defines threads that are organised in blocks and a grid. When we refer to such threads, blocks and grid, we shall use the terms CUDA thread, CUDA block and CUDA grid. If the term CUDA is dropped, then we are referring to something else.

A complete piece of execution code that is intended for consumption by a CUDA GPU is called a CUDA kernel. The entity within the GPU that executes a CUDA kernel is known as a Streaming Multiprocessor (SM). The P100 has 56 of such SMs.

A CUDA kernel is executed with a configurable CUDA grid layout. The layout consists of several CUDA blocks made of a configurable number of CUDA threads. From a hardware perspective, CUDA threads in a CUDA block are organised in 32 thread-wide warps. SMs schedule warps not CUDA threads, and by design, the hardware concurrently executes the same instruction for all the CUDA threads in a warp. The CUDA block is therefore made up of a configurable number of warps that gets dispatched to one SM, whereby dedicated resources such as  shared memory get assigned, and the block is launched in full for execution. All warps in a CUDA block will reside on the SM until the CUDA block execution finishes.

\subsection{Registers}
\label{sec:methodology:registers}
Each CUDA thread avails itself of a set of dedicated registers. These are on the SM, and the amount allocated per CUDA thread depends on the needs of the CUDA kernel set out during code compilation. The compiler provides a way to control the number of registers assigned per-thread by controlling the minimum amount of concurrent CUDA blocks executing on each SM, for which the compiler will have to control the number of registers allocated per CUDA thread to comply.

\subsection{Latency and Instruction Level Parallelism}
\label{sec:methodology:latency}
A given instruction needs some time to be fully executed, and such time is loosely referred to as latency. The SM is made up of many modules able to execute different type of instructions in parallel. As some module in the SM is executing an instruction for a given warp, the SM is capable of scheduling and executing further instructions of the same warp. Such scheduling and execution are possible if resources to execute the instruction are available, and all dependencies to execute the new instructions are fulfilled. Dependencies include results from previous instructions which might not yet be available due to latency. The warp is said to be stalled if for any reason a new instruction cannot be executed.   

The SM mitigates the effects of warp stalls by having a pool of warps from which it can try to schedule the execution of a new instruction. However, for sufficient mitigation, there must be a certain amount of Instruction Level Parallelism.

Instruction Level Parallelism is a measure of how many instructions can be executed simultaneously for a CUDA thread, possibly because such instructions do not depend on the results of each other. Increasing Instruction Level Parallelism helps in reducing stalls related to latency.

\subsection{Memory}
\label{sec:methodology:cudamemory}
The CUDA programming model defines various memory spaces that the CUDA kernel can avail of and here we are interested in three of these spaces which are shared memory, global memory and local memory.

Before discussing the three memory spaces, we clarify two terms used in the context of memory, which are the \textit{request} and the \textit{transaction}. A request is an instruction to access some memory module, while a transaction is the movement of a unit of memory data generally from one memory module to an other.

Shared memory is a dedicated on-chip per SM memory resource. It is dedicated to a given CUDA block throughout the lifetime of the CUDA block. Since shared memory is on-chip, it is quite fast to access and much faster than device memory. 

Global memory is a memory shared by all the CUDA grid and persists across CUDA kernel launches. In the P100, global memory resides in the device memory which is outside the GPU chip and is exclusively accessed by a two-layered cache system composed of the Unified Cache and the L2 Cache.

The Unified Cache resides on the SM while the L2 Cache is shared with all SMs. Memory requested to global memory are always transacted via  the Unified Cache. If the Unified Cache cannot deliver a transaction (a MISS) it will forward the transaction to the L2 Cache which in turn forwards it to device memory if there is also a MISS on the L2 Cache.

We do note that in our implementations, we instruct the P100 to load input record data without caching in the Unified Cache, as to reserve the Unified Cache store for the GCF. In such circumstances, the requests will still pass through the Unified Cache where temporary storage is allocated for the use of coalescing (Pascal Tuning Guide \cite{NVIDIATuningPascal}).

Local memory is local to a CUDA thread and used only when the compiler could not find a way to use registers. For example, registers cannot be referenced dynamically, and when arrays are used in a way that the compiler cannot determine a constant way to access them, then the compiler is forced to use local memory. Registry spilling is also another reason why the compiler is forced to use local memory. It happens when there are not enough registers available for a CUDA thread due to limits set either by the hardware, or set programmatically such as when setting a minimum for the CUDA blocks per SM, as discussed previously.   

In the P100, local memory is cached on the Unified Cache and uses the same structures used for global memory. In general, the so-called \textit{optimal solutions} (refer to Section \ref{sec:methodology:tuning})  for the Gridders and Pruners do not make use of local memory.   

\subsection{Shared memory access}

There are some best practice guidelines to adhere to, for efficient access of a warp to shared memory. Shared memory in the P100 is composed of thirty-two four-bytes wide banks. Best practice mandates that the access patterns to shared memory should be such that bank conflicts are avoided. A bank conflict happens when two or more CUDA threads in a warp try to access distinct data stored on the same bank. When access to shared memory causes a bank conflict, the GPU is forced to serialise such access.

\subsection{Global memory access and the optimal access pattern}
\label{sec:methodology:optimalaccesspattern}

In the P100, the two caches and device memory do transactions only with thirty-two-byte memory aligned words. The access pattern of the request dictates the Performance of a warp request to access global memory. We define the \textit{optimal access pattern} as that access pattern of the request that leads to best Performance. There are two important requirements that an access pattern has to abide by  in order to be optimal, which we call the \textit{coalescing requirement} and the \textit{vicinity requirement}. We now define the stated requirements:  
\begin{enumerate}
    \item \textbf{Coalescing requirement}: The Best Practices Guide \cite{NVIDIAbestpractice} states that for the P100, the most efficiently executed requests are those which access whole thirty-two-byte memory aligned words where neighbour threads request data from the word in a sequential manner. With such a request, the P100 will coalesce (merge) the request over the warp in thirty-two-byte transactions containing only needed data.
    \item  \textbf{Vicinity requirement}: Through experimentation, we found out there is a degradation in Performance if the words requested by the warp are far apart in memory. Therefore, an optimal access pattern has to generate transactions with words in the vicinity of each other but not necessarily adjacent to each other.
\end{enumerate}

We will state that an access pattern is \textit{degraded} if that access pattern does not adhere to any of the two requirements. 

A degradation in access pattern is very likely to penalise Performance by injecting latency in the CUDA kernel execution. This fact is crucial in what we call grid committing since in-general such committing generates severely degraded optimal patterns that visibly and unequivocally push down Performance.  \textit{Grid committing} is that action in a Gridder or Pruner that updates the output grid. Such updates are executed using atomic reductions for which the optimal access pattern still applies.

\subsection{Utilisation and boundedness}

As mentioned before, the GPU is composed of various modules that execute in parallel instructions meant for the different modules. For example, in a single SM, the GPU is capable of executing a memory-related instruction on the Unified Cache, and at the same time execute an integer arithmetic instruction on that module that handles integer arithmetic instructions. We find it useful to measure the utilisation of some of these modules, since it indicates whether GPU resources are being well used. We say that a module or set of modules bound the CUDA kernel execution if the utilisation of such module or set of modules is so high that is inhibiting further increase in Performance. 

Let us describe further boundedness. In our analyses, we will measure memory bandwidth utilisation of the shared memory, Unified Cache, L2 Cache and device memory. We will also present utilisation measures of compute. In general, a CUDA kernel execution is considered memory-bound if the memory bandwidth of any memory module is utilised over 60\% of the peak. If compute utilisation is above 60\% then we consider the execution CUDA kernel as compute-bound. It is possible to have CUDA kernel executions that are  neither compute-bound nor memory-bound, and in such a case the CUDA kernel execution is probably riddled with latency.

We do have to point out that the Performance Metrics (refer to Section \ref{sec:methodology:performance}) related to memory utilisation are integers that scale from 0 to 10. Therefore, we consider any memory utilisation equal to or above the value of 6 as memory-bound.  

\subsection{Occupancy}

In Section \ref{sec:methodology:latency}, we briefly mentioned that the SM manages a pool of warps which are being actively executed. The maximum size of this pool depends on an imposed hardware limit and resources requested by the CUDA kernel. In the P100, the hardware limit is at 64 warps per SM (See Table 15 of the Programming Guide \cite{NvidiaProgGuide}), but the maximum pool size will be reduced if the SM does not have enough shared memory or registers to handle the needs for all the warps in the pool.

Theoretical Occupancy, which in this thesis, shall be referred  to  as just \textit{Occupancy}, is a measure of the maximum size of the pool of active warps as imposed by the CUDA kernel against the hardware limit. Occupancy with a  value of 100\% implies that the SM has enough registers and shared memory to have the warp pool size reach the maximum set by hardware.

Occupancy is not directly related to Performance, though a too low Occupancy might lead to latency issues. Volkov \cite{Volkov2010} showed that, in CUDA kernels with a high-level of global memory access, lowering Occupancy is beneficial, and in discussing our implementations in the next chapters, we shall refer to this matter various times. 

\subsection{Profiling}

The CUDA toolkit provides two tools that we used extensively to profile our implementations in order to do in-depth analyses.  

The first tool is \texttt{nvprof} \cite{NvidiaProfiler}, which is a text-based profiler. We used it to collect data needed for Performance Metrics. 

The second tool is the \texttt{NVIDIA Visual Profiler} \cite{NvidiaProfiler}, which we used to visualise some of the data generated by \texttt{nvprof} to help us report better some of the findings.

NVIDIA is deprecating the two tools, but they served very well to our needs.

\section{Notes on implementation}
\label{sec:methodology:implimentation}
In this section, we discuss various aspects of our novel implementations. We shall also define and discuss some terminology that will be used from now on.

\subsection{Use of \textit{mt-imager}}
We implemented all our studied algorithms on the \textit{mt-imager} such that they can be tested and analysed. 

The \textit{mt-imager} is a GPU-based Radio Interferometric Imager we developed for Muscat \cite{Muscat2014}, which, during the tenure of the doctoral studies we restructured and enhanced, with Double-Precision arithmetic and the implementation of CLEAN \cite{Hogbom1974}, MS-MFS \cite{Rau2011} and other algorithms. We reported some of the stated enhancements in the Transfer Report (Muscat \cite{Muscat2015}). 

The \textit{mt-imager} provides all needed infrastructure such that we could implement and test all the algorithms with ease. In particular, the \textit{mt-imager} provides for the loading of Radio Interferometric observation data from MeasurmentSets (Kemball and Wieringa \cite{Kemball2000}), movement of data, GPU management via the General Array FrameWork (Muscat \cite{Muscat2014}), measurement of execution for a CUDA kernel, and saving of the output images in a FITS file (Pence \etal \cite{Pence2010}).

\subsection{Precision}
\label{sec:methodology:arithmeticprecision}
We developed the studied implementations, to work with Single and Double Precision. Input Visibility values available in MeasurmentSets are generally Single-Precision and on enabling Double-Precision, all steps in the studied implementations are computed using Double-Precision. The input Visibility values are converted to Double-Precision in a pre-processing phase, before being presented to a given Gridder for consumption. 

\subsection{Multi-polarised channels - $N_{\text{pol}}$}
\label{sec:methodology:multipolarisationchannels}
All our Gridders take advantage of multi-polarisation to maximise Performance. Radio Interferometers measure all Visibility polarisations simultaneously, implying equal $(u,v,w)$ coordinates. Therefore, all polarised images are generated together to reduce some compute common to all polarisations. We define $N_{\text{pol}}$ as the number of polarisations handled simultaneously, by the Gridder or Pruner. $N_{\text{pol}}$ can take values from one to four.

We note that Taylor coefficients in MS-MFS (Rau and Cornwell \cite{Rau2011}) also have equal $(u,v,w)$  coordinates. Therefore all our implementations can handle Taylor Coefficients in the same way they handle multi-polarisation.

\subsection{Pre-processing}

When implementing the studied algorithms, we followed a similar structure to the implementation of the \textit{w-projection} algorithm in Muscat \cite{Muscat2014}. The input Visibility records are subject to a pre-processing step that alters the input record stream in a way that is best suitable for digestion by the Gridder.

We designed the pre-processing step to handle work that can be done once in a Deconvolution process with many major cycles. Therefore, we regard pre-processing as part of an initialisation step that will not run for every major cycle and therefore not subject to a study of its Performance in this thesis. Similar reasoning was made by Merry \cite{Merry2016} in analysing an enhanced Convolutional Gridder for \textit{w-projection}.

The main inputs to the implemented algorithms are a stream containing $(u,v,w)$ coordinates of the multi-polarised Visibility records and another stream containing the multi-polarised Visibility values, obtained from a MeasurementSet. Such input is pre-processed before presenting it to a Gridder for processing. The  $(u,v,w)$ coordinates of the input are transformed\footnote{The $w$ coordinate is ignored since our algorithms do not cater for $w$.} into a set of integers usable by the Gridder to efficiently find the position of the record on the output grid and where applicable the right choice of the GCF. Pre-processing also does a full compression of the input which is discussed in one of the succeeding sub-sections. 

\subsubsection{Ordering Modes}
\label{sec:methodology:recordorder}
The pre-processor, in conjunction with other infrastructure within \textit{mt-imager}, sorts records in the streams in three different configurable Ordering Modes, which we now list and explain.

\begin{enumerate}
\item \textbf{\textit{No-Interleave (NI)}}: In NI mode, records are grouped by Antenna Pair, and Frequency Channel and each group is ordered in increasing time of measurement. This mode of ordering is the same as that given in Muscat \cite{Muscat2014} when frequency channel interleaving is disabled. 

\item \textbf{\textit{Interleave-Forward-Back (IFB)}}: In IFB mode, records are grouped by Antenna Pair and ordered similarly to NI mode, with the exception that there is interleaving over the Frequency Channels as shown in Figure \ref{fig:methodology:ifbordering}. 
\begin{figure}
\centerline{\includegraphics[scale=0.68]{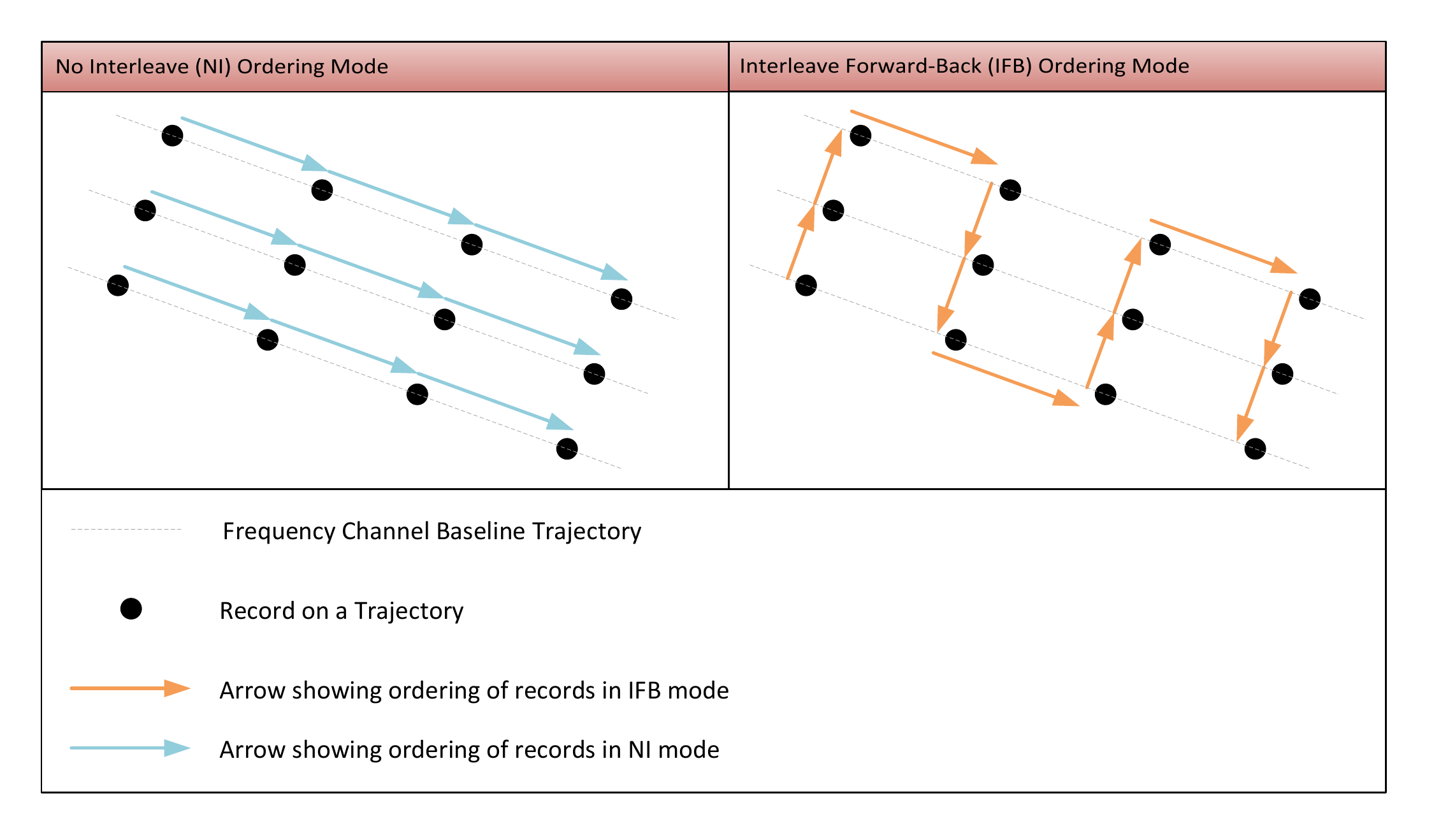}}

\caption[The NI and IFB Ordering Modes]{A sketch of  12 records sampled in 3 frequency channels in UV-space, showing how the records get ordered in the \textit{Interleave-Forward-Back (IFB)} (left sketch) and the \textit{No-Interleave (NI)} (right sketch) Ordering Modes. The arrows indicate the order of sorting. In NI mode, records from different frequency channels are not interleaved. In IFB mode, records are interleaved, and the ordering of records from different frequency channels is such that the ordering will go back and forth through the channels as illustrated by the arrows.
}
\label{fig:methodology:ifbordering}
\end{figure}

\item \textbf{\textit{NN Grid sorted (SS)}}: In SS mode all records are ordered based on the position of records on the virtual NN Grid as would be stored in memory. The virtual NN Grid would be stored in a row-major order where every row contains pixels with constant $v$ values and columns contain pixels with constant $u$ values. \end{enumerate}

\subsubsection{Full Compression}
\label{sec:methodology:fullcompression}
In Muscat \cite{Muscat2014}, we argued that if two Visibility records fall on the same grid pixel in the virtual NN Grid, then they can be added together prior to gridding with no effect on aliasing. We called this process of addition of records \textit{compression} in view that this reduces (compresses) the input record stream to one with fewer elements. In Muscat \cite{Muscat2014} we implemented compression in a way that only neighbour records on the stream that falls on same NN Grid pixel are added. The stream would have been sorted either in NI or a frequency channel interleaved Ordering Mode similar to IFB. 

In this thesis, we enhanced compression, whereby any records in the stream that falls on the same NN Grid pixel gets added. We call this \textit{full compression}. Except when otherwise stated, full compression is enabled in all experiments we present in this thesis.

We note that, by its nature, full compression ensures that the input number of records to the Gridder is constant for a given size of the NN Grid, irrespective of the Ordering Mode, and varies with varying size of the NN Grid.

\subsection{Gridding Step}
\label{sec:methodology:griddingstep}
As we stated in previous chapters, the implementation of a Gridding Step is referred to as a \textit{Gridder}. The exact name of each Gridder will be according to which algorithm it is to be applied. Therefore we have the Convolutional Gridder for Convolutional Gridding (Algorithm \ref{algo:maths:convgriddingoversampling}), the Hybrid Gridder for Hybrid Gridding (Algorithm \ref{algo:maths:hybrid}) and the NN Gridder for Pruned NN Interpolation (Algorithm \ref{algo:maths:purenninterpolation})

All Gridders are composed of one CUDA kernel, which in our experiments, grids all input records in one execution of the CUDA kernel.

The output of any Gridder is a multi-polarised grid, where its layout in memory depends on the Gridder. The Convolutional Gridder outputs a non-interleaved, multi-polarised grid with origin set at the centre. In contrast, the Hybrid Gridder outputs a non-interleaved, multi-polarised grid with the origin at the corner, and the NN Gridder outputs a polarisation-interleaved grid with the origin at the corner.

As pointed out in Section \ref{sec:methodology:optimalaccesspattern}, all Gridders access the output grid via atomic reductions. We are calling the process of updating a pixel as committing, in view that in most implementations, the value that will be added is generally the result of an accumulation in registers that will be added (committed) to the grid.

The Convolutional and Hybrid Gridders make use of what we call \textit{Sub-Warp Gridders}. In the stated Gridders, the number of CUDA threads required to grid one record is less than the total CUDA threads per warp. Therefore, we sub-divide the warp in Sub-Warp Gridders that grid in full a subset of the input records,  with no dependency on the other Sub-Warp Gridders.

Specifics of the three studied Gridders are discussed in detail in Chapters \ref{chap:2dgridding},\ref{chap:hybrid} and \ref{chap:purenn}

\subsection{Pruning Step}

In Chapter \ref{chap:pruning}  we deliver a detailed discussion on our implementation of the Pruning Step. In this sub-section, we just note that in the implementation of Pruned NN Interpolation, the Pruner de-interleaves the grid as to make it easier for the IFFT step to handle its input grid.

\subsection{IFFT Step}
\label{sec:methodology:ifftstep}
The IFFT Step is implemented using the cuFFT library \cite{Nvidiacufft} supplied with CUDA. The library works only with multi-polarised grids with the origin at the corner, and supports polarised-interleaved grids and also non-interleaved grids. For any value of $N_{\text{pol}}$,  IFFTs of the polarised grids, are executed as a batch to guarantee minimal execution time on GPU for the IFFT.

We note that there is a Performance penalty in cuFFT when handling polarisation-interleaved grids and therefore the Pruning Step ability to de-interleave helps cuFFT to be more performant.  

We also note that since the Convolutional Gridder outputs a grid with the origin at the centre, we implemented an extra pre-FFT step to convert the grid accordingly, which is similar to the one documented in Muscat \cite{Muscat2014}.

\subsection{Correction Step}
The Correction Step is implemented as a simple CUDA kernel which re-layouts the IFFT  output with the origin set at the centre, multiples with a pre-calculated corrector and truncates the image to its meant size. The \textit{mt-imager} generates the corrector as pre-processing, and its Performance will not be analysed.

In general, in Radio Interferometry the output of the Correction Step is expected to be real-valued and therefore the said CUDA kernel discards the imaginary while correcting.

In this thesis we shall not deliver any Performance analyses specific to the Correction Step. Its implementation is trivial, and in general, its effects on the overall Performance for  any of the studied implementations is minimal.

\section{Tuning and Brute Force Search} 
\label{sec:methodology:tuning}
When implementing the Gridding and Pruning Steps, we did whatever we could to achieve the highest Performance possible. We did so through common-sense, best-practice,  and by tuning the use of GPU resources, such as registries per threads, shared memory size and memory I/O. 

Tuning is a challenging task since the solution is dependent on many factors such as Precision, $N_{\text{pol}}$ and even the input $uv$-profile of the observation controls memory access to the output grid. 
Tuning in this thesis is done on parameters that we refer to as \textit{Tuning Parameters}. In coding our implementations, we do extensive use of C++ templates \cite{cppstandard} where most of the Tuning Parameters are implemented as C++ template parameters. The use of C++ Templates  enabled us to have one unique code for all combinations of selected values of the Tuning Parameters with Precision and $N_{\text{pol}}$, the latter two being also generalised as C++ template parameters.  

We achieve fine-tuning of the Tuning Parameters via a Brute Force Search. Such a search gave us more insights on the Performance of the studied implementations. For each Tuning Parameter, we select a set of reasonably good values and run performance tests over the implementations, instantiated with all possible combinations of the selected values of all Tuning Parameters. We then do a manual analysis to find a combination (a \textit{solution}) of Tuning Parameters values that perform well. We do note that many times in our test scenarios, there were no unique solutions, and our selection is a manually chosen \textit{best fit} achieving optimal Performance, for some group of test scenarios. We refer to such solutions as the \textit{optimal solutions}, and we used the term optimal Performance to mean some Performance which is near the maximum reported by the Brute Force Search. The Performance Metric \optimalityfactor defined in Section \ref{sec:methodology:optimalsolutionsperformancemetrics} will measure how optimal is an optimal solution.

For the Gridders, the Brute Force Search is also used to measure various Performance Metrics which are defined in Section \ref{sec:methodology:perfromancemetricsbrute}.

\section{Experimental setup}
\label{sec:methodology:experimentalsetup}
Let us now describe the setup used to perform experiments geared to understand the Performance behaviour of the Gridding and Pruning Steps of the studied implementations.  

\subsection{Hardware setup}
Our Performance-related experiments are run on a High-Performance server with two "Intel(R) Xeon(R) CPU E5-2640 v4 @ 2.40GHz" \cite{IntelCorporation2016} and four P100s \cite{Corporation2016} with clocks set to 1480 MHz.
Some specifications worth to note for the said GPUs are given in Table \ref{tab:methodology:p100specs}. A given experiment is exclusively run on one and only one GPU. When possible, multiple experiments were executed concurrently on different GPUs.

\begin{table}[]
\centering
\begin{tabular}{@{}l@{}c@{}l@{}}
\toprule
Specification       &\rule{2.5cm}{0cm}         & Value \\ \midrule
Product Name             &    & Tesla P100-SXM2-16GB\\

Clock Frequency (\texttt{$\lambda$})& &1480MHz \\
Memory Size &                 & 15.899GiB\\
L1 Cache Size &                & 24KB per SM \\
L2 Cache Size  &              & 4MiB \\
Single Precision Performance & & 10.612 TeraFLOPS\\
Double Precision Performance & & 5.306 TeraFLOPS\\
CUDA compute ability    &     & 6.0\\
Number of SMs (\texttt{N\subscript{SM}})& & 56\\
\begin{minipage}[]{6cm}Maximum Instructions Per Cycle for each SM (\texttt{MaxIPC\subscript{SM}})\end{minipage} & & 3 \\
 \bottomrule
\end{tabular}
\caption[P100 specifications]{List of selected specifications of the four P100s, over which we executed  most of the experiments presented in this thesis. These specifications were in general retrieved from profiling (displayed by the Visual Profiler) except for the L1 Cache Size, which was retrieved from the Programming Guide  \cite{NvidiaProgGuide}  and the Clock Frequency ($\lambda$), which was manually set.}
\label{tab:methodology:p100specs}
\end{table}

\subsection{Input and Simulation Parameters}
Unless otherwise specified, input for all simulations related to gridders is a quad-polarised 16-frequency channel LOFAR observation, with 6501600 multi-polarised records per frequency channel. Figure \ref{fig:introduction:uvcoverage} gives the $(u,v)$-coverage of the input observation. The observation is fully compressed by \textit{mt-imager}, and Table \ref{tab:methodology:compressedinputecords} gives the number of input multi-polarised records presented to any Gridder after full compression. 

\begin{table}[]
\centering
\begin{tabular}{@{}l@{\rule{2cm}{0cm}}r@{}}
\toprule
$N_{\text{nn}}$ &  Number of Records \\ \midrule
$2^{16}=65,536$      &  19,754,479          \\
$2^{14}=16,384$        & 3,198,831        \\ 
$2^{13}=8,192$     &  976,536  \\ \bottomrule
\end{tabular}
\caption[Number of gridded records in our experiments]{Table giving the input number of multi-polarised records presented to any Gridder after the 16-frequency channel LOFAR observation with 6501600 multi-polarised records per frequency channel is fully compressed. In all experiments, the virtual NN Grid is always of equal length and width and therefore we need only give the size in pixels of one of its dimensions ($N_{\text{nn}}$).}
\label{tab:methodology:compressedinputecords}
\end{table}

\textit{mt-imager} is instructed to produce an output image with a constant pixel width and length of 4.7arcsec. The pixel size ensures that all records from the observation fit within the central region of the output $uv$-grid of any Gridder, which is a common practice in Radio Interferometry.

Most of our Performance analyses is based on tracking various measured Performance Metrics (See Section \ref{sec:methodology:performance}) while we change values of various parameters, which we call \textit{Simulation Parameters}. In this way, we can identify changes happening within the Gridder or Pruner execution that are  varying Performance.

There are in all five Simulation Parameters which we now list and discuss next. 

\subsubsection{Precision}

We discussed Precision in Section \ref{sec:methodology:arithmeticprecision} and in our experiments, we shall consider Single and Double Precision.

\subsubsection{Number Of Polarisations ($N_{\text{pol}}$)}
We discussed the handling of multi-polarised data in Section \ref{sec:methodology:multipolarisationchannels}. In our experiments and subsequent analyses, we will consider imaging with $N_{\text{pol}}$ equal to one, two, three or four polarisations. 

\subsubsection{NN Grid Size ($N_{\text{nn}}$)}

 In all our experiments, the NN Grid is always with equal length and width, and we, therefore, quote one dimension of the NN grid size denoted by $N_{\text{nn}}$. For our analyses, we considered experiments with $N_{\text{nn}}=2^n , n\in\mathbb{N}, 6\le n \ge 16$, but we only chose to present results for two values for each Gridder in this thesis. For the Convolutional and Hybrid Gridders, we chose to present results for $N_{\text{nn}}=65536$ and $N_{\text{nn}}=16384$ while for the NN Gridder we chose to present results for $N_{\text{nn}}=8192$ and $N_{\text{nn}}=16384$. The difference in choice was necessary since simulations for NN gridder with $N_{\text{nn}}=65536$ were not possible due to memory limitations on the P100. Our choices of $N_{\text{nn}}$ ensure that full compression leaves enough records to let the studied Gridders saturate the GPU.

\subsubsection{Oversampling factor $\beta$}

In our experiments, we consider oversampling factor ($\beta$) values of $2^n$, $n\in\mathbb{N}, n<10, N_{\text{nn}}/\beta\ge128$. We note that the output image size in pixels is equal to $N_{\text{nn}}/\beta$.

\subsubsection{Record Ordering Mode}
We discussed the way records are ordered and presented to the studied Gridders in Section \ref{sec:methodology:recordorder}. In our analyses, we will consider all described modes: INO, IFB and SS, but we will not always provide detailed analyses for every Ordering Mode.

\subsection{The GCF}
\label{sec:methodology:thegcf}
In all experiments, we set the GCF used by the Convolutional and Hybrid Gridders to the Prolate Spheroidal of order one. We chose the stated GCF since it is the GCF we know of, with the smallest support and acceptable anti-aliasing properties for use in Radio Interferometry. We are assuming that the smaller the GCF is, the more performant is  the Convolutional Gridder. We thrive in making the Convolutional Gridder as performant as possible to ensure its validity as reference in the comparative analyses given in Chapter \ref{chap:comparative}. 

We discussed the use of the Prolate Spheroidal in Section \ref{sec:maths:gcfform}, and we note that in our experiments, the function is generated using the same mathematical method as used in CASA.

We also note that different GCFs are considered for the Pruning Step, and we detail this matter in Chapter \ref{chap:pruning}. 

\subsection{Maximum Performance Experiments}
For all Gridders, we shall perform special experiments that we call the Maximum Performance Experiments. These experiments try to estimate the maximum Performance, a given Gridder using the optimal solution, can deliver for specific values of Precision and $N_{\text{pol}}$. We perform such experiments by hacking in the pre-processing phase to vary the coordinates of the input records to make the Gridder give the highest Performance.  We do this exercise after acquiring knowledge on the given Gridder from other results.  

We give Performance results of the Maximum Performance Experiments in tables. 

\subsection{Layout for plotted results}
\label{sec:methodology:graphlayouts}
As we shall detail in Section  \ref{sec:methodology:performance}, we use what we call Performance Metrics to measure and analyse Performance. Except for the results of the Maximum Performance Experiments, we sub-divided these metrics into groups that are plotted in separate figures. These groups are listed and detailed as sub-sections in Section  \ref{sec:methodology:performance}.

We plot each Performance Metrics group using line or bar plots in figures containing an array of graphs. Each row in the array is dedicated to one Performance Metric. For results related to the studied Gridders, each column represents a given value of $N_{\text{pol}}$. For Performance results related to the Pruners, what a column represents is defined in the applicable figure. 

Note that for the studied Gridders, we generally report results for Single and Double Precision, in separate figures, but when convenient we do amalgamate the Single and Double Precision results as two sub-figures in one figure.

We use line plots wherever we report measurements of Performance Metrics against $\beta$ or $\alpha$. In Chapter \ref{chap:purenn} we use bar plots since plotting against ($\beta$) for the NN Gridder is meaningless.

All given line plots for the Convolutional and Hybrid Gridders has the value of $N_{\text{nn}}$ and Ordering Mode constant. Since $\beta$ varies over the line plot, then the output grid of the  Gridder will also vary in size.

Various plots have missing data at the head and tail because we were unable to perform related experiments either due to memory limitations imposed by the P100 or because of $N_{\text{nn}}/\beta<128$.

Unless otherwise specified, line and bar plots for the Gridders are labelled with the format \texttt{Ordering Mode-$N_{\text{nn}}$}. For example, a line plot for results with Ordering Mode set to IFB and $N_{\text{nn}}=65536$ is labelled \texttt{IFB-65536}.  When considering the Convolutional Gridder an extra suffix (either \texttt{-L} or \texttt{-H}) is sometimes added to a plot's label as to distinguish between the so-called \textit{Low Commit Rate} solutions (\texttt{-L}) and the \textit{High Commit Rate} solutions (\texttt{-H}). For example, a line plot labelled with \texttt{IFB-65536-L} gives results for a Low Commit Rate solution.   

\section{Measuring Performance}
\label{sec:methodology:performance}
As alluded previously, we measure and analyse Performance through \textit{Performance Metrics}. These metrics are mostly derived from direct measurement of quantities made via profiling using \texttt{nvprof} combined with other known information. 

All Performance metric and related quantity names are written with typewriter typeface. If a quantity is measured directly by \texttt{nvprof}, then the name of the Performance metric is the name given by \texttt{nvprof}, pre-appended with the text "\texttt{P100:}". Documentation of all quantities measured by \texttt{nvprof} is available in the Profiling Guide \cite{NvidiaProfiler}.

As discussed before, we divide all Performance Metrics into five groups, where each group is graphed exclusively in its own figure. Each group is discussed in the next sub-sections, but in the first sub-section, we define some useful quantities that will be used by many of the metrics.

\subsection{Useful quantities}

\subsubsection{\executiontime}

\executiontime means the time taken by a given CUDA kernel or a group of CUDA kernels to execute.
The \executiontime is one of the few quantities that is not measured through profiling and instead measured using CUDA events. The CUDA Runtime API Reference Guide \cite{cudareference} gives the impression that CUDA events can measure execution time on GPU with a resolution of 0.5 microseconds,\footnote{Refer to the documentation of function \texttt{cudaEventElapsedTime()} in \url{https://docs.nvidia.com/cuda/cuda-runtime-api/group__CUDART__EVENT.html##group__CUDART__EVENT} } but warns that the practice can induce some noise in the measurement. Such noise was evident in our experimental setup, and we mitigate the error by measured the execution duration of the CUDA kernel/s executed serially twenty times at one go using identical input. We repeat experiments three times and pick the lowest reported execution time as the measured value of the \executiontimestop

\subsubsection{\griddedrecords}

Definition of \griddedrecords slightly varies from when used in the context of a Gridder or Pruner.

A \textit{Gridded Record} for a Gridder is that multi-polarised record presented to the Gridder after full compression is applied. Therefore in a Gridder context, \griddedrecords is equal to what is given in Table \ref{tab:methodology:compressedinputecords}.

For the Pruners, a Gridded Record is one single-polarised pixel in the multi-polarised input grid. Therefore an input multi-polarised grid of dimensions $G \times H \times N_{\text{pol}}$ provides $GH\cdot N_{\text{pol}}$ \griddedrecordsstop

Note that the term record is equivalent to the term  \textit{Gridded Record} in this thesis.

\subsection{The Optimal Solution Performance Group }
\label{sec:methodology:optimalsolutionsperformancemetrics}
Performance Metrics in the  Optimal Solution Performance Group are used to measure aspects of Performance of a given optimal solution. 
\subsubsection{\gridrate}
The \gridrate is the main metric measuring Performance of Gridders and Pruners. It measures how fast a Gridder or Pruner is to execute in a given experiment using Equation \ref{equ:gridrate}:
\begin{equation}
\label{equ:gridrate}
    \text{\gridrate}=\frac{\texttt{Gridded Records}}{\executiontime}
\end{equation}

All measurements for the \gridrate are given in \texttt{Giga Records per second (GRecs/s)}.

\subsubsection{\efficiency}

\efficiency measures how efficient a Gridder is to tap the GPU's arithmetic power in order to make grid point updates. We measure this metric only for the Maximum Performance Experiments of the Convolutional Gridder. 

A record updates $S^2$ multi-polarised grid points when gridded by the Convolutional Gridder, and every update consumes four floating-point operations (FLOPs) for each polarisation. Therefore, \efficiency is computed by Equation \ref{equ:2dgridding:efficiency}. We remind the reader that $S$ denotes support of the GCF as defined in Section \ref{sec:maths:gcfform} and that in all our experiments such support is set to six ($S=6$) since we are using the Prolate Spheroidal of order one as per Section \ref{sec:maths:gcfform}.

\begin{equation}
\label{equ:2dgridding:efficiency}
\texttt{Efficiency}=\frac{4S^2\cdot \gridrate \cdot N_{\text{pol}}}{\texttt{P100 Maximum FLOPS}}
\end{equation}

Values for the \texttt{P100 Maximum FLOPS} are available in Table \ref{tab:methodology:p100specs}.

\subsubsection{\optimalityfactor}
The \optimalityfactor is a ratio given in percentage defined by Equation \ref{equ:methodology:optimalityfactor}. It measures how optimal the solution is in a given experiment.
\begin{equation}
\label{equ:methodology:optimalityfactor}
    \optimalityfactor=\frac{\gridrate}{\bestgridrate}\cdot 100\%
\end{equation}

where \bestgridrate is a Performance Metric defined in Section \ref{sec:methodology:perfromancemetricsbrute}.

Note that the \optimalityfactor is not measured for the Maximum Performance Experiments. 

\subsubsection{\computerate}
We want to measure how much compute is required on average to grid a record in the Gridders and Pruners. We do so by measuring the average number of warp-level non-memory related instructions executed per record, as given by Equation \ref{equ:methodology:computerate}.

\begin{equation}
\label{equ:methodology:computerate}
    \computerate=\frac{\gpumetric{inst_executed} - \texttt{Total Executed Memory Instructions}}{\texttt{Gridded Records}}
\end{equation}

where the \texttt{Total Executed Memory Instructions} is equal to the summation of all the P100 metrics listed in Table \ref{tab:methodology:memorymetrics}

\begin{table}[]
    \centering
    \begin{tabular}{l}
    \toprule
    Metric \\
    \midrule
    \gpumetric{inst\_executed\_global\_loads} \\
    \gpumetric{inst\_executed\_local\_loads}\\
    \gpumetric{inst\_executed\_shared\_loads}\\
    \gpumetric{inst\_executed\_surface\_loads}\\
    \gpumetric{inst\_executed\_global\_stores}\\
    \gpumetric{inst\_executed\_local\_stores}\\
    \gpumetric{inst\_executed\_shared\_stores}\\
    \gpumetric{inst\_executed\_surface\_stores}\\
    \gpumetric{inst\_executed\_global\_atomics}\\
    \gpumetric{inst\_executed\_global\_reductions}\\
    \gpumetric{inst\_executed\_surface\_atomics}\\
    \gpumetric{inst\_executed\_surface\_reductions}\\
    \gpumetric{inst\_executed\_shared\_atomics}\\ \gpumetric{inst\_executed\_tex\_ops}  \\
    \bottomrule
    \end{tabular}
    \caption[List of profiler metrics reporting on memory related  executed instructions.] {Table listing all metrics reported by the profiler that report memory related executed instructions.}
    \label{tab:methodology:memorymetrics}
\end{table}

Care must be taken when interpreting this metric vis-a-vis Performance, since the metric is void from various factors that affect Performance, such as the different amount of cycles different instructions need to execute as shown in Table 3 of the Programming Guide \cite{NvidiaProgGuide}. 

Measurements of the \computerate are given in \texttt{Instructions per record} (\texttt{Inst/rec}). 

\subsubsection {\ltexrate and \globalhitrate}
In the implementations of the Convolutional Gridder and Hybrid Gridder, we would like the P100 caching system to deliver GCF Data from the cache-store (which will cause a HIT on the Cache) rather than retrieving it from Device Memory (which is a MISS on the two caches).
We use the metrics \ltexrate and \globalhitrate to measure the HIT rate of the cache system. The  \globalhitrate gives the HIT rate of the Unified Cache for transactions related to loading, while the \ltexrate gives the HIT rate of transactions at L2 level which will see traffic that did not HIT on the Unified Cache.
The two metrics are expresses as ratios with values between 0 (everything was a MISS) and 1 (everything was a HIT). It is important to note that these two metrics, factor in the loading of input record data, which in general is expected to MISS on the Unified and L2 Cache.

\subsubsection{\commitrate}
We measure grid committing via the \commitrate metric defined in Equation \ref{equ:methodology:commitrate}.   

\begin{equation}
\label{equ:methodology:commitrate}
\text{\commitrate}=\frac{\text{\gpumetric{atomic_transactions}}}{\texttt{Gridded Records}}
\end{equation}

\gpumetric{atomic_transactions} is equal to the total atomic transactions made during the profiled CUDA kernel execution.  Since the metric measures transactions, it partially factors in coalescence and memory alignment, meaning, that a lack of the coalescence requirement for optimal memory access patterns will increase the \gpumetric{atomic_transactions.}  

All measurements for the \commitrate are given in \texttt{Transactions per Gridded Record (Trans/Rec)}.

We need to report unexpected behaviour in the \gpumetric{atomic_transactions} metric. We did an experiment whereby a CUDA kernel is executed with just one warp and commits only once in Single-Precision to global memory using an optimal access pattern where the transactions generated are not adjacent in memory to each other. In repeated execution of this experiment  \texttt{nvprof} aperiodically reports \gpumetric{atomic_transactions} with a value of 6 rather than 4. We have filed a bug report on the matter, but up to the time of writing this thesis, we had no clarification if this is a bug or something else. Note that we did not reproduce such behaviour when repeating the experiment with Double-Precision.

\subsubsection{\dramreadtransactionsrate}
The \dramreadtransactionsrate measures how many read transactions are made to the device memory per gridded record as defined by Equation \ref{equ:methodology:dramreadtransactionsrate}. Any read transaction that HITs the cache system is not counted.     

\begin{equation}
\label{equ:methodology:dramreadtransactionsrate}
    \text{\dramreadtransactionsrate}=\frac{\text{\gpumetric{dram\_read\_transactions}}}{\texttt{Gridded Records}}
\end{equation}

All measurements for the \dramreadtransactionsrate are given in \texttt{Transactions per Gridded Record (Trans/rec)}.

\subsection{The Utilisation Group}
\label{sec:methodology:performancemetricutilisation}
Performance Metrics in the  Utilisation Group measure the utilisation of various modules or group of modules in the P100. For better readability of figures that plot the Utilisation group the \gridrate metric is also plotted in the said figures.

\subsubsection{\computeutilisation}

In this thesis, we define the \computeutilisation metric as the ratio between the executed instructions related to compute in a given CUDA kernel execution against the theoretical maximum instructions the P100 could execute. Equation \ref{equ:methodology:computeutilisation} shows how the metric is calculated.

\begin{equation}
\label{equ:methodology:computeutilisation}
\computeutilisation=\frac{\computerate \times \griddedrecords} {\texttt{MaxIPC\subscript{SM}}\times\texttt{N\subscript{SM}} \times \lambda \times \tau } \times 100
\end{equation}

where $\tau$ is the \executiontimecomma while \texttt{MaxIPC\subscript{SM}}, \texttt{N\subscript{SM}}, \texttt{$\lambda$} are given in Table \ref{tab:methodology:p100specs}. 

We note that the Visual Profiler provides a measurement similar to our \computeutilisation Performance Metric. Unfortunately, the measurement is only available in a bar chart, and we were unable to find trustworthy documentation on the metric so that we can reproduce the measurement in our plots. This issue forced us to define our own \computeutilisation metric, on which we did factual verification that it tracks well the metric presented in the Visual Profiler.  

\subsubsection{\texutilisation}
The \texutilisation metric measures the utilisation of the Unified Cache expressed in terms of the peak utilisation. The measurement is an integer ranging from 0 to 10, where 0 implies the Unified Cache was idle while 10 implies full utilisation.

\subsubsection{\ltwoutilisation}
The \ltwoutilisation metric measures the utilisation of the L2 Cache expressed in terms of the peak utilisation. The measurement is an integer ranging from 0 to 10, where 0 implies the L2 Cache was idle while 10 implies full utilisation.

\subsubsection{\dramutilisation}
The \dramutilisation metric measures the utilisation of the device memory, expressed in terms of the peak utilisation. The measurement is an integer ranging from 0 to 10, where 0 implies the device memory was idle while 10 implies full utilisation.

\subsubsection{\sharedutilisation}
The \sharedutilisation metric measures the utilisation of the shared memory, expressed in terms of the peak utilisation. The measurement is an integer ranging from 0 to 10, where 0 implies shared memory was idle while 10 implies full utilisation.

\subsection{The Brute Force Search Group}
\label{sec:methodology:perfromancemetricsbrute}
Performance Metrics in the  Brute Force Search Group are metrics measured through a Brute Force Search.

\subsubsection{\bestgridrate}

The \bestgridrate is equal to the \gridrate of the most performant solution considered by the Brute Force Search for a given combination of the Simulation Parameters. In the Maximum Performance Experiments, since the Brute Force Search is not done, the \bestgridrate is equal to the \gridrate.

\subsubsection{\maxbestgridrate}
The \maxbestgridrate is equal to the \gridrate of the most performant solution considered by the Brute Force Search for a given combination of the Simulation Parameters excluding the Ordering Mode.

This metric is not plotted anywhere, but we use it to calculate other metrics.

\subsubsection{\polgain}
\polgain measures how much we gain by having the Gridder handling a number of polarisations together. It is defined by Equation \ref{equ:methodology:polgain}.

\begin{equation}
\label{equ:methodology:polgain}
\polgain=\frac{\bestgridrate\times \texttt{N_{\text{pol}}}}{\bestgridrate\text{ at }\texttt{N_{\text{pol}}=1}}
\end{equation}

\subsubsection{\wsplitoptimalityfactor{x}}

The Convolutional and Hybrid Gridders split the warp into independent Sub-Warp Gridders counted by $W_{\text{split}}$. $W_{\text{split}}$ is a Tuning Parameter which we call the \textit{Warp Split Factor}. We use the metric \wsplitoptimalityfactor{x} to measure the effect of $W_{\text{split}}$ on the Performance of the Gridders. \texttt{x} is an integer with possible values from 1 to 5 that represents the value of $W_{\text{split}}$ the metric is measuring for. The metric is measured for a given combination of Simulation Parameters, whereby through Brute Force Search the maximum \gridrate of all runs with $W_{\text{split}}=\texttt{x}$ is found. This maximum is compared with the \bestgridrate as per Equation \ref{equ:methodology:wsplitoptimal}.

\begin{multline}
    \wsplitoptimalityfactor{x}=\\
    \frac{\text{\gridrate for fastest run with } W_{\text{split}}=\texttt{x}}{\bestgridrate}\times 100\%
\label{equ:methodology:wsplitoptimal}
\end{multline}

\subsection{The Gridder Advantage Group}
\label{sec:methodology:performancemetricgridderadvantage}

The \gridderadvantage of the Hybrid or the NN Gridder measures by what factor the stated Gridder performs better than the Convolutional Gridder, for a given combination of the Simulation Parameters excluding the Ordering Mode. It is defined by Equation \ref{equ:methodology:advantage}.
\begin{multline}
\label{equ:methodology:advantage}
\gridderadvantage=\\ 
\frac{\maxbestgridrate \text{of Hybrid or NN Gridder} }{\maxbestgridrate \text{ of the Convolutional Gridder}}
\end{multline}

When analysing the NN Gridder, we will also measure the Gridder Advantage of the NN Gridder over the Hybrid Gridder. In this case, the Performance Metric will be called \gridderadvantageonhybrid and calculated by Equation \ref{equ:methodology:advantagehybrid}.  
\begin{multline}
\label{equ:methodology:advantagehybrid}
\gridderadvantageonhybrid =\\ \frac{\maxbestgridrate \text{ of the NN Gridder} }{\maxbestgridrate \text{ of the Hybrid Gridder}}
\end{multline}

\subsection{The Pruning Gain Group}
\label{sec:methodology:pruninggainmetrics}
Performance Metrics in the FFT Pruning Gain Group measure Performance gain caused by the Pruning Step in the inversion of the grid to an image. We only use this group of metrics in Chapter \ref{chap:pruning}.

Let us define some quantities which will be used to define Performance Metrics in this group. Let Grid \texttt{Inp} be the input for the Pruning Step. The grid will be downsampled to Grid \texttt{Out1}  by a Column Pruner\footnote{Details of the Column and Row Pruners are given in Chapter \ref{chap:pruning}.}. In Pruned NN Interpolation Grid \texttt{Out1} will be further reduced to Grid \texttt{Out2} by a Row Pruner. The Output Grids \texttt{Out1} or \texttt{Out2} are then IFFTed and corrected. We define the quantity \metric{FFT\subscript{Inp}} as the execution time taken to IFFT Grid \texttt{Inp}. We also define \metric{FFT\subscript{Out1}} and \metric{FFT\subscript{Out2}} as the execution time taken to IFFT Grid \texttt{Out1} and \texttt{Out2} respectively.

Note that in this group of metrics, we do not factor in the execution time of the Correction Step, since our implementation of the Corrector Step has an invariant execution time over the implementations of the different algorithms.   

\subsubsection{\columnpruninggain}
The \columnpruninggain measures the gain in Performance in the Fourier inversion when applying the Column Pruner as defined in Equation \ref{equ:methodology:colpruninggain}.

\begin{equation}
\label{equ:methodology:colpruninggain}
    \columnpruninggain=\frac{\metric{FFT\subscript{Inp}}}{\metric{Pruner\subscript{Col}}+\metric{FFT\subscript{Out1}}}
\end{equation}
\subsubsection{\rowpruninggain}
The \rowpruninggain measures the gain in Performance in the Fourier inversion on applying the Row Pruner as defined in Equation \ref{equ:methodology:rowpruninggain}.

\begin{equation}
\label{equ:methodology:rowpruninggain}
    \rowpruninggain=\frac{\metric{FFT\subscript{Out1}}}{\metric{Pruner\subscript{Row}}+\metric{FFT\subscript{Out2}}}
\end{equation}
\subsubsection{\accumulatedpruninggain}
The \accumulatedpruninggain measures the gain in Performance in the Fourier inversion on applying the Column Pruner and subsequently the Row Pruner as defined in Equation \ref{equ:methodology:accumulatepruninggain}.

\begin{equation}
\label{equ:methodology:accumulatepruninggain}
    \rowpruninggain=\frac{\metric{FFT\subscript{Inp}}}{\metric{Pruner\subscript{Col}}+\metric{Pruner\subscript{Row}}+\metric{FFT\subscript{Out2}}}
\end{equation}
\subsubsection{\columnprunerfootprint}
The \columnprunerfootprint is the ratio between the execution time of the Column Pruner and the total execution of the Fourier inversion of Grid \texttt{Inp} to Grid \texttt{Out1} as defined by Equation \ref{equ:methodology:colfootprint}.

\begin{equation}
\label{equ:methodology:colfootprint}
    \columnprunerfootprint=\frac{\metric{Pruner\subscript{Col}}}{\metric{Pruner\subscript{Col}}+\metric{FFT\subscript{Out1}}}
\end{equation}

\section{Conclusion}
This chapter introduced all the work we are to present in the subsequent chapters. Terminology was defined, and various concepts that are used in subsequent chapters were discussed.

CUDA programming and GPUs are the first topics discussed and proceeded with common information to all implementations. Afterwards, tuning and the experimental setup were discussed, and the chapter concludes with a list of Performance metrics measured in the many experiments we are to report for studying Performance.

\chapter{The Convolutional Gridder}
\label{chap:2dgridding}
This chapter discusses and provides analyses to our original implementation of the Convolutional Gridder for the P100. Our Convolutional Gridder implementation is based on a moving window strategy proposed by Romein \cite{Romein2012}, which we refer to as Romein's strategy. We will modify and enhance the original implementation by Romein \cite{Romein2012} to make it perform at its best for GCFs of support $6\times 6$.

This chapter is organised as follows: The first section reviews Romein's strategy and related literature, while the second section details our novel implementation of the Convolutional Gridder. Right afterwards, in another section, we disclose and discuss results related to Brute Force Search and progress to yet another section where we report more detailed results, together with an in-depth analysis. We conclude in Section \ref{sec:2dgridding:conclusion} by giving a summary of the main results. 

We remind the reader that validation of our Convolutional Gridding implementation is discussed in Chapter \ref{chap:comparative}, including that of the Convolutional Gridder.

\section{Romein's moving window strategy}

Enhanced by Muscat \cite{Muscat2014} and Merry \cite {Merry2016}, Romein's strategy is the fastest algorithm which we are aware of, able to grid on GPUs. It acquires high-performance by taking advantage of the trajectory of records for a given baseline. Neighbour records on the same trajectory are very near each other, facilitating the accumulation of grid-point updates on the fast on-chip registers rather than directly added to the grid via the much slower global memory.

\subsection{High-level description}
We can view Romein's strategy as an enhanced version of a simple scatter strategy (NVIDIA \cite{Nvidia2018}). In an easy simple multi-threaded scatter strategy, an independent group of threads equal to the convolution function affected pixel area ($S\times S$ pixels), grid full records. Each thread sequentially reads the coordinates and value of records to grid from a stream, calculates the convolution for a given pixel and immediately commits (adds) to the grid. Since a GPU is a highly parallel device, there will be many groups of such threads sharing the workload, and collisions during committing are probable, mandating the need for atomic additions (reductions) to evade race conditions.

The main disadvantage of a scatter strategy is its intense grid committing, whereby every record will generate $S\times S$ complex atomic reductions. Romein's strategy reduces such atomic reductions by giving the threads responsibility of a fixed set of points on the grid, enabling the possibility of accumulating updates rather than committing for every record. Such a strategy works for Radio Interferometric Imaging because a baseline measures Visibility records over a trajectory on the $uv$-grid, and the records tend to be near each other, making said accumulations probable. 

\subsubsection{Thread configuration}
\label{sec:2dgridding:threadconf}
\begin{figure}

\includegraphics[page=1,width=\linewidth]{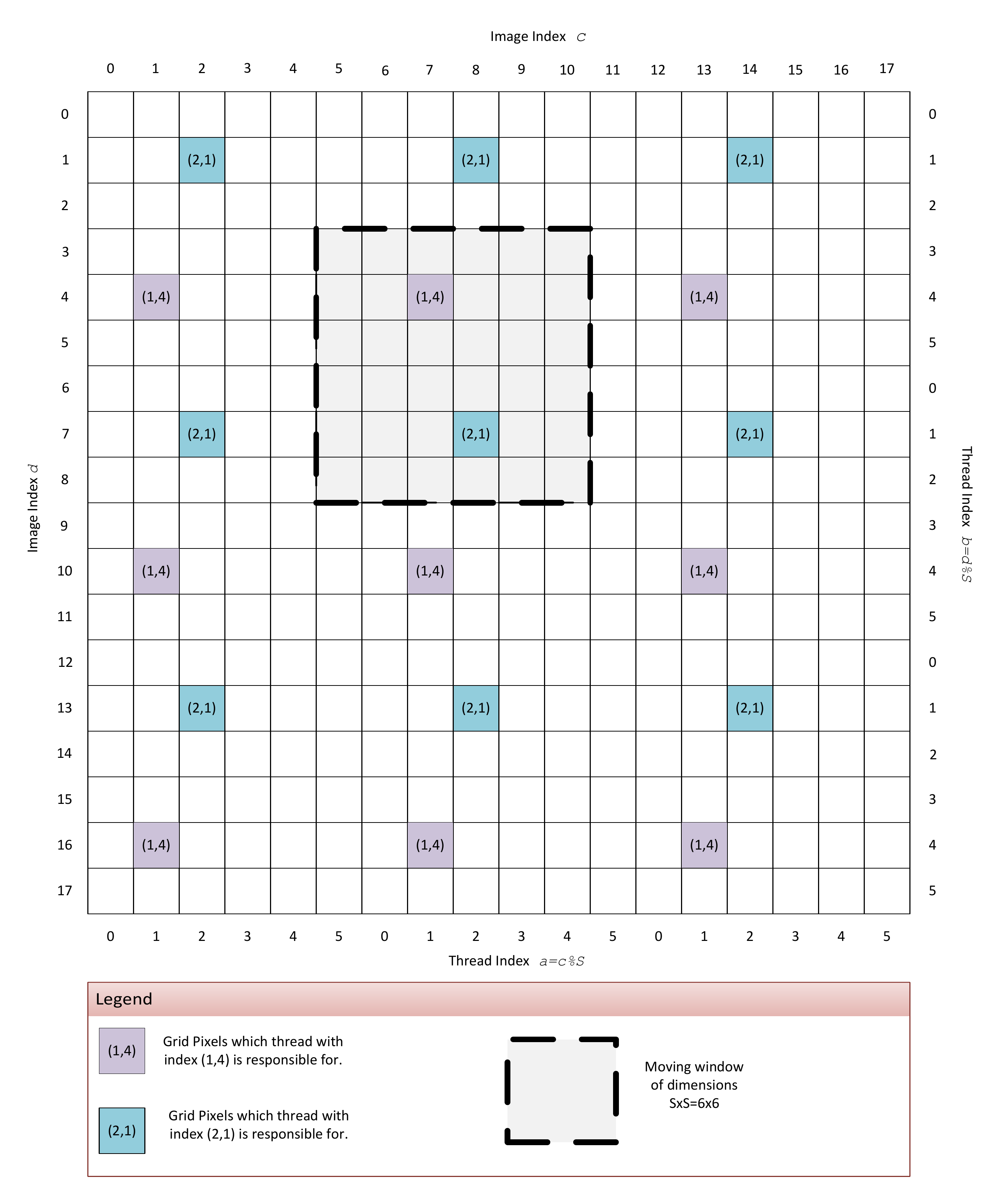}

\caption[Romein's strategy thread configuration for Convolutional Gridding]{Diagram showing the thread configuration of Romein's strategy. The diagram features a grid of size 18x18 pixels, to which a record is gridded using a GCF of size 6x6. The moving window shows the area on the grid the record will affect for gridding. A thread such as that indexed by (1,4) is responsible for updating several pixels on the grid, but only one pixel is inside the window for any given record.}
\label{fig:2dgridding:romeinmovingwindow}
\end{figure}

Figure \ref{fig:2dgridding:romeinmovingwindow} illustrates the thread configuration for Romein's strategy.
Given a grid of dimensions $N\times N$ pixels, $S\times S$ threads are allocated for gridding a stream of records. Let us index each thread using a two-dimensional index $(a,b)$ where $0\le a,b <S$. We also index the grid using another two-dimensional index $(c,d)$ where $0\le c,d< N $. A thread is given the responsibility to accumulate updates on grid pixels with index $(a+k_1S,b+k_2S)$ where $k_1,k_2 \in \mathbb{Z}$, which ensures that the thread updates one and only one pixel for a given record. Given records on a given baseline trajectory are fed to the strategy one after the other, the given thread will likely need to update the same grid point for neighbouring records, making it doable for the thread to accumulate the updates internally. When the neighbouring records move enough such that the given thread has to update a different grid point, then and only then the thread commits to the grid using atomic reductions.

In some literature (NVIDIA \cite{Nvidia2018}), Romein's strategy is referred to as a moving window strategy since there is accumulation for grid pixels within a window of size $S\times S$ moving in-sync with the $(u,v)$ co-ordinates of the input record stream. 

\subsection{Past work and results}

Before discussing our implementation of the Convolutional Gridder, we here review published work related to Romein's strategy.

Romein \cite{Romein2012} introduces the strategy and in his analysis, considers a quad-polarised ($N_{\text{pol}} = 4$) LOFAR observation gridded using Single-Precision with complex-valued GCFs and support equal or greater to $16 \times 16$.  Input records were not compressed.

In Muscat \cite{Muscat2014}, we introduced trajectory compression, and analysed a modified Convolutional Gridder implementation based on Romein's \cite{Romein2012} implementation.  We considered various  values of $N_{\text{pol}}$, and delivered the first analyses of gridding with a  complex $6\times6$ GCF. All analyses were made using Single-Precision, with no consideration for Double-Precision.

Merry \cite{Merry2016} proposed a performance enhancement to Romein's implementation by combining several threads into one. Such a combination of threads is known as thread coarsening. Strategically, Merry \cite{Merry2016} chose to combine neighbour thread on the grid in a bid to reduce Logic. He reports a best case of 90\% improvement for a large complex GCF of dimensions $128\times 128$, but does not deliver any results for GCFs of support near or equal to $6\times 6$. However, his results hint that if there is any gain using his implementation of thread coarsening, it should be small due to the introduction of some waste.

In an SDP Memo, NVIDIA \cite{Nvidia2018} compared Romein's strategy with a scatter and gather strategy and delivered a Performance analysis for Single and Double Precision gridding over the NVIDIA Tesla K40 GPU Accelerator. NVIDIA reported Romein's strategy to perform best for small GCFs (smaller or equal to $32 \times 32$) and lacked behind the gather strategy for larger GCFs. No study for the $6 \times 6$ GCFs was presented. NVIDIA \cite{Nvidia2018} also argued that thread coarsening should increase Performance of Romein's strategy.

We finalise this short review by pointing out that there are other studies that focus on alternative Convolutional Gridding deployments on GPU not based on Romein's strategy. In general, they focus on complex GCFs with varying support. Examples of such studies are  Edgar \etal \cite{Edgar2010}, Luo \etal \cite{Luo2018}, Lao \etal \cite{Lao2019}, van Amesfoort \etal \cite{Amesfoort2009}, Dingle \etal \cite{Dingle2017} and Antao  \cite{Antao2018}.    

\subsection{Contribution}

To the best of our knowledge, up to the writing of this thesis, there has been no published study of a  Convolutional Gridder using Romein's strategy that runs over GPUs using Single or Double Precision with a real-valued GCF of small dimensions such as $6\times6$.  This chapter is dedicated to such a Gridder.

\section{Implementation details}
\label{sec:2dgridding:implimentation}

We now explain our novel implementation of the Convolution Gridder based on Romein's strategy.  The implementation targets gridding using a real-valued GCF of support $6\times6$,  and it can easily be modified to cater for real or complex valued GCFs with support of $8\times 8$.

\subsection{CUDA grid layout}
\label{sec:methodology:cudagridlayout}
The CUDA grid layout is subject to tuning via a Brute Force Search discussed in Section \ref{sec:methodology:tuning}, whereby we define two Tuning Parameters, $B_{\text{Warps}}$ and $G_{\text{Blocks}}$, to control the way warps are organised over the whole CUDA Grid. $B_{\text{Warps}}$  configures the number of warps CUDA blocks contain, while $G_{\text{Blocks}}$ defines the maximum number of blocks the Convolutional Gridder CUDA kernel gets launched with. In case $G_{\text{Blocks}}$ is too large for the workload given, the CUDA kernel is launched with fewer blocks than defined by $G_{\text{Blocks}}$. 

The value of $B_{\text{Warps}}$   impacts performance. As clarified in Section \ref{sec:2dgridding:sharedmemory}, $B_{\text{Warps}}$ has a control on Occupancy and shared memory size that in turn controls the choice of $X_{\text{Load}}$ (discussed in Section \ref{sec:2dgridding:sharedmemory}). 

The value of $G_{\text{Blocks}}$ can also impact Performance. While in general it is recommended to have $G_{\text{Blocks}}$ as high as possible, our Convolutional Gridder can benefit with an increase in Performance by lowering $G_{\text{Blocks}}$ (and also lowering $B_{\text{Warps}}$)  since such a decrease can reduce grid committing.

\subsection{Sub-Warp Gridders and thread coarsening}

In our implementation, we split the warp into several Sub-Warp Gridders equal to the Tuning Parameter Warp Splitting Factor ($W_{\text{split}}$), with possible values of 1,2,4 and 5. The Sub-Warp Gridder\footnote{We introduced the Sub-Warp Gridder in Section \ref{sec:methodology:griddingstep}.} is an independent Gridder that grids in full a group of neighbour records in the input streams. It also loads input data using its CUDA threads, eliminating the need for synchronisation between warps.

The use of Sub-Warp Gridders makes our implementation much different from that of Romein \cite{Romein2012}. Romein focused on gridding with GCFs considerably larger than $6\times 6$, and as a consequence, he chose a CUDA block of warps to make the independent Gridder.

Sub-Warp Gridders have fewer CUDA threads available than Romein's strategy mandates, and to reduce the number of needed threads, we merge threads together through thread coarsening. The merged threads are programmatically interleaved on the CUDA thread, to increase Instruction Level Parallelism. Such Interleaving caused some Logic to become redundant, and therefore removed. 

We do stress that our adoption of thread coarsening is different from that of Merry \cite{Merry2016}. Merry focused on reducing Logic, which requires the GCF to be padded, differing from our implementation.

\begin{figure}
\includegraphics[page=1,width=\linewidth,trim=0.5cm 0.5cm 0.5cm 0.5cm,clip]{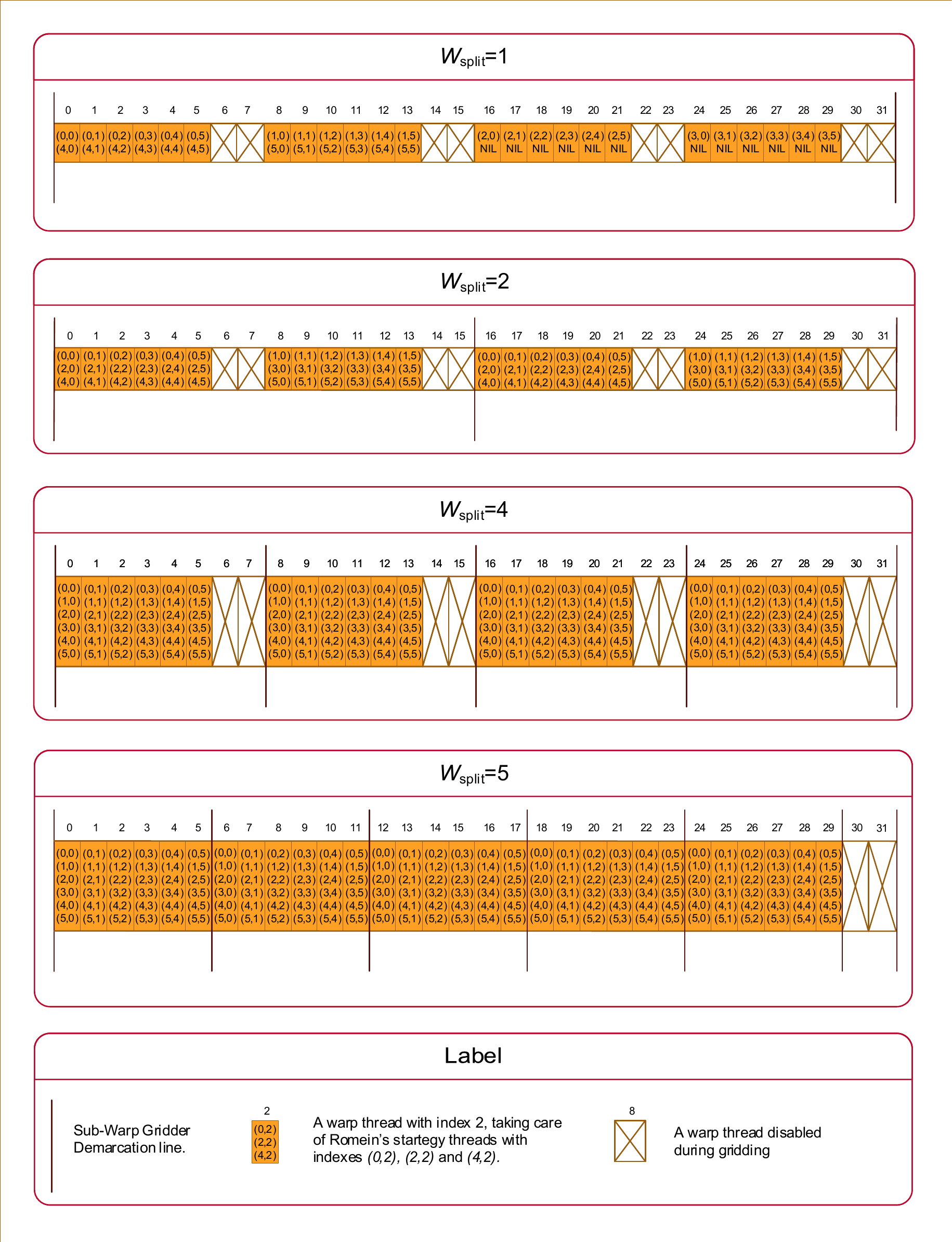}
\caption[CUDA threads layout for the Convolutional Gridder.]{Diagram showing how CUDA threads are organised within a warp in the Convolutional Gridder for the different values of the Warp Split Factor ($W_{\text{split}}$).}
\label{fig:2dgridding:warpgridder}
\end{figure}

Figure \ref{fig:2dgridding:warpgridder} illustrates the way threads are merged and organised over the warp for different values of $W_{\text{split}}$. One easily notices that some of the CUDA threads are disabled, which is unavoidable for $6 \times 6$ GCFs.\footnote{The issue vanishes when modifying the implementation for $8\times 8$ GCF using the same layout strategy of $6\times 6$ GCFs.} In the case of $W_{\text{split}} = 1$, two CUDA threads are disabled for every eight CUDA threads, and half of the active CUDA threads only merge one thread instead of two. When the  CUDA threads are executing for the second merged thread, the CUDA threads with no second merged thread get automatically disabled. For $W_{\text{split}}$ equal to two or four, the situation is somehow better, since all active CUDA threads have merged the same number of threads, and in $W_{\text{split}} = 5$ only two CUDA threads out of thirty-two are disabled.

An increase in $W_{\text{split}}$ brings in some issues whose effects can overcome the advantages brought in by thread coarsening. First of all, an increase in thread coarsening reduces the parallelism of the Convolution Gridder. Every record requires fewer CUDA threads for gridding, giving rise for the need of larger workloads to saturate the GPU. Merry \cite{Merry2016} already raised such an issue in his work,  but in our implementation, this issue is much more taxing, since the GCF is very small. Another issue is that when threads are merged, register pressure increases, that is, a given CUDA thread requires more registers to do the job. Higher register pressure can lead to a reduction in Occupancy, and we control register pressure through the Tuning Parameter $SM$, which sets the minimum concurrent CUDA blocks actively executing on each Streaming Multiprocessor (SM), as discussed in Section \ref{sec:methodology:registers}.

A final disadvantage arising from thread coarsening and unique to our implementation, is that as $W_{\text{split}}$ increases, more shared memory is needed per warp. Shared memory is a limited resource, and it affects Occupancy too. We will discuss this argument further in the next sub-section.

\subsection{Input record data, and use of shared memory}
\label{sec:2dgridding:sharedmemory}

Input record data is split into various input streams  listed hereunder:
\begin{enumerate}
    \item A 128-bit memory aligned  index made of four 32-bit integers.
    \item One 32-bit integer type index acting as a pointer to the selected GCF data. 
    \item $N_{\text{pol}}$ streams with each stream containing Visibility data of one polarisation.
\end{enumerate}

The first and second streams are generated in the pre-processing phase  from the $(u,v)$-coordinates of the input records. They are the same indexes used in Romein's original implementation, with the only difference being that Romein choose to load the $(u,v)$-coordinates and convert them to these indexes within the Gridder itself.

Shared memory is used as an intermediate store of record data. Each Sub-Warp Gridder loops between loading $X_{\text{Load}}$ records  in shared memory and then gridding the chunk up. In stark difference with Romein’s original implementation, our Convolutional Gridder retains accumulated grid point updates in the registers while switching between gridding and data loading.

 $X_{\text{Load}}$ is a Tuning Parameter subject to our Brute Force Search. Its value, together with the values of $W_{\text{split}}$ and $B_{\text{Warps}}$ defines the size of shared memory allocated per CUDA block as per:
\begin{equation}
\label{equ:2dgridding:sharedmemorysize}
\texttt{Shared memory size per CUDA block}=\texttt{Record size}\times W_{\text{split}}\times B_{\text{Warps}}
\end{equation}

The record size in shared memory is dependent on Precision and $N_{\text{pol}}$ and is given in Table \ref{tab:2dgridding:recordchunksplit}. The size of the allocated shared memory per CUDA block can affect Occupancy, and therefore  the choice of $X_{\text{Load}}$ combined with $W_{\text{split}}$ and $B_{\text{Warps}}$ can affect Performance as they exert control on shared memory.

\begin{figure}{}
\includegraphics[page=1,width=\linewidth,trim=0.5cm 0.5cm 0.5cm 0.5cm , clip]{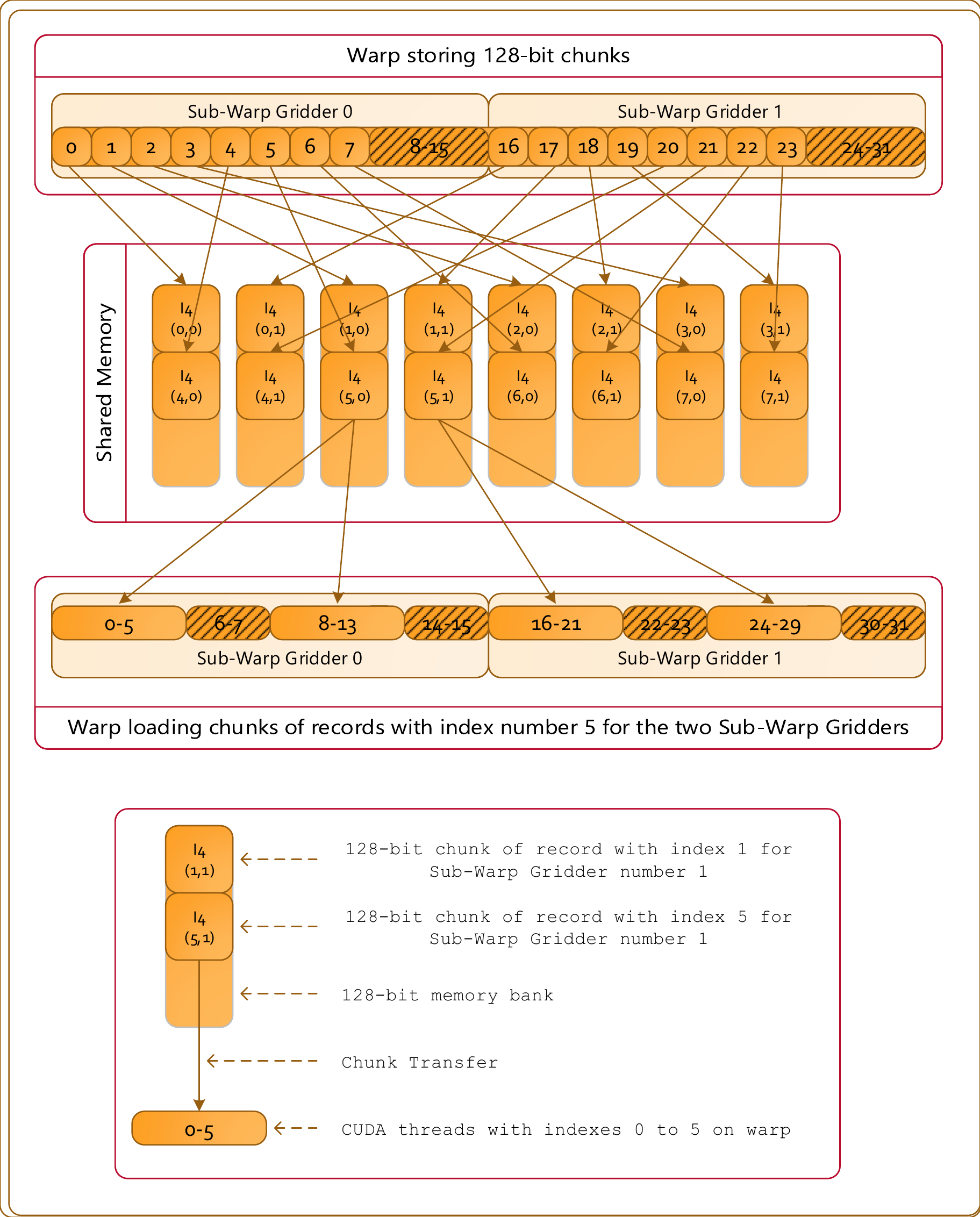}
\caption[Layout of chunks in shared memory for the Convolutional Gridder]{Figure depicting the layout of 128-bit chunks of the same type (for example I4) in shared memory for the Convolutional Gridder. The warp shown in the figure is split into two Sub-Warp Gridders ($W_{\text{split}}=2)$ with $X_{\text{Load}}=8$. With this layout, a warp request to store chunks in shared memory is executed with two transactions with each transaction utilising all the shared memory banks. Furthermore, a warp request to retrieve chunks from shared memory is executed with one transaction and no bank-conflicts.}
\label{fig:2dgridding:shared_memory_layout}
\end{figure}

Let us now explain how the record data is laid out in shared memory.
Since CUDA threads are able to load or store a maximum of 128 bits at one go, we divide a record in chunks of 128 bits, with the final chunk possibly smaller as given in Table \ref{tab:2dgridding:recordchunksplit}. Each different chunk is assembled during loading and stored in shared memory with a 128-bit store instruction. It is also retrieved from shared memory as a chunk with a 128-bit load instruction. 

Shared memory is divided in such a way that each warp has an exclusive area in shared memory where to load and store a particular chunk type. The layout of such areas is depicted in Figure \ref{fig:2dgridding:shared_memory_layout}, explaining how Sub-Warp Gridders in a given warp store and retrieve chunks. The layout ensures storage of chunks in shared memory with a good access pattern and  retrieval of chunks from shared memory with no bank conflicts.

\begin{table}[]
\centering
\begin{tabular}{@{}c@{}c@{}c@{}c@{}c@{}c@{}c@{}c@{}cc@{}}
\toprule
Precision & \rule{0.4cm}{0cm}$N_{\text{pol}}$\rule{0.4cm}{0cm} & \rule{0.7cm}{0cm} &\begin{minipage}[]{1cm}\centering C1\end{minipage} & \begin{minipage}[]{2.1cm}\centering C2\end{minipage}  & \begin{minipage}[]{2.1cm}\centering C3\end{minipage}     & \begin{minipage}[]{1cm}\centering C4\end{minipage}     & \begin{minipage}[]{1cm}\centering C5\end{minipage}    & \begin{minipage}[]{1cm}\centering C6\end{minipage} & Record size  \\ \midrule
\multirow{4}{*}{single}    & 1   &       & I4      & IC+P+V1 &             &             &             &        & 32     \\
    & 2  &        & I4      & V1+V2    & IC$^\dagger$ &             &             &         & 36    \\
    & 3   &       & I4      & IC+P+V1  & V2+V3       &             &             &      & 48        \\
    & 4   &       & I4      & V1+V2    & V3+V4       & IC$^\dagger$ &             &       & 52      \\ \midrule
\multirow{4}{*}{double}   & 1  &        & I4      & V1       & IC$^\dagger$ &             &             &      & 36         \\
   & 2     &     & I4      & V1       & V2          & IC$^\dagger$ &             &        & 52     \\
   & 3      &    & I4      & V1       & V2          & V3          & IC$^\dagger$ &        & 68     \\
    & 4      &    & I4      & V1       & V2          & V3          & V4          & IC$^\dagger$ & 84\\ \bottomrule
\end{tabular}
\caption[Splitting of records in 128-bit chunks for the Convolutional Gridder]{This table shows how record data is split in 128-bit chunks such that they can be laid out efficiently in shared memory. The last column gives the amount of space in bytes a record consumes in shared memory. \\\textbf{C\textit{x}}: Chunk No \textit{x} \\\textbf{I4}: 4-Integer Index \\ \textbf{IC}: Convolution Index \\ \textbf{IC$^\dagger$}: Chunk made up of the GCF index only. Such a chunk is only 32 bits wide \\ \textbf{P}: A 32-bit pad, needed to secure a 128-bit chunk when a Single-Precision Visibility and the Convolution Index form a chunk.\\ \textbf{V\textit{x}} Visibility value of the \textit{x}$^{th}$ polarisation. Single-Precision Visibility values are 64-bit wide while Double-Precision Visibility values are 128-bit wide.}
\label{tab:2dgridding:recordchunksplit}
\end{table}

\subsection{Loading of the GCF}

Gridding a record requires the loading of a set of GCF values from global memory. We rely on the P100 cache system to retrieve such data as fast as possible, where we use the \texttt{\_\_ldg()} function to explicitly instruct the compiler to cache GCF data in the Unified Cache. At the same time, we instruct the compiler for a default caching policy that does not cache in the Unified Cache, such that record data is not cached by the Unified Cache while being loaded. With this setup, we maximise the availability of cache memory in the Unified Cache for the storage of GCF data. 

\subsection{Output grid}
As already discussed in Section \ref{sec:methodology:griddingstep}, the Convolutional Gridder outputs a non-interleaved multi-polarised grid with its origin at the centre. 

\section{Brute Force Search Results}
\label{sec:2dgridding:bruteforce}

\begin{figure}
\includegraphics[page=1,width=\linewidth]{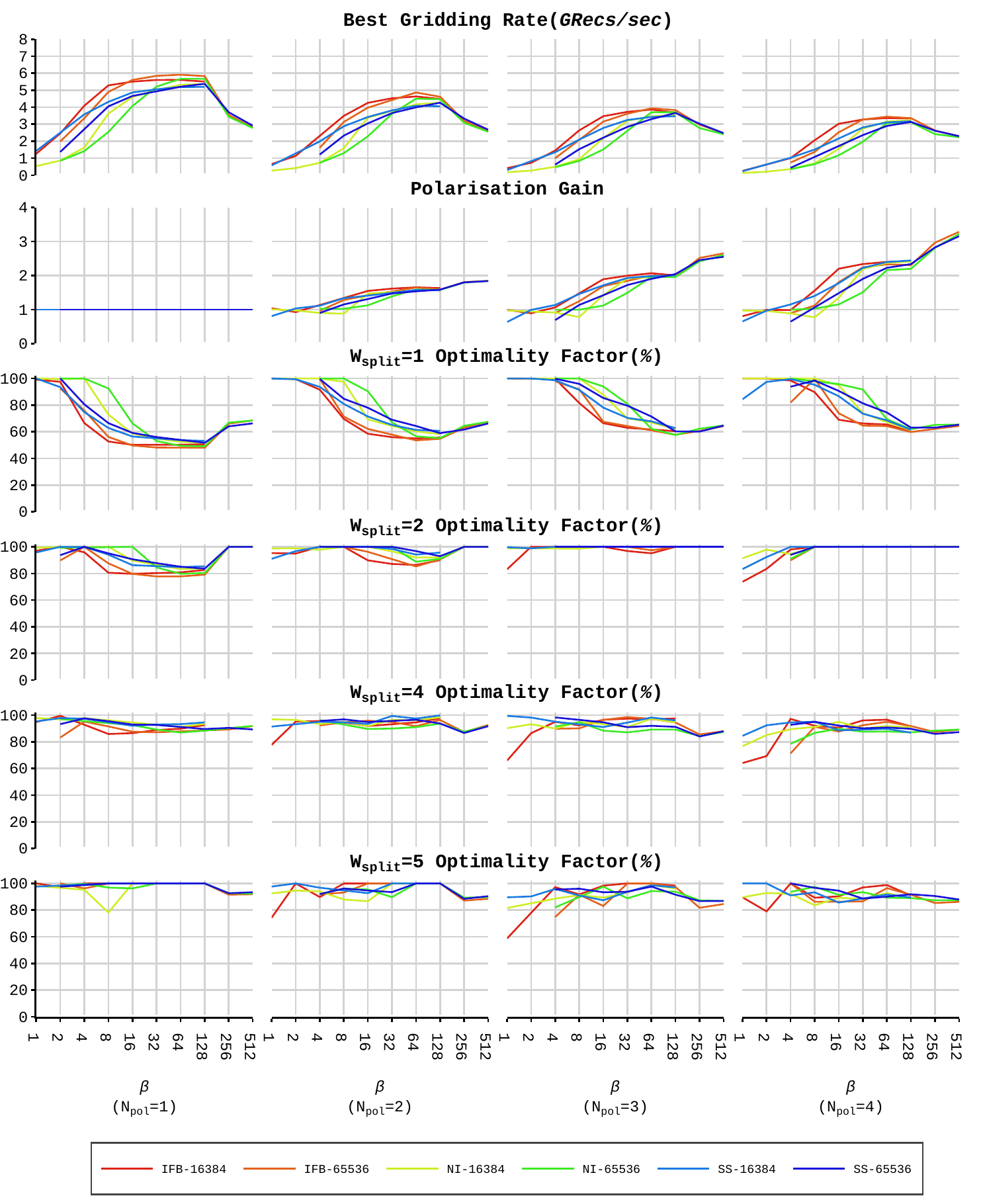}
\caption[Brute Force Search results for the Single-Precision Convolutional Gridder]{Brute Force Search results for the Single-Precision Convolutional Gridder. The layout is as described in Section \ref{sec:methodology:graphlayouts}. Performance metrics plotted in this figure are defined in Section \ref{sec:methodology:perfromancemetricsbrute}.}
\label{fig:2dgridding:convbrute_single}
\end{figure}
\begin{figure}

\includegraphics[page=1,width=\linewidth]{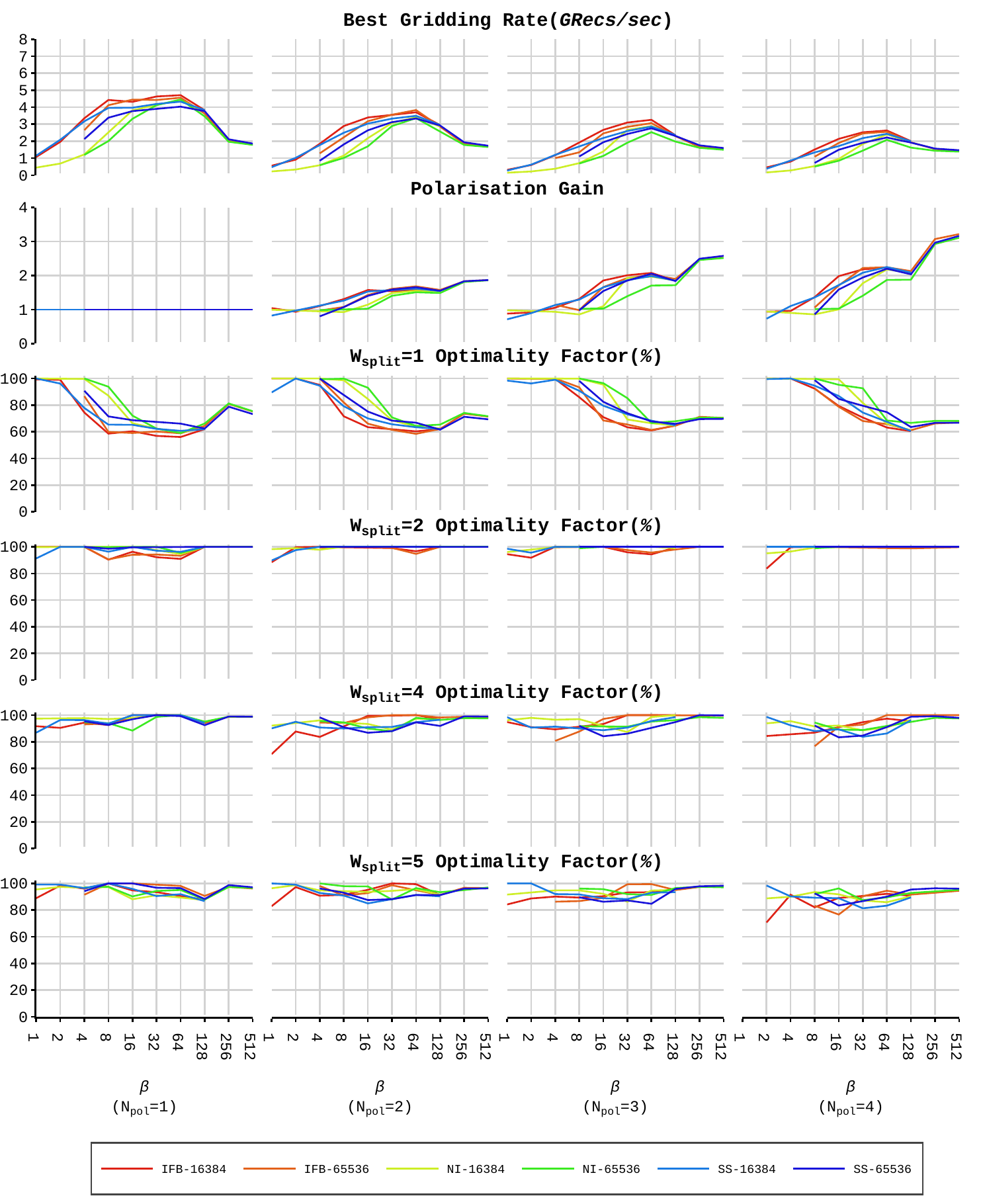}

\caption[Brute Force Search results for the Double-Precision Convolutional Gridder]{Brute Force Search results for the Double-Precision Convolutional Gridder. The layout is as described in Section \ref{sec:methodology:graphlayouts}. Performance metrics plotted in this figure are defined in Section \ref{sec:methodology:perfromancemetricsbrute}.}
\label{fig:2dgridding:convbrute_double}
\end{figure}

We execute a Brute Force Search on the Convolutional Gridder over the stated Tuning Parameters as discussed in Section \ref{sec:methodology:tuning}. Measured Brute Force Search Performance metrics described in Section \ref{sec:methodology:perfromancemetricsbrute} are plotted in Figure \ref{fig:2dgridding:convbrute_single} for Single-Precision and Figure \ref{fig:2dgridding:convbrute_double} for Double-Precision. 

From the Brute Force Search, we discovered optimal solutions which we report and analyse in Section \ref{sec:2dgridding:optimalsolutions} and Table \ref{tab:2dgridding:optimisationparameters}.

In the next sub-sections, we analyse the stated Brute Force Search results.

\subsection{\bestgridrate and \polgain}

Results of the Brute Force Search shows that the Performance of the Convolutional Gridder varies with the Ordering Mode and $\beta$ suggesting that Performance is dependent on the $uv$-profile as inputted to the Gridder, which is similar to what Merry \cite{Merry2016} observed. Performance also varies with $N_{\text{pol}}$ and Precision, which is due to a change in the required GPU. It is interesting to note that the Convolution Gridder performed relatively well in our experiments when the SS Ordering Mode was used.

\polgain is clearly dependent on $\beta$, whereby little gain or even a loss is registered for the lowest values of $\beta$, but then increases with an increased value of $\beta$. This behaviour is primarily due to grid committing, which we discuss further on in Section \ref{sec:2dgridding:gridcommiting}.

\subsection{The Warp Split Factor and Thread Coarsening}

We now discuss the effects of $W_{\text{split}}$ on Performance and remind the reader that in the Convolutional Gridder, the value of $W_{\text{split}}$ directly controls the level of thread coarsening.

From the results given in Figures \ref{fig:2dgridding:convbrute_single}  and \ref{fig:2dgridding:convbrute_double}, it is clear that except for the various Single-Precision experiments with $N_{\text{pol}}=1$, $W_{\text{split}}=2$, is in general the right optimal choice. For $W_{\text{split}}=2$, three threads are merged. The maximum increase in Performance when increasing $W_{\text{split}}$ from $1$ to $2$ is well over 60\%. On increasing further $W_{\text{split}}$ from 2 to values of 4 and 5, Performance is generally affected negatively, except for the Single-Precision, $N_{\text{pol}}=1$ experiments which generally reach optimal Performance at $W_{\text{split}}=5$.

\subsection{Optimal Solutions}
\label{sec:2dgridding:optimalsolutions}
Using Brute Force Search, we did not find any optimal solutions that are independent of $\beta$ and Ordering Mode, suggesting that tuning is dependent on the \commitratestop Such behaviour is in agreement with the work of Merry \cite{Merry2016}, whereby he observes that tuning of his Gridder is dependent on the input's $uv$-profile.

\begin{figure}
\centering
\includegraphics[page=1,scale=1]{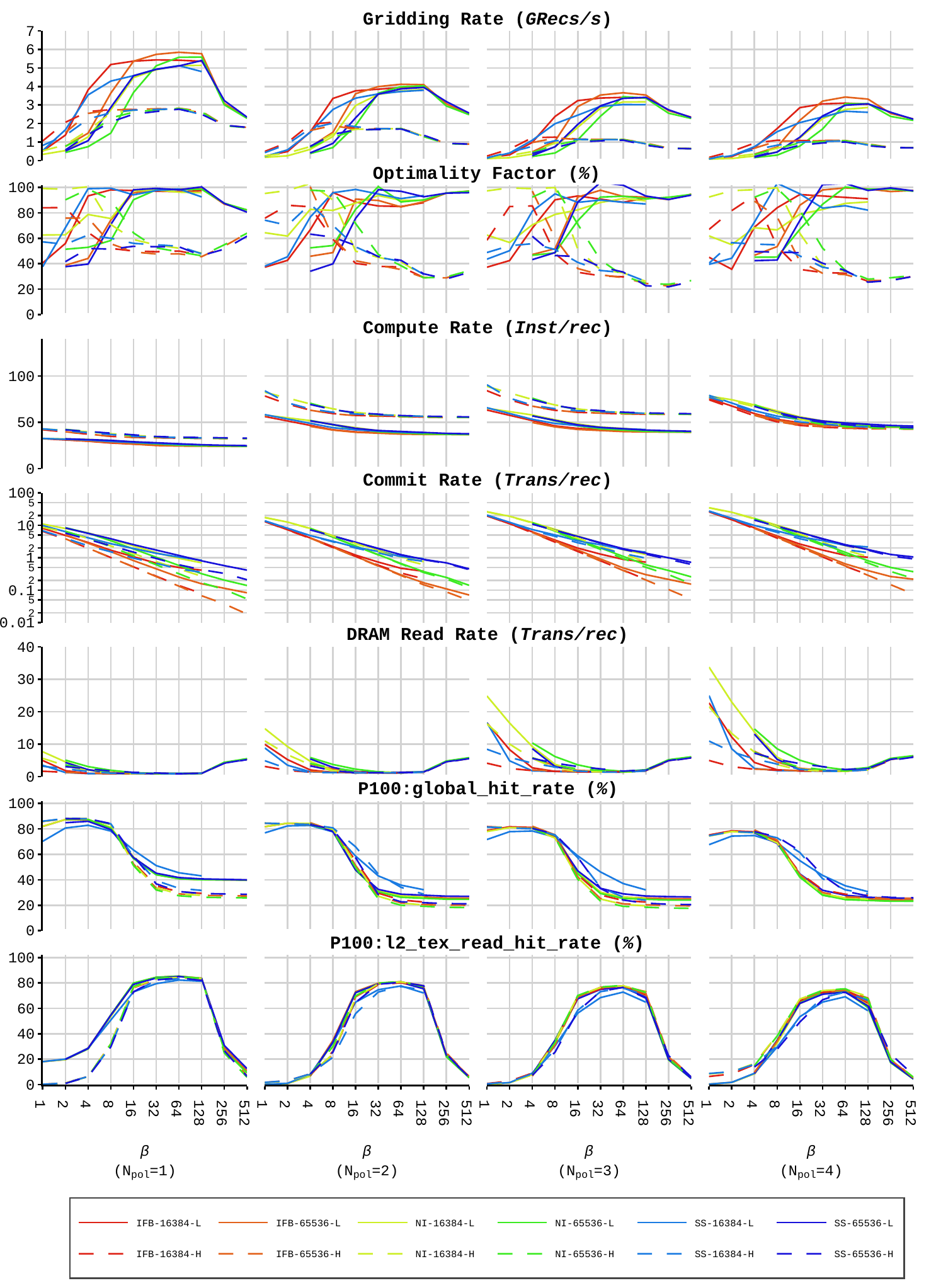}

\caption[Optimal Solutions Performance results for the Single-Precision Convolutional Gridder]{Optimal Solutions Performance results for the Single-Precision Convolutional Gridder. The layout is as described in Section \ref{sec:methodology:graphlayouts}. Performance metrics plotted in this figure are defined in Section \ref{sec:methodology:optimalsolutionsperformancemetrics}.}

\label{fig:2dgridding:conv_single}
\end{figure}

\begin{figure}

\includegraphics[page=1,width=\linewidth]{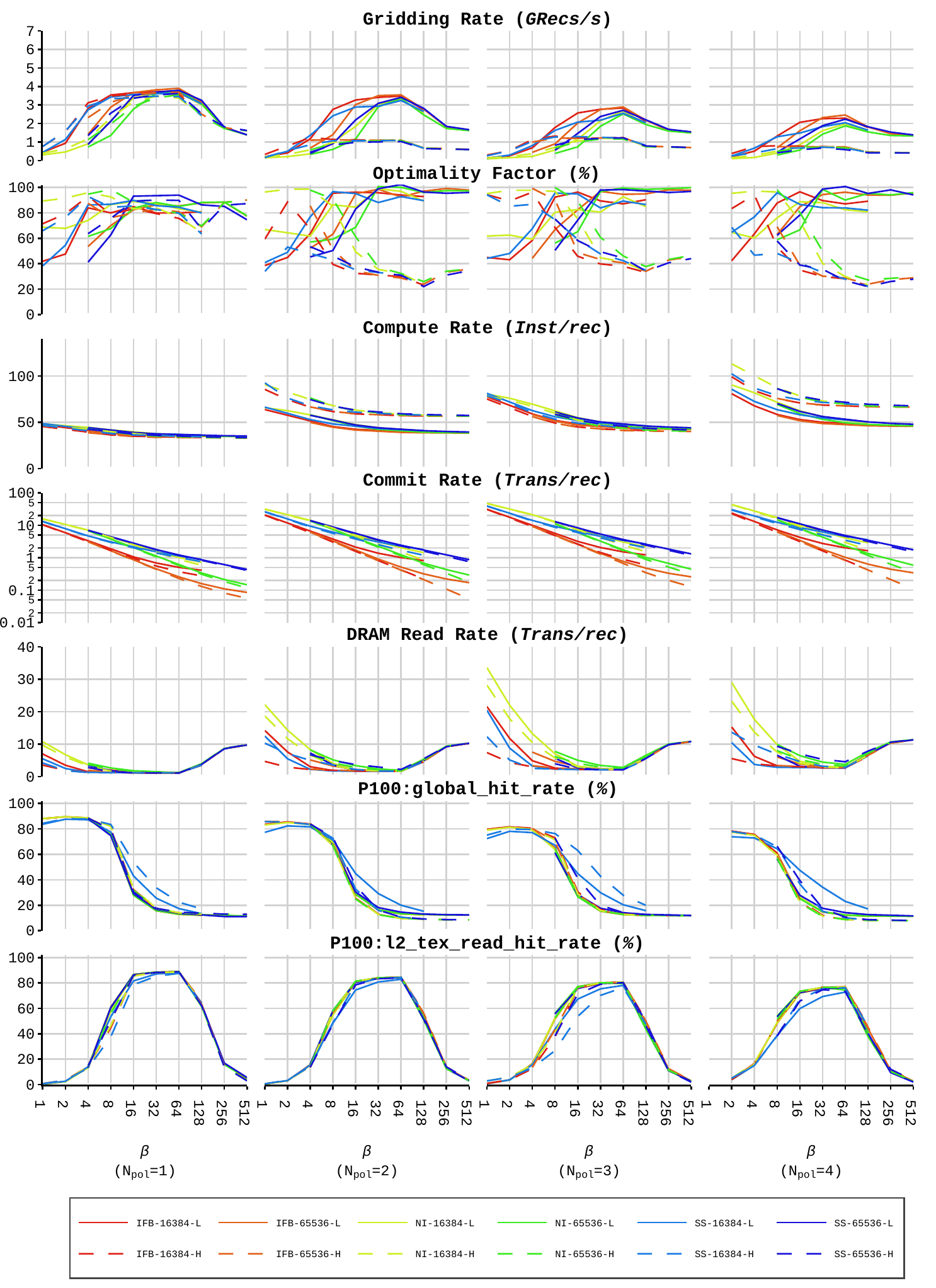}

\caption[Optimal Solutions Performance results for the Double-Precision Convolutional Gridder]{Optimal Solutions Performance results for the Double-Precision Convolutional Gridder. The layout is as described in Section \ref{sec:methodology:graphlayouts}. Performance metrics plotted in this figure are defined in Section \ref{sec:methodology:optimalsolutionsperformancemetrics}.}

\label{fig:2dgridding:conv_double}
\end{figure}

We choose two optimal solutions for every combination of Precision and $N_{\text{pol}}$, that we will analyse in-depth in Section \ref{sec:2dgridding:performance}. These optimal solutions are described in Table \ref{tab:2dgridding:optimisationparameters} and will be referred to as the \textit{Low Commit Rate} and \textit{High Commit Rate} solutions. The Low Commit Rate solutions deliver optimal Performance for those experiments with oversampling factors of roughly 16 and over, where the \commitrate is relatively low. On the other hand, the chosen High Commit Rate solutions are optimal for those experiments with the lowest $\beta$ using the NI Ordering Mode, where the \commitrate is at its highest (refer to Figures \ref{fig:2dgridding:conv_single} and \ref{fig:2dgridding:conv_double}). It is interesting to note that, excluding three exceptions, the High Commit Rate solutions have a lower Occupancy than their counter-part Low Commit Rate solutions. The High Commit Rate solutions are likely adapting to higher memory usage from grid committing by reducing Occupancy.
 
\begin{table}[]
\centering
\begin{tabular}{@{}c@{\rule{0.2cm}{0cm}}cccccccc@{}}
\toprule
Solution  & Precision                                    & $N_{\text{pol}}$ & $W_{\text{split}}$ & $X_{\text{Load}}$ & $SM$ & $B_{\text{Warps}}$  & $G_{\text{Blocks}}$& Occupancy  \\ \midrule
\multirow{8}{*}{%
\begin{minipage}[c]{1.7cm}%
\begin{tabular}{@{}c@{}}%
Low\\ 
Commit\\ 
Rate\\
\end{tabular}%
\end{minipage}}                          & \multicolumn{1}{@{}c}{\multirow{4}{*}{single}} & 1      & 2           & 16         & 4    & 16          & 1600   & 37.5     \\
                   & \multicolumn{1}{@{}c}{}                        & 2      & 2           & 16         & 1    & 8           & 1600     & 62.5   \\
                      & \multicolumn{1}{c}{}                        & 3      & 2           & 16         & 1    & 8           & 3000    & 50    \\
                      & \multicolumn{1}{c}{}                        & 4      & 2           & 16         & 4    & 8           & 3000 & 50        \\ \cmidrule(l){2-9} 
                      & \multicolumn{1}{@{}c}{\multirow{4}{*}{double}} & 1      & 2           & 16         & 6    & 8           & 3000 & 75        \\
                      & \multicolumn{1}{c}{}                        & 2      & 2           & 16         & 2    & 8           & 3000  & 50      \\
                      & \multicolumn{1}{c}{}                        & 3      & 2           & 16         & 3    & 8           & 3000  & 37.5        \\
                      & \multicolumn{1}{c}{}                        & 4      & 2           & 16         & 1    & 8           & 3000  & 25        \\ \midrule
\multirow{8}{*}{%
\begin{minipage}[c]{1.7cm}%
\begin{tabular}{@{}c@{}}%
High\\ 
Commit\\ 
Rate\\
\end{tabular}%
\end{minipage}} & \multicolumn{1}{@{}c}{\multirow{4}{*}{single}} & 1      & 2           & 16         & 1    & 1           & 600  & 50        \\
                      & \multicolumn{1}{c}{}                        & 2      & 1           & 64         & 1    & 16          & 600    & 25     \\
                      & \multicolumn{1}{c}{}                        & 3      & 1           & 32         & 2    & 1           & 600     & 50     \\
                      & \multicolumn{1}{c}{}                        & 4      & 2           & 64         & 1    & 1           & 200      & 14.1   \\ \cmidrule(l){2-9} 
                      & \multicolumn{1}{@{}c}{\multirow{4}{*}{double}} & 1      & 2           & 48         & 1    & 8           & 1600    & 25    \\
                      & \multicolumn{1}{c}{}                        & 2      & 1           & 32         & 1    & 1           & 600      & 50   \\
                      & \multicolumn{1}{c}{}                        & 3      & 2           & 64         & 1    & 4           & 1600    & 6.2    \\
                      & \multicolumn{1}{c}{}                        & 4      & 1           & 64         & 1    & 2           & 200     & 18.5    \\ \bottomrule
\end{tabular}
\caption[Optimal Solutions for the Convolutional Gridder]{Optimal Solutions for the Convolutional Gridder that we discovered and chose through Brute Force Search. For each value of $N_{\text{pol}}$ and Precision we study two solutions known as the Low Commit Rate solution and the High Commit Rate solution. The last column gives the theoretical Occupancy.}
\label{tab:2dgridding:optimisationparameters}
\end{table}


\begin{figure}

\includegraphics[page=1,width=\linewidth]{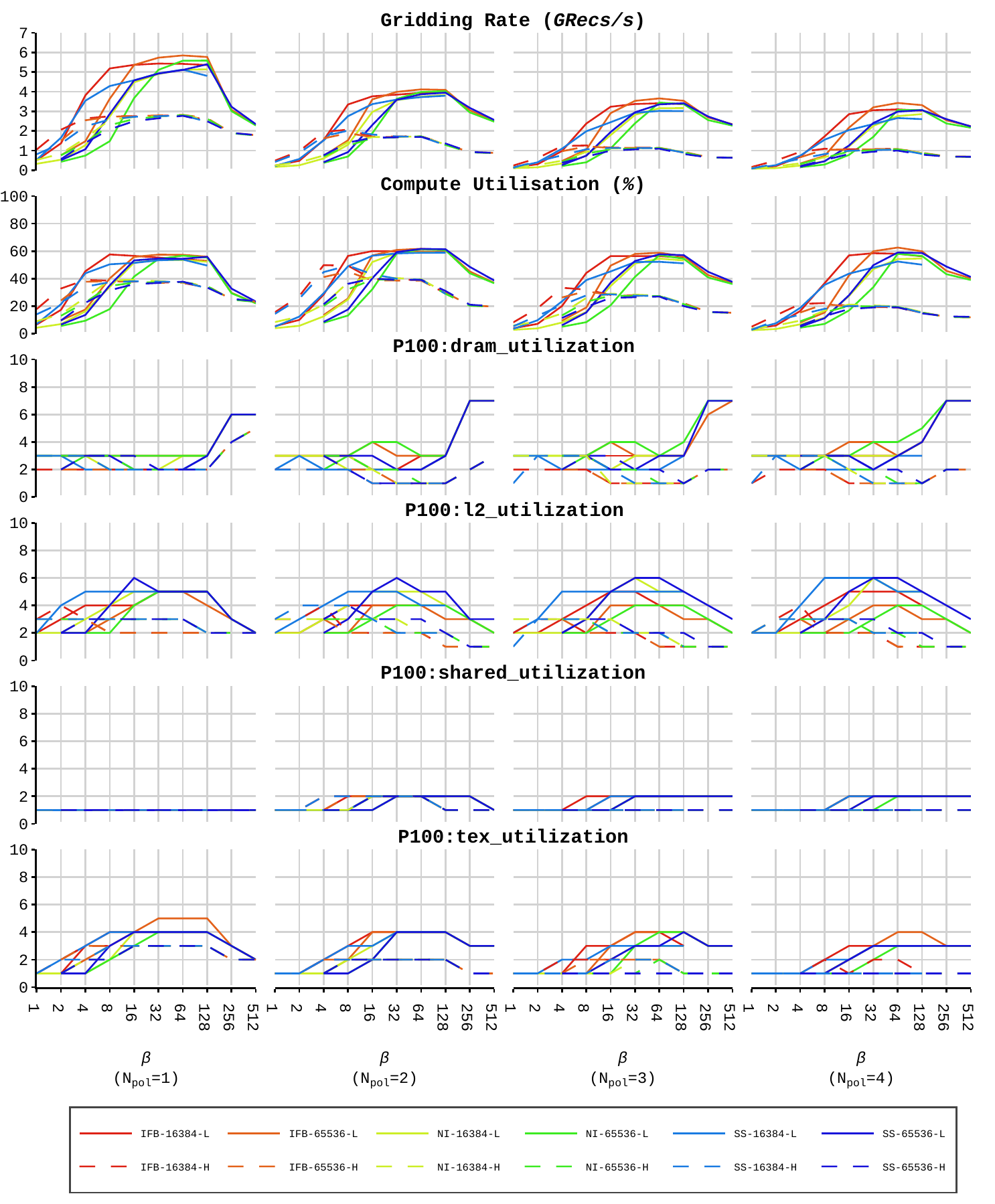}

\caption[Utilisation results for the Single-Precision Convolutional Gridder]{Utilisation results for the Single-Precision Convolutional Gridder. The layout is as described in Section \ref{sec:methodology:graphlayouts}. Performance metrics plotted in this figure are defined in Section  \ref{sec:methodology:performancemetricutilisation}.}

\label{fig:2dgridding:conv_single_utilisation}
\end{figure}

\begin{figure}

\includegraphics[page=1,width=\linewidth]{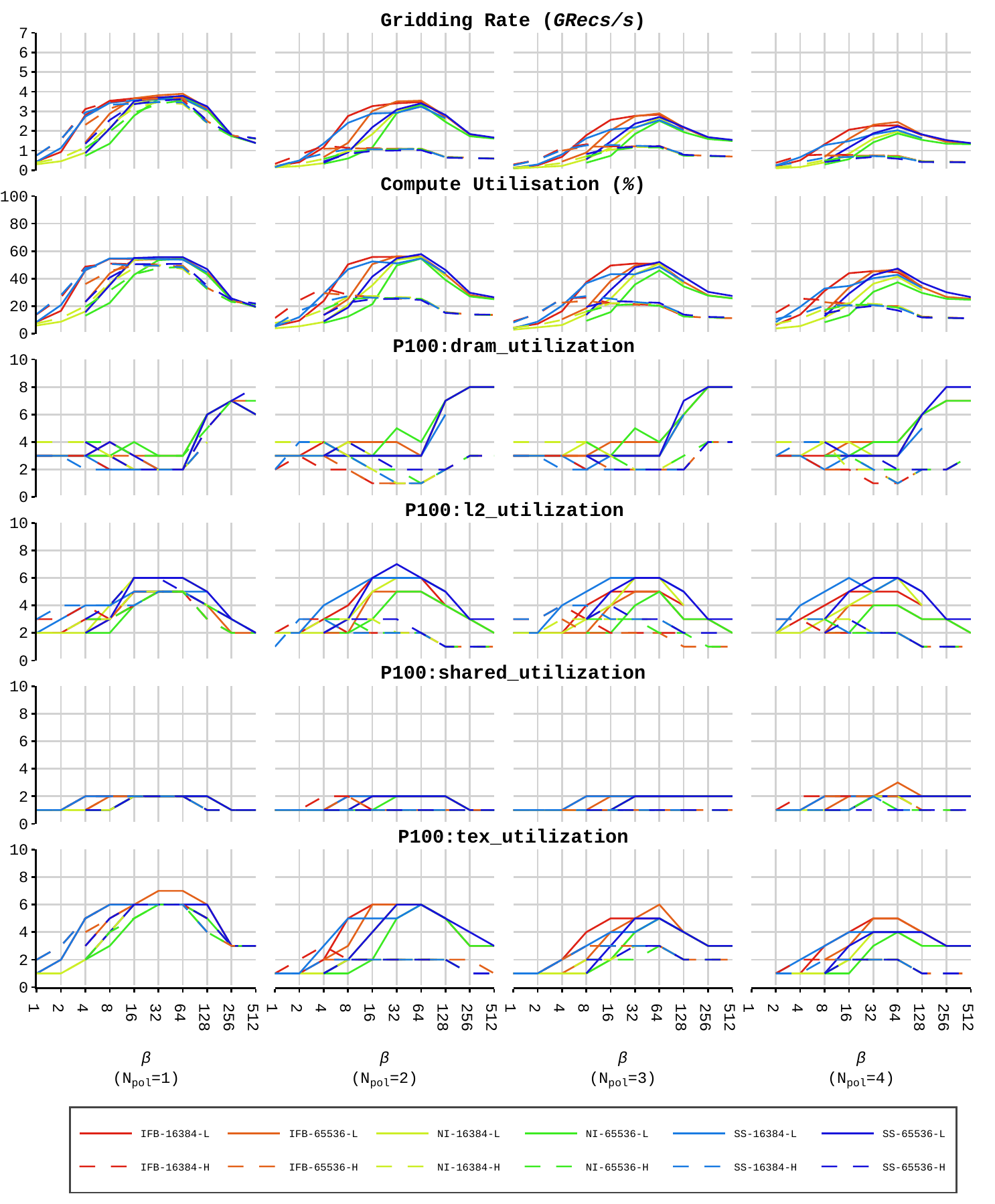}

\caption[Utilisation results for the Double-Precision Convolutional Gridder]{Utilisation results for the Double-Precision Convolutional Gridder. The layout is as described in Section \ref{sec:methodology:graphlayouts}. Performance metrics plotted in this figure are defined in Section  \ref{sec:methodology:performancemetricutilisation}.}
\label{fig:2dgridding:conv_double_utilisation}
\end{figure}

\section{ In-depth analyses}
\label{sec:2dgridding:performance}

 Optimal Solution Performance results are plotted in Figures \ref{fig:2dgridding:conv_single} and \ref{fig:2dgridding:conv_double} while Utilisation  results are plotted in Figures \ref{fig:2dgridding:conv_single_utilisation} and \ref{fig:2dgridding:conv_double_utilisation}. Results given are all about the optimal solutions stated in Table \ref{tab:2dgridding:optimisationparameters}.

Table \ref{tab:2dgridding:convmaxtheoratical}  gives the results for the Maximum Performance Experiments, whereby \efficiency is included. The Maximum Performance Experiments were only executed on the Low Commit Rate solutions whereby we modified the pre-processing phase and forced all records to have co-ordinates mapping to the same NN Grid pixel. In this way we reduced to a minimum the adverse effects on Performance caused by GCF data retrieval and grid committing.  Compression is also switched off, so as to increase the workload and minimise the \commitratestop

In order to help us in the detailed analyses that follows, we split results into three regions dependent on $\beta$. The \textit{Low-$\beta$ Region} includes all experiment results with values of $\beta$ ranging from 1 to that value where Performance stops to increase in a fast way and stabilises to values near the peak.  The second range known as the \textit{Mid-$\beta$ Region} includes results with values of $\beta$ ranging between where the previous range stopped and until Performance peaks at $\beta=64$. The \textit{High-$\beta$ Region} includes experimental results with $\beta\ge 128$. 

We now provide detailed analyses of the stated results in the next sub-sections.

\begin{table}[]
\centering
\begin{tabular}{@{}l@{}c@{}cccc@{}}
\toprule
 
$N_{\text{pol}}$ &\rule{1.7cm}{0cm}&\begin{minipage}[]{1.2cm}\centering1\end{minipage} &\begin{minipage}[]{1.2cm}\centering2\end{minipage}& \begin{minipage}[]{1.2cm}\centering3\end{minipage}& \begin{minipage}[]{1.2cm}\centering4\end{minipage}\\ \midrule
\rowcolor{mygrey}\multicolumn{6}{c}{Single-Precision}\\
\midrule
\gridrate \texttt{(GRecs/s)} & & 6.33 & 4.29 & 3.83 & 3.62  \\
\efficiency \texttt{(\%)} & & 8.59 & 11.6 & 15.6 & 19.7 \\
\polgain & & 1 &1.35 & 1.81 & 2.29 \\
\computerate \texttt{(Inst/rec)}& & 23.4 & 35 & 37.5 & 42.2\\
\globalhitrate \texttt{(\%)}& & 62.5 & 72.6 & 68.5 & 64.8  \\
\ltexrate \texttt{(\%)} & & 18.4 &  0.321 & 0.294 & 0.314 \\
\commitrate \texttt{(Trans/rec)} & & 1.35e-02  & 7.25e-03 & 2.03e-02 & 2.70e-02  \\
\dramreadtransactionsrate \texttt{(Trans/rec)} & & 0.875 & 1.13 & 1.38 & 1.63  \\
\computeutilisation \texttt{(\%)} & &59.5 & 60.4 & 57.7 & 61.4 \\
\texutilisation & & 5 & 4&4 &4 \\
\ltwoutilisation & &1 &1 &1 &1 \\
\dramutilisation & & 3 &3 &3&3 \\
\sharedutilisation & & 2 &2 &2&2 \\
\midrule
\rowcolor{mygrey}\multicolumn{6}{c}{Double-Precision}\\
\midrule
\gridrate \texttt{(GRecs/s)} & & 4.00 & 3.81 & 3.23 & 2.84 \\
\efficiency \texttt{(\%)} & & 10.8 & 20.7 & 26.3 & 30.8 \\
\polgain & & 1 &1.35 & 1.81 & 2.29 \\
\computerate \texttt{(Inst/rec)} & & 32.9 & 36.1 & 39.6 & 42.2 \\
\globalhitrate \texttt{(\%)} & & 79.9 & 73.3 & 67.7 & 62.9 \\
\ltexrate \texttt{(\%)} & & 0.66 & 0.696 & 0.757 & 0.807 \\
\commitrate \texttt{(Trans/rec)}&  & 1.35e-02 & 2.70e-02 & 4.05e-02 & 5.41e-02 \\
\dramreadtransactionsrate \texttt{(Trans/rec)} &  & 1.13 & 1.63 & 2.13 & 2.63 \\
\computeutilisation \texttt{(\%)} & &53 & 55.2 & 51.5 & 48.2 \\
\texutilisation & & 7 & 7&6 &6 \\
\ltwoutilisation  & &1 &1 &1 &1 \\
\dramutilisation & & 3 &3 &4&4 \\
\sharedutilisation & & 2 &2 &3&3 \\
\bottomrule
\end{tabular}
\caption[Results of the Maximum Performance Experiments for the Convolutional Gridder.]{Results of the Maximum Performance Experiments for the Convolutional Gridder.}
\label{tab:2dgridding:convmaxtheoratical}
\end{table}

\subsection{Utilisation and boundedness}
\label{sec:2dgridding:boundedness}
The Single-Precision Maximum Performance Experiments are compute-bound since the \computeutilisation metric nears or exceeds 60\%. In contrast, the Double-Precision Maximum Performance Experiments are memory-bound at the Unified Cache since \texutilisation is at a value of 6 and above. We realise that for all the Maximum Performance Experiments \computeutilisation and \texutilisation are substantially high. Probably the change in boundedness between Single and Double-Precision happened because the increased utilisation of memory to handle Double-Precision records and GCF data surpassed the increase in compute.

Experiments in the Mid-$\beta$ Region using the Low Commit Rate solutions showed similar \computeutilisation (Refer to  Figures \ref{fig:2dgridding:conv_single_utilisation} and \ref{fig:2dgridding:conv_double_utilisation}) to the Maximum Performance Experiments, but lower  \texutilisation and a considerable increase in \ltwoutilisation that bounds the Double-Precision experiments. The increased utilisation of the L2 Cache is due to pressure on the Unified Cache caused by the loading of the GCF data, which we discuss in Section \ref{sec:2dgridding:gcfdataretrieval}. 

Experiments in the High-$\beta$ Region using the Low Commit Rate solutions are memory-bound at the device memory (refer to \dramutilisation in Figures \ref{fig:2dgridding:conv_single_utilisation} and \ref{fig:2dgridding:conv_double_utilisation}), which is to due to more pressure on the caching system caused by loading of GCF data, which we also discuss in Section \ref{sec:2dgridding:gcfdataretrieval}.  

Without doubt, experiments in the Low-$\beta$ Region are in general neither compute nor memory bound and suffer from latency, probably injected by the elevated level of grid committing which we discuss in Section \ref{sec:2dgridding:gridcommiting}. 

Experiments using the High Commit Rate solutions are neither compute nor memory bound and suffer from latency. It is interesting to note that when they deliver higher Performance, than their counter-part Low Commit Rate solutions, both memory and compute utilisation increase. 

\subsection{Compute}
\label{sec:2dgridding:compute}
We argue that compute has a considerable impact on the Maximum Performance Experiments and the Mid-$\beta$ Range experiments using the Low Commit Rate solutions. Suggesting so are the high values of the \computeutilisation metric. Nevertheless, the high utilisation of the caching modules hints that a hypothetical reduction in the \computerate would not bring any substantial Performance enhancements. 

For all the other experiments not mentioned in the previous paragraph, we note that compute has minimal impact on Performance since \computeutilisation is considerably low, and is not the factor that bounds Performance.

It is worth doing a more in-depth analysis of compute. We split compute between FMA instructions, which take care of the accumulation in registers and all other instructions, which we collectively refer to as Logic. For $W_{\text{split}}=2$, the FMA Warp-level instructions per record are equal to 3$\times N_{\text{pol}}$. Using the measurements for the \computerate metric (ref to Figures \ref{fig:2dgridding:conv_single} and \ref{fig:2dgridding:conv_double} and Table \ref{tab:2dgridding:convmaxtheoratical}) we find out that  the ratio between instructions related to Logic and all the compute instructions is at minimum 71\%, implying that impact of compute on Performance is dominated by Logic. 

It is worth to note that the increase in \computerate due to an increase in $N_{\text{pol}}$ in the Maximum Performance Experiments is mostly FMA instructions and the \computerate is not too far from being inversely proportional to the \gridratestop Such behaviour suggests that the \polgain measured for the Maximum Performance Experiments is limited by Logic, and to attain higher gains, Logic has to be reduced. In this suggestion, we are assuming that there is no way to reduce FMA instructions, and that the schedulers in the P100 SMs are working at nearly full capacity, which we verified to be true by means of profiling.

We now finalise our deliberation on compute by having a look at the \efficiency metric given for the Maximum Performance Experiments in Table \ref{tab:2dgridding:convmaxtheoratical}. Values stated are by far much lower than what Merry \cite{Merry2016} achieved with thread coarsening. However, one needs to consider that Merry's work is for complex-valued GCFs and not real-valued GCFs, where handling complex-valued GCFs generally doubles the FMA warp-level instructions. We again point out that Merry did not report on gridding with $6 \times 6$ GCFs.

It is interesting to note that on comparing Double-Precision Maximum Performance Experiments with their counter-part Single-Precision Maximum Performance Experiments, we find out that \efficiency is higher in Double-Precision. At the same time, \computeutilisation is lower and \computerate is nearly invariant except for $N_{\text{pol}}=1$ where $W_{\text{split}}$ differs. We see it likely that this behaviour has to do with the fact that the P100 computes integer and Single-Precision arithmetic on the same module, but computes Double-Precision on a different module, implying that the P100 can parallelise better in Double-Precision experiments.

\subsection{GCF data retrieval}
\label{sec:2dgridding:gcfdataretrieval}
We observed that GCF data retrieval generates around six memory  transactions per record  for Single-Precision\footnote{For the Single-Precision, $N_{\text{pol}}=1$ case the values are smaller due to a different value of $W_{\text{split}}$.} and twelve memory transactions per record for Double-Precision. These values are much higher than the amount needed to load data of one record.\footnote{The memory transactions needed to load a record for a given Precision and $N_{\text{pol}}$ is approximately equal to the measured \dramreadtransactionsrate in the Maximum Theoretical Performance Experiments given in Table \ref{tab:2dgridding:convmaxtheoratical} and is constant on varying $\beta$.} Furthermore, they are a lot higher than the memory transactions per record generated by grid committing, implying that  GCF data retrieval has a substantial share in memory utilisation. Therefore, in experiments where there is  high utilisation of some memory modules, GCF data retrieval should tangibly impact Performance.

We find it reasonable to expect that the level of impact on Performance by GCF data retrieval is related to the ability of the caching system to retrieve data from its memory store (a HIT) rather than retrieving it from device memory (a MISS). Such ability is dependent on the full size of the GCF data stored in device memory equal to $\beta^2S^2$ Single or Double Precision floating point numbers where $S=6$. The said ability is also tight to the distribution of the requested GCF data which we verified to be nearly flat. 

As discussed in Section \ref{sec:methodology:cudamemory}, the cache system in the P100 is made up of two cascading layers of cache known as the Unified Cache and L2 Cache. We consult the Performance Metrics \globalhitrate and \ltexrate to understand the cache behaviour while clarifying that these metrics do not just measure HIT rates for GCF data retrieval but also include MISSed record data loading. 

All experiments at a given $\beta$ value show similar patterns for the stated metrics. Initially, at $\beta=1$, \globalhitrate is at a near peak, and \ltexrate is nearly 0. We can assume that GCF data HIT rate is at a near 100\% since the GCF data size is small and \ltexrate is near 0, indicating minimal GCF data related traffic is reaching the L2 Cache, which would otherwise result in higher values of \ltexratestop The Single-Precision, $N_{\text{pol}}=1$, Low Commit Rate experiments are an exception whereby \ltexrate is at 20\%. The culprit of such a behaviour is because when $W_{\text{split}} = 5$, the loading of records data from global memory is not fully aligned. Such a misalignment will cause some transactions related to record data loading to repeat and HIT the L2 Cache. 

We took caution in stating that at $\beta=1$, the \globalhitrate is \textit{at a near peak} rather than \textit{at the peak}.  Sometimes the peak is reached at some low value of $\beta$ but not at $\beta=1$.  There is a simple explanation for this behaviour, whereby at low values of $\beta$, there is a high probability that Sub-Warp Gridders in a given warp request the same GCF data, which the GPU will coalesce (merge) in the same group of transactions. Such merging lowers the average memory transactions executed per record for loading  GCF data, which quantifies itself as a decrease in \globalhitratecomma since the transactions needed for loading record data for each record is constant. The stated probability should generally decrease with increasing $\beta$,  implying that at low $\beta$, it is possible to have a rising \globalhitrate with a rising $\beta$ for a constant HIT ratio of the loading of GCF data. We note that in the Maximum Performance Experiments, the stated probability is at 100\% and in fact, the reported values of \globalhitrate are lower than for experiments with $\beta=1$.

Results show that increasing $\beta$ from one to about eight, the \globalhitrate and \ltexrate vary slightly, implying no significant changes in the behaviour of the caching system. The \gridrate increases, but this is due to less grid committing which we discuss in Section \ref{sec:2dgridding:gridcommiting}. 

A further increase of $\beta$ up to a value of $128$, pushes down the \globalhitrate and raises the \ltexratestop The Unified Cache is now under pressure and begins to MISS  GCF related transactions. L2 is backing up the Unified Cache and serves MISSed GCF data with HITs on its cache memory. At $\beta=8$, the size of the GCF data is 2KiB for Single-Precision, similar to the size of the Unified Cache. Therefore, the pressure on the Unified Cache is expected. Interestingly enough, all experiments reach a peak in \gridrate at $\beta=64$, and the peak values of the \gridrate are not too far from the Maximum Performance Experiments. We see such  behaviour as indicative that in the region of $\beta$ we are discussing, the change in behaviour of the caching system has only a minor impact on the \gridratestop 

Finally, upon increasing $\beta$ from 128 to 256 and further, the \ltexrate goes down. The \gridrate also goes down, and, as we stated in Section \ref{sec:2dgridding:boundedness}, experiments become memory bound.  It is clear that now the L2 Cache is under pressure and, as also indicated by the \dramreadtransactionsrate metric, a substantial amount of GCF data is transacted from the device memory. At $\beta=256$, the full Single-Precision stored GCF data has a size of around 8MB, which is two times bigger than the L2 Cache size, so the pressure on L2 Cache is understandable.  It is clear that in this discussed region of $\beta$, which we previously defined as the High-$\beta$ Region, GCF data retrieval has a substantial impact on the Performance of the Convolutional Gridder, and is the main culprit pushing down Performance. 

\subsubsection{Chocking effects of the GCF}

We suspect that the GCF data retrieval might be \textit{choking up} other memory operations. Feeding our suspicion is the nearly constant \dramutilisation for the Low-$\beta$ Region experiments, whereby in this region, the \commitrate metric increases with decreasing $\beta$. By design, in the P100, all memory-related operations need to compete for  Unified Cache since there is no direct access to the device memory. In the Maximum Performance Experiments, the Unified Cache is very busy. While we do not regard any of the evidence we gave here as conclusive, we believe as likely that retrieving GCF data is impacting Performance of the other memory operations, by forcing a ceiling on the \dramutilisationstop

\subsection{Loading of records data and use of shared memory}
\label{sec:2dgridding:recordloading}
The loading of record data contributes less than 20\% of the total memory transactions, and therefore, we see no concern in regards to the impact on Performance by such loading. However, the chocking effect discussed before might be slowing down record data loading.   

Shared memory usage is not limiting Performance, in view that the \sharedutilisation metric is always less or equal to 3.

\subsection{Grid committing}
\label{sec:2dgridding:gridcommiting}
We expect grid committing to be an inefficient and relatively slow process, since it requires to read and write on global memory with degraded access patterns. Possibly, grid committing might also be \textit{choked} by the GCF data retrieval, as discussed in Section \ref{sec:2dgridding:gcfdataretrieval}.

Evidence shows that grid committing is the main reason for Performance variation upon changes in $\beta$ and Ordering Mode and controls  \polgainstop In the following paragraphs, we discuss the evidence.    

Grid committing is the only step in the Convolutional Gridder that is controlled by a  predicate. Therefore, Performance should have a dependency on the frequency that Grid Committing executes, measured by the \commitrate Performance Metric. Results for the Low and Mid-$\beta$ Region experiments clearly show a correlation between \commitrate and Performance, whereby an increase in \commitrate results in loss of Performance. The High-$\beta$ Region experiments do not follow suit, but this is due to GCF Data Retrieval biting in. 

By design, the \commitrate is expected to obey the approximate relationship:
\begin{equation}
    \commitrate \approx (\commitrate \texttt{ at } N_{\text{pol}}=1)\times N_{\text{pol}}
\end{equation}

Minor variations in the relationship are due to different tuning. The relationship implies that Grid Committing does not contribute to an increase in \polgaincomma but rather pushes the metric towards a value of one. Therefore, increased \commitrate values imply lower \polgaincomma as validated by results at low $\beta$. They have the highest \commitrate values and lowest \polgainstop

Results show that a high \commitrate pushes down the utilisation of compute and all considered memory modules, causing the Convolutional Gridder to be crippled with latency. In our view, this is caused by the degraded memory access patterns of grid committing, whereby they  neither meet the vicinity requirement nor the coalescence requirement (refer to Section \ref{sec:methodology:optimalaccesspattern}).

Finally, we point out that the increase in \computerate with a decrease in $\beta$ is due to an increase in the \commitrate since grid committing, by design,  requires the execution of some extra compute related instructions.

\section{Summary}
\label{sec:2dgridding:conclusion}

This chapter discussed our implementation of a Convolutional Gridder on the P100 for real-valued GCFs of support 6 × 6. We are here summarising in point form the main Performance results.
\begin{itemize}
\item In our experiments, the Convolutional Gridder delivered acceptable Performance for all Ordering Modes, and the highest Performance with the IFB Ordering Mode.   
\item Maximum Performance of the Convolutional Gridder in ideal conditions per Precision and $N_{\text{pol}}$ is stated in the below table.

\begin{tabular}{@{}c@{}c@{}cccc@{}}
\toprule
 $N_{\text{pol}}$                                      &\rule{1cm}{0cm}& 1    & 2    & 3    & 4    \\ 
 \midrule
\begin{minipage}{8cm}\begin{spacing}{1}Single-Precision Maximum \gridrate (GRecs/s)\\[-20pt]\end{spacing}\end{minipage} & \rule{0cm}{25pt}& 6.33 & 4.29 & 3.83 & 3.62 \\
\begin{minipage}{8cm}\begin{spacing}{1}Double-Precision Maximum \gridrate (GRecs/s)\\[-20pt]\end{spacing}\end{minipage} & \rule{0cm}{25pt}& 4    & 3.81 & 3.23 & 2.84 \\ \bottomrule
\end{tabular}

\item When running with maximum Performance, the Convolutional Gridder has a high level of Compute Utilisation and Memory Utilisation at the Unified Cache. Compute bounds the Convolutional Gridder when running in Single-Precision, while utilisation of the  Unified Cache bounds the Convolutional Gridder when running in Double-Precision.
\item The two main limiters that push down Performance from its maximum are the loading of GCF data, and grid committing.  
Loading of GCF data will considerably degrade Performance when there is pressure on the L2 Cache such that it cannot hold all the GCF data, in its store as to provide the best HIT rates. In our experiments, such pressure was felt for $\beta\ge128$, where the experiments turned Memory-bound on device memory.
\item Grid committing degrades Performance according to the frequency at which it is predicated to occur. In our experiments, elevated levels of grid committing occured for low values of $\beta$, whereby the Convolutional Gridder was not anymore bound by memory or compute. Instead, The Convolutional Gridder was riddled with latency injected by the degraded access patterns of grid committing.
\item The \polgain of the Convolutional Gridder when delivering at maximum Performance is limited by Logic.
\end{itemize}

\chapter{The Hybrid Gridder}
\label{chap:hybrid}

This chapter describes and provides analyses to our novel implementation of the Hybrid Gridder targeted for the P100. It is laid out similarly to Chapter \ref{chap:2dgridding}. Validation of the implemented Hybrid Gridder is discussed in Chapter \ref{chap:comparative}.

\section{Modified Romein's strategy}
\label{sec:hybrid:romeinstrategy}
We implement the Hybrid Gridder using the same Romein's strategy used for the Convolution Gridder, but cosmetically modified as to reflect the use of GCFs with support $S\times1$.

\begin{figure}

\includegraphics[page=1,width=\linewidth]{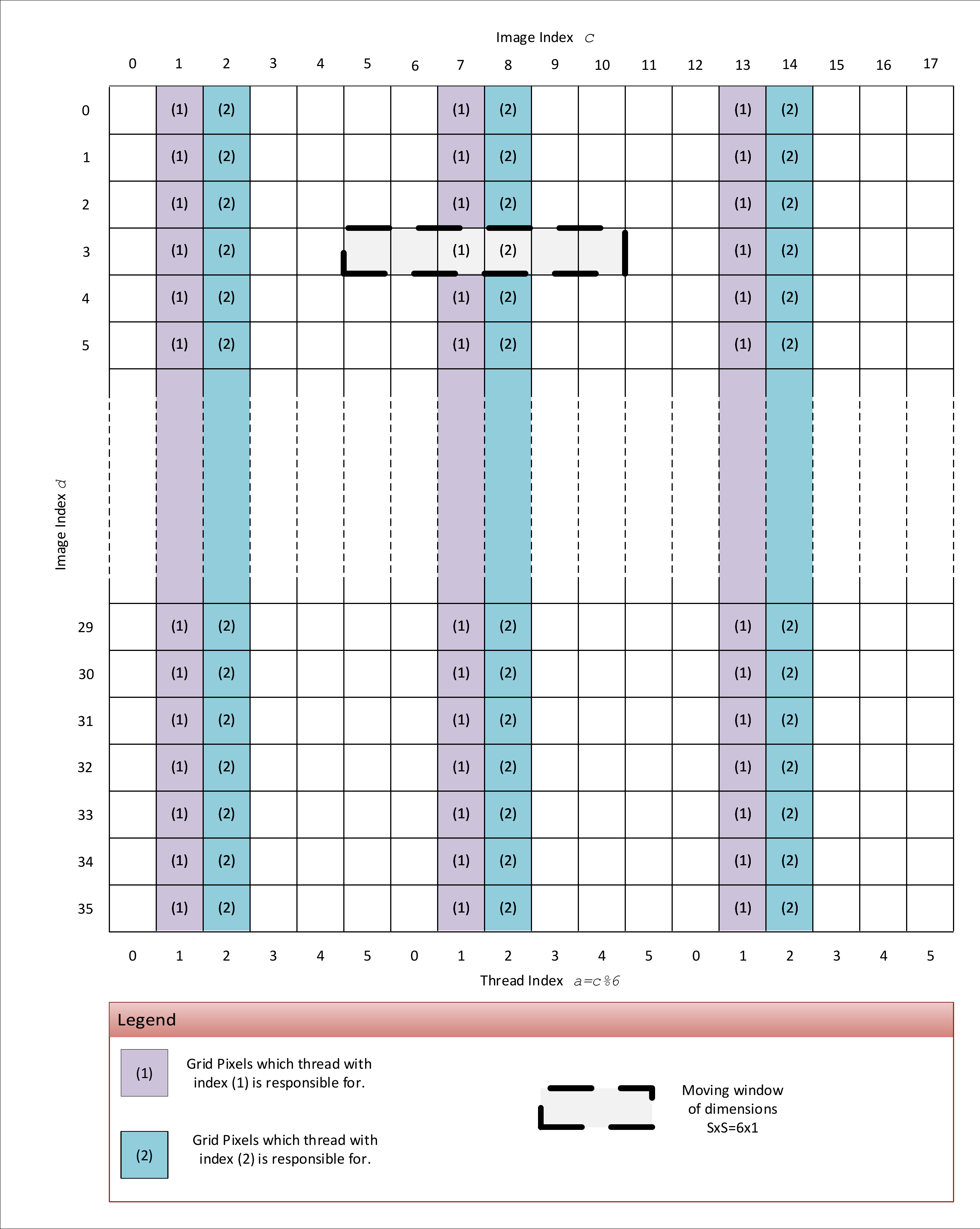}

\caption[Romein's strategy thread configuration for Hybrid Gridding]{Diagram showing the thread configuration of the Romein's strategy as cosmetically modified for the Hybrid Gridder. We are showing a grid of size $18\times 36$ pixels, over which a record is gridded using a GCF of support $6\times 1$. The moving window shows the area on the grid a gridded record updates. A thread such as that indexed by (2) is responsible for updating a set of pixels on the grid. Nevertheless, any given gridded record only updates one pixel from the set.}
\label{fig:hybrid:romeinhybridmovingwindow}
\end{figure}

Figure \ref{fig:hybrid:romeinhybridmovingwindow} illustrates the modified thread configuration of Romein's strategy for Hybrid Gridding. We use the same line of thought made in Section \ref{sec:2dgridding:threadconf}.  Given a grid size of $N\times\beta N$, $S$ threads are allocated for gridding a stream of records. Let's index each thread using a 1D index $(a)$ where $0\le a < S$ and index the two-dimensional grid using a two-dimensional index $(c,d)$ where $0\le c < N$ and $0\le d < \beta N$. Each thread is given the responsibility to update grid pixels with index $(a+k_1 S,k_2)$, where $k_1, k_2 \in \mathbb{Z}$.

The shape of the GCF forces us to reconsider which Ordering Mode for the input records is best suited for a performant Hybrid Gridder. Romein's strategy obtains Performance through the accumulation of grid pixel updates with no grid committing, where such accumulation is only possible when the window moves sideways. Therefore, it is likely that SS Ordering Mode delivers better Performance results than NI or IFB.

\section{Implementation}
Our implementation of the Hybrid Gridder is based on the Convolutional Gridder. In an attempt to simplify this text, we describe the Hybrid Gridder using the same sub-sections, and flow of ideas given in Section \ref{sec:2dgridding:implimentation}, while stating what differs from the Convolutional Gridder. We note that our implementation is focused on delivering Performance for a real-valued  GCF of support $6\times 1$. However, the implementation can easily be modified to cater for GCFs with support up to $32 \times 1$.

\subsection{CUDA grid layout}

The CUDA grid layout is equivalent to that of the Convolutional Gridder. Therefore the Hybrid Gridder is tuned with the Tuning Parameters $B_{\text{Warps}}$  and $G_{\text{Blocks}}$, which retain the same definition stated for the Convolutional Gridder. We expect the parameters to influence Performance in the same way discussed in Section \ref{sec:methodology:cudagridlayout}.

\subsection{Sub-Warp Gridders and Instruction Level Parallelism}

\begin{figure}
\includegraphics[page=1,width=\textwidth,trim=0.5cm 0.5cm 0.5cm 0.5cm,clip]{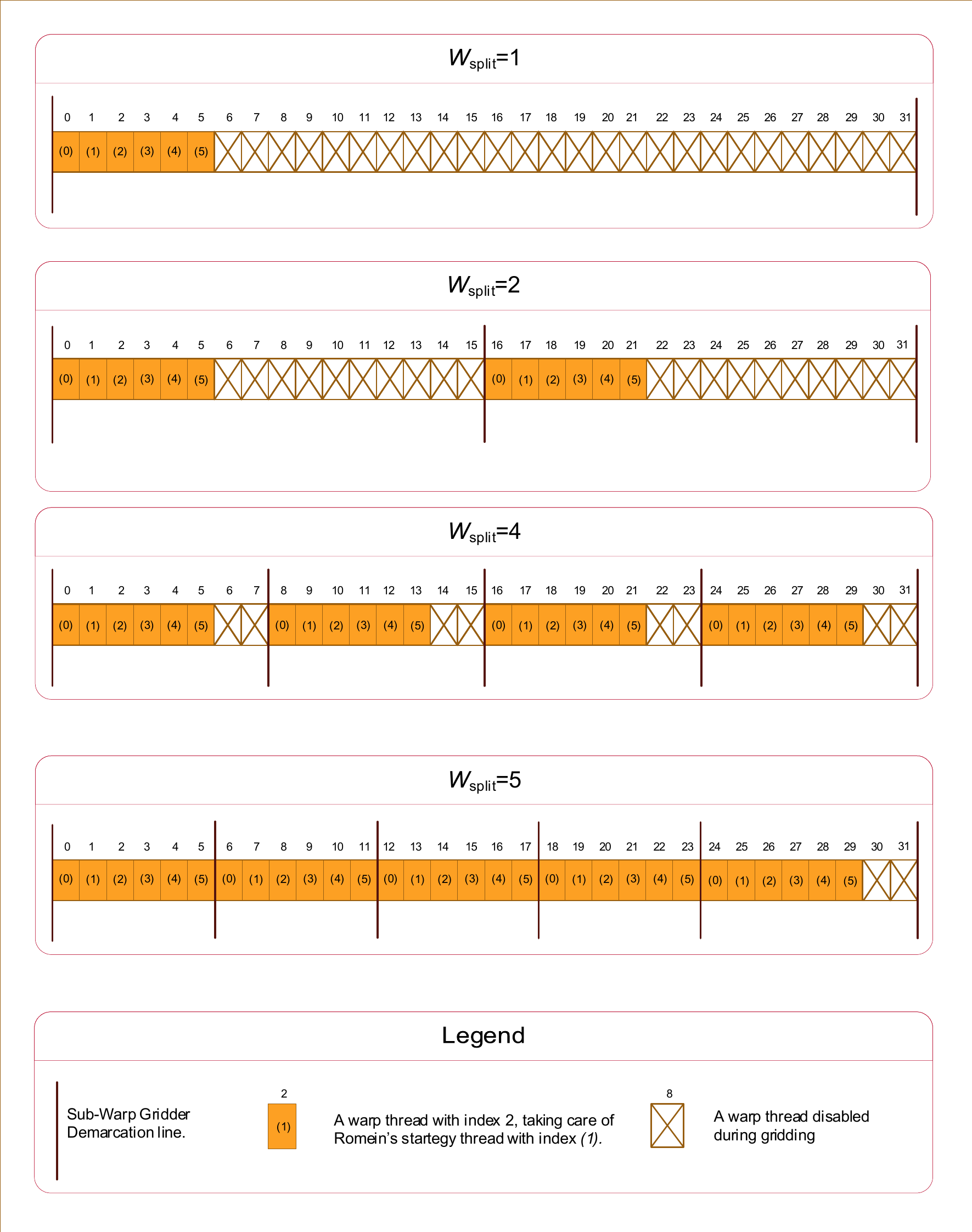}
\caption[CUDA thread layout for the Hybrid Gridder]{Diagram showing how CUDA threads are organised within a warp for the different values of the Warp Split Factor ($W_{\text{split}}$). Except for $W_{\text{split}}=5$, CUDA threads crossed out are not used for gridding but still used to load record data from global memory. }
\label{fig:hybrid:gridderthreadconfiguration}
\end{figure}

Similar to the Convolutional Gridder, a warp in the  Hybrid Gridder is split into several Sub-Warp Gridders according to the Tuning Parameter $W_{\text{split}}$. The values considered for $W_{\text{split}}$ in the Brute Force Search are 1,2,4 and 5, and the respective CUDA thread configuration for each value of $W_{\text{split}}$ value is illustrated in Figure \ref{fig:hybrid:gridderthreadconfiguration}.

Since there are few threads involved in gridding a record, we did not consider the use of thread coarsening in the Hybrid Gridder, and took a different approach to obtain an increase in Instruction Level Parallelism and mitigate latency. The Sub-Warp Gridder is programmed in such a way that it grids two records at one go by interleaving the execution of Logic and retrieval of GCF data. Such programming is possible because much of the gridding Logic for a given record is independent of the gridding Logic of other records.

Our method for increasing Instruction Level Parallelism in the Hybrid Gridder requires more registers per thread, which, similar to thread coarsening in the Convolutional Gridder,  can impact Performance. We again use $\text{SM}$ as a Tuning parameter to control the number of registers used per thread.

\subsection{Input record data and use of shared memory}

Input Record Data is split in various streams as listed hereunder.
\begin{enumerate}
    \item A 128-bit memory aligned four integer index.
    \item $N_{\text{pol}}$ streams with each stream containing Visibility data of one polarisation.
\end{enumerate}

The first stream is generated from the records' $(u,v)$-coordinates and is not equivalent to the four-integer index input stream used in the Convolutional Gridder. 

Shared Memory is used as interim storage, much like how it is used in the Convolutional Gridder. The loading of record data from global memory is the same, whereby the Tunning Parameter $X_{\text{Load}}$ defines how many records are loaded at one and then immediately  gridded. Equation \ref{equ:2dgridding:sharedmemorysize} is still valid for the Hybrid Gridder, but the record size differs and is tabulated in Table \ref{tab:hybrid:recordchunksplit}.

Shared memory is laid out using the same methods used in the Convolutional Gridder, but chunks are defined differently, and are always 128-bit long. Table \ref{tab:hybrid:recordchunksplit} tabulates the contents in the chunks. We note that, since the Hybrid Gridder grids two records at one go, some chunks have Visibility data of two records to ensure 128-bit chunks. 
\begin{table}[]
\centering
\begin{tabular}{@{}c@{}c@{}c@{}c@{}c@{}c@{}c@{}c@{}c@{}c@{}}
\toprule
Precision & \rule{0.4cm}{0cm}$N_{\text{pol}}$\rule{0.4cm}{0cm} & \rule{0.7cm}{0cm} &\begin{minipage}[]{1.7cm}\centering C1\end{minipage} & \begin{minipage}[]{1.7cm}\centering C2\end{minipage}  & \begin{minipage}[]{1.7cm}\centering C3\end{minipage}     & \begin{minipage}[]{1.7cm}\centering C4\end{minipage}     & \begin{minipage}[]{1.7cm}\centering C5\end{minipage}    & \rule{0.0cm}{0cm} & Record size  \\ \midrule
\multirow{4}{*}{single}   & 1         & & I4      & V1($2\times$) &             &       &      &    &  24           \\
    & 2         & & I4      & V1+V2    &    &             &    & & 32             \\
    & 3         & & I4      & V1+V2  & V3(2$\times$)       &           &  &  & 40                    \\
    & 4        &  & I4      & V1+V2    & V3+V4       &  &  &  & 48                 \\ \midrule
\multirow{4}{*}{double}    & 1         & & I4      & V1       & &             &                  &     & 32     \\
  & 2         & & I4      & V1       & V2          &  &             & & 48        \\
   & 3         & & I4      & V1       & V2          & V3          &    & & 56         \\
   & 4         & & I4      & V1       & V2          & V3          & V4    & & 64    \\ \bottomrule
\end{tabular}
\caption[Splitting of records in 128-bit chunks for the Hybrid Gridder.]{This table shows how record data is split in 128-bit chunks for the Hybrid Gridder. The split is in such a way that it can be laid out efficiently in shared memory. The last column gives the amount of space in bytes a record consumes in shared and device memory.    
\\\textbf{C\textit{x}}: Chunk No \textit{x} \\\textbf{I4}: A four-integer index (128 bits long) \\\textbf{V\textit{x}}: Visibility value of the \textit{x}$^{th}$ polarisation. \\\textbf{V\textit{x}(2$\times$)}: Visibility values of the \textit{x}$^{th}$ polarisation of two consecutive records, ensuring a chunk length of 128 bits.}
\label{tab:hybrid:recordchunksplit}
\end{table}

\subsection{Loading of the GCF}
Our implementation of the Hybrid Gridder uses the same methods used in the Convolutional Gridder implementation for loading GCF data. Caching of the GCF data is forced using the \texttt{__ldg()} function, and loading of record data is set not to be cached in the Unified Cache.  

\subsection{Output}
\label{sec:hybrid:output}

The Hybrid Gridder outputs non-interleaved multi-polarised grids with the origin set at the corner of the grid. Internal indexes assume that the origin of the grid is at the centre, and conversion of indexes takes place during Grid Committing. As we shall show through experiments, the Hybrid Gridder is not compute-intensive, and the extra Logic needed for the  conversion has only a minor impact on Performance.

\section{Brute Force Search results}
\begin{figure}
\includegraphics[page=1,width=\linewidth]{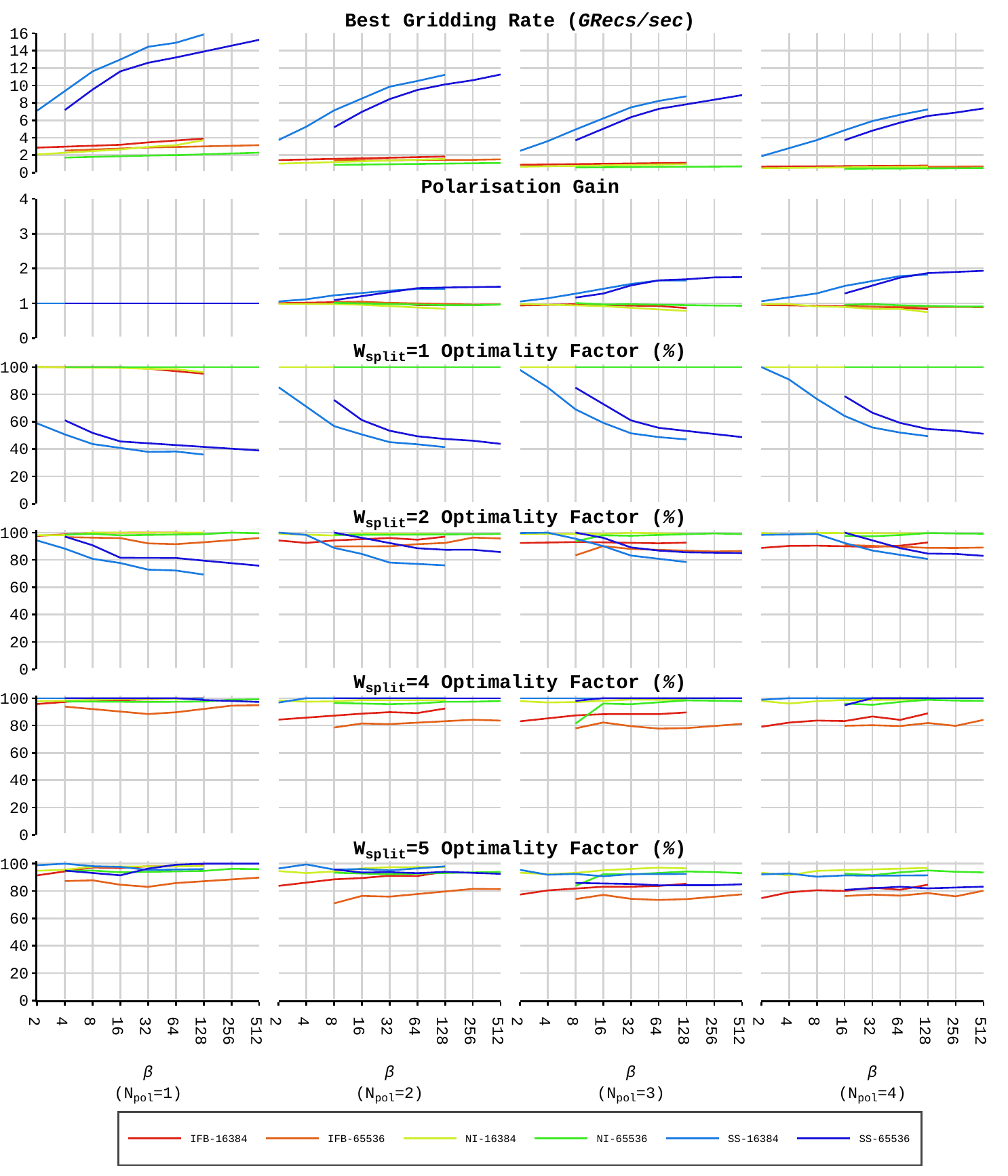}
\caption[Brute Force Search results for the Single-Precision Hybrid Gridder]{Brute Force Search results for the Single-Precision Hybrid Gridder. The layout is as described in Section \ref{sec:methodology:graphlayouts}. Performance metrics plotted in this figure are defined in Section \ref{sec:methodology:perfromancemetricsbrute}.}
\label{fig:hybrid:brute-single}
\end{figure}

\begin{figure}

\includegraphics[page=1,width=\linewidth]{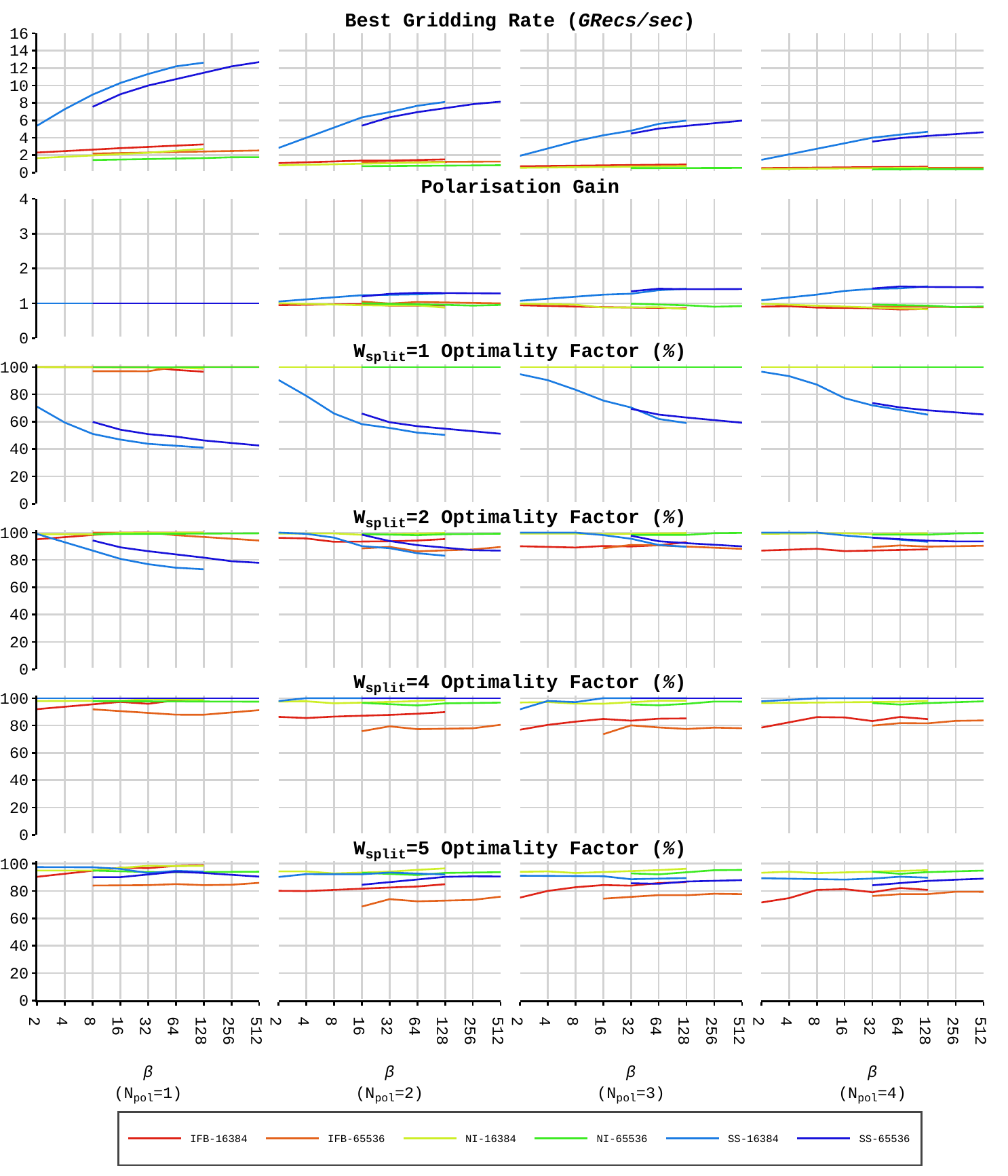}
\caption[Brute Force Search results for the Double-Precision Hybrid Gridder]{Brute Force Search results for the Double-Precision Hybrid Gridder. The layout is as described in Section \ref{sec:methodology:graphlayouts}. Performance metrics plotted in this figure are defined in Section \ref{sec:methodology:perfromancemetricsbrute}.}
\label{fig:hybrid:brute-double}
\end{figure}

\begin{figure}[h]
\centering
\includegraphics[page=1,width=\linewidth]{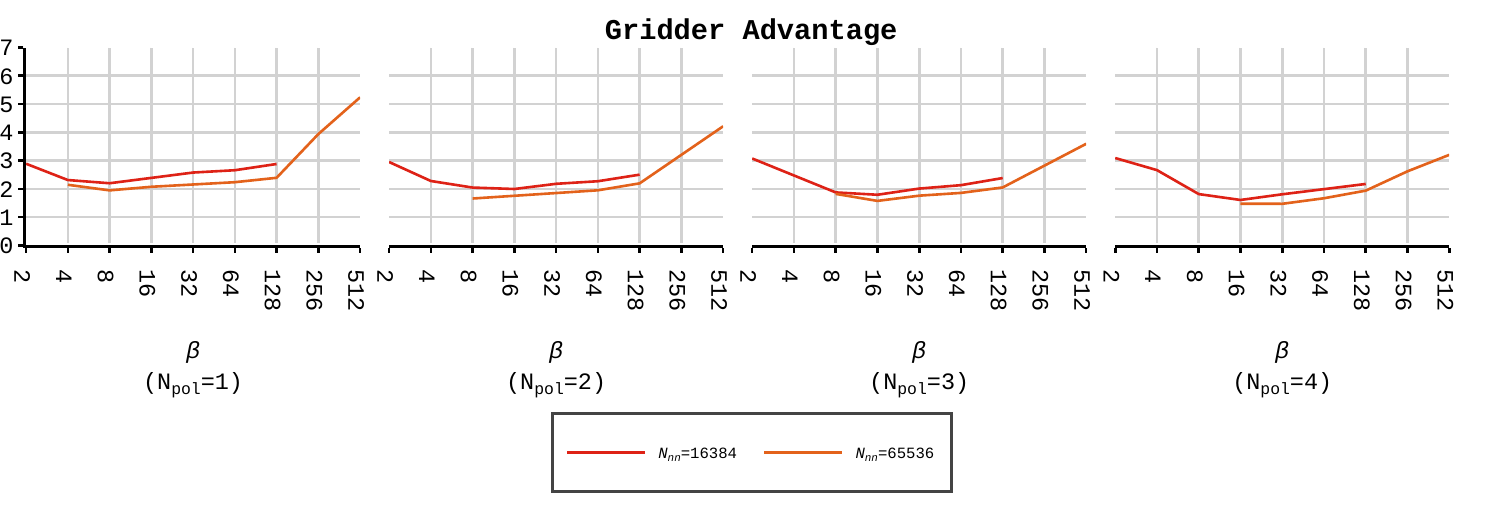}
\begin{center}
    \smaller{\textbf{(a)} Single-Precision}
\end{center}
\includegraphics[page=1,width=\linewidth]{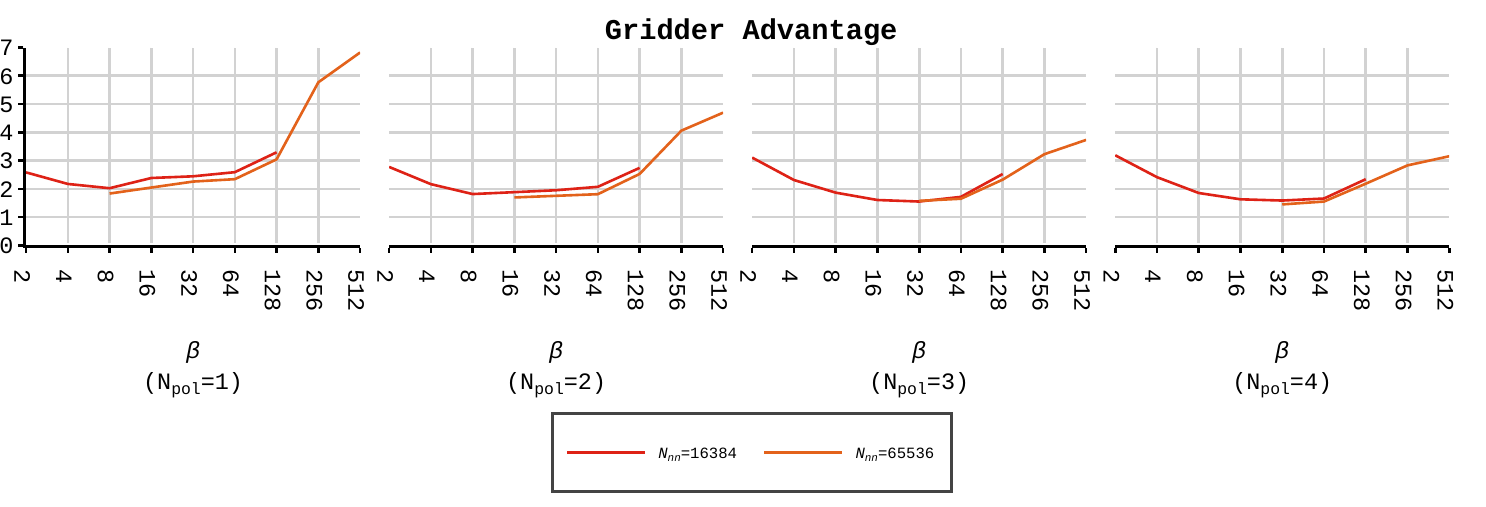}
\begin{center}
\smaller{\textbf{(b)} Double-Precision}
\end{center}
\caption[Gridder Advantage results for the Hybrid Gridder]{Gridder Advantage results for the Single and Double Precision Hybrid Gridders. The layout is as described in Section \ref{sec:methodology:graphlayouts}. The \gridderadvantage metric is defined in Section \ref{sec:methodology:performancemetricgridderadvantage}.}
\label{fig:hybrid:gridderadvantage}
\end{figure}

We execute a Brute Force Search on the Hybrid Gridder over the stated Tuning Parameters, in the way discussed in Section \ref{sec:methodology:tuning}. Measurements for the related group of Performance Metrics described in Section \ref{sec:methodology:perfromancemetricsbrute} are plotted in Figure \ref{fig:hybrid:brute-single} for Single-Precision and Figure \ref{fig:hybrid:brute-double} for Double-Precision.  Results for the \gridderadvantage metric are plotted in Figure \ref{fig:hybrid:gridderadvantage}. 

From the Brute Force Search, we discover optimal solutions, which we report and analyse in Section \ref{sec:hybrid:optimalsolutions}.

In the next sub-sections, we analyse the stated Brute Force Search results.


\subsection{\bestgridratecomma \polgain and \gridderadvantage}
A look at the \bestgridrate shows excellent Performance for  experiments done with SS Ordering Mode, but poor Performance otherwise. Based on what we discussed in the previous chapter, such a variation in Performance is due to the \commitratecomma which we will discuss further on.

The \polgain metric shows that the Hybrid Gridder is less capable than the Convolution Gridder to take advantage of gridding multi-polarisation together. The reasons why are discussed in the detailed analyses.

The measured \gridderadvantage shows that the Hybrid Gridder Performs faster than the Convolutional Gridder, but only when the SS Ordering Mode is used. \gridderadvantage goes down with an increase in $N_{\text{pol}}$.

\subsection{Effects of the \textit{Warp Splitting Factor}}

$W_{\text{split}}=4$ is clearly the optimal choice for experiments using the SS Ordering Mode. Splitting the warp in five instead of four reduces Performance slightly, and we hypothesise that this is due to the loading of data which is not fully aligned in the $W_{\text{split}}=5$ case, which we leave it as a hypothesis. 

Interestingly enough, $W_{\text{split}}=1$ is optimal for the IFB and NI Ordering Modes and the gain in Performance when compared to $W_{\text{split}}=4$ is at maximum around 37\%. In this thesis, we will not seek to understand the nature of this gain, in view that the Hybrid Gridder does not perform well with NI and IFB Ordering Modes. Nevertheless, it is very likely that the Hybrid Gridder is adapting to a high level of grid committing.  

\subsection{Optimal solutions}
\label{sec:hybrid:optimalsolutions}
Table \ref{tab:hybrid:bruteforceresults} lists solutions we found through Brute Force Search that are optimal for SS and NI Ordering Modes. They are unsuitable for experiments using the IFB Ordering Mode. 

\begin{table}[]
\centering
\begin{tabular}{@{}cc@{}c@{}ccccc@{}}
\toprule
\multirow{2}{*}{Precision} & \multirow{2}{*}{$N_{\text{pol}}$} & \rule{1.1cm}{0cm} & \multicolumn{5}{c}{Tuning Parameters} \\ 
\cmidrule{4-8} & & & $W_{\text{split}}$ &$X_{\text{Load}}$ & $\text{SM}$ & $B_{\text{Warps}}$  & $G_{\text{Block}}$ \\ \midrule
\multirow{4}{*}{single}   & 1 & & 4  & 16  & 1 & 8  & 1600 \\
                          & 2 & & 4  & 16  & 1 & 16  & 800 \\
                          & 3 & & 4  & 16  & 1 & 16  & 800 \\
                          & 4 & & 4  & 16  & 1 & 8   & 1600\\ \midrule
\multirow{4}{*}{double}   & 1 & & 4  & 16  & 1 & 16  & 1600\\
                          & 2 & & 4  & 16  & 1 & 16  & 800 \\
                          & 3 & & 4  & 16  & 1 & 8   & 1600\\
                          & 4 & & 4  & 16  & 1 & 4   & 1600\\ \bottomrule
\end{tabular}
\caption[Optimal Solutions for the Hybrid Gridder]{Optimal Solutions for the Hybrid Gridder that we discovered and chose through Brute Force Search.}
\label{tab:hybrid:bruteforceresults}
\end{table}

\begin{figure}

\includegraphics[page=1,width=\linewidth]{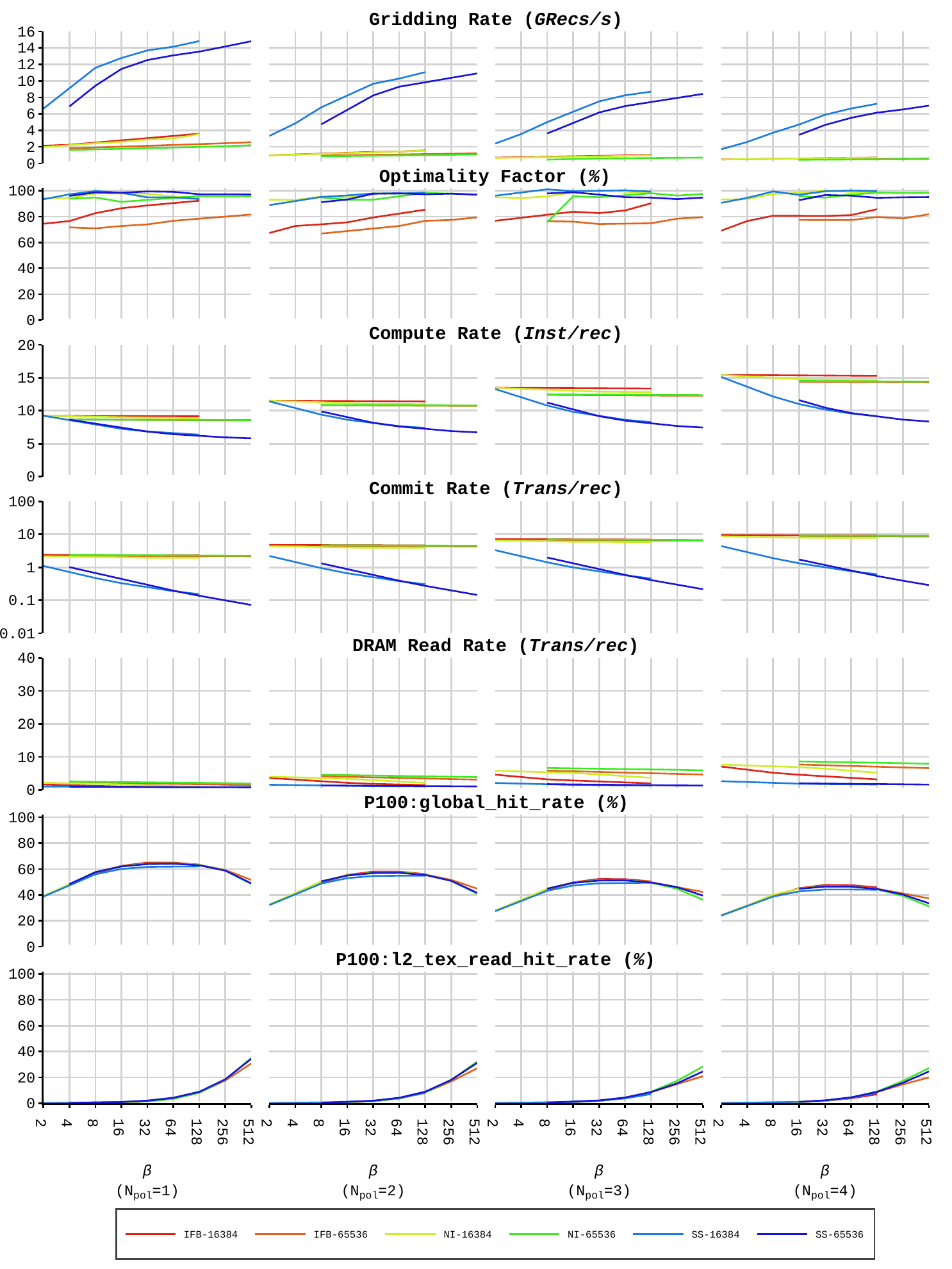}
\caption[Optimal Solutions Performance results for the Single-Precision Hybrid Gridder]{Optimal Solutions Performance results for the Single-Precision Hybrid Gridder. The layout is as described in Section \ref{sec:methodology:graphlayouts}. Performance metrics plotted in this figure are defined in Section \ref{sec:methodology:optimalsolutionsperformancemetrics}.}
\label{fig:hybrid:sims-single}
\end{figure}
\begin{figure}
\includegraphics[page=1,width=\linewidth]{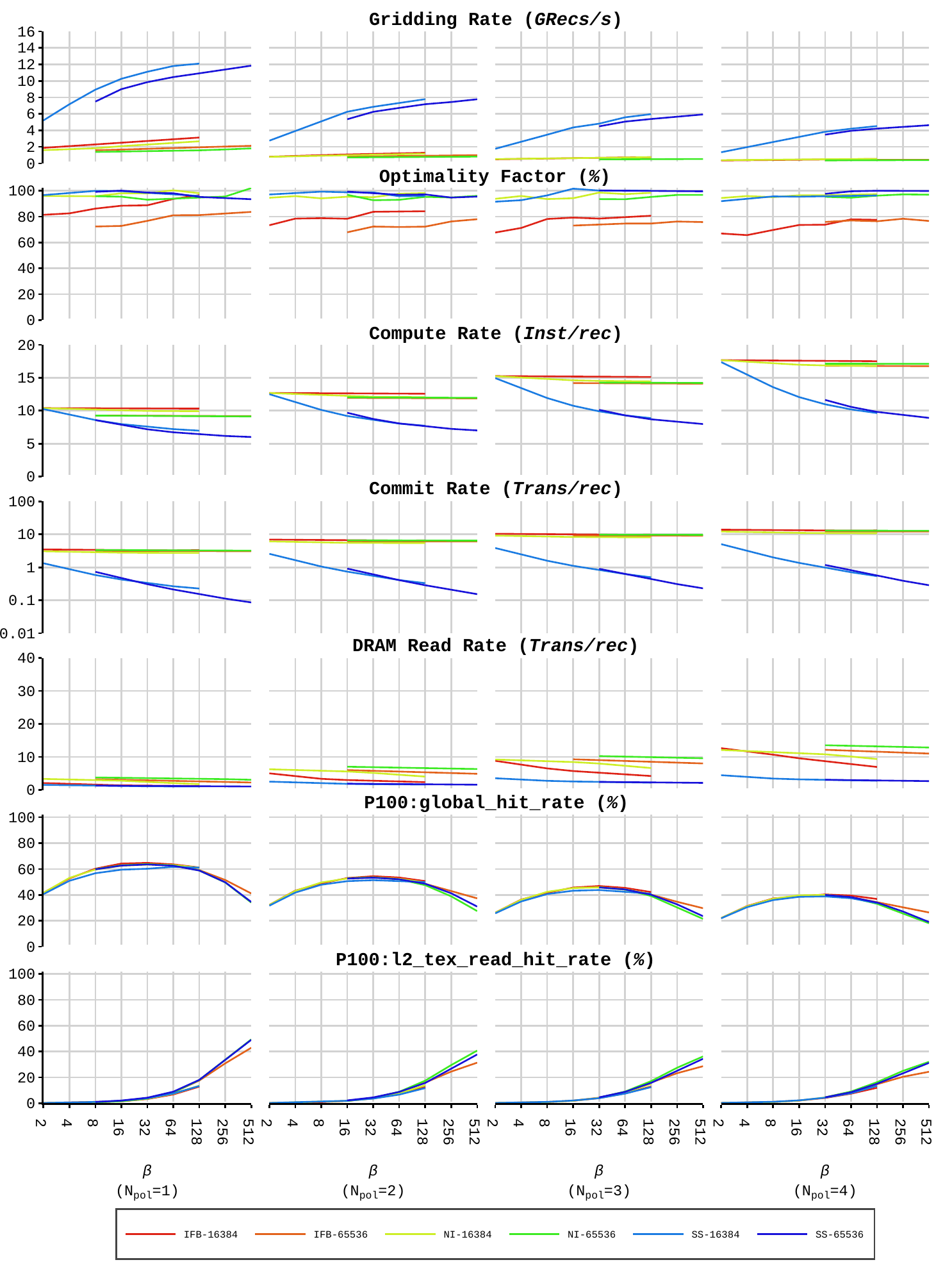}
\caption[Optimal Solutions Performance results for the Double-Precision Hybrid Gridder]{Optimal Solutions Performance results for the Double-Precision Hybrid Gridder. The layout is as described in Section \ref{sec:methodology:graphlayouts}. Performance metrics plotted in this figure are defined in Section \ref{sec:methodology:optimalsolutionsperformancemetrics}.}
\label{fig:hybrid:sims-double}
\end{figure}
\begin{figure}

\includegraphics[page=1,width=\linewidth]{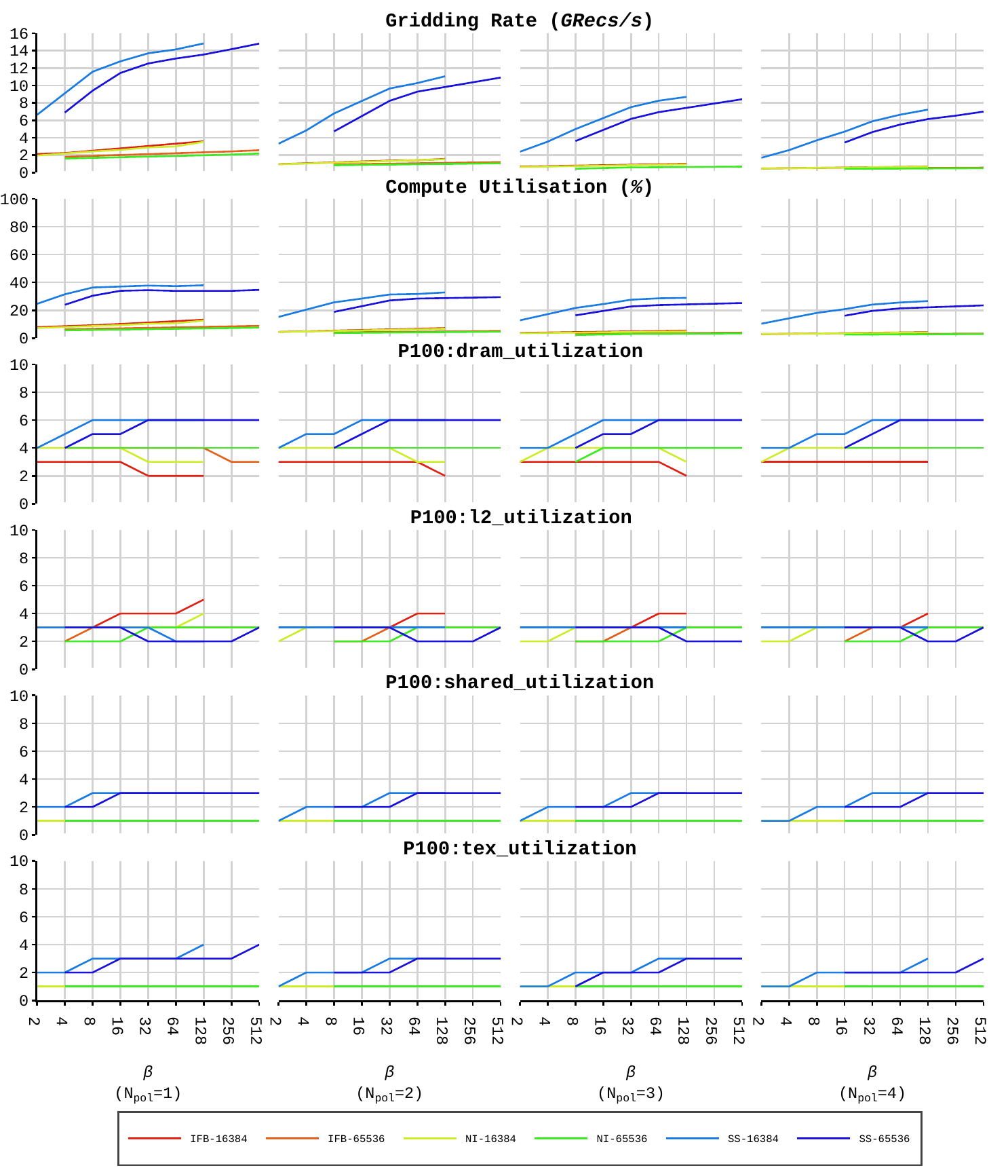}
\caption[Utilisation results for the Single-Precision Hybrid Gridder]{Utilisation results for the Single-Precision Hybrid Gridder. The layout is as described in Section \ref{sec:methodology:graphlayouts}. Performance metrics plotted in this figure are defined in Section \ref{sec:methodology:performancemetricutilisation}.}
\label{fig:hybrid:sims-single-utilisation}
\end{figure}
\begin{figure}
\includegraphics[page=1,width=\linewidth]{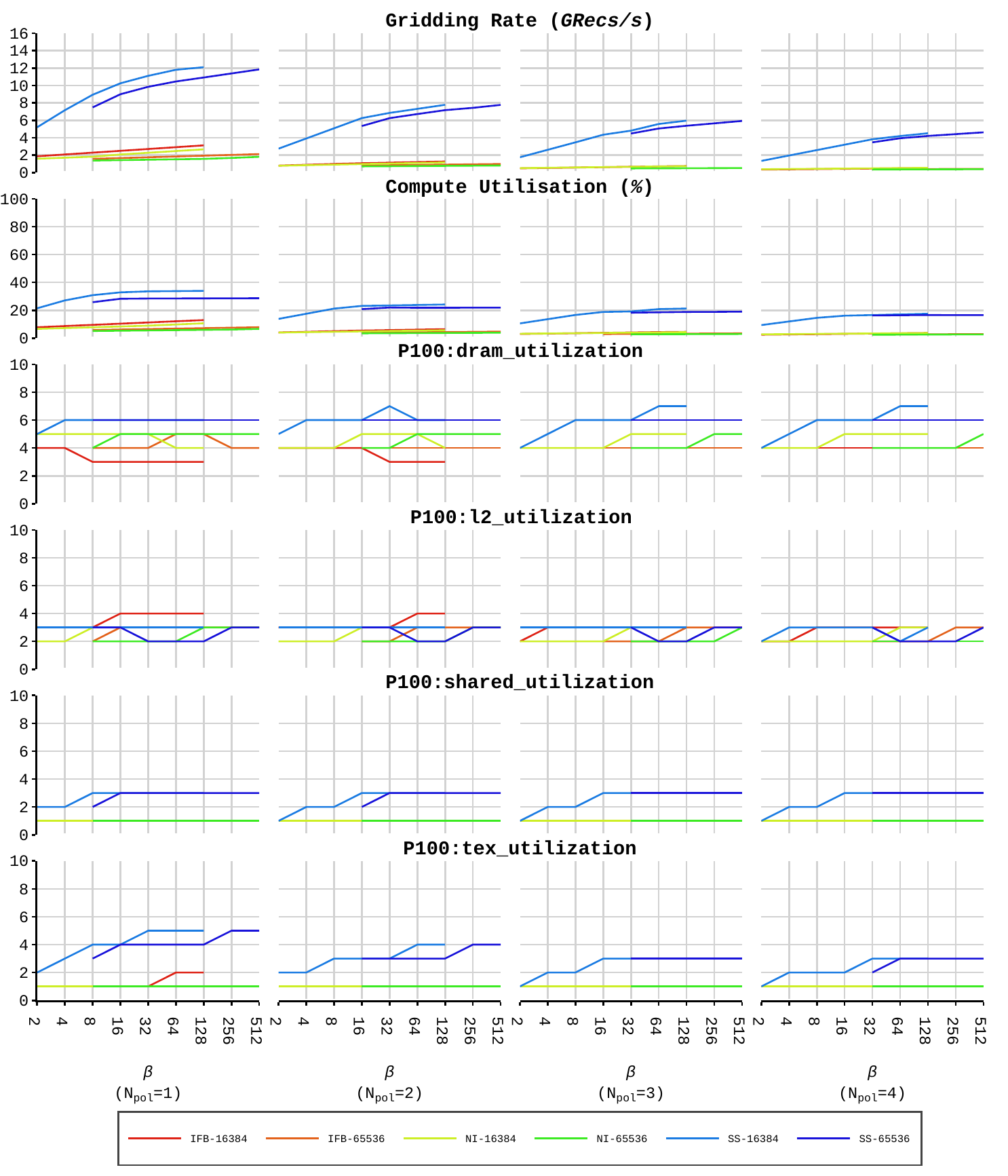}
\caption[Utilisation results for the Double-Precision Hybrid Gridder]{Utilisation results for the Double-Precision Hybrid Gridder. The layout is as described in Section \ref{sec:methodology:graphlayouts}. Performance metrics plotted in this figure are defined in Section \ref{sec:methodology:performancemetricutilisation}.}
\label{fig:hybrid:sims-double-utilisation}
\end{figure}

\section{In-depth analyses}

In Figures \ref{fig:hybrid:sims-single} and \ref{fig:hybrid:sims-double},
we plot the Optimal Solution Performance results, and in Figures \ref{fig:hybrid:sims-single-utilisation} and \ref{fig:hybrid:sims-double-utilisation}, we plot the Utilisation results for experiments on the optimal solutions given in Table \ref{tab:hybrid:bruteforceresults}.

We tabulate the results of the Maximum Performance Experiments in Table \ref{tab:hybrid:hybridmaxtheoratical}. We conduct these experiments using the same rationale we used for the Convolutional Gridder, whereby we modify the pre-processing phase and force the coordinates of all records to point to the same NN Grid Pixel. We also disabled compression, but due to memory limitations, we only gridded 12 frequency channels from the input $uv$-profile.

\begin{table}[]
\centering
\begin{tabular}{@{}l@{}c@{}cccc@{}}
\toprule
 
$N_{\text{pol}}$ &\rule{1.7cm}{0cm}&\begin{minipage}[]{1.2cm}\centering1\end{minipage} &\begin{minipage}[]{1.2cm}\centering2\end{minipage}& \begin{minipage}[]{1.2cm}\centering3\end{minipage}& \begin{minipage}[]{1.2cm}\centering4\end{minipage}\\ \midrule
\rowcolor{mygrey}\multicolumn{6}{c}{Single-Precision} \\ \midrule
\gridrate \texttt{(GRecs/s)} & & 17.6 & 13.6 & 11 & 9.55  \\
\gridderadvantage & & 2.78 & 3.17 &  3.58 & 2.63 \\
\polgain & & 1 & 1.54 & 1.87 & 2.17 \\
\computerate \texttt{(Inst/rec)} & & 5.44 & 6.19 & 6.82 & 7.63 \\
\globalhitrate \texttt{(\%)} & & 24.9 & 19.9 & 16.6 & 14.2  \\
\ltexrate \texttt{(\%)}& & 0.0728 & 0.0734 & 0.0891 & 0.0918 \\
\commitrate \texttt{(Trans/rec)}& & 2.4e-03 & 4.81e-03 & 7.21e-03 & 9.62e-03  \\
\dramreadtransactionsrate \texttt{(Trans/rec)} & & 0.75 & 1 & 1.25 & 1.5  \\
\computeutilisation \texttt{(\%)} & & 38.6 & 33.8 & 30.1 & 29.3 \\
\texutilisation & & 4 & 4 & 4 & 3 \\
\ltwoutilisation & & 2 & 2 & 2 & 2 \\
\dramutilisation & & 6 & 6 & 7 & 7 \\
\sharedutilisation & & 4 & 4 & 4 & 4 \\\midrule
\multicolumn{6}{c}{\cellcolor{mygrey}Double-Precision} \\ \midrule
\gridrate \texttt{(GRecs/s)} & & 14.3 & 10.5 & 7.73 & 6.32  \\
\gridderadvantage & & 3.58 &  2.76 & 2.39 & 2.23 \\
\polgain & & 1 & 1.47 & 1.62 & 1.76 \\
\computerate \texttt{(Inst/rec)} & & 5.53 & 6.39 & 7.21 & 8 \\
\globalhitrate \texttt{(\%)} & & 33.2 & 24.9 & 19.8 & 16.5  \\
\ltexrate \texttt{(\%)}& & 0.161 & 0.169 & 0.188 & 0.188 \\
\commitrate \texttt{(Trans/rec)} & & 4.81e-03 & 4.81e-03 & 7.21e-03 & 4.83e-03   \\
\dramreadtransactionsrate \texttt{(Trans/rec)} & & 1 & 1.5 & 2 & 2.5  \\
\computeutilisation \texttt{(\%)} & & 31.8 & 27 & 22.4 & 20.3 \\
\texutilisation & & 6 & 5 & 4 & 4 \\
\ltwoutilisation & & 2 & 2 & 2 & 2 \\
\dramutilisation &  & 7 & 7 & 7 & 7 \\
\sharedutilisation & & 4 & 4 & 4 & 4 \\
 \bottomrule

\end{tabular}
\caption[Results of the Maximum Performance Experiments for the Hybrid Gridder]{Results of the Maximum Performance Experiments for the Hybrid Gridder.}
\label{tab:hybrid:hybridmaxtheoratical}
\end{table}

We now present an in-depth analysis of the Hybrid Gridder's Performance based on the results obtained. We also include comparisons with the Convolutional Gridder.

\subsection{Boundedness and utilisation}

Utilisation Results show that the Hybrid Gridder is in general memory-bound on the device memory, since the \dramutilisation metric is in general equal to 6 or higher. The only exception is  when $\beta$ is low, whereby such experiments are neither compute nor memory bound, since none of the utilisation metrics reach the 60\% threshold. In stark difference with the Convolutional Gridding, utilisation of the L2 Cache and the Unified Cache never bound the Hybrid Gridder, and the  \computeutilisation is always lower than 40\%, implying that the Hybrid Gridder is far from being compute-bound. In-line with what we have discussed in Section \ref{sec:methodology:optimalaccesspattern}, the loss of boundedness and increase in latency at low value of $\beta$ is due to a high level of grid committing with degraded access patterns.

\subsection{Compute}

Based on the findings on utilisation we just discussed, it is reasonable to conclude that the level of compute, as measured by the \computerate metric, has only a minor impact on the Performance of the Hybrid Gridder.

Let us give some observations regarding the \computeratestop

\begin{itemize}
    \item As stated in Section \ref{sec:hybrid:output}, grid committing requires some extra Logic, which is responsible for the increase in \computerate when there is an increase in \commitratestop  
    \item There is a slight increase in \computerate with a decrease in $N_{nn}$. We attribute the increase to instructions related to initialisation that average out less in the metric because of a decrease in workload. 

\end{itemize}

\subsection{GCF data retrieval}

Loading of the GCF through the Unified Cache presented a similar behaviour to that of the Convolution Gridder. However, since the Hybrid Gridder deals with a smaller sized GCF, there are some differences worth to note.

\begin{itemize}
\item Trends in \ltexrate and \dramreadtransactionsrate clearly show that the L2 Cache is never under pressure, even at the highest $\beta$ considered. We can explain this behaviour by pointing out that the L2 Cache in the Hybrid Gridder has enough space to store all the GCF data even at $\beta=512$, since the GCF data size is $6\beta$ times  smaller than the GCF data size of the Convolutional Gridder.

\item We find no evidence in our results showing any chocking effects. The Hybrid Gridder requires far fewer transactions per record to load GCF data than the Convolutional Gridder, and therefore GCF data loading should be competing less for Unified Cache resources.

\item The \globalhitrate is visibly lower for the Hybrid Gridder when compared to the Convolutional Gridder. This behaviour happens because the loading transactions per record ratio between GCF data loading and record data loading is much lower for the Hybrid Gridder than for the Convolutional Gridder.

\item The \globalhitrate never peaks at $\beta=1$, since, by design, all Sub-Warp Gridders will request the same GCF data at $\beta=1$, which will be coalesced in the same transactions. Using argumentation made for a similar observation in the Convolutional Gridder  (Section \ref{sec:2dgridding:gcfdataretrieval}) we can explain why \globalhitrate never peaks at $\beta=1$ in the Hybrid Gridder.
\end{itemize}

\subsection{Loading of records from global memory and use of shared memory}

We state that loading of records from global memory controls the Performance of the Hybrid Gridder, particularly in experiments where the \commitrate is relatively low. The fact that the Maximum Performance Experiments are device memory bound corroborates our statement, in view that \dramutilisation in the Maximum Performance Experiments is mostly related to record data loading. The reader can validate our last statement by noticing that the \dramreadtransactionsrate in the Maximum Performance Experiments is always equal to the record size given in Table \ref{tab:hybrid:recordchunksplit} when expressed in terms of 32-bytes transactions.  

In experiments with a high \commitratecomma loading of record data needs to compete for device memory utilisation against grid committing. Therefore,  an interplay between loading of record data and grid committing is controlling Performance.

As a direct consequence of our statement, \polgain is controlled by the loading of record data and grid committing. When \commitrate is low enough to cause a negligible impact on Performance, \polgain is mostly controlled by the savings made in record size, while increasing $N_{\text{pol}}$.\footnote{There is a saving in record size of 4 integers for a unit increase of $N_{\text{pol}}$.} 

In regards to shared memory usage, the measured \sharedutilisation metric is sometimes higher than what we measured for the Convolutional Gridder, but in no way high enough to indicate any considerable limitations in Performance caused by the use of shared memory. 

\subsection{Grid committing}

As expected, grid committing controls Performance of the Hybrid Gridder in a similar way to the Convolutional Gridder. Results show  a good correlation between the \commitrate and \gridratecomma with some exceptions when changing  the value of $N_{nn}$. Based on experience, these exceptions are probably due to a deteriorating vicinity requirement.

The Hybrid Gridder did not perform well for NI and IFB Ordering Modes, because the \commitrate is high enough to dominate Performance over the loading of record data. 

When decreasing $\beta$ in the SS Ordering Mode experiments, \commitrate increases and begins to dominate Performance in such a way that the \gridrate and \polgain decrease with decreasing $\beta$.  The \polgain is set to drop with an increase in \commitratecomma because  atomic reductions of grid committing push \polgain towards 1 when increasing $N_{\text{pol}}$.

\section{Summary}
This chapter discussed our implementation of a Hybrid Gridder on the P100 for real-valued GCFs of support $6 \times 1$. We are here summarising in point form the main Performance results.

\begin{itemize}
    \item In our experiments, the Hybrid Gridder delivered acceptable Performance for the SS Ordering Mode only.  
    \item Maximum Performance of the Hybrid Gridder in ideal conditions per Precision and $N_{\text{pol}}$ is stated in the below table.

\begin{tabular}{@{}c@{}c@{}cccc@{}}
\toprule
 $N_{\text{pol}}$                                      &\rule{1cm}{0cm}& 1    & 2    & 3    & 4    \\ 
 \midrule
\begin{minipage}{8cm}\begin{spacing}{1}Single-Precision Maximum \gridrate (GRecs/s)\\[-20pt]\end{spacing}\end{minipage} & \rule{0cm}{25pt}&17.6 & 13.6 & 11 & 9.55 \\
\begin{minipage}{8cm}\begin{spacing}{1}Double-Precision Maximum \gridrate (GRecs/s)\\[-20pt]\end{spacing}\end{minipage} & \rule{0cm}{25pt}& 14.3    & 10.5 & 7.73 & 6.32 \\ \bottomrule
\end{tabular}

    \item When delivering maximum Performance, the Hybrid Gridder is memory-bound on device memory, where device memory utilisation is dominated at near 100\% by the loading of record data.
    \item The main limiter that pushes down Performance from the Hybrid Gridder maximum is grid committing.  
    \item In stark difference from the Convolutional Gridder, GCF data retrieval does not considerably degrade Performance in any of our experiments. In our experiments, there was never pressure on L2 Cache, since it was always able to hold all the GCF data in its store as to provide the best HIT rates. 
    \item Grid committing degrades Performance according to the frequency at which it is predicated to occur. In our experiments, elevated levels of grid committing occurred for low values of $\beta$ in the SS Ordering-Mode, and for all values of $\beta$ in the IFB and NI Ordering Modes. At elevated values of grid committing, the Hybrid Gridder loses memory-bound status, since grid committing injects latency due to its degraded pattern in accessing global memory. 
    \item When the Hybrid Gridder is delivering at maximum Performance, \polgain is limited by the loading of record data from device memory. 
    \item In our experiments \gridderadvantage varied from a low of 1.45 to a peak of 6.8.
\end{itemize}

\chapter{The NN Gridder}
\label{chap:purenn}
This chapter discusses and provides analyses to our naive and straightforward implementation of the NN Gridder.

This chapter is organised into four sections. The first section describes the implementation of the NN Gridder and the following section states and provides analyses on Brute Force Search results. Another section progresses towards an in-depth analyses of results for experiments on the optimal solutions. The last section is a simple summary of the main results.

Note that validation of the NN Gridder is discussed in Chapter \ref{chap:comparative}.

\section {Gridder implementation}
Our implementation of the NN Gridder is simple, since all that is required is to add input Visibility records to a pixel on the output grid. The main trick for good Performance is to seek ways how to access global memory with access patterns that try to adhere as much as possible to the coalescing and vicinity requirements. 

\subsection{Inputs and output}
The NN Gridder gets record data in two input streams as given hereunder:
\begin{enumerate}
    \item Position of records on the output NN Grid expressed by a 32-bit unsigned integer.
    \item A stream containing polarisation-interleaved Visibility data.
\end{enumerate}

The NN Gridder grids on a multi-polarised interleaved Grid with the origin at the corner.

\subsection{Occupancy reduction}
We expect the NN Gridder to be memory bound, and try to control Occupancy via two Tuning Parameters, $B_{\text{Warps}}$  and $M_{\text{shared}}$. $B_{\text{Warps}}$  sets the number of warps in a CUDA block, similar to how it is defined for the other Gridders.  $M_{\text{shared}}$ states the size in bytes of shared memory to be allocated per CUDA block. The NN Gridder does not use shared memory, but the allocation of shared memory sets up a ceiling on Occupancy. 

We also define the Tuning Parameter $G_{\text{Blocks}}$ for the NN Gridder, in the same way we did for the two other Gridders. In general, we do not expect $G_{\text{Blocks}}$ to have any effect on Performance, and we just included it to keep some similarities with the other Gridders.
\begin{figure}[h]
\includegraphics[page=1,width=\linewidth]{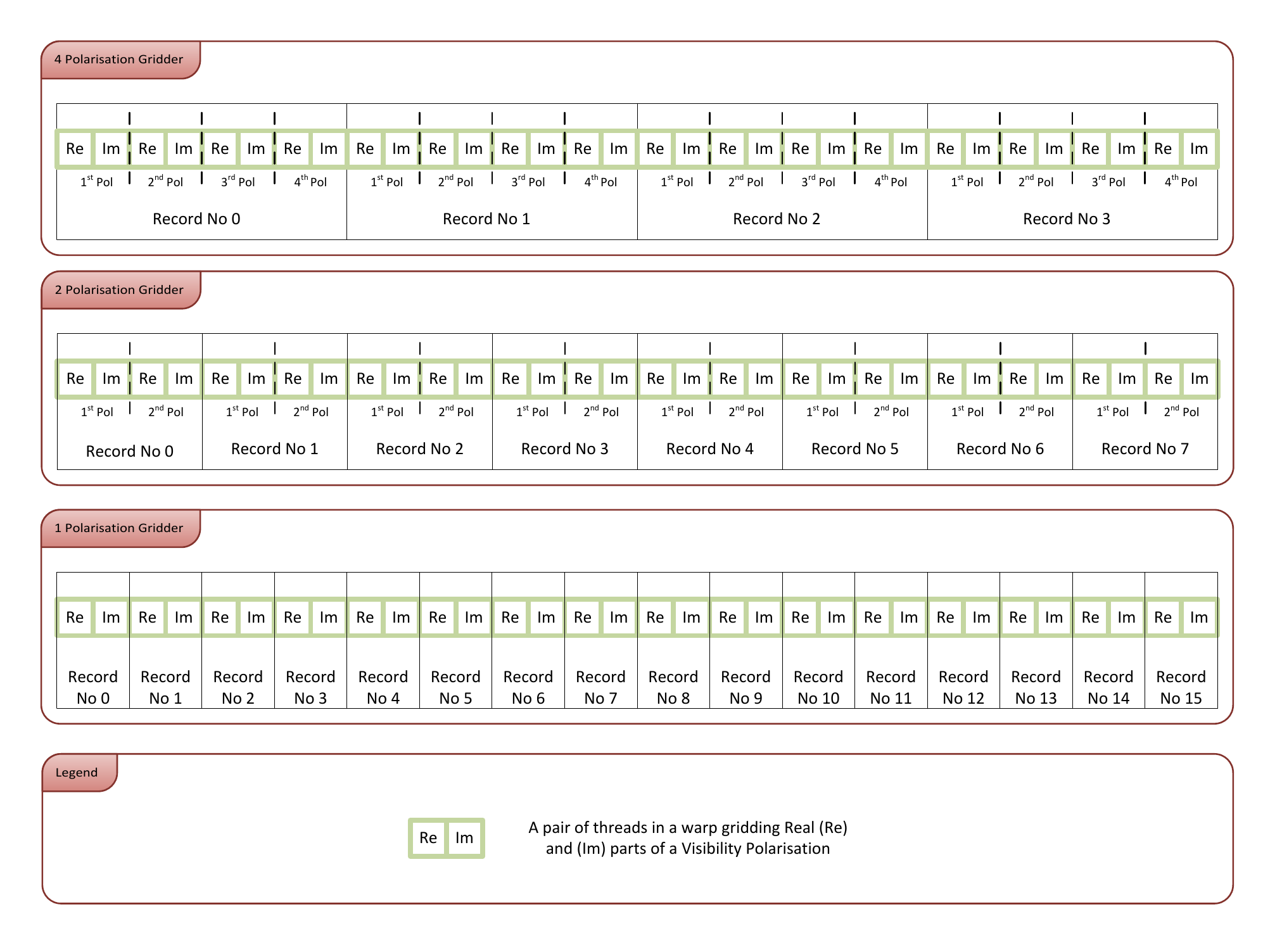}

\caption[Layout of the CUDA threads in the NN Gridder]{Layout of the CUDA threads in a warp for the NN Gridder.}
\label{fig:purenn:threadconfigueration}
\end{figure} 
\subsection{Thread configuration}
In the NN Gridder, we consider the whole CUDA Grid as just one big Gridder. All CUDA threads in the CUDA Grid are indexed in their natural order, with every $2 N_{\text{pol}}$ CUDA threads grouped and assigned to grid a full record as depicted in Figure \ref{fig:purenn:threadconfigueration}. Consecutive groups of CUDA threads are assigned to grid consecutive input records, such that a warp grids $32/(2\cdot N_{\text{pol}})$ records in parallel, given $N_{\text{pol}}=1,2$ or $4$. For $N_{\text{pol}}=3$, some records will be gridded by two warps. All threads will load the data required from the input streams and immediately commit to the output grid. If the CUDA grid provides fewer CUDA threads than needed to grid all records, then the CUDA threads will cycle through more than one record, in a way that the previously stated layout is kept intact.  

The layout ensures access to global memory to load input record data with optimal access patterns. The layout also helps to have access patterns for grid committing to be as close as possible to optimal, but at the end of the day, such optimally will depend on the input $uv$-profile. Interleaving the output grid helps to enhance the degraded access patterns, in particular the coalescence requirement for optimal access patterns. 

An increase in $N_{\text{pol}}$ will generally force an increase in coalescence for a warp grid commit request, since all polarisations of a given record are adjacent to each other. There is a guarantee that the NN Gridder  fully adheres to the coalescence requirement if run in Single-Precision with $N_{\text{pol}}=4$ or Double-Precision with $N_{\text{pol}}=2$ or $N_{\text{pol}}=4$. We would also like to get as close as possible to reach the vicinity requirement, for which we see that our best chances are when the input record data is ordered in SS Mode.

Finally, we note that the layout sets a minimum for the \commitratecomma which is obtained when grid committing adheres to the coalescence requirement. This minimum is described in Equation \ref{equ:pruning:mincommitrate}

\begin{equation}
\label{equ:pruning:mincommitrate}
    \texttt{Minimum \commitrate}=\frac{\texttt{sizeof(Precision)}\times 2N_{\text{pol}}}{32}
\end{equation}

where $\texttt{sizeof(Single-Precision)}=4$ and $\texttt{sizeof(Double-Precision)}=8$

\section{Brute Force Search}
\begin{figure}
\centering
\includegraphics[page=1]{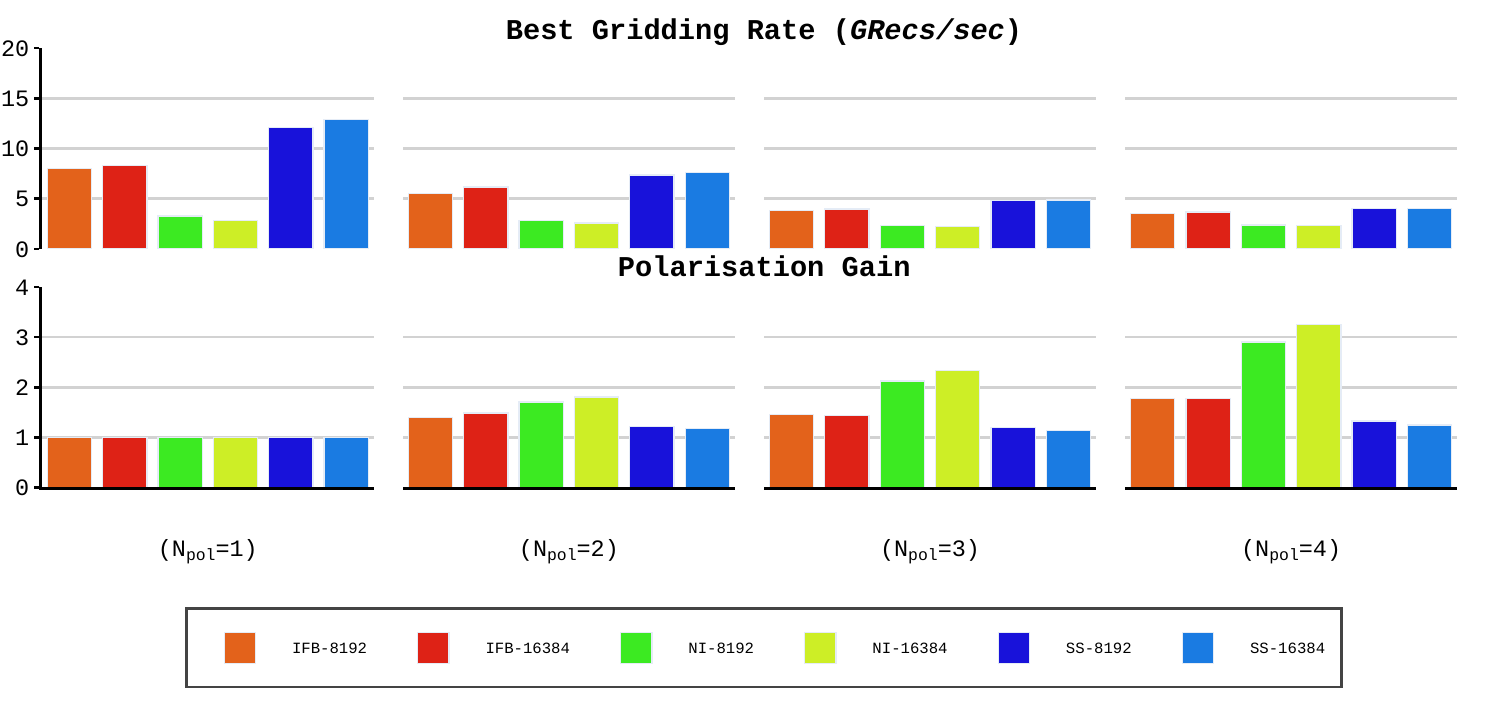}
\begin{center}
    \smaller{\textbf{(a)} Single-Precision}
\end{center}
\includegraphics[page=1]{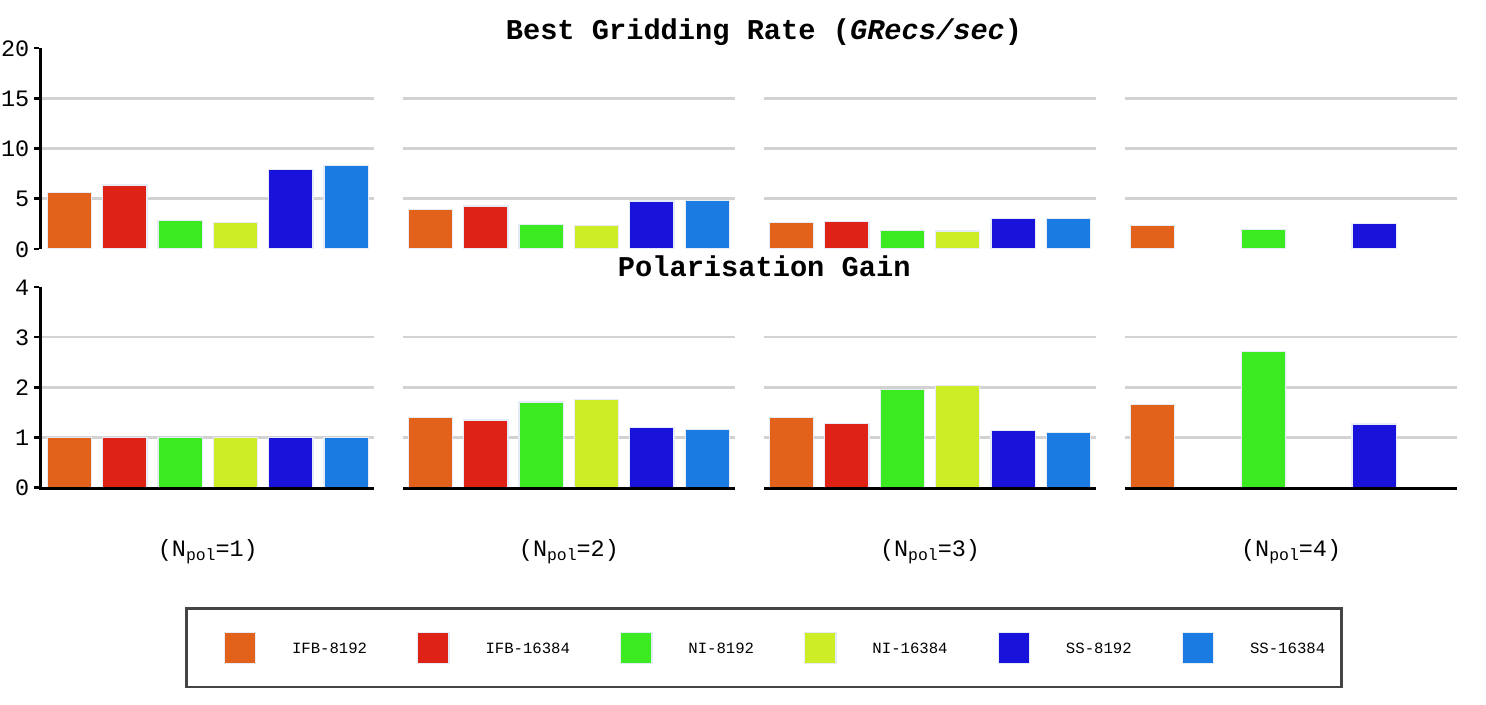}
\begin{center}
    \smaller{\textbf{(b)} Double-Precision}
\end{center}
\caption[Brute Force Search results for the NN Gridder]{Brute Force Search results for the Single and Double Precision NN Gridders. The layout is as described in Section \ref{sec:methodology:graphlayouts}. Performance metrics plotted in this figure are defined in Section \ref{sec:methodology:perfromancemetricsbrute}.}
\label{fig:purenn:brute_nn}
\end{figure}

\begin{figure}
\includegraphics[page=1]{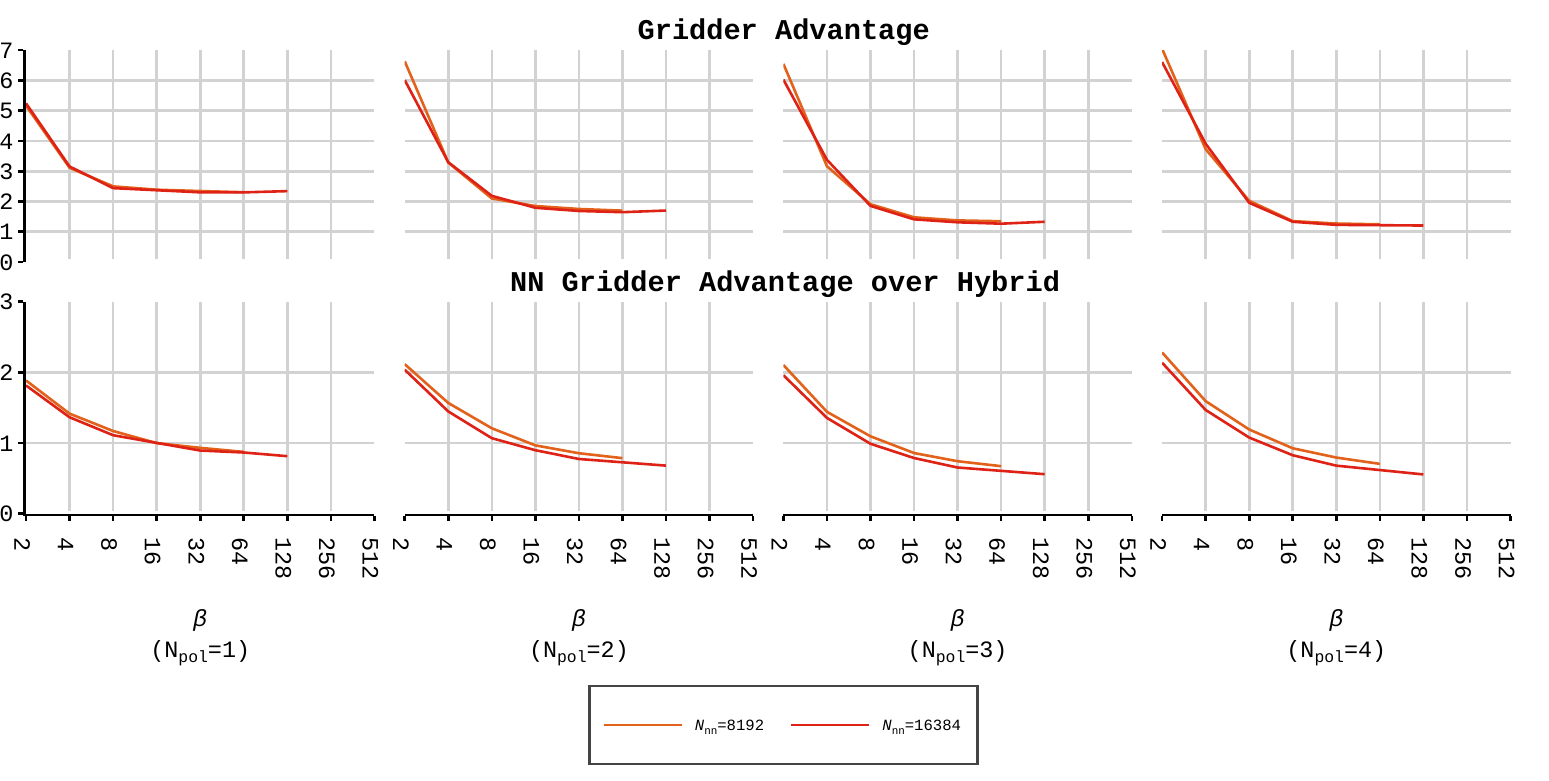}
\begin{center}
    \smaller{\textbf{(a)} Single-Precision}
\end{center}
\includegraphics[page=1]{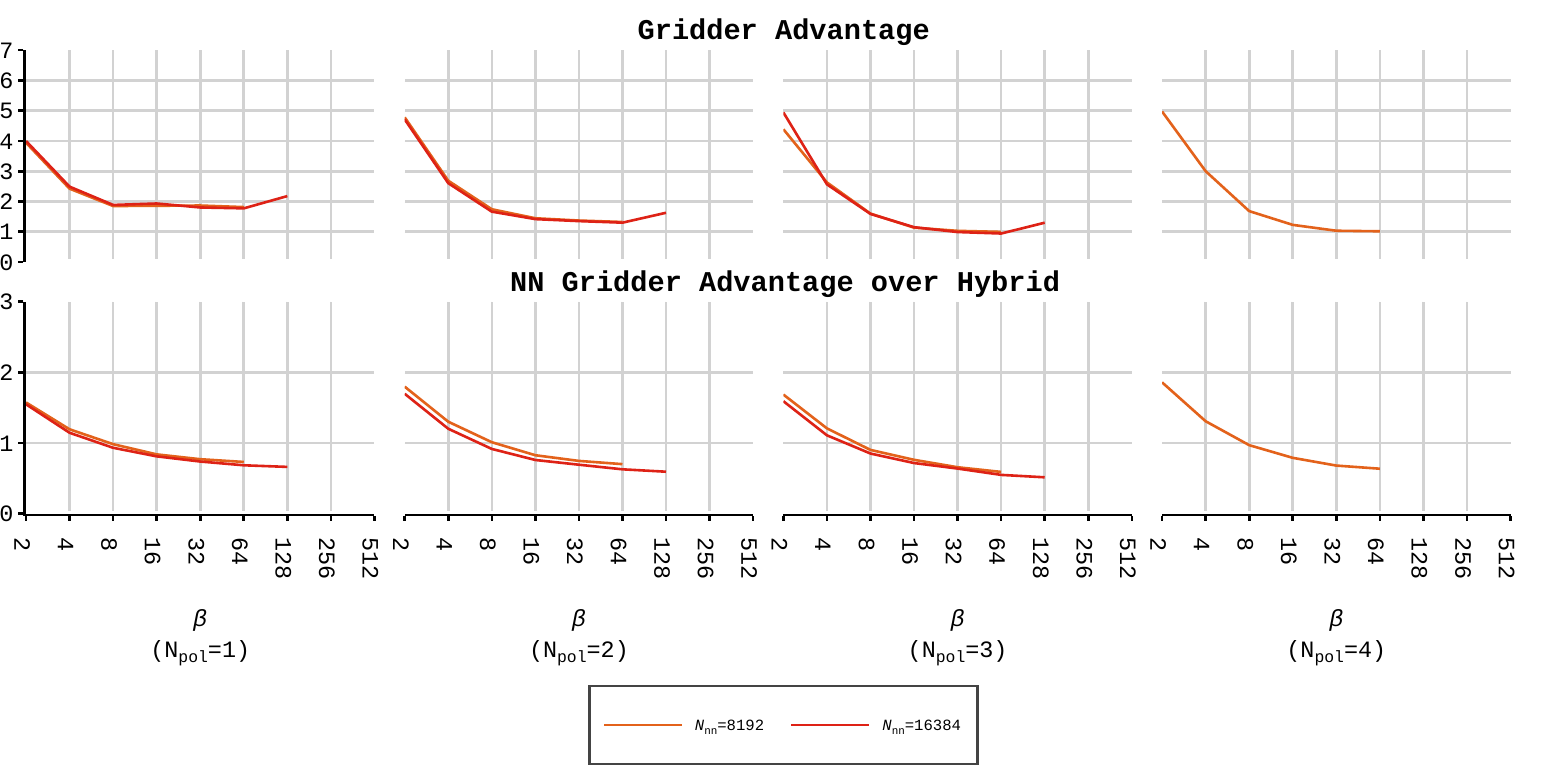}
\begin{center}
    \smaller{\textbf{(a)} Double-Precision}
\end{center}
\caption[Gridder Advantage results for the NN Gridder]{Gridder Advantage results for the Single and Double Precision NN Gridders. The layout is as described in Section \ref{sec:methodology:graphlayouts}. Performance metrics plotted in this figure are defined in Section \ref{sec:methodology:performancemetricgridderadvantage}.}
\label{fig:purenn:advantage_nn}
\end{figure}

We execute a Brute Force Search on the NN Gridder over the stated Tuning Parameters. Measurements for  the \bestgridrate and \polgain described in Section \ref{sec:methodology:perfromancemetricsbrute} are plotted in Figure \ref{fig:purenn:brute_nn}. Gridder Advantage results are plotted in Figure \ref{fig:purenn:advantage_nn}.

We discovered optimal solutions which we report and analyse in Section \ref{sec:purenn:optimalsolutions}. 

Due to memory limitations, we could not analyse the behaviour of the NN Gridder for $N_{nn}=65536$, and instead presented analyses on the behaviour of the NN Gridder with $N_{nn}=8192$.

In the next sub-sections we analyse the stated Brute Force Search results.

\subsection{\bestgridratecomma \polgain and \gridderadvantage}
\label{sec:purenn:bruteforcepolgain}
As expected, in our test scenarios, the NN Gridder grids fastest when the SS Ordering Mode is used, and Performance decreases when changing to the IFB and  NI Ordering Modes. There is a trend whereby the NN Gridder has the best \polgain for the  Ordering Mode that has the least Performance. Detailed analyses showed that the varying behaviour observed is related to grid committing, which we discuss in Section \ref{sec:purenn:gridcommiting}.

The Gridder Advantage results show that the NN Gridder grids faster than the Convolutional Gridder but the Hybrid Gridder beats the NN Gridder for all $\beta \ge 16$.

\subsection{Optimal solutions}
\label{sec:purenn:optimalsolutions}
Through the Brute Force Search, we discovered a solution that is optimal for all combinations of Precision and $N_{\text{pol}}$ for the experiments analysed in this chapter.  This solution is tabulated in Table \ref{tab:purenn:optimalsolutions}.

\begin{table}[]
\centering
\begin{tabular}{ccccc}
\toprule
Precision   & $N_{\text{pol}}$ & $M_{\text{shared}}$  & $B_{\text{Warps}}$  & $G_{\text{Blocks}}$  \\ \midrule
\multirow{4}{*}{single} & 1 & \multirow{4}{*}{49152} & \multirow{4}{*}{32} & \multirow{4}{*}{1600} \\
    & 2  &   &     &  \\
    & 3  &   &     &  \\
    & 4  &   &     &  \\ \midrule
\multirow{4}{*}{double} & 1  & \multirow{4}{*}{49152} & \multirow{4}{*}{32} & \multirow{4}{*}{1600} \\
    & 2  &   &     &  \\
    & 3  &   &     &   \\
    & 4  &   &     &  \\ \bottomrule        
\end{tabular}
\caption[Optimal Solutions for the NN Gridder]{Optimal Solutions for the NN Gridder that we discovered through Brute Force Search.}
\label{tab:purenn:optimalsolutions}
\end{table}

\begin{figure}
\includegraphics[page=1]{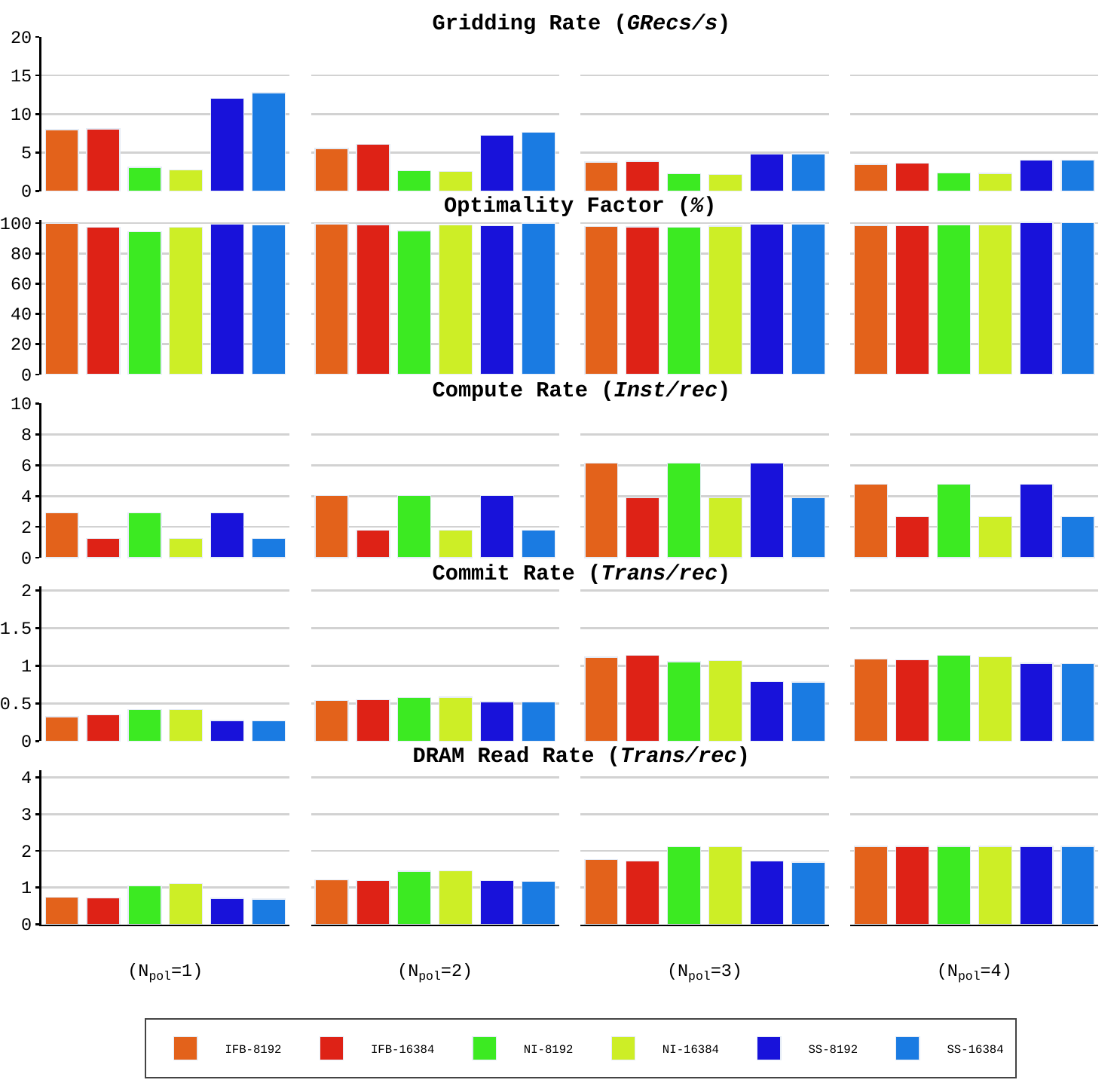}
\caption[Optimal Solutions Performance results for the Single-Precision NN Gridder]{Optimal Solutions Performance results for the Single-Precision NN Gridder.  The layout is as described in Section \ref{sec:methodology:graphlayouts}. Performance metrics plotted in this figure are defined in Section \ref{sec:methodology:optimalsolutionsperformancemetrics}.}
\label{fig:purenn:single_nn}
\end{figure}

\begin{figure}
\includegraphics[page=1]{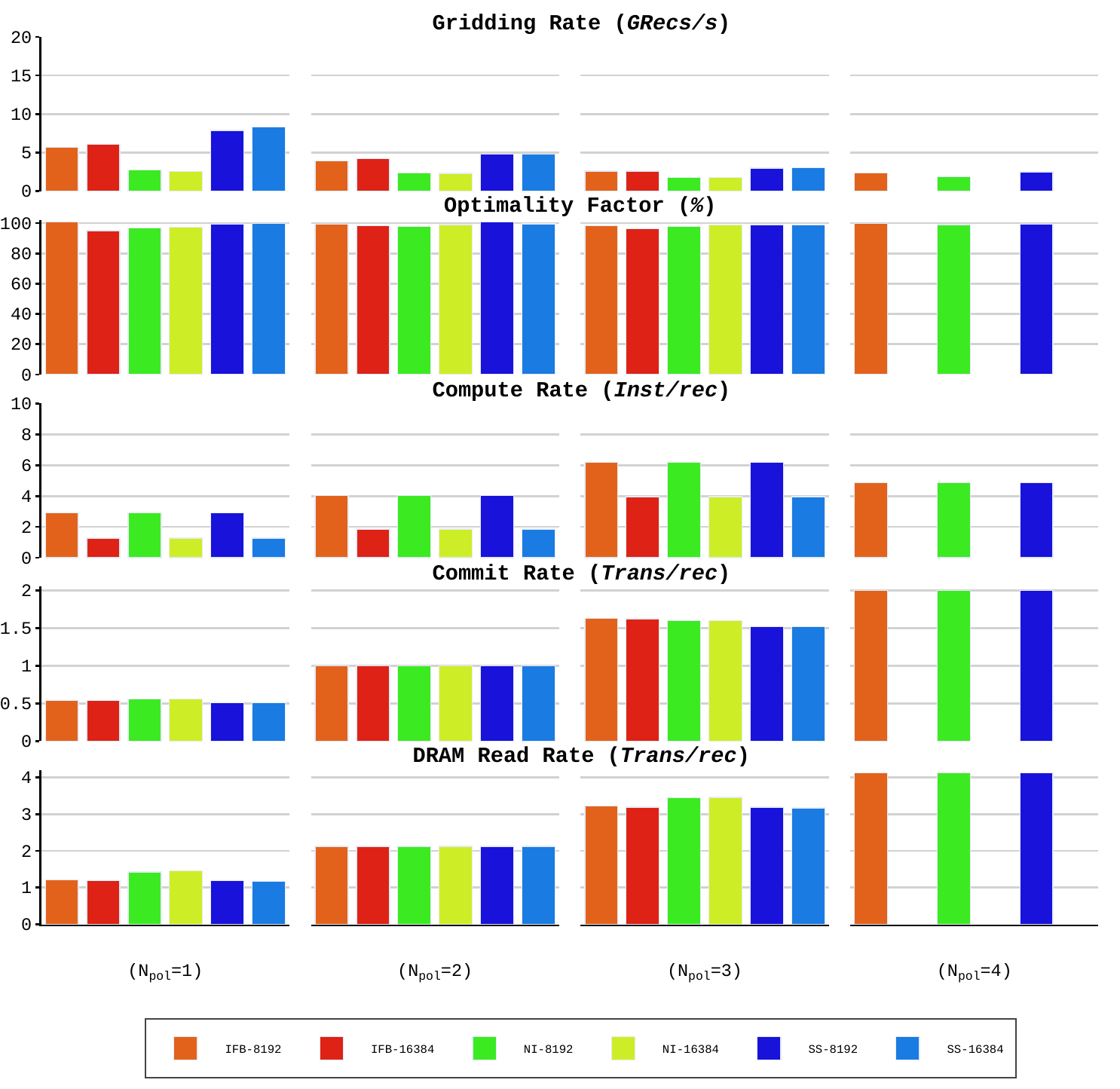}
\caption[Optimal Solutions Performance results for the Double-Precision NN Gridder]{Optimal Solutions Performance results for the Double-Precision NN Gridder.  The layout is as described in Section \ref{sec:methodology:graphlayouts}. Performance metrics plotted in this figure are defined in Section \ref{sec:methodology:optimalsolutionsperformancemetrics}.}
\label{fig:purenn:double_nn}
\end{figure}

\begin{figure}
\includegraphics[page=1]{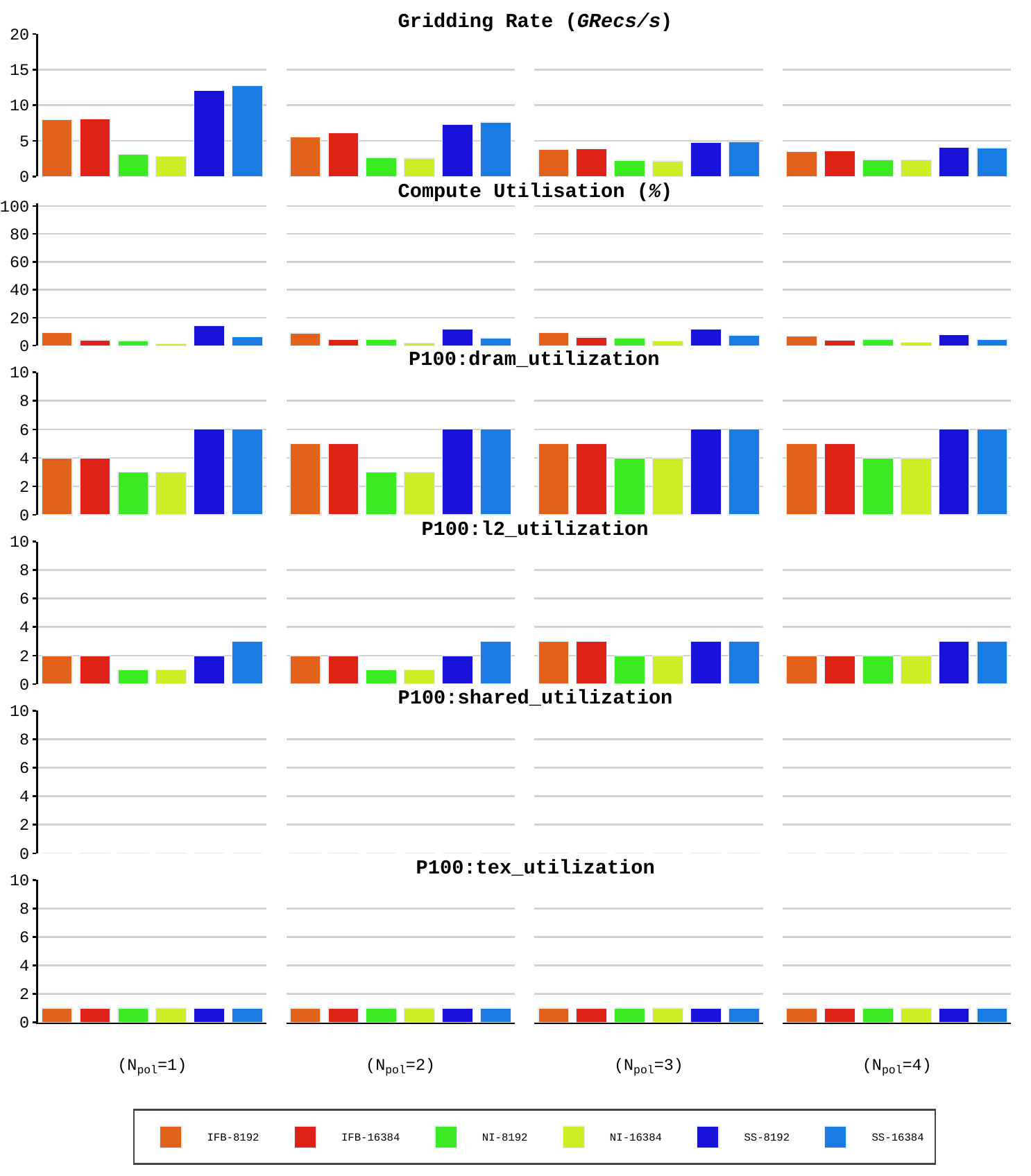}
\caption[Utilisation results for the Single-Precision NN Gridder]{Utilisation results for the Single-Precision NN Gridder.  The layout is as described in Section \ref{sec:methodology:graphlayouts}. Performance metrics plotted in this figure are defined in Section \ref{sec:methodology:performancemetricutilisation}.}
\label{fig:purenn:single_nn_utilisation}
\end{figure}

\begin{figure}
\includegraphics[page=1]{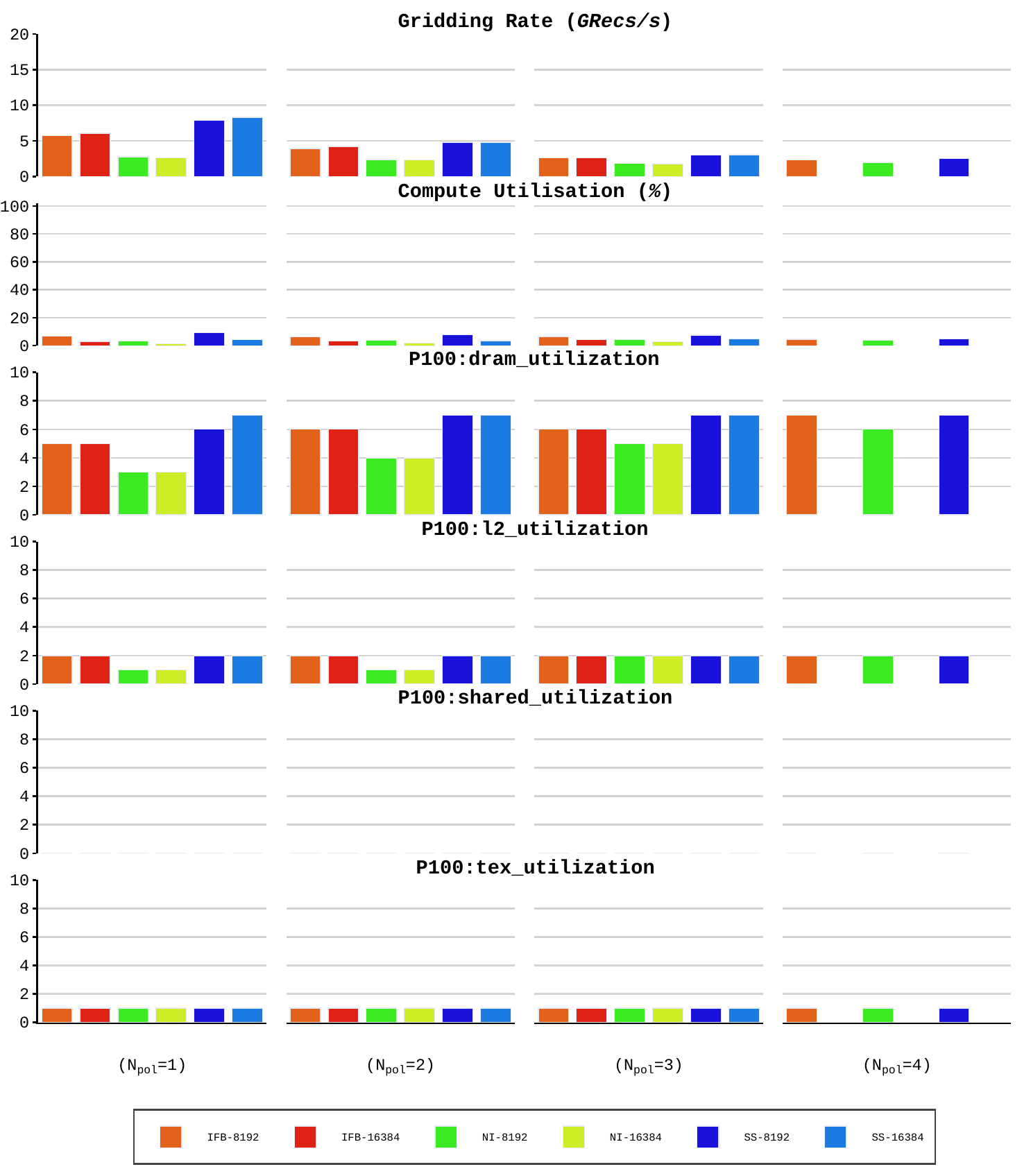}
\caption[Utilisation results for the Double-Precision NN Gridder]{Utilisation results for the Double-Precision NN Gridder.  The layout is as described in Section \ref{sec:methodology:graphlayouts}. Performance metrics plotted in this figure are defined in Section \ref{sec:methodology:performancemetricutilisation}.}
\label{fig:purenn:double_nn_utilisation}
\end{figure}

\section{Detailed analyses}
In Figures \ref{fig:purenn:single_nn} and \ref{fig:purenn:double_nn}, we plot the Optimal Solution Performance results and in Figures \ref{fig:purenn:single_nn_utilisation} and \ref{fig:purenn:double_nn_utilisation} we plot the Utilisation results. The results we plot are of experiments done with the discovered optimal solutions stated in Table \ref{tab:purenn:optimalsolutions}.

We tabulate the results of the Maximum Performance Experiments in Table \ref{tab:purenn:nnmaxtheoratical}. In these experiments, we modify the pre-processing phase to force input record data to have an incremental position on the NN Grid to ensure an optimal access pattern for grid committing.

\begin{table}[]
\centering
\begin{tabular}{@{}l@{}c@{}cccc@{}}
\toprule
 
$N_{\text{pol}}$ &\rule{1.7cm}{0cm}&\begin{minipage}[]{1.2cm}\centering1\end{minipage} &\begin{minipage}[]{1.2cm}\centering2\end{minipage}& \begin{minipage}[]{1.2cm}\centering3\end{minipage}& \begin{minipage}[]{1.2cm}\centering4\end{minipage}\\ \midrule
\rowcolor{mygrey}\multicolumn{6}{c}{Single-Precision}\\
\midrule
\gridrate \texttt{(GRecs/s)}  & & 15.3 & 7.85 & 5.4 & 4.06   \\
\gridderadvantage & &2.42 &	1.83 &	1.41 &	1.12 \\
\gridderadvantageonhybrid & & 0.87 & 0.58 &	0.49 & 0.44 \\
\polgain & & 1 & 1.03 & 1.06  & 1.06 \\  
\computerate \texttt{(Inst/rec)}& & 0.467 & 0.867 & 2.76 & 1.65 \\
\commitrate \texttt{(Trans/rec)} & &  0.25 & 0.5 & 0.75 & 1  \\
\dramreadtransactionsrate \texttt{(Trans/rec}) & & 0.625 & 1.13 & 1.63 & 2.13   \\
\computeutilisation \texttt{(\%)} & & 2.87 & 2.74 & 5.98 & 2.7 \\
\texutilisation & & 1 & 1 & 1 &1 \\
\ltwoutilisation & & 3 & 3 & 3 & 3 \\
\dramutilisation & & 6 & 6 & 6 & 6 \\
\sharedutilisation & &0 & 0 & 0 & 0 \\
 \midrule
\rowcolor{mygrey}\multicolumn{6}{c}{Double-Precision}\\
\midrule
\gridrate  \texttt{(GRecs/s)} & & 9.98 & 5.14 & 3.47 & 2.63   \\
\gridderadvantage & & 2.50 & 1.35 &	1.07 &	0.93 \\
\gridderadvantageonhybrid & & 0.70 &	0.49 &	0.45  &	0.42 \\
\polgain & & 1 & 1.03 & 1.04 & 1.05 \\
\computerate \texttt{(Inst/rec)} & & 0.518 & 0.965 & 2.9 & 1.85 \\
\commitrate \texttt{(Trans/rec)} &  & 0.5 & 1 & 1.5 & 2  \\
\dramreadtransactionsrate  \texttt{(Trans/rec)} & & 1.13 & 2.13 & 3.13 & 4.13    \\
\computeutilisation \texttt{(\%)}& & 2.08 & 2 & 4.06 & 1.96 \\
\texutilisation & & 1 & 1 & 1 & 1 \\
\ltwoutilisation & & 2 & 2 & 2 & 2 \\
\dramutilisation & & 8 & 8 & 8 & 8 \\
\sharedutilisation & & 0 & 0 & 0 & 0 \\
\bottomrule
\end{tabular}
\caption[Results of the Maximum Performance Experiments for the NN Gridder ]{Results of the Maximum Performance Experiments for the NN Gridder.}
\label{tab:purenn:nnmaxtheoratical}
\end{table}

\subsection{Boundedness and utilisation}

The Maximum Performance Experiments are memory-bound on the device memory, whereby the measured \dramutilisation metrics are at a value of six for Single-Precision and eight for Double-Precision. Utilisation values for the two caches (Unified and L2) are low. 
In the other experiments \dramutilisationcomma degrades. We attribute such degradation to the degrading access patterns of grid committing, which we discuss in Section \ref{sec:purenn:gridcommiting}.

All results show that compute is utilised minimally in the NN Gridder, whereby \computeutilisation is always below 20\%.

\subsection{Compute}

Based on the discussion made in the previous sub-section,  compute should have no major impact on the NN Gridder's Performance. We now explain some of the behaviour of the \computerate metric.
\begin{enumerate}
    \item The variation in \computerate with a change in $N_{nn}$ is due to initialisation logic averaging out differently.
    \item There is no variation in the \computerate for a change in Ordering Mode, which is expected, since there are no predicates based on the record that can vary execution. In particular, grid committing is guaranteed to happen for every record processed. 
    
  \item By design, \computerate increases with an increase in $N_{\text{pol}}$, but for $N_{\text{pol}}=3$, the increase does not follow the trend. The compute for $N_{\text{pol}}=3$ is much larger than suggested by the trend since Logic related to division and modulus with the integer 3 need more instructions to compute than with integers 1,2 and 4.  
\end{enumerate}

\subsection{Loading of input record data}
The loading of input record data is responsible for a maximum of 44\% of device memory transaction which happens in the Maximum Performance Experiment for Single Precision at $N_{\text{pol}}=1$. Variations in the percentage are dependent on the \commitratecomma $N_{\text{pol}}$ and Precision. Since Performance of the NN Gridder is impacted by memory usage, we conclude that the loading of input record data from memory has an impact on Performance. 

\subsection{Grid committing}
\label{sec:purenn:gridcommiting}

We first point out that the measurement of \commitrate taken for the Single-Precision experiments with $N_{\text{pol}}=4$ show the unexpected behaviour discussed in Section \ref{sec:methodology:optimalsolutionsperformancemetrics} when the \commitrate metric was being defined.  

Grid committing is the central player in limiting and varying Performance. The access patterns generated by grid committing play a central role in Performance. 

The Maximum Performance Experiments provide optimal access patterns which minimise the \commitrate and maximise \dramutilisationstop The \dramutilisation tracks very well the measured \gridrate when comparing experiments with equal $N_{\text{pol}}$ and Precision.  In the context of the NN Gridder, a decrease in \dramutilisation implies an increase in latency brought up by degrading grid committing access patterns.  

An increase in \commitrate seen in experiments with equal $N_{\text{pol}}$ and Precision is also another indication that access patterns are degrading. However, the \commitrate only changes when access patterns degrade for the coalescing requirement and not the vicinity requirement. For example, there is no change in commit rate for Double-Precision experiments with $N_{\text{pol}}=2$ and $N_{\text{pol}}=4$, where adherence to the coalescence requirement is guaranteed. 

We previously stated that the Maximum Performance Experiments minimise the \commitrate and we validate this  statement  by noting that the \commitrate of the Maximum Performance Experiments abide with Equation \ref{equ:pruning:mincommitrate}, which states what the minimum \commitrate should be. 

We finalise our analyses by commenting on the behaviour of the \polgainstop All Maximum Performance Experiments reported a \polgain of just a little bit more than one. Theoretically, if we had to consider the grid committing access pattern, we expect a value exactly equal to one. The extra small increase in \polgain is probably due to the increase in workload when increasing $N_{\text{pol}}$. Such an increase forces each CUDA thread to grid more records, which reduces the impact of initialisation Logic in all CUDA threads. 

As we stated in Section \ref{sec:purenn:bruteforcepolgain}, \polgain seems to be highest for the least performant Ordering Modes. It stands to reason that, since the worst Performing Ordering Modes should produce the most degraded grid committing access patterns, the mandatory improvement in the access patterns caused by an increase in $N_{\text{pol}}$ gives the best boost in Performance.

\section{Summary}
This chapter discussed our implementation of an NN Gridder on the P100. We are here summarising in point form the main Performance results.

\begin{itemize}
\item In our experiments, the NN Gridder delivered the highest level of Performance using the SS Ordering Mode. Performance decreases for the IFB and NI Ordering Modes in the order mentioned. 

\item Maximum Performance of the NN Gridder in ideal conditions per Precision and $N_{\text{pol}}$ is stated in the below table.

\begin{tabular}{@{}c@{}c@{}cccc@{}}
\toprule
 $N_{\text{pol}}$                                      &\rule{1cm}{0cm}& 1    & 2    & 3    & 4    \\ 
 \midrule
\begin{minipage}{8cm}\begin{spacing}{1}Single-Precision Maximum \gridrate (GRecs/s)\\[-20pt]\end{spacing}\end{minipage} & \rule{0cm}{25pt}&  15.3 & 7.85 & 5.4 & 4.06 \\
\begin{minipage}{8cm}\begin{spacing}{1}Double-Precision Maximum \gridrate (GRecs/s)\\[-20pt]\end{spacing}\end{minipage} & \rule{0cm}{25pt}& 9.98 & 5.14 & 3.47 & 2.63 \\ \bottomrule
\end{tabular}

\item When running with maximum Performance, the NN Gridder is memory-bound on device memory, which is utilised by the loading of record data and grid committing. The main limiter that pushes down Performance from the NN Gridder maximum is the quality in memory access pattern  of grid committing.
\item Polarisation Gain of the NN Gridder is controlled by the improvement made in the grid committing global memory access patterns while increasing $N_{\text{pol}}$.
\item  In our experiments, \gridderadvantage varied from a low of 0.94 to a peak of 7.05. Nevertheless, the \gridderadvantageonhybrid was below one for most of the  Single-Precision experiments with $\beta > 8$ and Double-Precision experiments with $\beta > 4$.
\end{itemize}

\chapter{The Pruners}
\label{chap:pruning}
This chapter discusses and provides analyses on two Pruners, called the Column Pruner and Row Pruner, used in the Pruning Step of Hybrid Gridding and Pruned NN Interpolation. Pruned NN Interpolation uses the Column and Row Pruners, with the Column Pruner entrusted to de-interleave the grid. In contrast, Hybrid Gridding makes use only of the Column Pruner, with de-interleaving disabled.

We do remind the reader that our Pruners are convolution-based, with the theory discussed in Section \ref{sec:maths:fftpruning}. They implement Equations \ref{equ:maths:hybridpruner}, \ref{equ:maths:nnpruner}, and \ref{equ:maths:prungrid} which apply a convolution and downsample the grid. They do not apply an IFFT, and in our implementations, the IFFT is still executed by the cuFFT library as explained in Section \ref{sec:methodology:ifftstep}.   We shall use $\alpha$ to represent the down-sampling factor, which in our studied implementations is always equal to $\beta/2$.

Our primary goal in this chapter is to investigate how much Performance and Pruning Gain the Pruners can deliver in the context of the studied implementations. We also have a secondary goal, which is to verify the aliasing suppression properties of the least-misfit gridding functions (Ye \etal \cite{Ye2019}) when used as GCFs for our Pruners. We shall investigate how well aliasing is suppressed below arithmetic noise.

This chapter is organised as follows: In the first section, we describe the implementation of the Row and Column Pruners. In Section \ref{sec:comparative:experiments} we explain the Performance-related experiments of which results are given and analysed in the subsequent four sections. In Section \ref{sec:comparative:aliasingexperiments} we deliver results for experiments related to aliasing, which serves as a validation of the Pruners, and then finalise the chapter in Section \ref{sec:comparative:summary} with a small summary of the important results.    
 
\section{Implementation details}
We implement the Row and Column Pruners using the same design, with the main difference being the layout of the CUDA threads within the CUDA grid.
In the next sub-sections, we shall explain the commonalities of the two Pruners by first making some mathematical formulation to help us define the \textit{Input Chunk} and then in various sub-sections, delve into the main concepts.  Finally, in the last two subsections, we give details specific to a given Pruner.

\subsection{Mathematical formulation}
Let us first do some mathematical formulation that serves as an aid to discuss the implementation of the Pruners.

Let $M$ be equal to the number of pixels in a row or column to be pruned. We define $\myvec{I}=\{i_{-\alpha S/2}, i_{-\alpha S/2 +1},...,i_0,i_1,...,i_{M-1},i_{M},i_{M+1},...,i_{M+\alpha S/2-1} \}$ to be a series of pixels, input to the Convolution-Based Pruner. $\myvec{I}$ is circularly convolved with a GCF of support $S_{\text{z}}$ in tandem with downsampling by a factor of $\alpha$. The output is the series $\myvec{O}=\{o_0,o_1,...,o_{M/\alpha-1}\}$. We assume $M$ to be always divisible by $\alpha$ and for convenience we included in $\myvec{I}$ extra elements at the head and tail to help in applying circular convolution. At the head of $\myvec{I}$, extra elements are with an index $m<0$ and they are equivalent to the elements with index $(M+m)$. At the tail, extra elements are with index $m>M-1$, and they are equivalent to elements with index $(m-M)$.

We conveniently order the GCF values that are used in the computation in a series of vectors $\myvec{C_{\alpha}}=\{\myvec{C'_{0}},\myvec{C'_{1}},...,\myvec{C'_{\alpha-1}}\}$ whereby a given element $\myvec{C'_{n}}$ is a vector defined by $\myvec{C'_{n}}=\{c_{n,0},c_{n,1},...,c_{n,S-1}\}$, which contains ordered values of the GCF such that $o_j \in \myvec{O}$ is computed using Equation \ref{equ:pruning:output_j}.

\begin{equation}
\label{equ:pruning:output_j}
    o_j=\sum_{s=0}^{S-1}\sum_{a=0}^{\alpha-1}c_{b,s}i_{\alpha(j-S/2+s)+a}
\end{equation}

For better readability we define the $S_{\text{z}}$ long vectors $\myvec{I_{j,b}}=\{i_{\alpha(j-S/2)+b},i_{\alpha(j-S/2)+b+\alpha},...,i_{\alpha(j-S/2)+b+(S-1)\alpha}\}$ such that Equation \ref{equ:pruning:output_j} can be re-written as: 
\begin{equation}
\label{equ:pruning:output_j_vec}
    o_j=\sum_{a=0}^{\alpha-1}\myvec{C'_{a}}\cdot\myvec{I_{j,a}}
\end{equation}
It is important to realise that $\myvec{I_{j+1,a}}$ becomes equal to $\myvec{I_{j,a}}$ after removing the first element of $\myvec{I_{j,a}}$ and adding a new element at the end. The main takeaway from Equation \ref{equ:pruning:output_j_vec} is that elements in $\myvec{I_{j,a}}$ will only need $\myvec{C'_{a}}$ to calculate the output and no other vector elements in $\myvec{C_{\alpha}}$

We want to parallelise the computation of $\myvec{O}$ and the first step we do is define the Tuning Parameter $T$ such that we independently calculate $\alpha/T$ partial output series  $\myvec{O_{p}}=\{o_{p,0},o_{p,1},...,o_{p,
M/\alpha-1}\}$, $0\le p <\alpha/T$, where $T\in \mathbb{N}, T\ne 0$ and $T$ divides $\alpha$. $\myvec{O_{p}}$ is described by the set of Equations \ref{equ:pruning:output_jq_vec}.

\begin{subequations}
\label{equ:pruning:output_jq_vec}
\begin{equation}
    o_j=\sum^{\alpha/T-1}_{p=0} \myvec{o_{p,j}}
\end{equation}
\begin{equation}
    o_{p,j}=\sum_{a=pT}^{pT+T-1}\myvec{C'_{b}}\cdot\myvec{I_{j,a}}
\end{equation}
\end{subequations}

We note that all elements in $\myvec{I_{j,a}}$ are all involved in calculating elements in one partial output series $\myvec{O_{p}}$ and are never involved in calculating other partial output series.

We now define $\myvec{I_{p}}$ as the series of all elements in $\myvec{I}$ that calculate $\myvec{O_{p}}$.
\begin{equation}
    \myvec{I_{p}}=\bigcup\limits_{b=pT}^{pT+T-1}\bigcup\limits_{j=0}^{M-1}\myvec{I_{j,b}}
\end{equation}
where the operator $\bigcup$ is the union of vectors or series, outputting a series containing all elements in the vectors and ordered by the value of their index in $\myvec{I}$. 

We can now split $\myvec{I_{p}}$ into \textit{Input Chunks} ($\myvec{Z_{p,x}}$), sometimes referred to just as \textit{chunks}. Each chunk except for the last one contains $S_{\text{z}}\alpha/T$ neighbouring elements from $\myvec{I_{p}}$, as described in Equation \ref{equ:pruning:chunkdef}.
\begin{equation}
\label{equ:pruning:chunkdef}
    \myvec{Z_{p,x}}=\begin{cases} 
   \bigcup\limits_{b=pT}^{pT+T-1}\bigcup\limits_{j=xS_{\text{z}}}^{xS_{\text{z}}+x-1}\myvec{I_{j,b}} & \text{if } (x+1)S_{\text{z}}<M-1 \\
   \bigcup\limits_{b=pT}^{pT+T-1}\bigcup\limits_{j=xS_{\text{z}}}^{M-1}\myvec{I_{j,b}} &  \text{if } (x+1)S_{\text{z}} \ge M
  \end{cases}
\end{equation}

The Input Chunk is defined such that we can apply Algorithm \ref{algo:pruning:ChunkCalculation}. A given Input Chunk "finalises" the calculation of $S$ neighbouring elements in $\myvec{O_{p}}$, while it prepares some calculations for the next $S$ elements in $\myvec{O_{p}}$.

We achieve parallelism in our implementations via the Input Chunks, whereby the processing of different Input Chunks is generally distributed over different CUDA threads. However, a given Input Chunk is guaranteed to be processed only by one CUDA thread.

\setlength{\algomargin}{0.1em}
\begin{algorithm}
\begin{mdframed}
\setstretch{1.5}
\KwIn{$\myvec{Z_{p,x}}=\{z_{0,0},z_{0,1},...,z_{0,T-1},z_{1,0},...,z_{1,T-1},...,z_{S-1,0},z_{S_{\text{z}}-1,1},...,z_{S_{\text{z}}-1,T-1}\}$}\par
\KwPar{GCF Support ($S_{\text{z}}$), Tuning Parameter $T$}
\SetKwInput{KwHelp}{Helper Input}
\KwHelp{\begin{minipage}[t]{11.2cm}\setstretch{1.5}GCF Data: $\{\myvec{C'_{pT}},\myvec{C'_{pT+1}},...,\myvec{C'_{pT+T-1}}\}$,\\$S_{\text{z}}$ Pre-filled Accumulators: $\text{Acc}[S]$\\\rule{10pt}{0cm} (From calculation of previous chunk $\myvec{Z_{p,x-1}}$)\\
\end{minipage}}%
\vskip -12pt
\KwOut{$S_{\text{z}}$ neighbour elements of $\myvec{O_{p}}$ denoted by $\{o_{p,0},o_{p,1},...,o_{p,S_{\text{z}}-1}\}$, $S_{\text{z}}$ Accumulators $\text{Acc}[S_{\text{z}}]$ for use in when calculating with the next chunk $\myvec{Z_{p,x+1}}$. }
\myalgoline
\Begin{
\For{a $\gets$ 0 to S-1}{
\For{t $\gets$ 0 to T-1}{
Load $z_{t,a}$\;
\ForAll{s $\gets$ 0 to S-1}{
$g \gets ( s + a ) \text{mod} S$\;

$\text{Acc}[g]=\text{Acc}[g]+z_{t,a}c_{pT+t,s}$\;
}

}
Save $o_{p,a}=\text{Acc}[a]$ \;
$\text{Acc}[a] \gets 0$\;
}
}

\end{mdframed}
\caption[Calculation of partial outputs using one Input Chunk]{Algorithm showing how an Input Chunk $\myvec{Z_{p,x}}$ is used to calculate $S_{\text{z}}$ partial outputs. It needs accumulated data generated through the previous Input Chunk $\myvec{Z_{p,x-1}}$ and will generate accumulated data for the use of the next Input Chunk $\myvec{Z_{p,x+1}}$. }
\label{algo:pruning:ChunkCalculation} 
\end{algorithm}

\subsection{Regularity and constant indexing}
\label{sec:pruning:constantindexing}
We use Algorithm \ref{algo:pruning:ChunkCalculation} to take advantage of the regularity of the input and output grids to reduce some of the Logic previously needed by the Convolutional and Hybrid Gridders and enable the use of constant indexing. Constant indexing is a type of indexing that can be evaluated at compile-time and therefore optimised out from execution. Constant indexing is necessary to store C++ arrays exclusively in registers, whereby we use such arrays to store $S_{\text{z}}$ accumulators and GCF data.

In implementing Algorithm \ref{algo:pruning:ChunkCalculation}, we explicitly request the compiler to unroll all the loops, in order to enable constant indexing and optimise out Logic. We apply a slightly different code to handle the first Input Chunk as to handle the elements with a negative index, and we modify Algorithm \ref{algo:pruning:ChunkCalculation} for the last chunk since the last chunk differs in size. The latter modification includes the introduction of predicates which breaks constant indexing of the accumulators. Through various experiments, we were able to conclude that loss of constant indexing in the processing of the last chunk has only a minor impact on Performance.  

\subsection{Distribution of chunks and use of shared memory}

We distribute calculation of chunks for specific rows or columns over different blocks, in a bid to increase parallelism.  Too few chunks processed by a CUDA block leads to a substantial increase in \commitratestop Therefore, as to control such increase, we define the Tuning Parameter \textit{Minimum Chunks Per Thread} (\texttt{Min\subscript{chunks}}) to state a minimum of chunks that a CUDA thread has to handle.

By design, we entrust  warps in a given CUDA block to calculate the same chunks of specific rows or columns but with different index $p$. In this way, we can use shared memory to add partial outputs together before committing to global memory. When it is time to save a partial output as specified in Algorithm \ref{algo:pruning:ChunkCalculation}, the partial output is written  to shared memory. Once $S_{\text{z}}$ partial outputs are written in shared memory, the block is synchronised, and one  warp performs additions over shared memory and commits the result to the output grid.

\subsection{Handling GCF data}

As we hinted before, on initialisation, we load and store GCF data into C++ arrays and use such data throughout the whole execution of a CUDA thread without repeated loading from global memory. Such loading is possible because of constant indexing that allows the C++ array to be stored in the CUDA thread registers. In this way, we get rid of latency reported for the Hybrid Gridder and the chocking issue reported for the Convolutional Gridder, due to loading of the GCF data.
Since registers are a scarce resource, it is often not viable to load all the GCF data in one CUDA thread, and this is the reason why $T$ was earlier defined as a Tuning Parameter, controlling the GCF data to be loaded by a given CUDA thread. We do point out that an increase in $T$ implies more registers being used to store GCF data in each CUDA thread, which drives Occupancy down. The Column Pruner tends to be memory bound, implying a reduction in Occupancy might be beneficial to Performance.   

\subsection{Column Pruner specifics}
In the Column Pruner, we consider the complex-valued column as two real-valued columns and assign consecutive CUDA threads in a  warp to handle input chunks ($\myvec{Z_{p,x}}$) from consecutive real-valued columns with equal $p$ and $x$ indexes. In this way we ensure that a warp loads input pixels from global memory with an optimal access pattern, honouring the coalescing and vicinity requirements.

As mentioned previously, the warps of a CUDA block work all on the same columns processing input chunks with equal $x$ but different $p$. We define the Tuning Parameter $B_{\text{Warps}}$  to control the number of warps in a CUDA block, and we do not allow $B_{\text{Warps}}>\alpha/T$.

As mentioned earlier, the Column Pruner can de-interleave an interleaved grid in tandem with downsampling. Such de-interleaving forces the Column Pruner to commit with a degraded access pattern that does not adhere to Coalescence and Vicinity requirements and therefore impacts Performance negatively.   

\subsection{Row Pruner specifics}
We found the implementation of a Row Pruner quite challenging, particularly in finding a way to have the loading of data and grid committing done with optimal access patterns. 

In order to attain optimal access patterns, we force the Tuning Parameter $T$ to be always equal to 1 and in general, have a CUDA warp assigned a part of a row to convolve.  A CUDA thread handles either the imaginary or real part of the chunk, and therefore we use a pair of consecutive CUDA threads to handle a complex-valued Input Chunk in a row.  Consecutive CUDA thread pairs on the warp are assigned to handle Input Chunks of same index $x$ but increasing index $p$. The discussed thread layout requires $2\alpha$ threads to handle all partial outputs for Input Chunks with equal index $x$.  If $\alpha < 16$ a warp provides more CUDA threads than needed, and in such a case, the warp is sub-divided to handle chunks with different $x$ at one go, thus breaking the vicinity requirement for the loading of record and grid committing.   For $\alpha > 16$ more CUDA threads are required than supplied by a warp, and in this case, the necessary number of warps (that is $\alpha/16$ warps) are assigned within the same CUDA block to calculate all chunks of the same row with equal index $x$.

\section{Experiments}
\label{sec:comparative:experiments}
The experiments on the Pruners we here report are different from what we reported for the Gridders.

We report results for experiments done using  quad-polarised (\texttt{N_{\text{pol}}}=4) grids only and consider GCFs with support $S_{\text{z}}$ of 6, 8, 10, 12 and 14. We do three different sets of experiments which we now list and describe.

\begin{enumerate}
    \item \textbf{Interleaved Square Grid Experiments}: In this set of experiments, the input to the Pruners is an $8192\times8192$ polarisation-interleaved grid with $N_{\text{pol}}=4$ which is downsampled to $8192\alpha^{-1} \times 8192\alpha^{-1}$ by first applying the Column Pruner and subsequently the Row Pruner. The Column Pruner de-interleaves the grid.
    
    We designed the Interleaved Square Grid Experiments to study the Pruning Step of Pruned NN Interpolation (Algorithm \ref{algo:maths:purenninterpolation}).
    
    \item \textbf{Non-Interleaved Square Grid Experiments}: This set of experiments is nearly identical to the Interleaved Square Grid Experiments, but the input grid is not interleaved, such that the Column Pruner does not de-interleave.
    
    We designed this set of experiments for comparison with the Interleaved Square Grid Experiments.

    \item \textbf{Non-Interleaved Rectangular Experiments}: In this set of experiments a quad-polarised non-interleaved grid of size $65536(2\alpha)^{-1} \times 65536$  is fed to the Column Pruner as to down-sample the Grid to a size of $65536(2\alpha)^{-1} \times 65536\alpha^{-1}$. 
    
    We designed this set of experiments to study the Pruning Step of  Hybrid Gridding (Algorithm \ref{algo:maths:hybrid}).

\end{enumerate}

\section {Brute Force Search results}
We executed a Brute Force Search over the stated Tuning Parameters for the Column and Row Pruners and discovered solutions whose Performance is with an \optimalityfactor well above 90\% for all the experiments we are reporting. Tables \ref{tab:pruning:colconv_brute} and \ref{tab:pruning:rowconv_brute} list our choice of optimal solutions  on which we deliver an in-depth analysis.   

\begin{table}[]
\centering
\begin{tabular}{@{}cccccccccc@{}}
\toprule
\multirow{2}{*}{\begin{minipage}[c]{0.5cm}\centering$\alpha$\end{minipage}} & \multirow{2}{*}{$S_{\text{z}}$} &\rule{0.6cm}{0cm} & \multicolumn{3}{c}{Single Precision} &  \rule{0.6cm}{0cm}&\multicolumn{3}{c}{Double Precision} \\ \cmidrule(l){4-6} \cmidrule(l){8-10}
 & & & T & $B_{\text{Warps}}$  &\texttt{Min\subscript{chunks}} & & T & $B_{\text{Warps}}$  & \texttt{Min\subscript{chunks}} \\ \midrule
2 & 6,8,10,12,14 & & 1 & 2 & 10 &  & 1 & 2 & 10 \\
4 & 6,8,10,12,14 & & 1 & 4 & 10 & & 1 & 4 & 10 \\
8 & 6,8,10,12,14 & & 1 & 8 & 10 & & 1 & 8 & 10 \\ \midrule
\multirow{3}{*}{16} & 6,8 & & 1 & 16 & 10 & & 2 & 8 & 10 \\
 & 10 & & 2 & 8 & 10 & & 1 & 16 & 10 \\
 & 12 & & 2 & 8 & 10 & & 4 & 4 & 10 \\
 & 14 & & 4 & 4 & 10 & & 2 & 8 & 10 \\ \midrule
\multirow{4}{*}{32} & 6 & & 8 & 4 & 10 & & 8 & 4 & 10 \\
 & 8 & & 8 & 4 & 10 & & 4 & 8 & 10 \\
 & 10 & & 2 & 16 & 5 & & 2 & 16 & 10 \\
 & 12 & & 4 & 8 & 10 & & 2 & 16 & 10 \\
 & 14 & & 4 & 8 & 10 & & 2 & 8 & 10 \\ \midrule
\multirow{4}{*}{64} & 6 & & 4 & 16 & 10 & & 8 & 8 & 10 \\
 & 8 & & 8 & 8 & 5 & & 4 & 8 & 10 \\
 & 10 & & 2 & 32 & 5 & & 2 & 16 & 10 \\
 & 12 & & 4 & 16 & 10 & & 4 & 8 & 10 \\
 & 14 & & 4 & 8 & 5 & & 2 & 8 & 5 \\ \midrule
\multirow{3}{*}{128} & 6,8 & & 1 & 8 & 5 & & 2 & 2 & 5 \\
 & 10 & & 1 & 8 & 5 & &  1 & 8 & 1 \\
 & 12 & & 1 & 8 & 5 & & 2 & 8 & 1 \\
 & 14 & & 1 & 8 & 5 & & 2 & 8 & 5 \\ \midrule
\multirow{3}{*}{256} & 6 & & 2 & 4 & 5 & & 2 & 4 & 5 \\
 & 8 & & 1 & 4 & 10 & & 2 & 4 & 5 \\
 & 10,12,14 & & 2 & 4 & 5 & & 2 & 4 & 5 \\ \midrule
512 & 6,8,10,12,14 & & 1 & 4 & 10 & & 1 & 4 & 10\\
\bottomrule
\end{tabular}
\caption[Optimal Solutions for the Column Pruner]{Optimal Solutions for the Column Pruner that we discovered and chose through Brute Force Search.}
\label{tab:pruning:colconv_brute}
\end{table}
\begin{table}[]
\centering
\begin{tabular}{@{}cccccccc@{}}
\toprule
\multirow{2}{*}{$\alpha$} & \multicolumn{1}{c}{\multirow{2}{*}{\begin{minipage}[c]{1cm}\centering$S_{\text{z}}$\end{minipage}}} & \rule{0.5cm}{0cm} &  \multicolumn{2}{c}{Single-Precision} &\rule{0.8cm}{0cm} &  \multicolumn{2}{c}{Double-Precision} \\ \cmidrule(l){4-5}\cmidrule(l){7-8} 
 & \multicolumn{1}{c}{} & & $B_{\text{Warps}}$  & \texttt{Min\subscript{chunks}} & & $B_{\text{Warps}}$  & \texttt{Min\subscript{chunks}} \\ \midrule
\multirow{4}{*}{2} & 6 & & 16 & 10 & & 8 & 10 \\
 & 8 & & 16 & 10 & & 8 & 10 \\
 & 10 & & 16 & 10 & &  8 & 10 \\
 & 12 & & 2 & 10 & & 8 & 10 \\
 & 14 & & 2 & 10 & & 8 & 10 \\ \midrule
\multirow{4}{*}{4} & 6 & & 4  & 10 & & 2 & 5 \\
 & 8 & & 8 & 5 & & 4 & 10 \\
 & 10 & & 4 & 5 & & 4 & 5 \\
 & 12 & & 4 & 5 & & 4 & 5 \\
 & 14 & & 4 & 5 & & 2 & 5 \\ \midrule
\multirow{4}{*}{8} & 6 & & 4 & 10 & & 4 & 10 \\
 & 8 & & 8 & 5 & & 4 & 5 \\
 & 10 & & 2 & 5 & & 2 & 5 \\
 & 12 & & 4 & 10 & &  8 & 5 \\
 & 14 & & 4 & 10 & & 2 & 5 \\ \midrule
\multirow{4}{*}{16} & 6 & & 4 & 10 & & 4 & 10 \\
 & 8 & & 8 & 5 & & 4 & 5 \\
 & 10 & & 2 & 10 & & 4 & 10 \\
 & 12 & & 4 & 10 & & 8 & 5 \\
 & 14 & & 4 & 10 & & 2 & 5 \\ \midrule
\multirow{4}{*}{32} & 6 & & 4 & 10 & & 4 & 15 \\
 & 8 & & 4 & 10 & & 4 & 15 \\
 & 10 & & 4 & 5 & & 4 & 5 \\
 & 12 & & 4 & 10 & & 4 & 5 \\
 & 14 & & 4 & 5 & & 2 & 5 \\ \midrule
\multirow{4}{*}{64} & 6 &  & 4 & 10 & & 4 & 10 \\
 & 8 & & 4 & 5 & & 4 & 10 \\
 & 10 & & 4 & 5 & & 4 & 5 \\
 & 12 & & 4 & 5 & & 4 & 5 \\
 & 14 & & 4 & 5 & & 4 & 5 \\ \bottomrule 
\end{tabular}
\caption[Optimal Solutions for the Row Pruner]{Optimal Solutions for the Row Pruner that we discovered and chose through Brute Force Search.}
\label{tab:pruning:rowconv_brute}
\end{table}

\section{Performance of the Column Pruner}

\begin{figure}

\includegraphics[page=1,width=\linewidth]{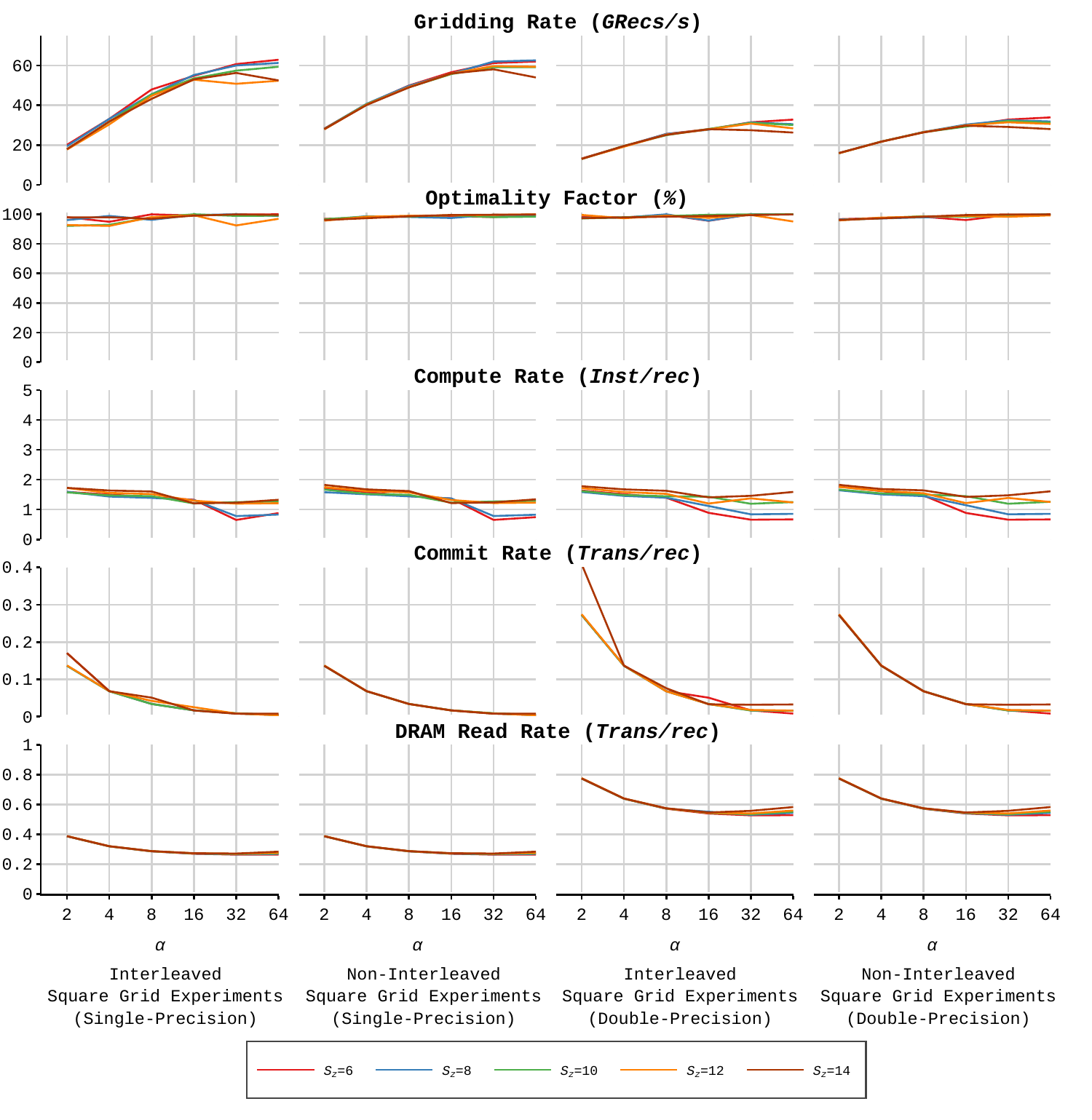}
\caption[Optimal Solutions Performance results of the Column Pruner in the  Interleaved and Non-Interleaved Square Grid Experiments]{Optimal Solutions Performance results of the Column Pruner, in the  Interleaved and Non-Interleaved Square Grid Experiments. The layout is as described in Section  \ref{sec:methodology:graphlayouts}. Performance metrics plotted in this figure are defined in Section \ref{sec:methodology:optimalsolutionsperformancemetrics}.}
\label{fig:pruning:pruningmetricexp13}
\end{figure}
\begin{figure}
\includegraphics[page=1,width=\linewidth]{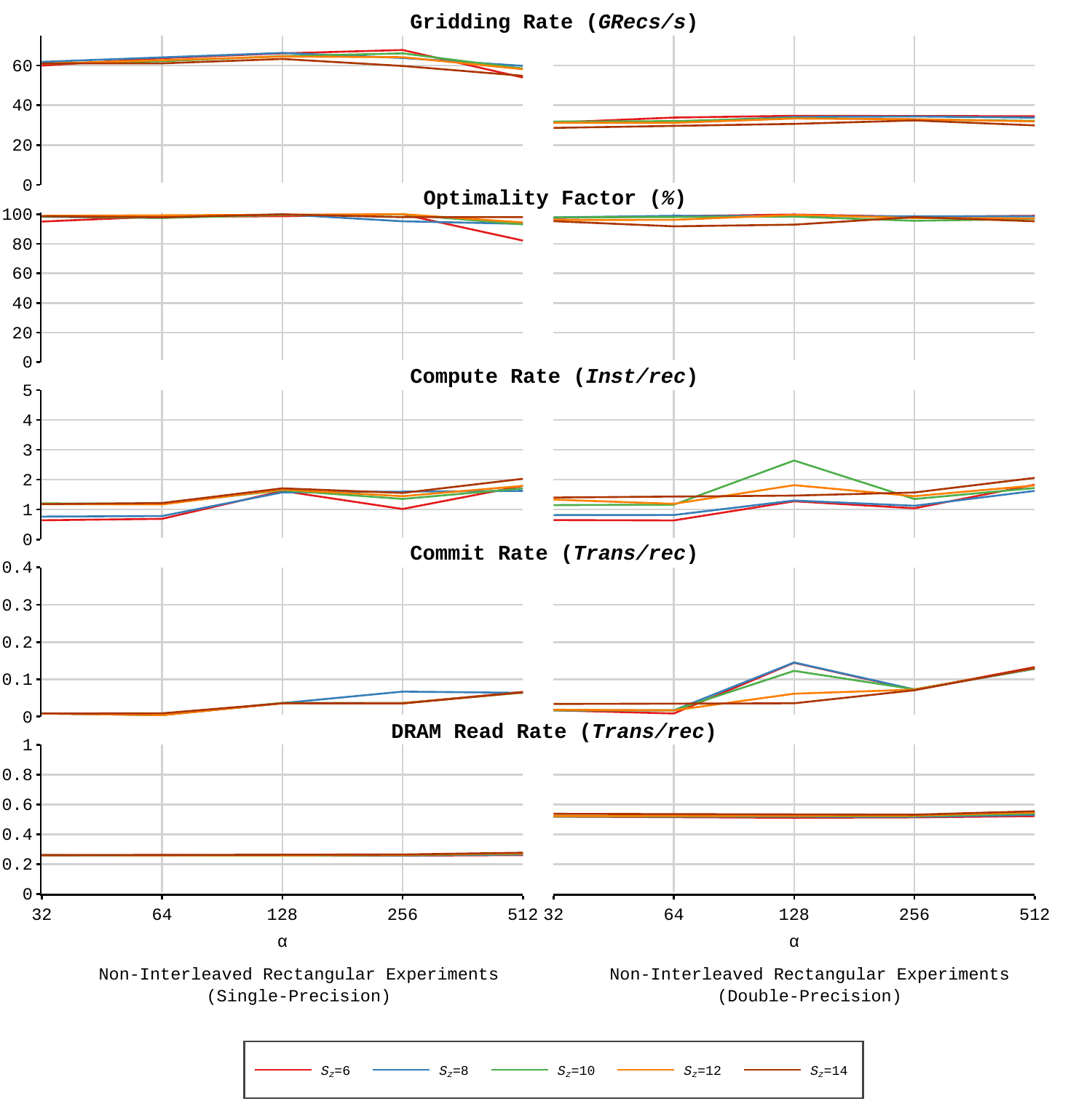}
\caption[Optimal Solutions Performance results of the Column Pruner in the Non-Interleaved Rectangular Grid Experiments]{Optimal Solutions Performance results of the Column Pruner, in the Non-Interleaved Rectangular Grid Experiments. The layout is as described in Section  \ref{sec:methodology:graphlayouts}. Performance metrics plotted in this figure are defined in Section \ref{sec:methodology:optimalsolutionsperformancemetrics}.}
\label{fig:pruning:pruningmetricexp4}
\end{figure}
\begin{figure}

\includegraphics[page=1,width=\linewidth]{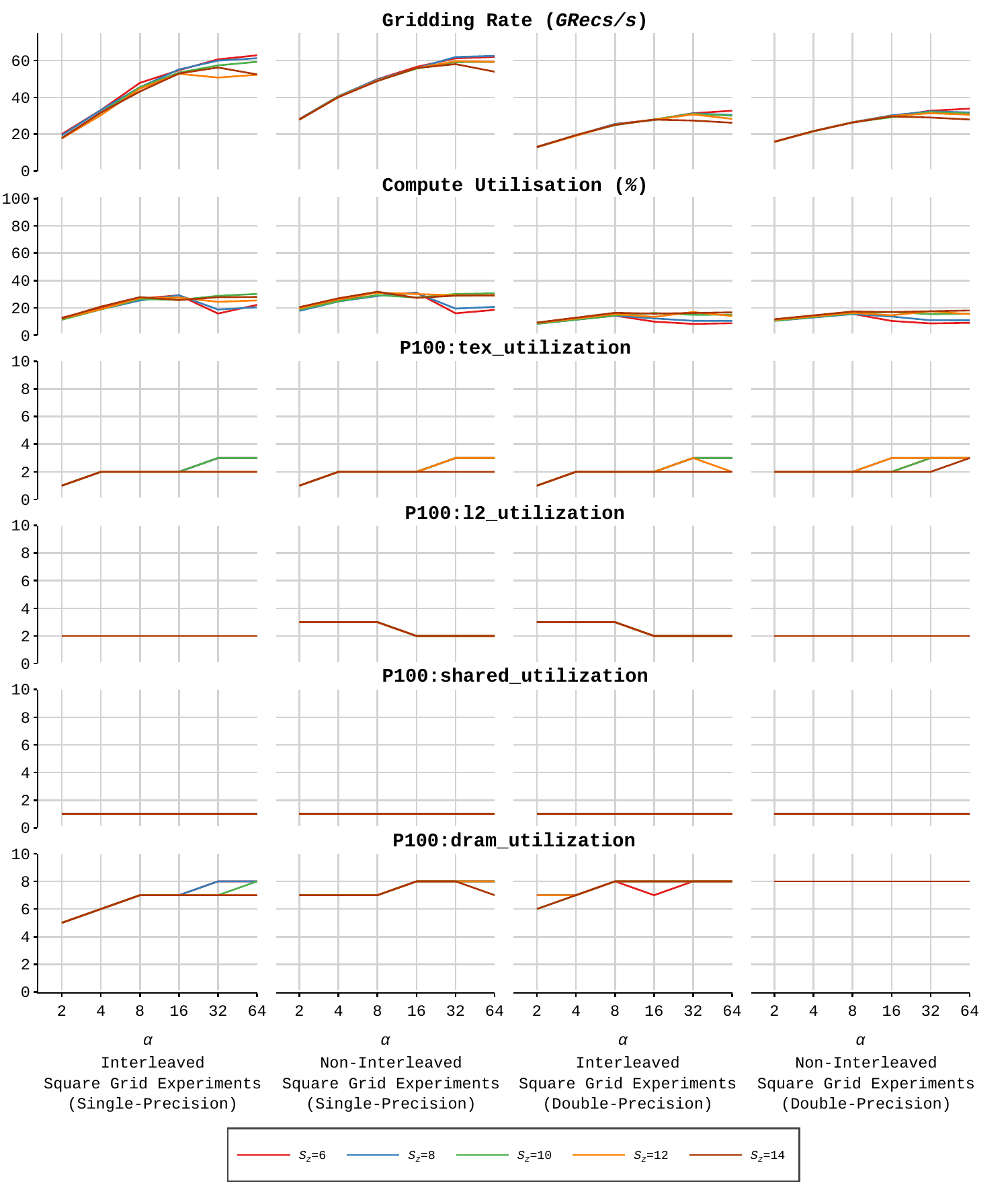}
\caption[Utilisation results of the Column Pruner in the  Interleaved and Non-Interleaved Square Grid Experiments]{Utilisation results of the Column Pruner, in the  Interleaved and Non-Interleaved Square Grid Experiments. The layout is as described in Section  \ref{sec:methodology:graphlayouts}. Performance metrics plotted in this figure are defined in Section \ref{sec:methodology:performancemetricutilisation}. }
\label{fig:pruning:pruningmetricexp13_utilisation}
\end{figure}
\begin{figure}
\includegraphics[page=1,width=\linewidth]{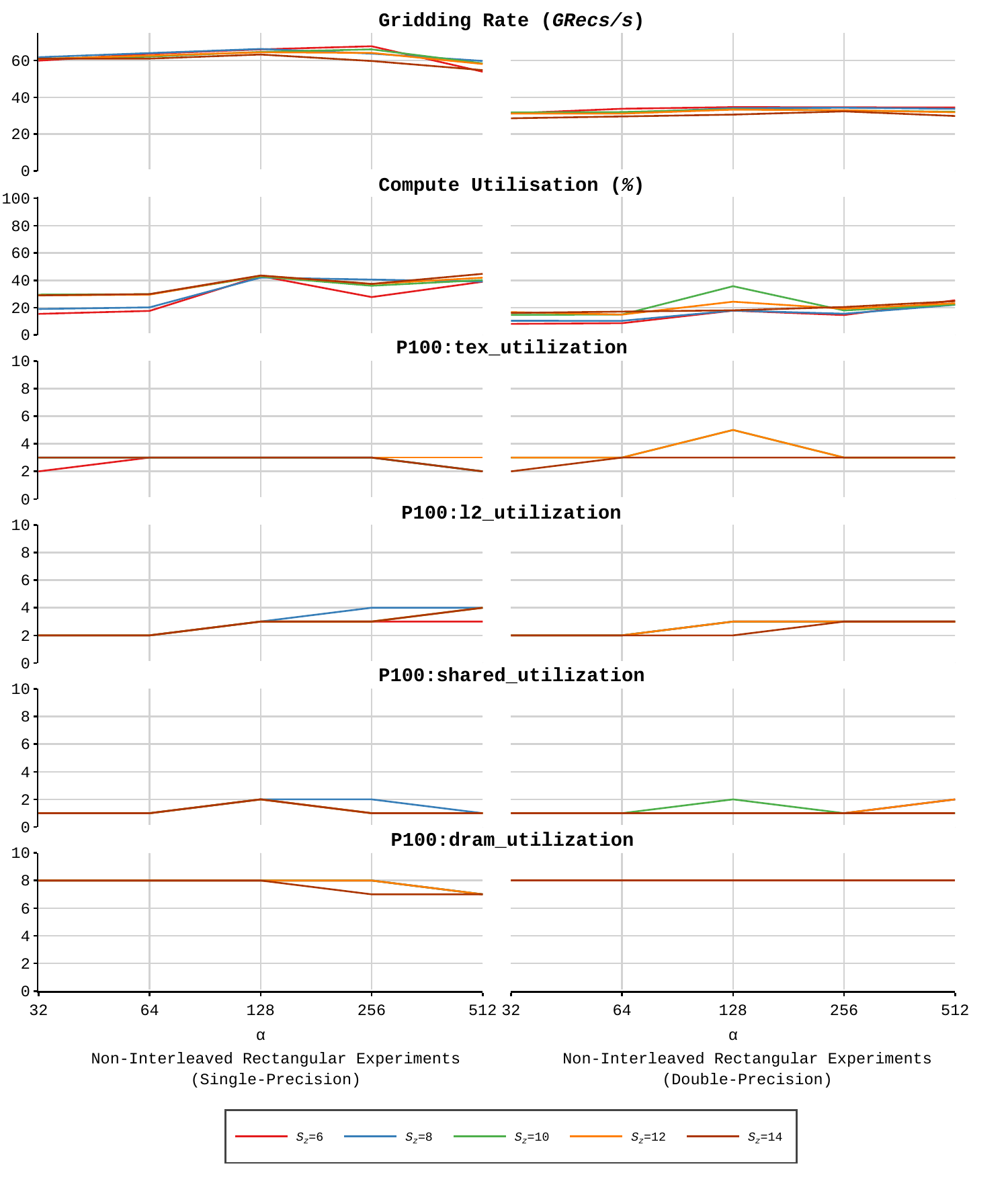}
\caption[Utilisation results of the Column Pruner in the Non-Interleaved Rectangular Grid Experiments]{Utilisation results of the Column Pruner, in the Non-Interleaved Rectangular Grid Experiments. The layout is as described in Section  \ref{sec:methodology:graphlayouts}. Performance metrics plotted in this figure are defined in Section \ref{sec:methodology:performancemetricutilisation}.}
\label{fig:pruning:pruningmetricexp4_utilisation}
\end{figure}

We plot the Optimal Solution Performance results of the Column Pruner in Figures \ref{fig:pruning:pruningmetricexp13} and \ref{fig:pruning:pruningmetricexp4} and Utilisation results in Figures \ref{fig:pruning:pruningmetricexp13_utilisation} and \ref{fig:pruning:pruningmetricexp4_utilisation}. We remind the reader that in context of the Pruners and the measured Performance Metrics, a \texttt{Gridded Record}, many times referred to just as a \texttt{record}, is one single-polarised complex-valued pixel, in the input grid.

Performance obtained for the Column Pruner can be summarised with the following five statements: 
\begin{itemize}
    \item Maximum Performance reported in the results for the Column Pruner is 67.7 GRecs/s   for Single-Precision and 34.8 GRecs/s for Double-Precision.
    \item Minimum Performance reported in the results for the Column Pruner is 17.7 GRecs/s for Single-Precision and 13.0 GRecs/s for Double-Precision.
    \item Performance of the Column Pruner depends on $\alpha$. The trend is that Performance increases with increasing $\alpha$ until reaching a peak, whereby afterwards there is a slight decrease with further increase in $\alpha$.
    \item Performance of the Column Pruner degrades when de-interleaving is enabled. Such degradation is mostly felt at low $\alpha$ for Single-Precision.
    \item In general, the Column Pruner is memory-bound on device memory with \computeutilisation always less than 50\%. 
\end{itemize}

Let us now analyse in further detail the Performance behaviour of the Column Pruner.

A look at the Utilisation metrics show that in general, the experiments are memory-bound on device memory, with the \dramutilisation metric usually well above 6,  implying that access to device memory is a central player in controlling Performance. On the other hand, \computeutilisation is well below 50\%, making it unlikely that compute has any significant effects on Performance. Clearly, in order to understand the Performance of the Column Pruner, we need to assess the utilisation of device memory.

Two processes require access to the device memory: the loading of records from the input grid, and the committing to the output grid. In general, much as expected, \commitrate reduces with increasing $\alpha$, implying less access to device memory, and thus promoting the noted trend of increased Performance with an increase in $\alpha$. The trend tends to break from $\alpha=32$ and higher, due to a change in the optimal solution, which is adapting for significantly higher demand on GPU resources (such as registers and shared memory). The same argument stands to explain the visible changes in Performance with changing values of $S_{\text{z}}$ in experiments with $\alpha\ge 32$. 

We now explain why de-interleaving causes a decrease in Performance, particularly for  the Single-Precision experiment with $\alpha=2$. In de-interleaving,  grid committing accesses memory with a degraded pattern, whereby neither the coalescence nor the vicinity requirements are honoured. In the results, the worst degradation happened for Single-Precision where only 8 bytes are merged in one 32-byte transaction. In Double-Precision, 16 bytes are merged in one 32-byte transaction, implying an access pattern that is less degraded than that for Single-Precision. 

As it is always the case in our analyses, the \commitrate indicates the level of Performance penalisation caused by the degraded access pattern. \commitrate is highest for the smallest $\alpha$, and therefore we expect  Performance to be mostly penalised for the lowest values of $\alpha$, particularly for Single-Precision as stated before.

We finalise this discussion by analysing compute. Let us compare the maximum \computerate measured for the Column Pruner using $S_{\text{z}}=6$ GCFs with the least measured value of \computerate for the Hybrid Gridder at $N_{\text{pol}}=1$. In all Column Pruner experiments with $S_{\text{z}}=6$, the measured \computerate was always lower than 2 Inst/rec, which we shall take as a maximum.  The minimum measured \computerate for the stated Hybrid Gridder is 5.53 Inst/rec, which is $2.7\times$ larger than the maximum Compute Rate of the Column Pruner. The stated values show that Logic in the Column Pruner is optimised out as suggested in Section \ref{sec:pruning:constantindexing}, which we corroborated through a visual inspection of the generated assembly code for the Column Pruner.

\section{Performance of the Row Pruner}

\begin{figure}

\includegraphics[page=1,width=\linewidth]{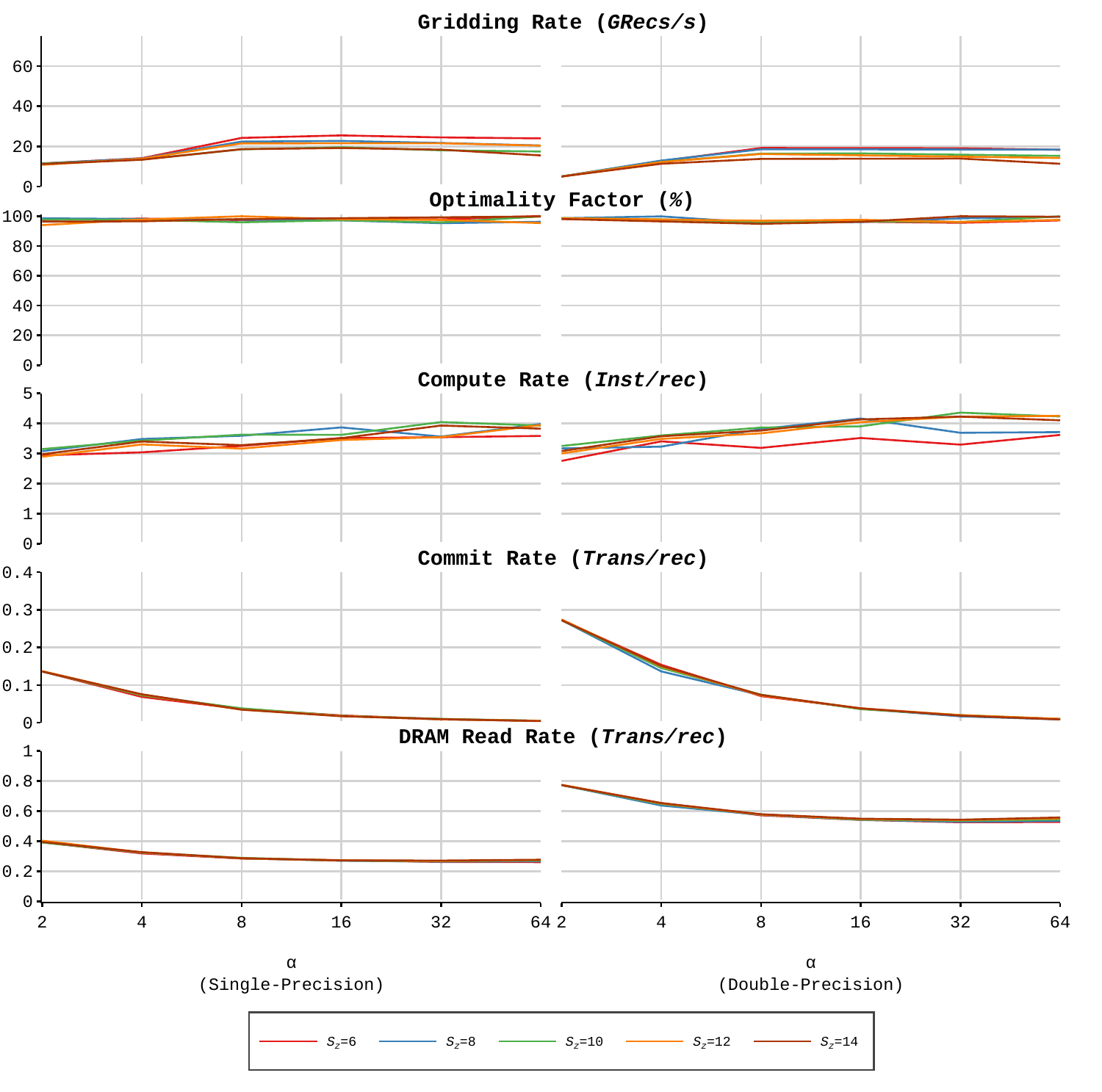}
\caption[Optimal Solutions Performance results for the Row Pruner]{Optimal Solutions Performance results for the the Row Pruner. The layout is as described in Section  \ref{sec:methodology:graphlayouts}. Performance metrics plotted in this figure are defined in Section \ref{sec:methodology:optimalsolutionsperformancemetrics}.}

\label{fig:pruning:pruningmetricexp2}
\end{figure}
\begin{figure}

\includegraphics[page=1,width=\linewidth]{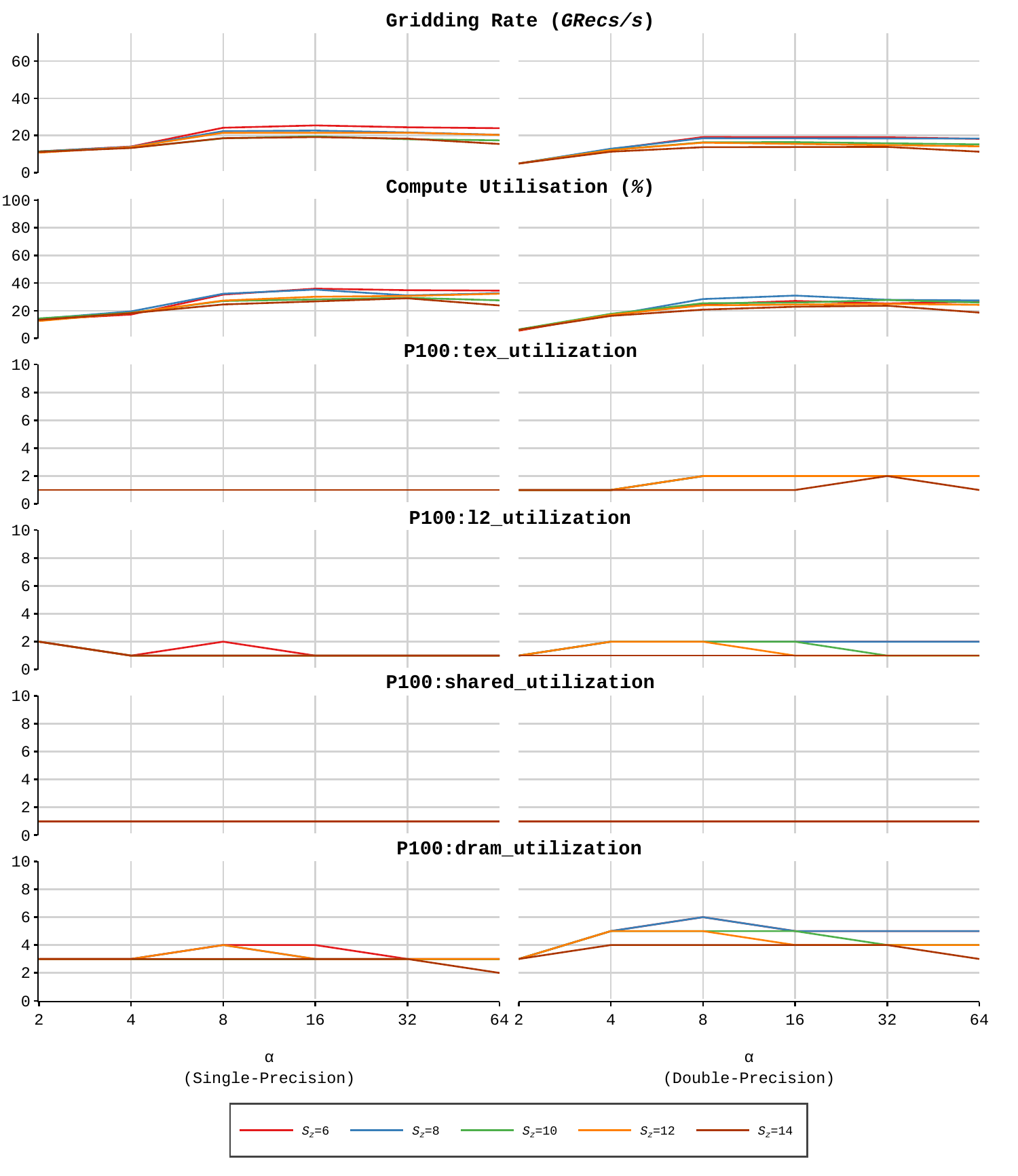}

\caption[Utilisation results for the Row Pruner]{Utilisation results for the Row Pruner. The layout is as described in Section  \ref{sec:methodology:graphlayouts}. Performance metrics plotted in this figure are defined in Section \ref{sec:methodology:performancemetricutilisation}.}
\label{fig:pruning:pruningmetricexp2_utilisation}
\end{figure}

We plot Optimal Solutions Performance results of the Row Pruner in Figure \ref{fig:pruning:pruningmetricexp2} and Utilisation results in Figure \ref{fig:pruning:pruningmetricexp2_utilisation}. The Row Pruner Performance is less than that of the Column Pruner by a factor of between 2 and 3, whereby the maximum \gridrate we are reporting for the Row Pruner is 25.4 GRecs/s for Single-Precision and 19.2 GRecs/s for Double-Precision. A look at utilisation suggests that except for one experiment, the Row Pruner is neither compute-bound nor memory-bound and riddled with latency. Based on the experience we achieved in examining Performance of the Column Pruner, we see it likely that forcing $T$ to 1 is the main cause of latency since it minimises Instruction Level Parallelism.

\begin{figure}

\includegraphics[page=1,width=\linewidth]{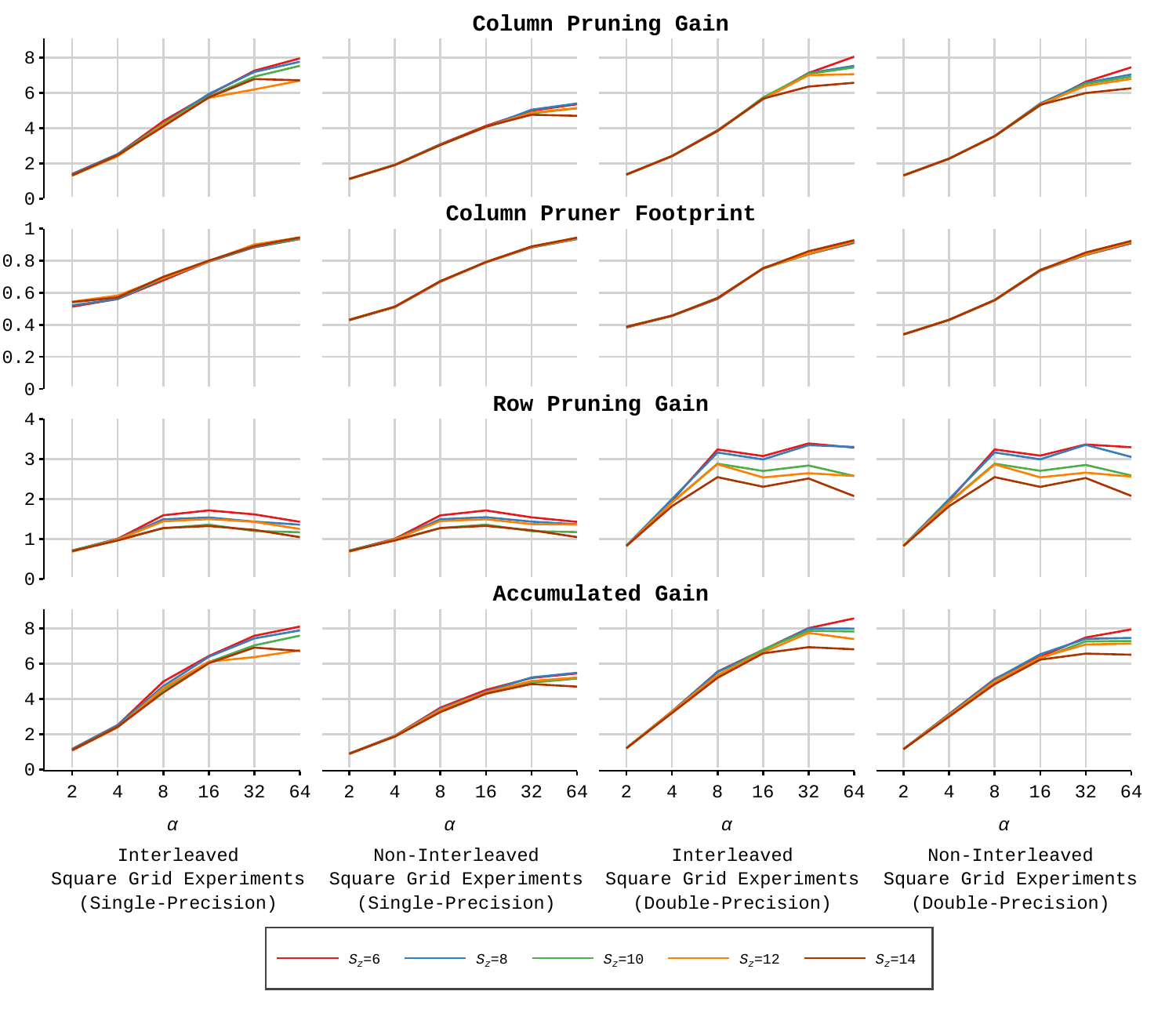}

\caption[Pruning Gain results of the Column and Row Pruners in the Interleaved and Non-Interleaved Square Grid Experiments]{Pruning Gain results for the Row and Column Pruners in the Interleaved and Non-Interleaved Square Grid Experiments.  The layout is as described in Section  \ref{sec:methodology:graphlayouts}. Performance metrics plotted in this figure are defined in Section \ref{sec:methodology:pruninggainmetrics}.}
\label{fig:pruning:pruninggainexp13}
\end{figure}
\begin{figure}[h]
\centering
\includegraphics[page=1]{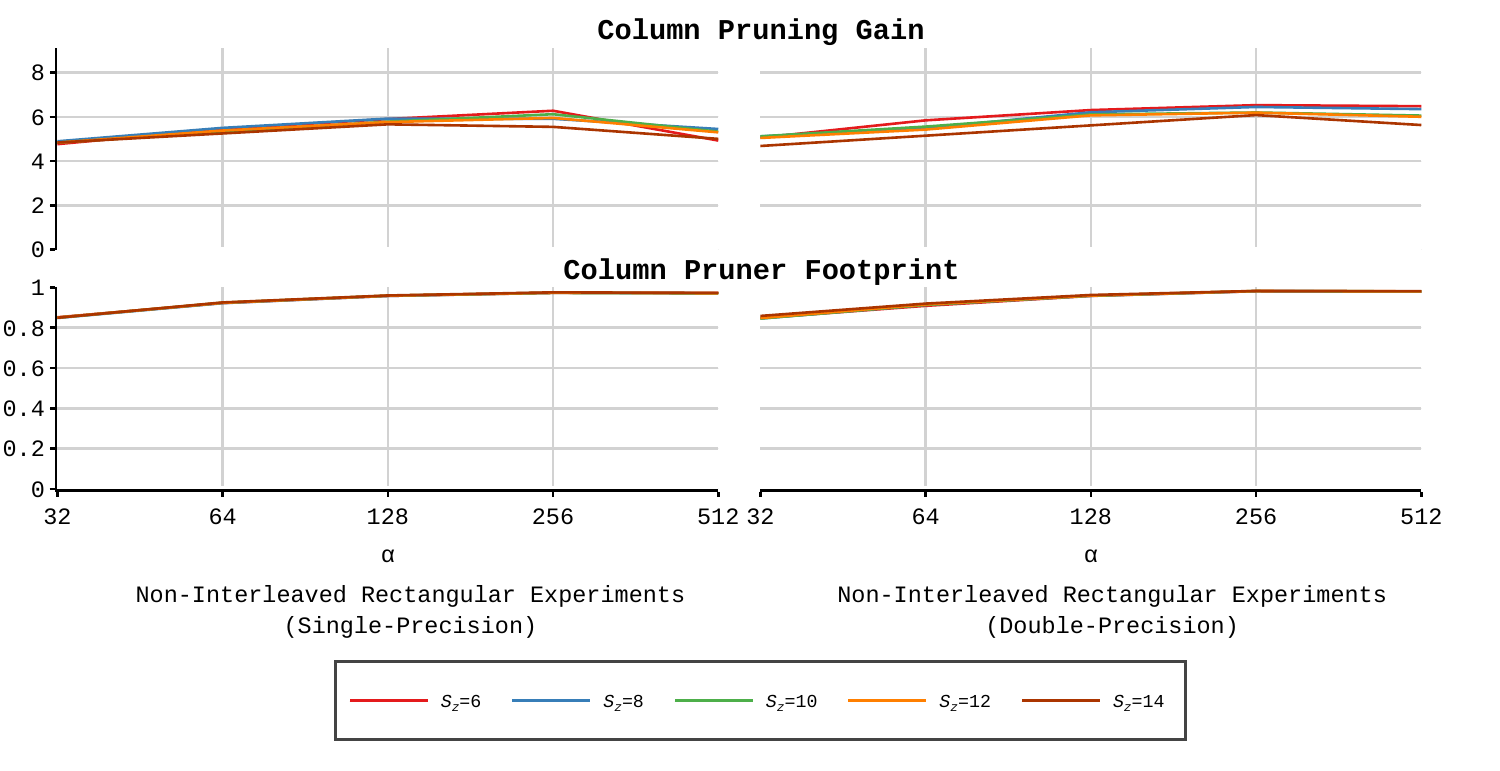}
\caption[Pruning Gain results of the Column Pruner in the Non-Interleaved Square Grid Experiments]{Pruning Gain results of the Column Pruner in the Non-Interleaved Square Grid Experiments. The layout is as described in Section  \ref{sec:methodology:graphlayouts}. Performance metrics plotted in this figure are defined in Section \ref{sec:methodology:pruninggainmetrics}.}
\label{fig:pruning:pruninggainexp4}
\end{figure}

\section{\pruninggain}

The main goal of the Pruners in this thesis is to reduce the time of inversion of the Hybrid and NN Gridders' output grid. We will now analyse such reductions using the Pruning Gain Group of metrics defined in Section \ref{sec:methodology:pruninggainmetrics}, of which measurements are plotted in Figures \ref{fig:pruning:pruninggainexp13} and \ref{fig:pruning:pruninggainexp4}. Note that for these results, we conducted more experiments to measure the execution time of IFFT that would be done if the Pruning Step was absent.

\subsection{\columnpruninggain}
Let us first review gains related to the Column Pruner. Results show a maximum \columnpruninggain of approximately 8 for the case of  Double-Precision Interleaved Square Grid Experiment  at $\alpha=64$ and a minimum \columnpruninggain of approximately 1.1 for the case of Single-Precision Non-Interleaved Square Grid Experiment at $\alpha=2$. We regard the results obtained for the \columnpruninggain as proof that Convolution-based Pruning, as applied in this thesis, can deliver good gains.

Let us now delve into some details related to the \columnpruninggainstop

\begin{itemize}

    \item \columnpruninggain is lowest at the lowest $\alpha$ and in general increases with an increase with $\alpha$.

    Two factors are controlling the \columnpruninggainstop These are the Performance of the Column Pruner and the reduced execution time of the IFFT due to downsampling. For the latter, we can assume that in general, the higher $\alpha$ is, the more IFFT execution time is reduced since IFFT will operate on smaller grids. When this assumption is coupled with the fact that the Performance of the Column Pruner generally increased with $\alpha$, we understand why the \columnpruninggain behaves in the way stated.

     \item \columnpruninggain is higher for the Interleaved Square Grid Experiments than for the Non-Interleaved Square Grid Experiments.
     
     We remind the reader that cuFFT, the library we are using to perform IFFTs, incurs some Performance loss when handling interleaved grids. The Column Pruner also incurs some penalties in Performance when handling interleaved grids. However, such penalties are less in such an amount that pruning is recouping some of the lost Performance in IFFTs  when such IFFTs handle de-interleaving of a grid with no pruning.
     
     \item In general, the \columnpruninggain metric shows little variation in Performance with a change in $S_{\text{z}}$.
     
     The result follows from the Performance of the Column Pruner where in general, there is little variation in Performance with a varying $S_{\text{z}}$. This result indicates that in real-life scenarios, it is worth considering a GCF for pruning with the largest $S_{\text{z}}$, when they have the best anti-aliasing properties since we expect only a minor impact on the Performance of the Pruners.
\end{itemize}

\subsection{\rowpruninggain and \accumulatedpruninggain}
In Figure \ref{fig:pruning:pruninggainexp13}, we report \rowpruninggain measurements. The maximum measured \rowpruninggain is around 3.3, for the  Double-Precision Interleaved/Non-Interleaved Square Experiments at $\alpha=64$. The lowest measured \rowpruninggain is at around 0.7 (implying a loss) for the  Single-Precision Interleaved/Non-Interleaved Square Experiments at $\alpha=2$,. 

Comparing the \accumulatedpruninggain with the \columnpruninggaincomma we deduce that the contribution of the Row Pruner towards the \accumulatedpruninggain is near to negligible. As shown by the measured \columnprunerfootprint for large values of $\alpha$, the Column Pruner dominates the execution time of the whole inversion, not leaving much for the Row Pruner to contribute to the \accumulatedpruninggainstop On the other hand, for the low values of $\alpha$, the Row Pruner can give a significant contribution to the \accumulatedpruninggain but lacks sufficient Performance to do so.

\section{Aliasing in Convolution-Based FFT Pruning }
\label{sec:comparative:aliasingexperiments}
We now present an experiment targeted to verify if applying the least-misfit gridding functions to the Pruners will lead to a Convolution-Based Pruning Algorithm that suppresses aliasing below arithmetic noise. 

\subsection{Experiment description and results}

In this experiment, we prune the 2\textsuperscript{nd} dimension (column) of a two-dimensional image $I(x,y,a)$ of initial size $256\times256\alpha$ to a size of $128\times128$. $x\in\mathbb{Z}$ is the 1\textsuperscript{st} dimension coordinate, and $y\in \mathbb{Z}$ is the 2\textsuperscript{nd} dimension coordinate of a pixel in the image. The third argument denoted as $a$, controls aliasing in a  way that will be evident further on in the text. The downsampling factor $\alpha$ is equal to 4.

We execute pruning of $I(x,y,a)$ using two different methods, which we call the Reference Method and the Pruning Method, described hereunder.

\textbf{Reference Method}:

\begin{enumerate}
    \item Apply Fourier Transform on image to get a Fourier grid of size $256\times256\alpha$.
    \item Apply Inverse Fourier Transform on grid to get back image of size $256\times256\alpha$.
    \item Remove outer region of the image to reduce the image to a size of $128\times128$.
\end{enumerate}

\textbf{Pruning Method}:

\begin{enumerate}
    \item Apply Fourier Transform on image to get a Fourier grid of size $256\times256\alpha$
    \item Apply Column Pruner to grid to reduce its size to $256\times256$.
    \item Apply Inverse Fourier Transform on the pruned grid to get an image of size $256\times256$.
    \item Remove outer region of the image to reduce the image to a size of $128\times128$.
\end{enumerate}

We repeat the two methods for Single and Double-Precision. We furtherly repeat the Pruning Method using five GCFs with different support ($S_{\text{z}}$) which are the Prolate Spheroidal used in CASA ($S_{\text{z}}=6$), and the least-misfit gridding functions ($x_0=0.25$) of support 8, 10, 12 and 14. The latter four GCFs were supplied by Ye \cite{YePrivate2020} through a python script.

Let us now describe  $I(x,y,a)$, whereby we will assume the origin is at the centre with $x=y=0$.

We are using a two-dimensional image, because the Column Pruner works on two-dimensional grids. In truth, we show no interest in the first dimension for this experiment, so much so that we set all columns to be identical to each other, that is  $I(x,y,a)=I(0,y,a)$ $\forall x$. 

\begin{figure}

\includegraphics[page=1]{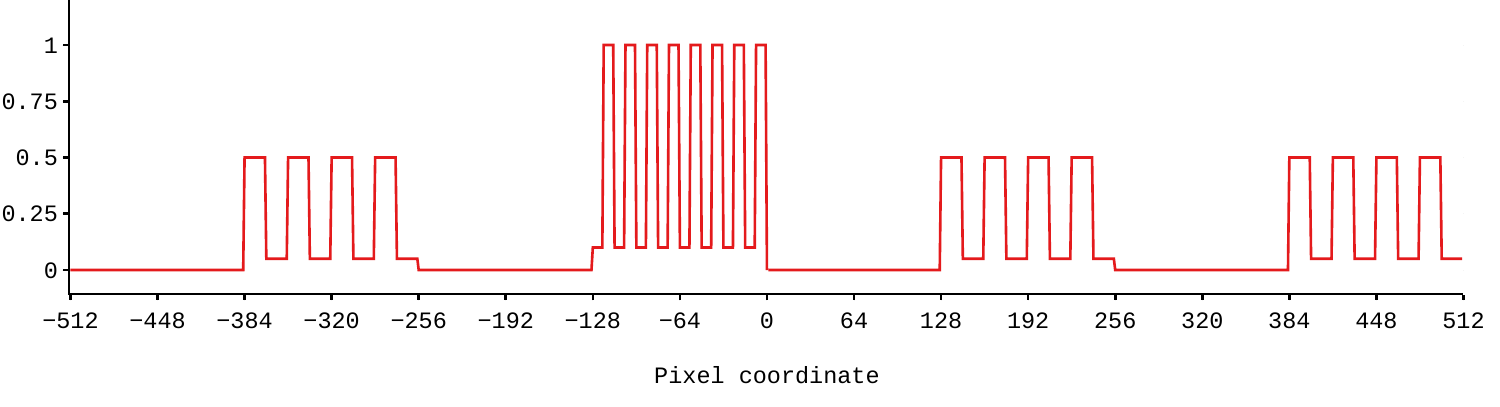}

\caption[Plot of $I(0,y,a)$ vs pixel coordinate $y$ for $a=0.5$]{Plot of $I(0,y,A)$ vs pixel coordinate $y$ for $a=0.5$ }
\label{fig:pruning:prunprecisionref}
\end{figure}

We now refer to Figure \ref{fig:pruning:prunprecisionref} containing a plot of $I(0,y,0.5)$ against $y$ for $\alpha=4$ and  meant as an aid for the below explanation.

We refer to the region in $I(0,y,a)$ within the interval $0\le y \le 64$ as the \textit{zero region} whereby $I(x,y,a)=0$. All regions that in the Pruning Method will distort through aliasing the zero region are set to zero. In this way, the zero region in the output image will only contain arithmetic noise.  

We refer to the region in $I(x,y,a)$ within the interval $-64\le y<0$ as the aliasing region. As we show in Figure \ref{fig:pruning:prunprecisionref}, the aliasing region is populated with a square function with a minimum of $0.1$, a maximum of $1$ and a wave period of 8 pixels. All regions in $I(x,y,a)$ that in the Pruning Method will cause distortion to the aliasing region, are populated by a square function with a wave period of 16 pixels, a minimum of $0.1a$, and a maximum of $a$. In this experiment, we will consider $a$ with values of 0.5, 10,  and 100 where such values control the level of aliasing distortion in the aliasing region, when the Pruning Method is used.

\begin{figure}[h]

\includegraphics[page=1]{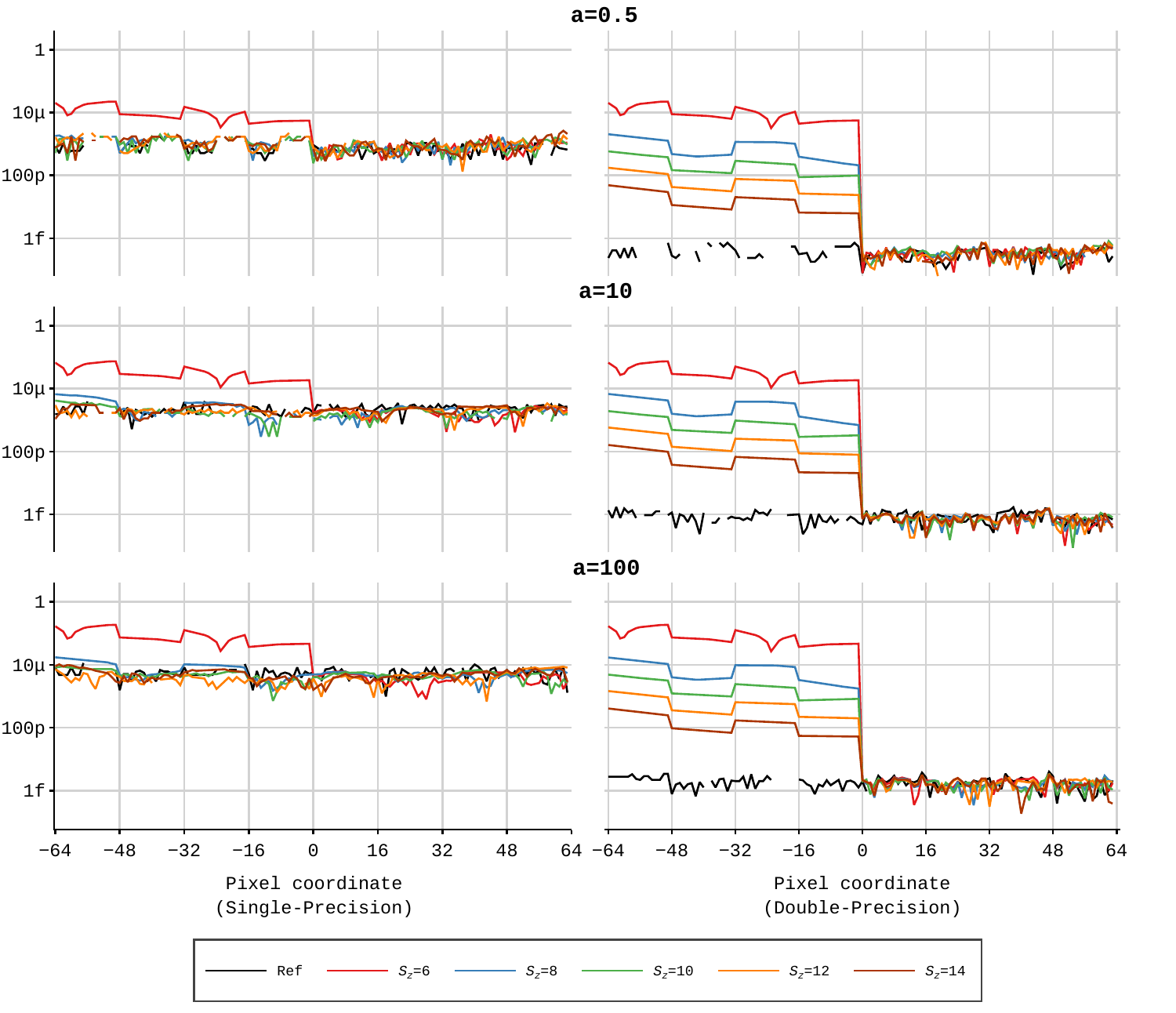}

\caption[Aliasing results for Convolution-Based FFT Pruning]{Results of aliasing experiment for Convolution-Based FFT Pruning as described in Section \ref{sec:comparative:aliasingexperiments}.}
\label{fig:pruning:prunprecision}
\end{figure}

Results are given in Figure \ref{fig:pruning:prunprecision} where we plot the absolute difference between the original $I(0,y,a)$ and the image we get after pruning through the two methods for the region $-64\le y<64$.

\subsection{Discussion}
Unfortunately, some of the results given in Figure \ref{fig:pruning:prunprecision}, though promising, do not reach the expectations set by Ye \etal  \cite{Ye2019}. For Single-Precision, the expectations were that all the four least-misfit gridding functions suppress aliasing below arithmetic noise. In our results the least-misfit gridding functions with $S_{\text{z}}=8$ and $S_{\text{z}}=10$ fail such expectations for $a=10$ and $a=100$. As for Double-Precision, expectations were that the least-misfit gridding function with $S_{\text{z}}=14$ would suppress aliasing below arithmetic noise. However, in our results, none of the functions reached such a level of aliasing suppression. Nevertheless, there is a trend in the results that suggest the existence of a least-misfit gridding function with $S_{\text{z}}=20$ that delivers the sought level of alias suppression.  

On the bright side, the least-misfit gridding functions delivered a much better aliasing suppression than the Prolate Spheroidal and their ability in suppressing aliasing increased with increasing $S_{\text{z}}$, which is as expected from Ye \etal \cite{Ye2019}.

Results show that arithmetic noise generated through the Pruning Method is at an acceptable level in the aliasing and zero regions. Sometimes such arithmetic noise is at a lower level than that produced by the Reference Method, mainly when \textit{a} is on the high side.

\section{Summary}
\label{sec:comparative:summary}
In this chapter, we discussed and analysed the Pruners for Performance and aliasing suppression.

Following is a list of the most important results.
\begin{itemize}
\item The following table gives the maximum Performance of the two Pruners together with their maximum Pruning Gain that resulted in our experiments. 

\begin{tabular}{@{}l@{\rule{0.8cm}{0cm}}cc@{}@{}c@{}cc@{}}
\toprule
Pruner & \multicolumn{2}{@{}c@{}}{Single-Precision} &\rule{0.8cm}{0cm} & \multicolumn{2}{@{}c@{}}{Double-Precision} \\
\cline{2-3}\cline{5-6}
& \begin{minipage}{2.5cm}%
\begin{spacing}{1}
\centering Maximum \gridrate (GRecs/s)%
\end{spacing}
\end{minipage}    & \begin{minipage}{2.5cm}\begin{center}\begin{spacing}{1} Maximum \pruninggain \\ \rule{0cm}{0cm} \end{spacing}\end{center} \end{minipage} & \rule{0cm}{42pt}& \begin{minipage}{2.5cm}\begin{spacing}{1}\centering Maximum \gridrate (GRecs/s)%
\end{spacing}\end{minipage}  & \begin{minipage}{2.5cm}\begin{spacing}{1}\centering Maximum \pruninggain  \\ \rule{0cm}{0cm} \end{spacing}\end{minipage}   \\  \midrule
Column  & 67.7 & 8 & & 34.8 & 8  \\
Row & 25.4 & 1.72 & & 19.2 &  3.4\\ \bottomrule
\end{tabular}

\item In most experiments, the Column Pruner was memory-bound while the Row Pruner was neither compute-bound nor memory-bound and riddled with latency.
\item The Row Pruner did not deliver any substantial contribution to the Accumulated Gain.
\item In our aliasing related experiments, using the least-misfit functions as GCFs, we were not able to reproduce in full the expected results by Ye \etal \cite{Ye2019}. In our test scenarios, the least-misfit functions did not always suppress aliasing below arithmetic noise.
\end{itemize}

\chapter{Comparative studies}
\label{chap:comparative}
This chapter finalises our study in this thesis by making final comparisons between the studied implementations to show if and how we reached the thesis's primary goal. We stated the primary goal at the beginning of Chapter \ref{chap:introduction}, which is to seek modifications to Convolutional Gridding that perform better on the P100, without causing any increase in aliasing. 

We divide our analysis into two main sections. In the first section, we validate our implementation of Convolutional Gridding in terms of Performance and aliasing, by comparing results with other well-known Gridding implementations. We require such validation since, in the second section, we do a comparative study between the three studied implementations in which the Convolutional Gridding implementation will be the reference. In the said study we show how our proposed modifications that morphed Convolutional Gridding to  Hybrid Gridding and Pruned NN Interpolation reach the primary goal.

\section{Validation of the Convolutional Gridding implementation}

In order to validate our implementation of Convolutional Gridding, we do experiments to compare Performance, aliasing and arithmetic noise of our implementation with other Gridding implementations. In the first sub-section, we will investigate how our implementation of the Convolutional Gridder fairs in terms of Performance with two other well known Gridders. In the sub-section that follows, we provide a comparative analysis of aliasing and arithmetic noise, to ensure that our Convolutional Gridding implementation does not cause aliasing distortion and arithmetic noise above what is expected. Such expectation is set by the implementation of Convolutional Gridding in CASA.

We clarify that unless otherwise stated, the terms \textit{Convolutional Gridder} and \textit{Convolutional Gridding} refer to our implementation of the Convolutional Gridder and Convolutional Gridding.

\subsection{Performance of the Convolutional Gridder}

We compare the Performance of the Convolutional Gridder with two well-known Gridder implementations. These are the CPU-based Convolutional Gridder of WSClean (Offringa \etal \cite{Offringa2014}) hereafter referred to as the WSClean Gridder and the GPU-based Image Domain Gridder (IDG) (van der Tol \etal \cite{VanderTol2019}, Veenboer \etal \cite{Veenboer2017} ), as integrated into WSClean, hereafter referred to as the IDG Gridder. The WSClean Gridder computes in Double-Precision and grids polarisation channels one at a time. On the other hand,  the IDG Gridder computes in Single-Precision solely for $N_{\text{pol}}=4$. In our experiments, the IDG Gridder is set to work in \textit{hybrid mode}, which is the recommended mode to use in running IDG over GPUs \cite{wscleanmanual1}.

We aim to estimate by what factor the Convolutional Gridder is more performant than the WSClean and IDG Gridders, through a comparison of the \gridratestop  The procedure we followed to make such an estimate is as follows: We performed a Brute Force Search over the WSClean Gridder and the IDG Gridder in order to measure a respective \bestgridratestop The Tuning Parameters for the WSClean Gridder were the output image size, oversampling factor ($\beta$), number of w-planes and various parameters that control parallelisation over the CPUs. The Tuning Parameters for the IDG Gridder were just the image size and number of w-planes\footnote{Note that the oversampling factor has no meaning in the IDG Gridder.}. Once we found the \bestgridrate of the WSClean and IDG Gridders, we calculated a Minimum and Maximum \textit{Gain} of the Convolution Gridder over the WSClean and IDG Gridders. With the term \textit{Gain}, we mean that factor by which the Convolutional Gridder performs better than the other Gridders in terms of \gridratestop We calculate the Minimum Gain by comparing the measured \bestgridrate of the WSClean and IDG Gridder with the Minimum \bestgridrate reported for the Convolutional Gridder, which is at $\beta=1$. We calculate the Maximum Gain by comparing the same \bestgridrate of the two well-known Gridders with the \gridrate reported by the Maximum Performance Experiments performed on the Convolutional Gridder.

We ran all experiments using a compiled from source WSClean version 2.8 with the IDG library integrated. We used the same hardware we used throughout the thesis and described in Chapter \ref{chap:methodology}. As input, we used the same LOFAR observation we used when experimenting with our Gridders in Chapters \ref{chap:2dgridding}, \ref{chap:hybrid} and \ref{chap:purenn} and we kept the pixel size of the output image to 4.7arcsec. We left the support of the GCF for the WSClean Gridder to the default. We took measurements of the execution time of the two well-known Gridders, by inspecting the output log of WSClean, which logs such measurements. It is worth to note that WSClean does not support compression, and therefore the WSClean and IDG Gridders had to grid the whole $6501600\times16$ records available in the LOFAR observation. 

We measured the \bestgridrate for the WSClean Gridder at 5.6MRec/s for single polarised-records. The measured value is higher by a factor of $1.3\times$ than what is reported by van der Tol \etal \cite{VanderTol2019}, possibly because of the different hardware used. We measured \bestgridrate for the IDG Gridder at 68.3MRec/s for quad-polarised records which is similar to what Veenboer \etal\cite{Veenboer2017} reported. Resultant Gains are calculated in Table \ref{tab:comparative:wscleangridder} and \ref{tab:comparative:idggridder}, where we report four-digit Maximum Gains for the Convolutional Gridder against the WSClean Gridder and two-digit Maximum Gains for the Convolutional Gridder against the IDG Gridder.   

Based on the just stated results and the many results we delivered in Chapter \ref{chap:2dgridding}, we see it reasonable to conclude that our implementation of the Convolutional Gridder performs well. Performance is at such a level that it is valid for the use as a reference in comparing Performance of the studied implementations. 

We do have to stress out that the experiments in this section were solely designed to validate the Convolutional Gridder for its use as a reference and in no way to discuss any possible superiority of the Convolutional Gridding against the two other gridders. One should consider more factors  to compare for the superiority of any gridder. For example, the IDG Gridder was designed to perform against scenarios that use W-projection and A-projections, which the Convolutional Gridder does not support. Also, the image sizes considered for the IDG Gridder and WSClean Gridder in the Brute Force search were at maximum $1024\times1024$ pixels.

\begin{table}[]
\centering
\begin{tabular}{@{}c|@{}c@{}|@{}c@{}c@{}|@{}c@{}c@{}}
\toprule
\multirow{2}{*}{$N_{\text{pol}}$}         & WSClean Gridder & \multicolumn{2}{c|}{Convolutional Gridder} & \multicolumn{2}{c}{Gain}  \\
\cmidrule{2-6}&%
\begin{minipage}[c]{3.8cm}%
  \vspace{4pt}%
  \centering%
  \bestgridrate \\%
  (GRec/s)\\%
  \vspace{4pt}%
  \end{minipage}&%
  \begin{minipage}[c]{2.6cm}%
  \vspace{4pt}%
  \centering%
  Maximum \gridrate\\%
  (GRec/s)\\%
  \vspace{4pt}%
  \end{minipage}&%
  \begin{minipage}[c]{2.6cm}%
  \vspace{4pt}%
  \centering%
  Minimum \gridrate\\%
  (GRec/s)\\%
  \vspace{4pt}%
  \end{minipage}&%
  \begin{minipage}[c]{2.2cm}%
  \vspace{4pt}%
  \centering%
  Maximum\\%
  \vspace{4pt}%
  \end{minipage}&%
  \begin{minipage}[c]{2.2cm}%
  \vspace{4pt}%
  \centering%
  Minimum\\%
  \vspace{4pt}%
  \end{minipage}\\
  \midrule
1  & 0.00530 & 4.00 & 0.45   & 754$\times$  & 84$\times$          \\
2  & 0.00265 & 3.81 & 0.33   & 1438$\times$ & 124$\times$   \\
3  & 0.00177 & 3.23 & 0.22   & 1825$\times$ & 124$\times$         \\
4  & 0.00133 & 2.84 & 0.16   & 2135$\times$ & 120$\times$         \\ \bottomrule
\end{tabular}
\caption[Convolutional Gridder Gain over the WSClean Gridder] {Table giving the estimated Minimum and Maximum Gains of the Convolution Gridder over the WSClean Gridder. The WSClean Gridder does not grid polarisations concurrently, and so we calculated the \bestgridrate  of the WSClean Gridder for $N_{\text{pol}}>1$ to be the  \bestgridrate at $N_{\text{pol}}=1$ divided by $N_{\text{pol}}$. We only consider Double-Precision results since the WSClean Gridder only computes in Double-Precision. The Maximum \gridrate values for the Convolutional Gridder is taken as the ones we reported for the  Double-Precision Maximum Performance Experiments available in Table \ref{tab:2dgridding:convmaxtheoratical}. The Minimum \gridrate values are taken as the minimum value of the \bestgridrate reported in Figure \ref{fig:2dgridding:convbrute_double}.}
\label{tab:comparative:wscleangridder}.
\end{table}

\begin{table}[]
\centering
\begin{tabular}{@{}c|@{}c@{}|@{}c@{}c@{}|@{}c@{}c@{}}
\toprule
\multirow{2}{*}{$N_{\text{pol}}$}         & IDG Gridder & \multicolumn{2}{c|}{Convolutional Gridder} & \multicolumn{2}{c}{Gain}  \\
\cmidrule{2-6}
  & \begin{minipage}[c]{2.8cm}%
  \vspace{4pt}%
  \centering%
  \bestgridrate \\%
  (GRec/s)\\%
  \vspace{4pt}%
  \end{minipage}&%
  \begin{minipage}[c]{2.8cm}%
  \vspace{4pt}%
  \centering%
  Maximum \gridrate\\%
  (GRec/s)\\%
  \vspace{4pt}%
  \end{minipage}&%
  \begin{minipage}[c]{2.8cm}%
  \vspace{4pt}%
  \centering%
  Minimum \gridrate\\%
  (GRec/s)\\%
  \vspace{4pt}%
  \end{minipage}&%
  \begin{minipage}[c]{2.4cm}%
  \vspace{4pt}%
  \centering%
  Maximum\\%
  \vspace{4pt}%
  \end{minipage}&%
  \begin{minipage}[c]{2.4cm}%
  \vspace{4pt}%
  \centering%
  Minimum\\%
  \vspace{4pt}%
  \end{minipage}\\
  \midrule

4  & 0.0683  & 3.62 & 0.129  & 53$\times$  & 1.88$\times$          \\
\bottomrule
\end{tabular}
\caption[Convolutional Gridder Gain over the IDG Gridder]{Table giving the estimated Minimum and Maximum Gains of the Convolution Gridder against the IDG Gridder. Only Single-Precision computation at $N_{\text{pol}}=4$ is considered because the IDG Gridder works only for $N_{\text{pol}}=4$ and Single-Precision. The Maximum \gridrate values for the Convolutional Gridder is taken as the ones we reported for the  Single-Precision Maximum Performance Experiments available in Table \ref{tab:2dgridding:convmaxtheoratical}. The Minimum \gridrate values are taken as the minimum value of the \bestgridrate reported in Figure \ref{fig:2dgridding:convbrute_single}.}
\label{tab:comparative:idggridder}
\end{table}

\subsection{Aliasing in the Convolutional Gridder}
\label{sec:comparative:convgridderaliasing}

We now discuss experiments meant to validate our Convolutional Gridding implementation in terms of aliasing suppression. We do so by comparing images outputted by our Convolutional Gridding implementation with the output of CASA, using its implementation of Convolutional Gridding henceforth referred to as \textit{CASA Gridding}. We do have to point out that any measurement of aliasing distortions will always include arithmetic noise. Therefore our experiments will inherently verify that the level of arithmetic noise generated by our Convolutional Gridding implementation is at acceptable levels and no higher than that of CASA Gridding.

The GCF used in the two Gridders is the Prolate Spheroidal of order 1 with support 6. It is the only supported GCF in CASA, and for our implementation, the GCF  is calculated similarly to how the GCF is calculated in CASA.

In all the experiments presented in this section, we use three observations, which we named after the telescopes that made the observation: the GMRT, EVLA, and LOFAR. The LOFAR observation is a different observation from what was in the other Chapters. For experiments in this section, we use the Visibility data measured in the stated observations and generate duplicates with all Visibility measurements set to unity, therefore generating PSFs. We differentiate the duplicates from the original by including the term PSF in the observation given name. For example, the duplicate of the VLA observation is called the VLA PSF observation.

In all experiments, we generate an output image of size $4096\times4096$ pixels with pixel intervals, set  as shown in Table \ref{tab:comparative:pixelinterval}. 

\begin{table}[]
\centering
\begin{tabular}{@{}ccccc@{}}
\toprule
Telescope &\rule{0.8cm}{0cm}&Pixel Interval (arcsec)\\ 
 \midrule
GMRT     &        & $5\times5$                   \\
VLA       &   & $0.1\times0.1$               \\
LOFAR     &   & $1.65\times1.65$               \\ \bottomrule
\end{tabular}
\caption[Pixel intervals and images sizes used for various experiments in Chapter \ref{chap:comparative}]{Table giving the pixel interval and image size used for different observations in experiments presented in Sections \ref{sec:comparative:convgridderaliasing}  and \ref{sec:comparative:aliasing}. The three different observations are identified by the telescope that made the particular observation.}
\label{tab:comparative:pixelinterval}
\end{table}

The observations we use in this section are also used in Section \ref{sec:comparative:aliasing}.

For the set of experiments discussed in this section, we generate an image from the six sets of data using the following three methods: 

\begin{enumerate}
\item \textbf{DFT method:} In this method, we use an in-house built GPU-based implementation that computes the output image with Double-Precision using the DFT Equation \ref{equ:maths:dft}. The generated output image is void from aliasing effects, and therefore we use it as a reference. We validated this implementation against a CPU-based DFT implementation in WSClean, whereby the two implementations returned identical results, once we programmatically disabled the final beam correction in the WSClean. We used our implementation instead of that of WSClean, since it is considerably faster to compute on the P100.

\item \textbf{CASA method:} In this procedure, we use the imager tool available in CASA, to generate the image using CASA Gridding. W-projection is switched off as to enable the \textit{Normal Interferometric Gridding mode}. In this mode, CASA automatically computes with Double-Precision using a hard-coded oversampling factor ($\beta$) of 100.

\item \textbf{Conv100 method}: In this method we generate the image using our implementation of Convolutional Gridding computing in Double-Precision, with $\beta$ set to 100. We will use the results outputted by this method to compare with the results of the CASA method.  
\end{enumerate}

We measure the sum of aliasing effects and arithmetic noise by computing the Root Mean Square (RMS) of the difference between the image produced by the reference DFT method and the image produced either by the CASA or Conv100 methods. We measure the RMS in two regions of the image, which are the Central Half ($64\times64$) Region and the Full Region, which is all the image. To help us understand better the difference between RMS readings of the two methods we define the ratio $\Gamma$ as per Equation \ref{equ:comparative:gamma}.
\begin{equation}
    \Gamma=\frac{\texttt{RMS of Conv100 Method} - \texttt{RMS of CASA method}}{\texttt{RMS of CASA method}}\times 100\%
    \label{equ:comparative:gamma}
\end{equation}

Note that a negative value of $\Gamma$ implies that the Conv100 method returned a lower RMS value than the CASA method.

\begin{table}[]
\centering
\begin{tabular}{@{}l@{\rule{1.7cm}{0cm}}c@{\rule{1.7cm}{0cm}}c@{\rule{1.7cm}{0cm}}r@{}}
\toprule
Observation & CASA & Conv100 & $\Gamma$ (\%) \\
\midrule
\multicolumn{4}{c}{\cellcolor{mygrey}Central Half Region}\\
\midrule
GMRT      & 6.909e-06 &      6.910e-06 &  1.540e-02 \\
GMRT PSF  & 8.999e-07 &      9.006e-07 &  8.172e-02 \\
VLA       & 3.583e-04 &      3.583e-04 & -1.019e-02 \\
VLA PSF   & 1.034e-06 &      1.046e-06 &  1.162 \\
LOFAR     & 7.331e-05 &      7.333e-05 &  2.532e-02 \\
LOFAR PSF & 7.788e-04 &      7.787e-04 & -1.175e-02 \\
\midrule
\multicolumn{4}{c}{\cellcolor{mygrey}Full Region}\\
\midrule
GMRT      & 2.185e-03 &      2.186e-03 & 4.827e-03 \\
GMRT PSF  & 3.713e-04 &      3.714e-04 & 7.382e-03 \\
VLA       & 5.618e-02 &      5.618e-02 & 6.635e-03 \\
VLA PSF   & 8.520e-04 &      8.521e-04 & 1.035e-02 \\
LOFAR     & 1.183e-02 &      1.183e-02 & 3.393e-03 \\
LOFAR PSF & 8.555e-04 &      8.555e-04 & 1.453e-03 \\
\bottomrule
\end{tabular}

\caption[Results for experiments validating our implementation of Convolutional Gridding]{Table giving the RMS and $\Gamma$ values measured by experiments validating our implementation of Convolutional Gridding against an implementation in CASA. Conv100 refers to our implementation of  Convolutional Gridding with $\beta$ set to 100.}
\label{tab:comparative:casaaliasresults}
\end{table}

\begin{figure}
\centering
\begin{tabular}{@{}c@{}c@{}c@{}}
\multicolumn{3}{c}{\small{\texttt{\bfseries GMRT}}} \\ 
\includegraphics[height=135pt]{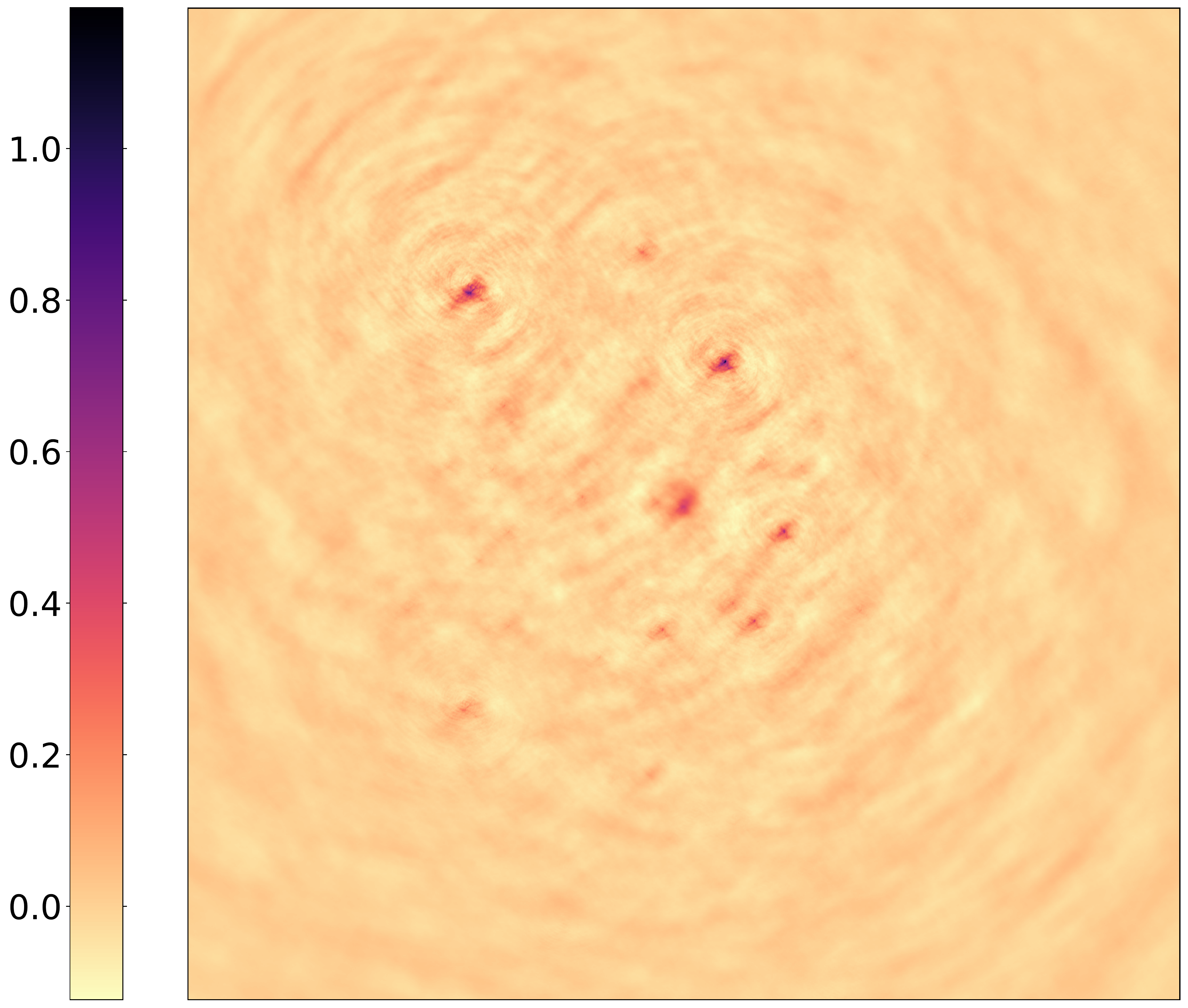} &
\includegraphics[height=135pt]{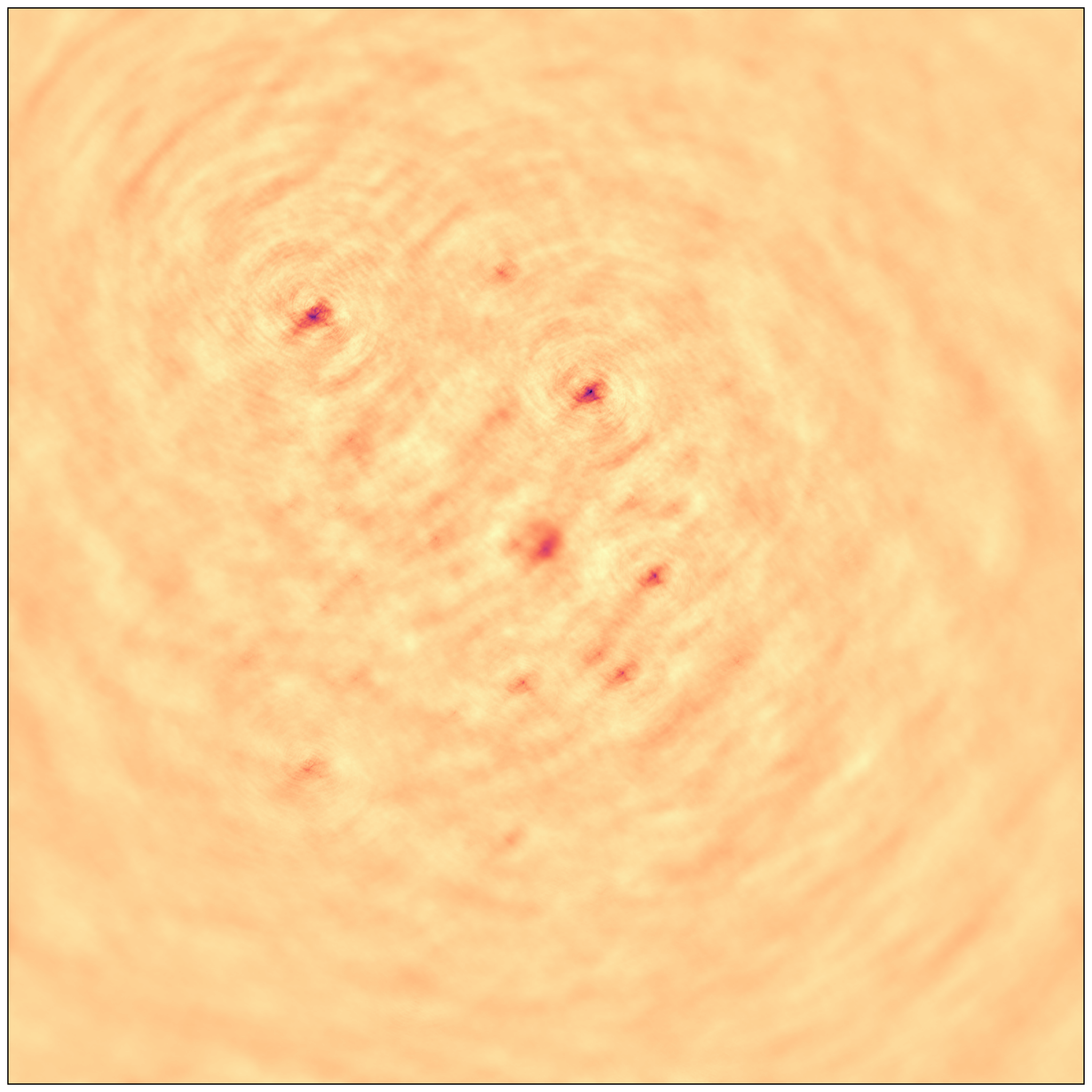} &
\includegraphics[height=135pt]{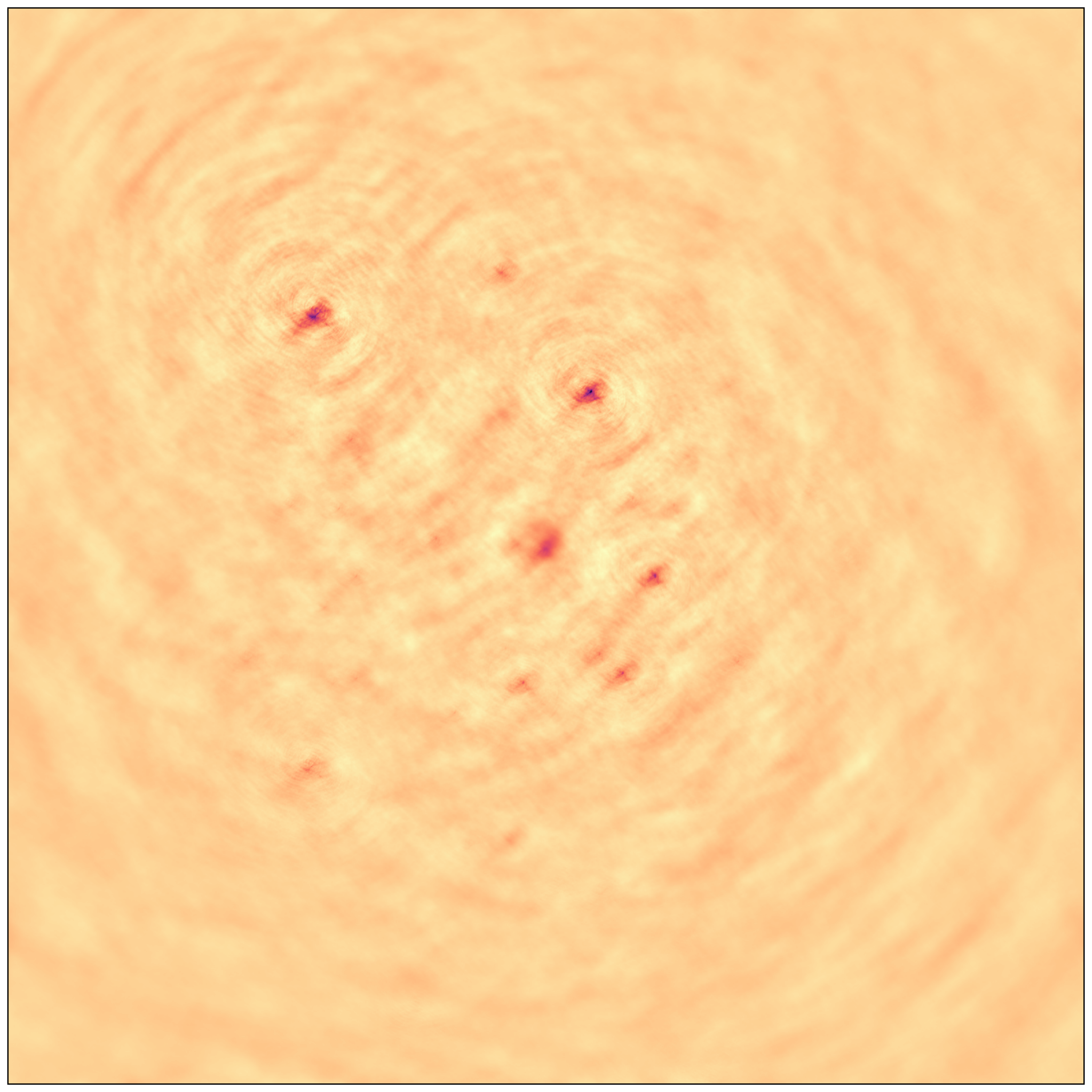} \\

\multicolumn{3}{c}{\small{\texttt{\bfseries VLA}}} \\ 
\includegraphics[height=135pt]{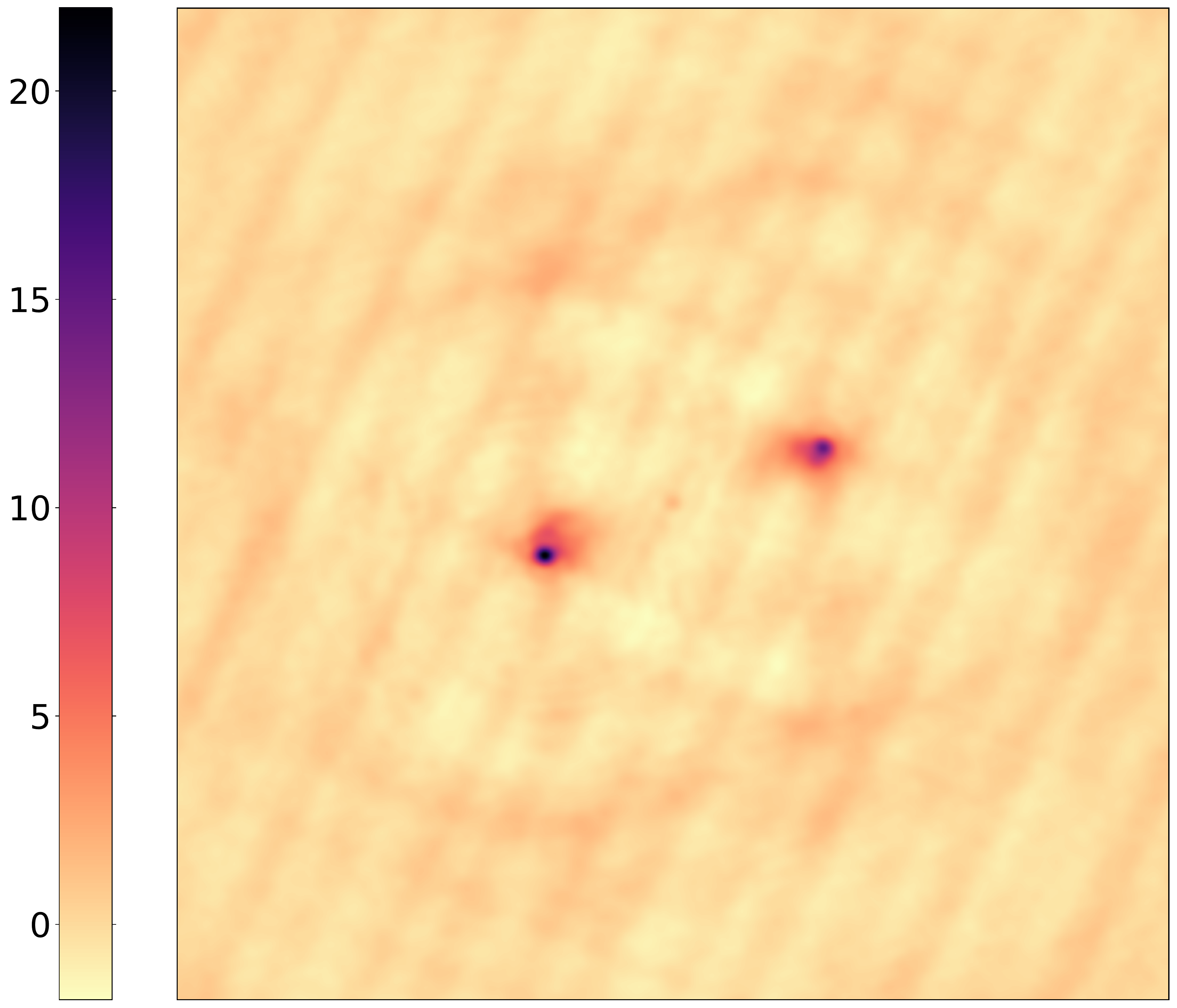} &
\includegraphics[height=135pt]{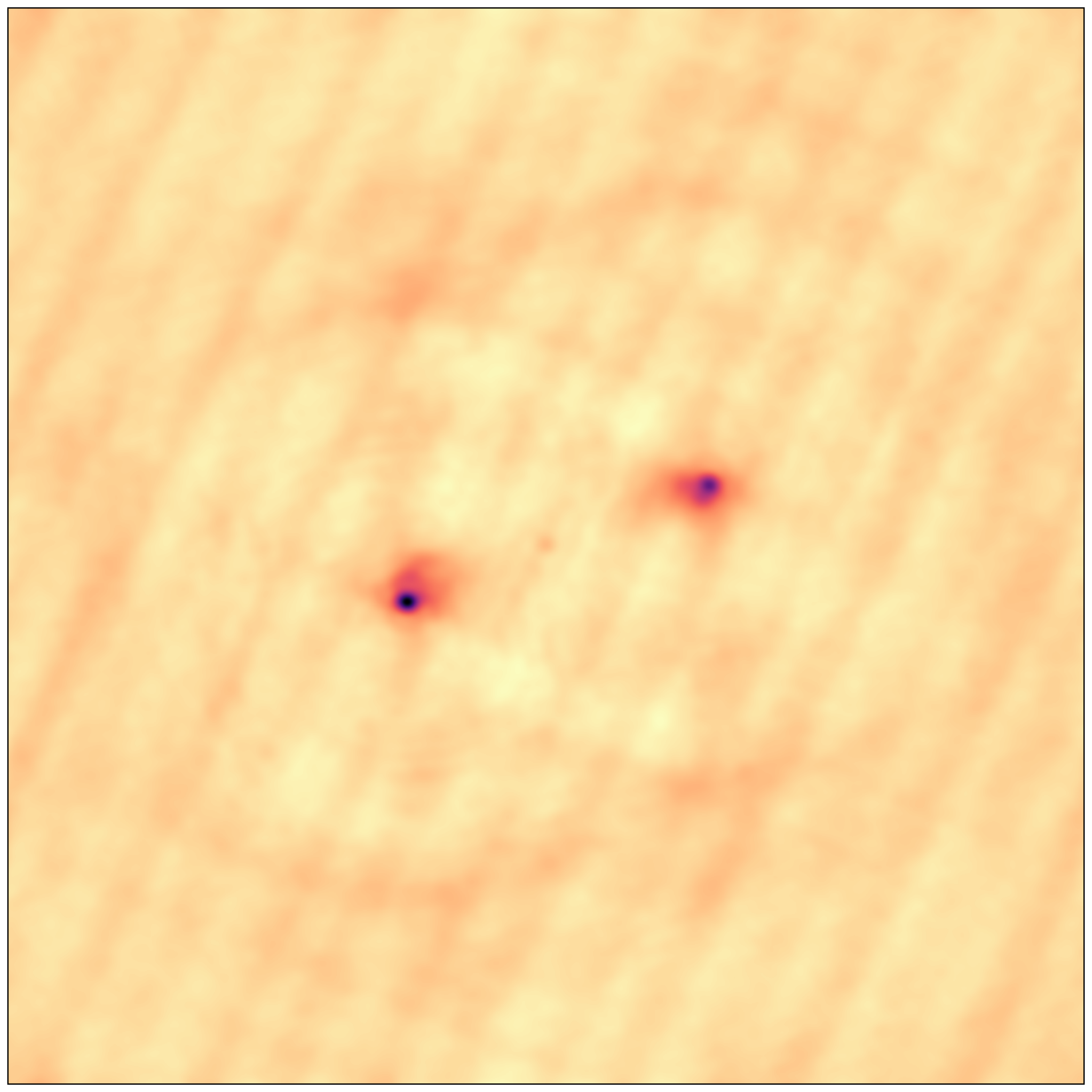} &
\includegraphics[height=135pt]{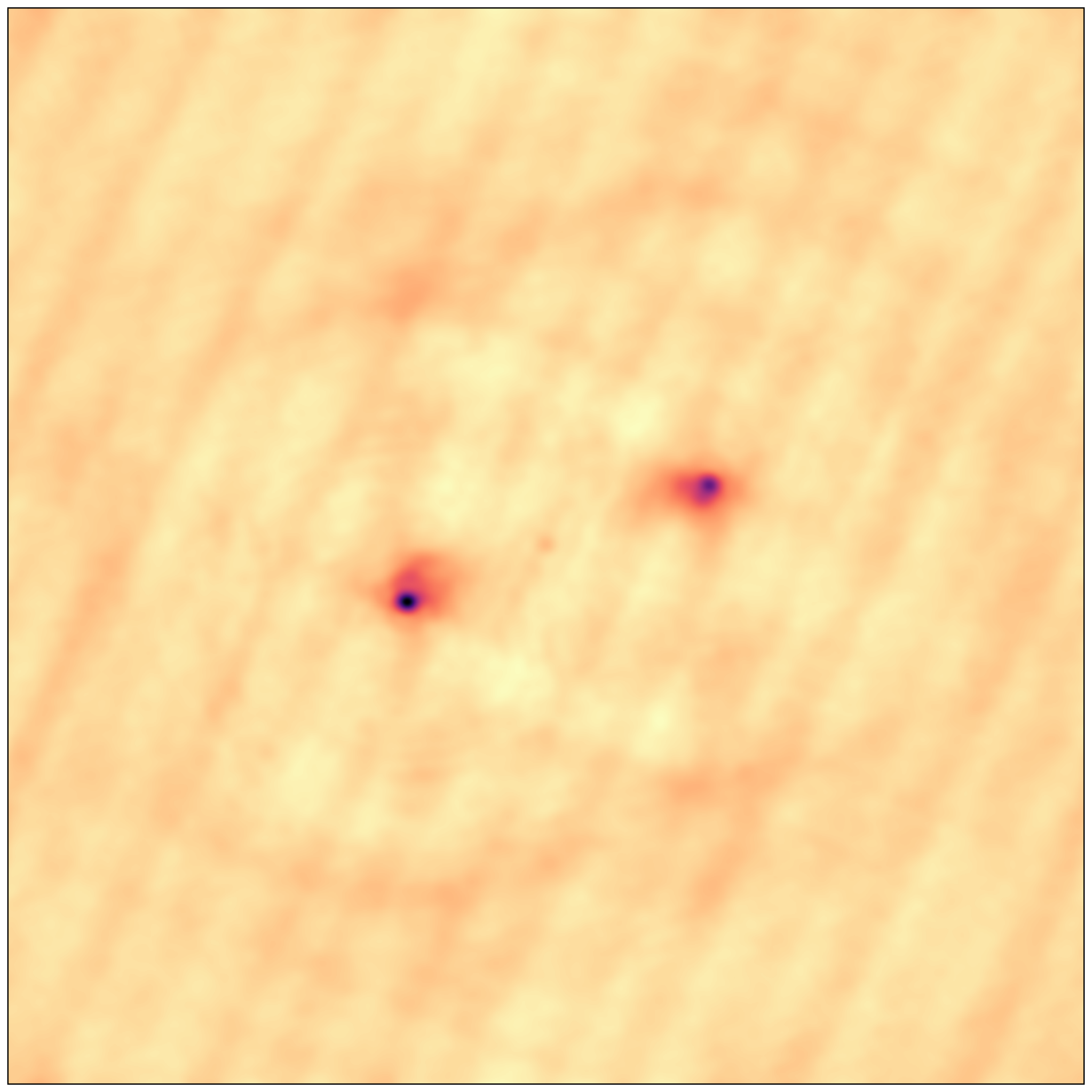} \\
\multicolumn{3}{c}{\small{\texttt{\bfseries LOFAR}}} \\ 
\includegraphics[height=135pt]{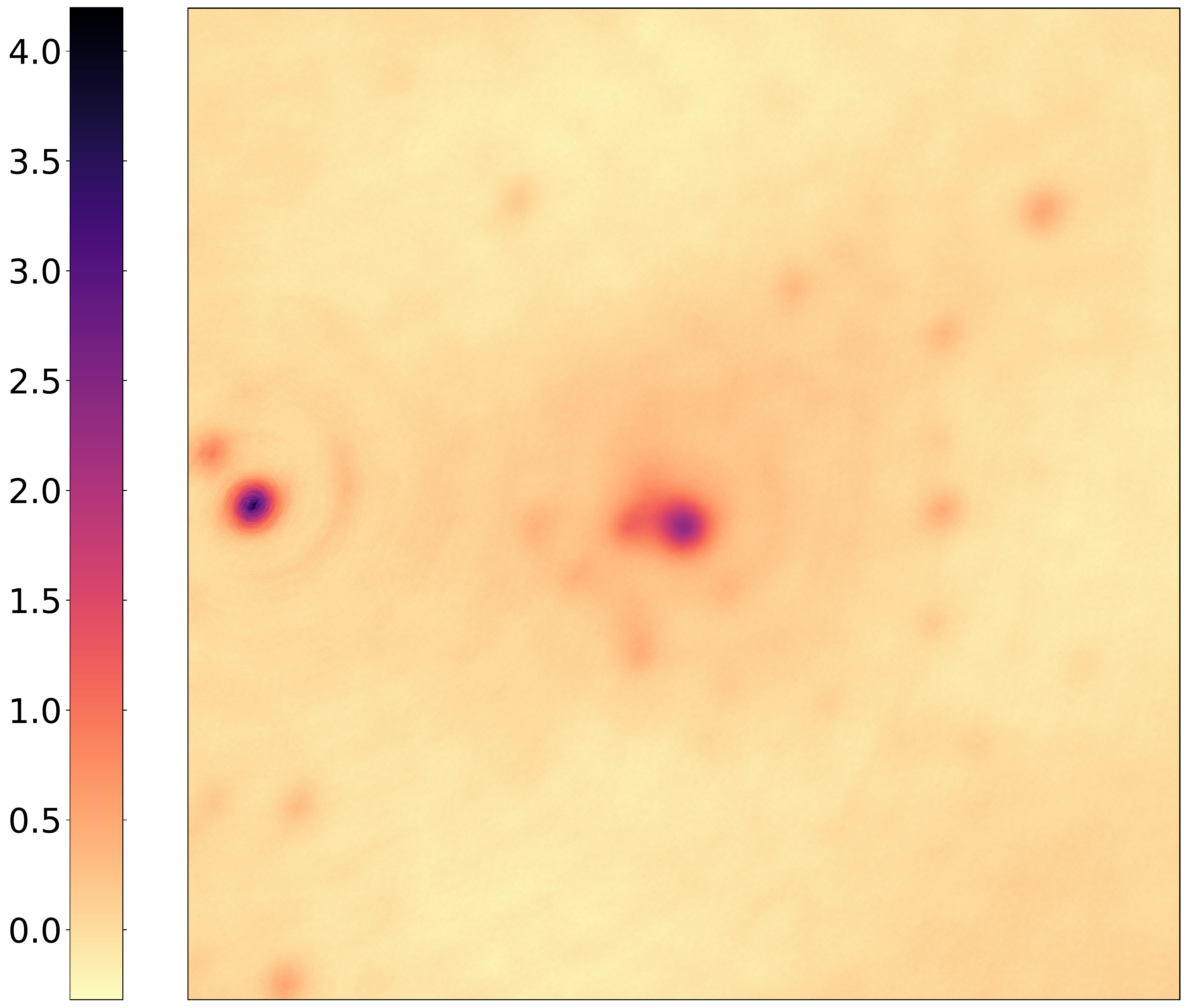} &
\includegraphics[height=135pt]{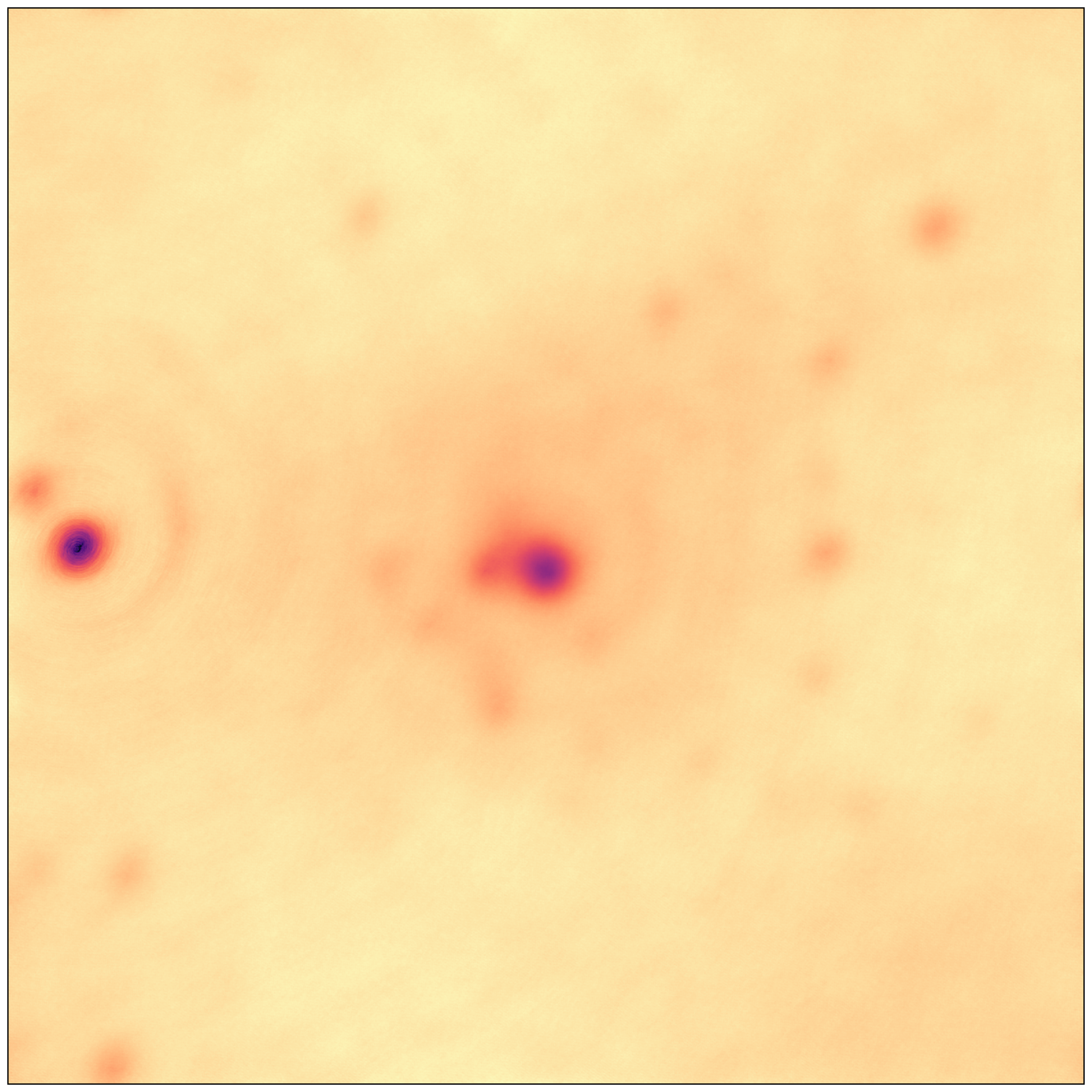} &
\includegraphics[height=135pt]{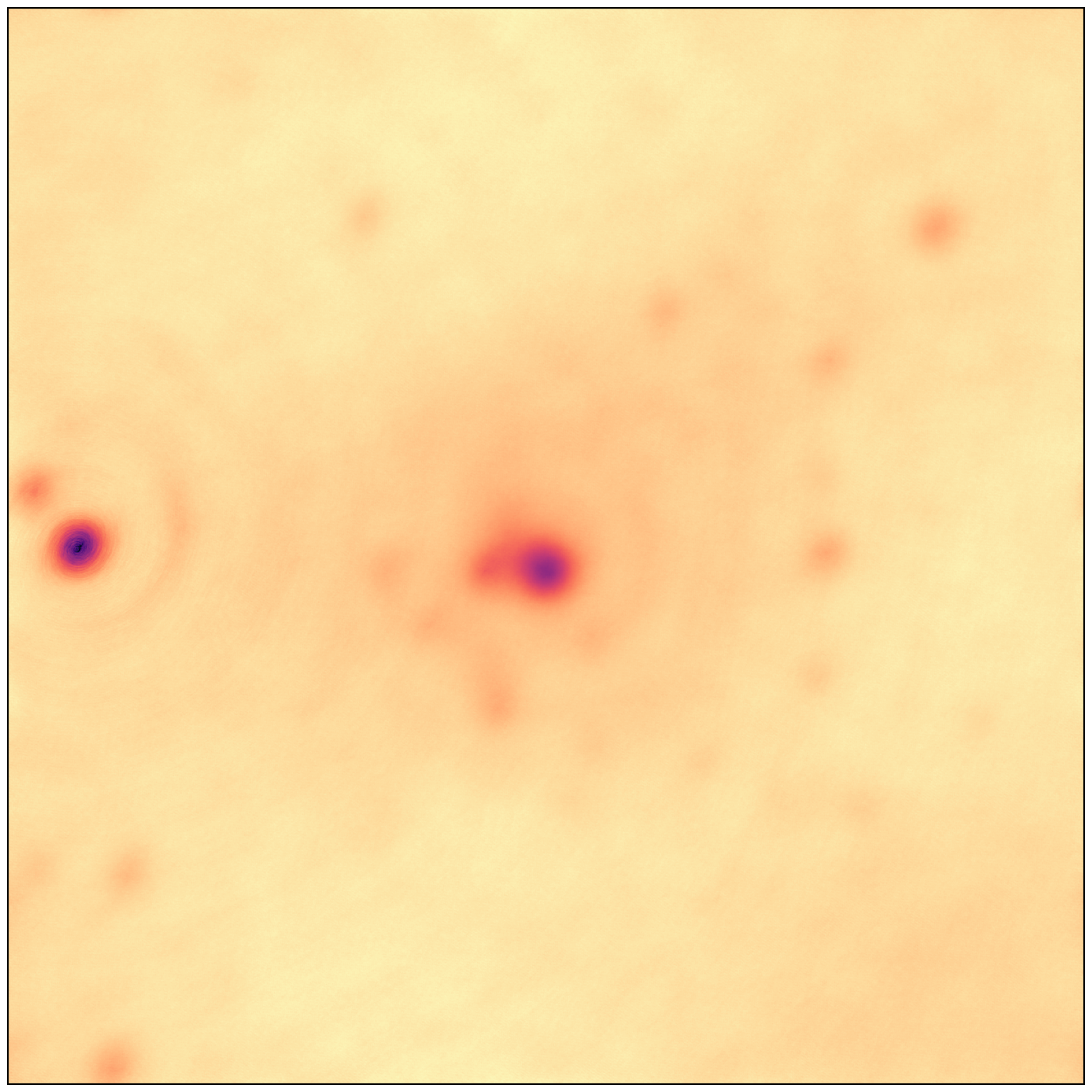} \\
\footnotesize\texttt{DFT method} & \footnotesize\texttt{CASA method} & \footnotesize\texttt{Conv100 method}\\
\\
\end{tabular}
\caption[Images generated by the set of  experiments in Section \ref{sec:comparative:convgridderaliasing}]{Images generated with the GMRT, VLA and LOFAR observations by the experiments of Section \ref{sec:comparative:convgridderaliasing}. There is no apparent visual difference between the images generated by the three different methods.}

\label{fig:pruning:convaliasing1}
\end{figure}

\begin{figure}
\centering
\begin{tabular}{@{}c@{}c@{}c@{}}

\multicolumn{3}{c}{\small{\texttt{\bfseries GMRT PSF}}} \\
\includegraphics[height=135pt]{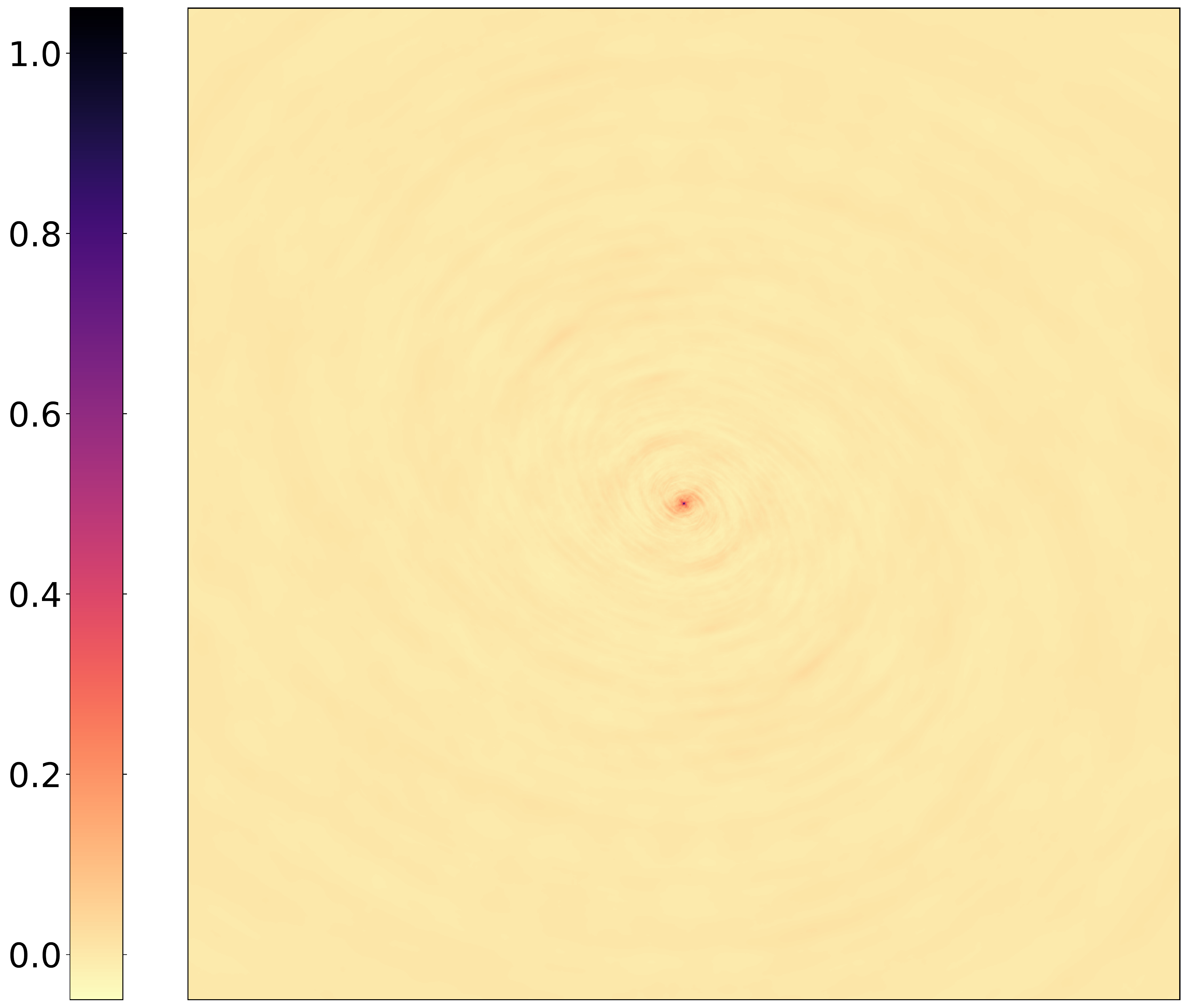} &
\includegraphics[height=135pt]{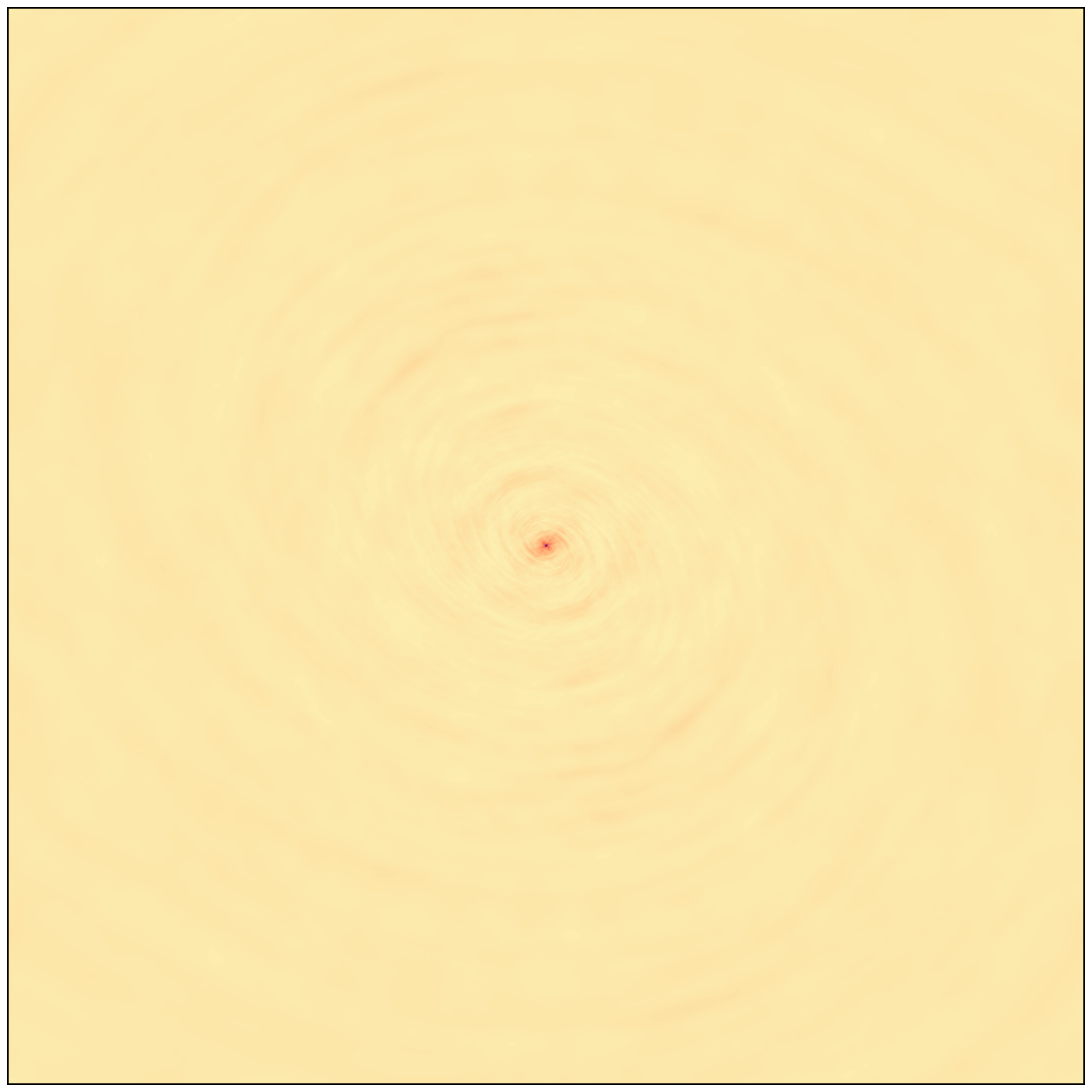} &
\includegraphics[height=135pt]{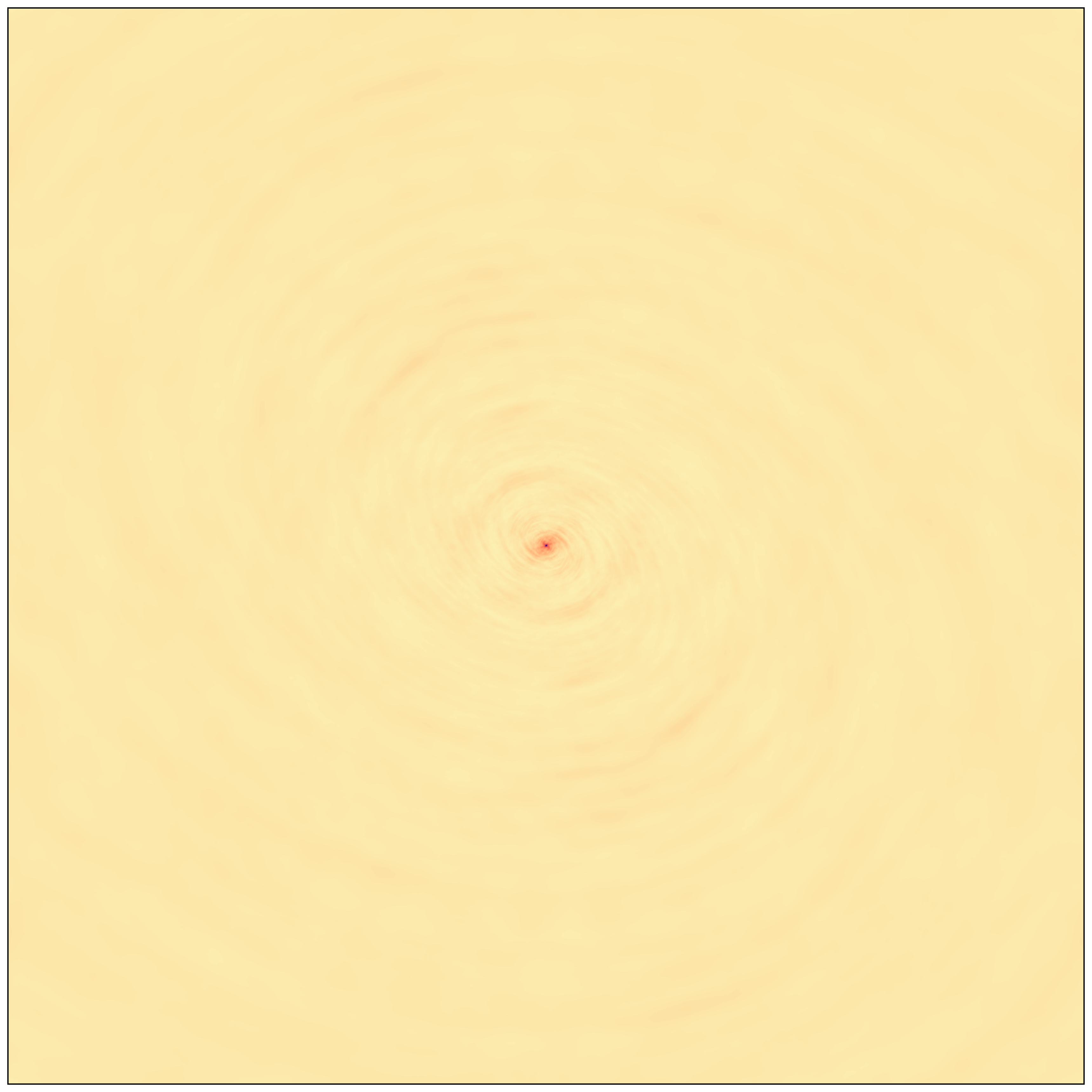}\\

\multicolumn{3}{c}{\small{\texttt{\bfseries VLA PSF}}} \\ 
\includegraphics[height=135pt]{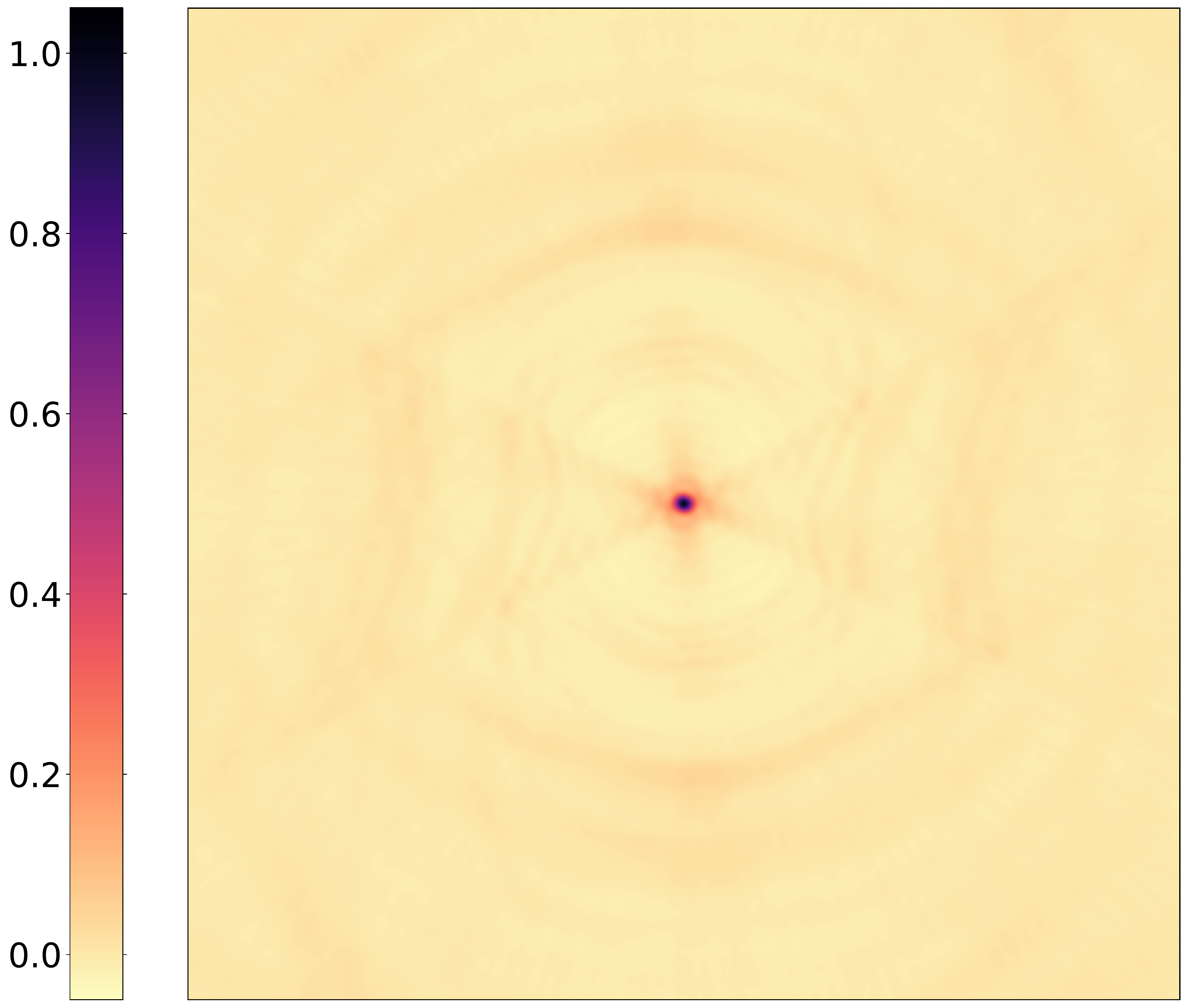} &
\includegraphics[height=135pt]{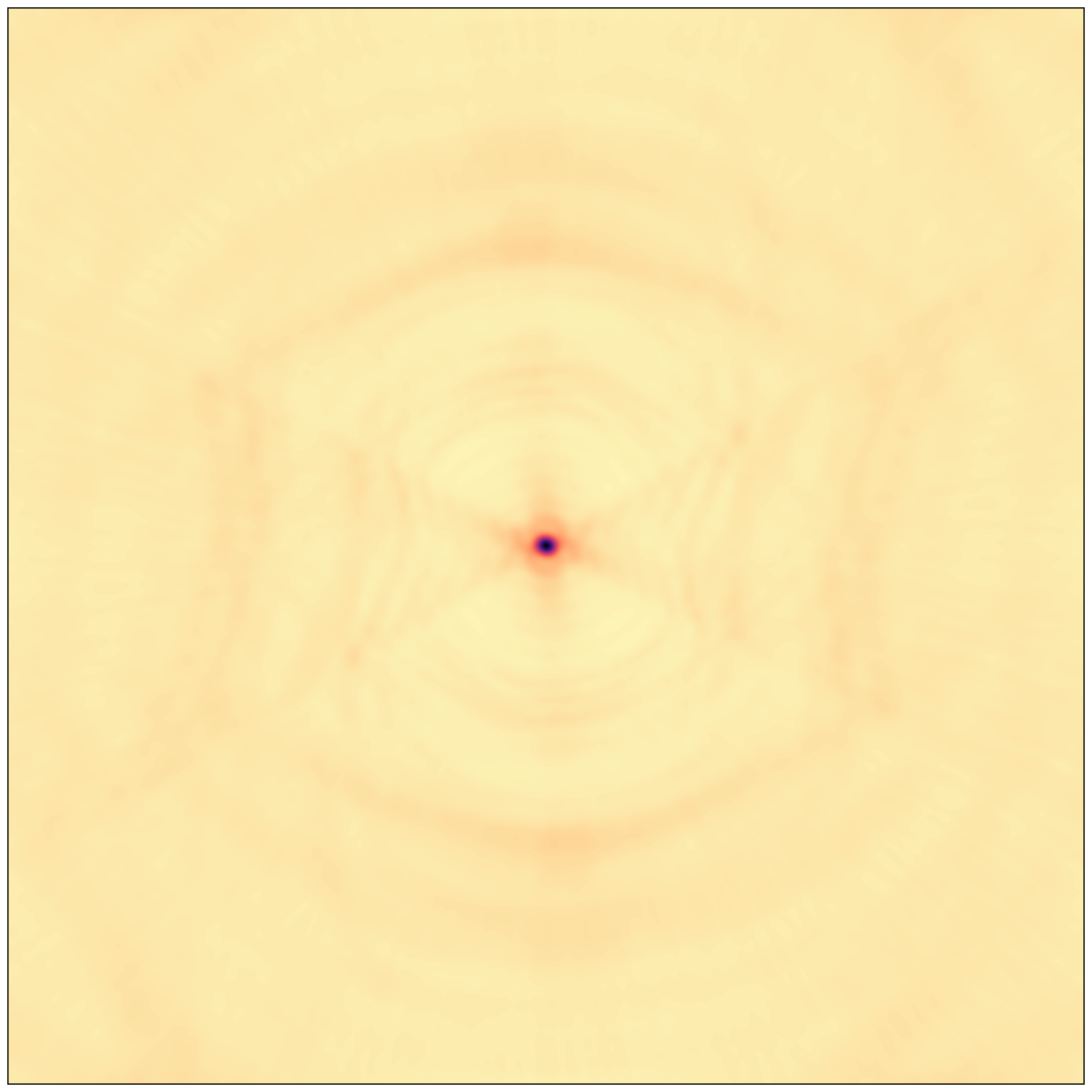} &
\includegraphics[height=135pt]{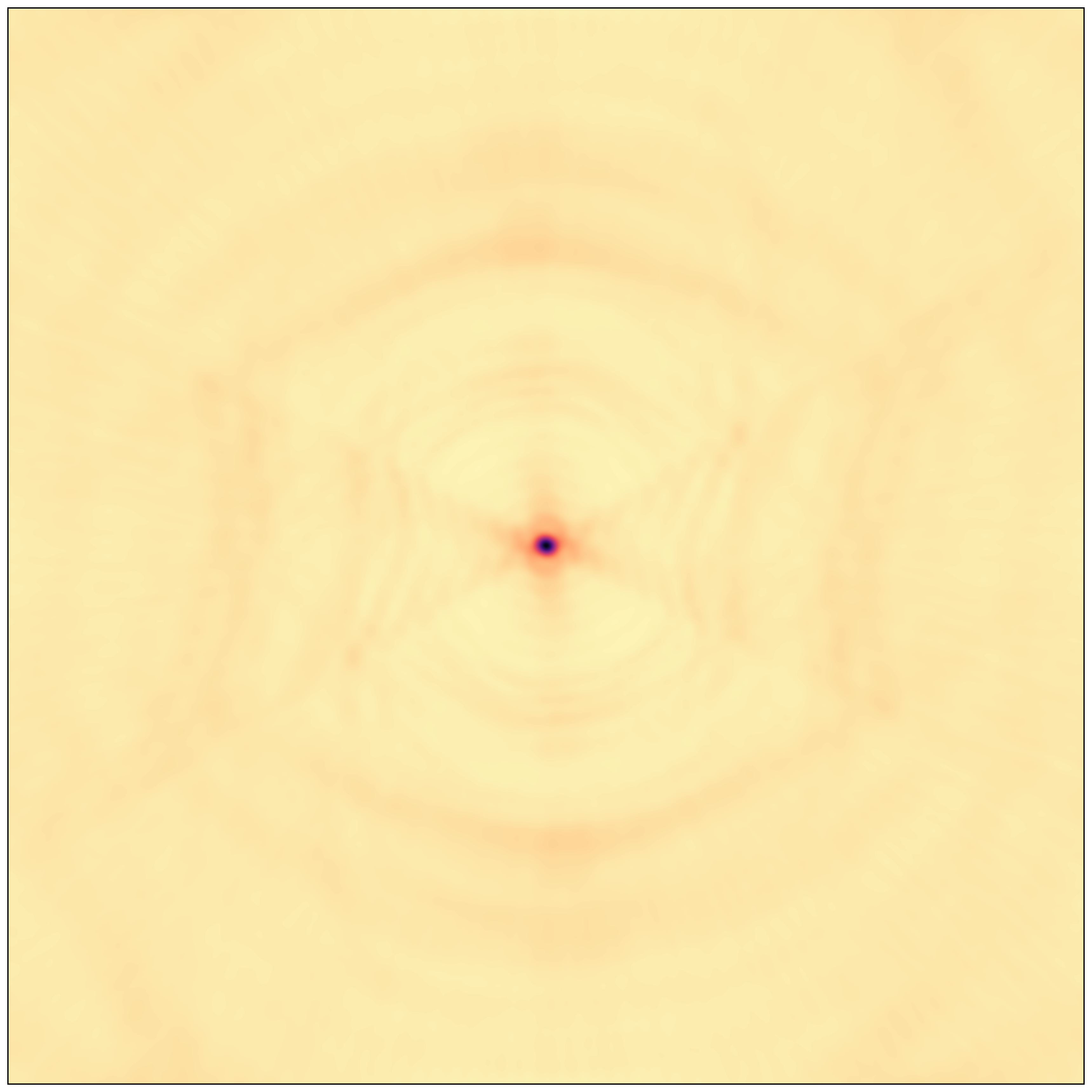} \\

\multicolumn{3}{c}{\small{\texttt{\bfseries LOFAR PSF}}} \\ 
\includegraphics[height=135pt]{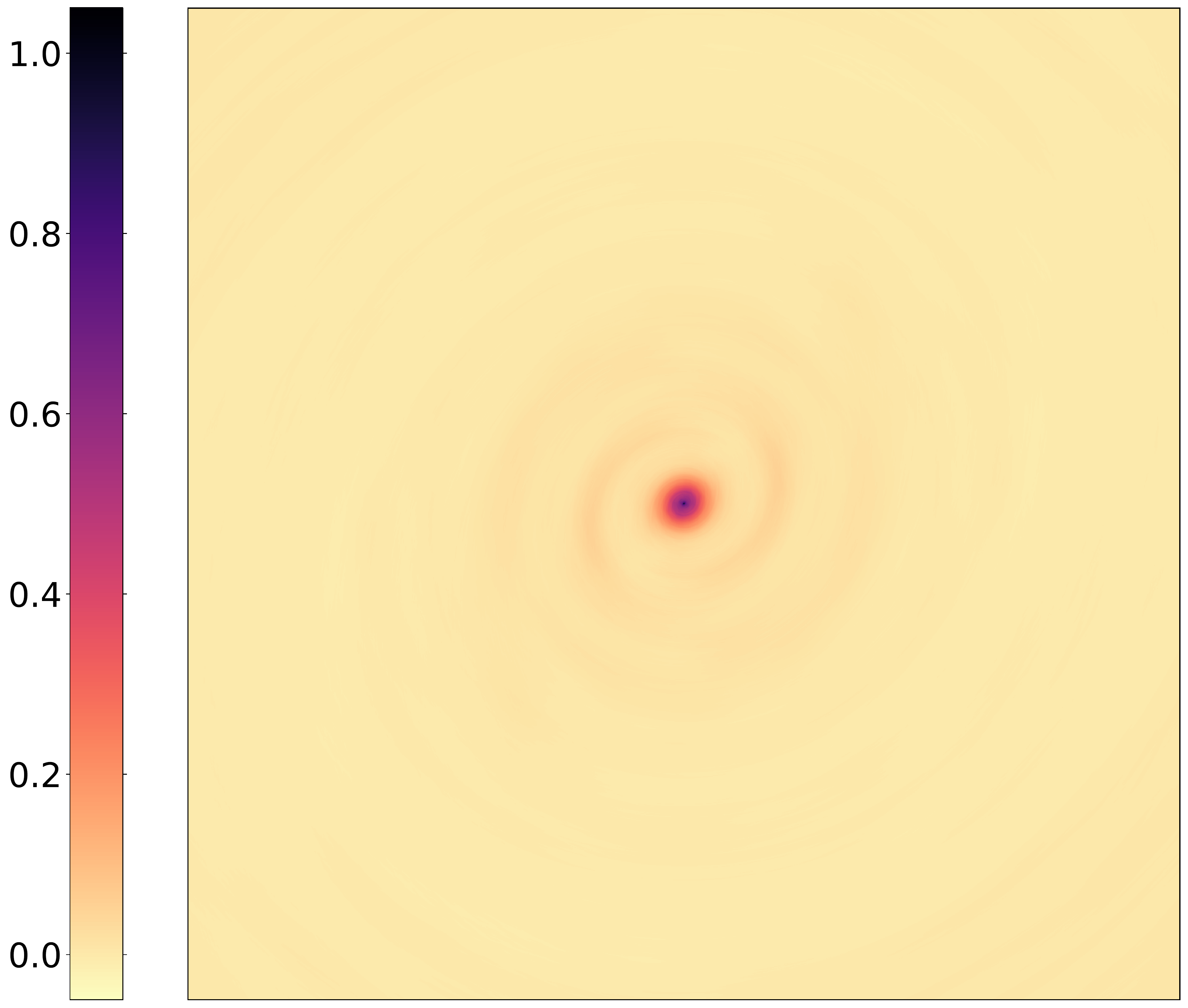} &
\includegraphics[height=135pt]{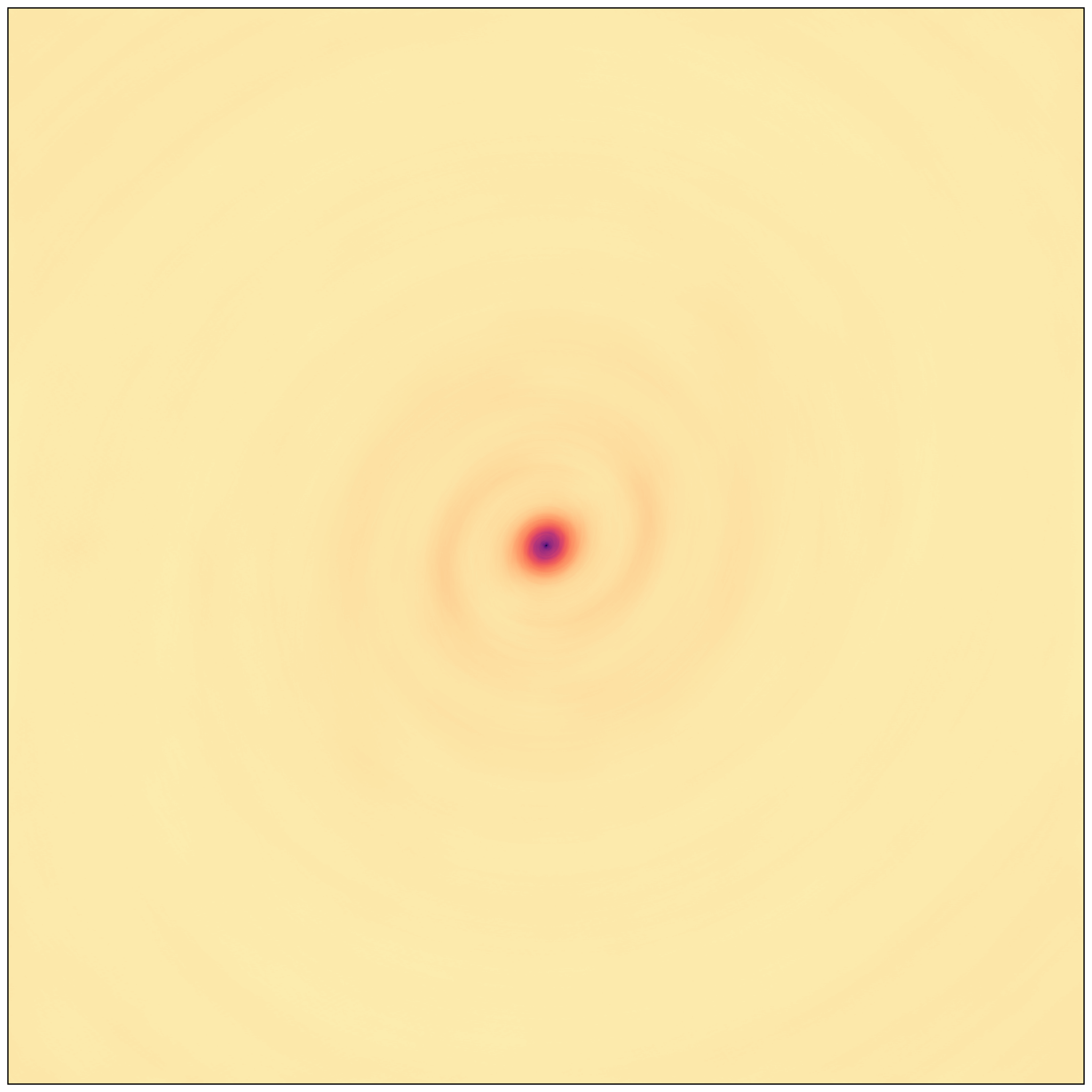} &
\includegraphics[height=135pt]{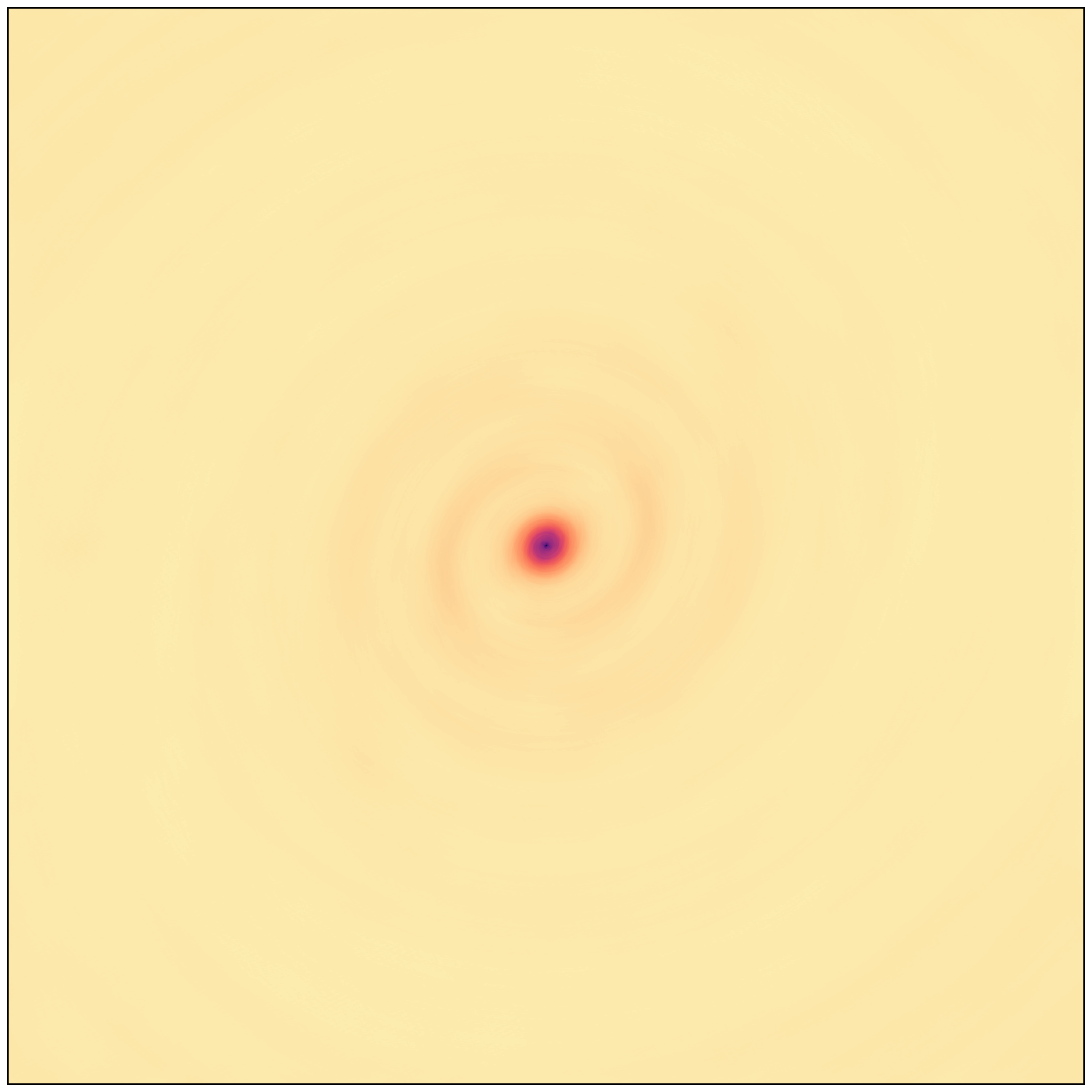} \\
\footnotesize\texttt{DFT method} & \footnotesize\texttt{CASA method} & \footnotesize\texttt{Conv100 method}\\
\\
\end{tabular}
\caption[Images generated by the set of  experiments in Section \ref{sec:comparative:convgridderaliasing}]{Images generated with the GMRT PSF, VLA PSF and LOFAR PSF observations by the experiments of Section \ref{sec:comparative:convgridderaliasing}. There is no apparent visual difference between the images generated by the three different methods.}
\label{fig:pruning:convaliasing2}
\end{figure}

All images generated by the experiments in this section are reproduced in Figures \ref{fig:pruning:convaliasing1} and \ref{fig:pruning:convaliasing2} while we tabulate the quantitative results in Table \ref{tab:comparative:casaaliasresults}.

A visual inspection of the images in Figures \ref{fig:pruning:convaliasing1} and \ref{fig:pruning:convaliasing2} do not show any visible difference between what is generated by the three methods. The quantitative results in Table \ref{tab:comparative:casaaliasresults} captured some slight variations with the absolute value of $\Gamma$ generally less than $0.1\%$. Results of the VLA PSF are an exception where $\Gamma=1.162$. Taking into account that for the VLA PSF the RMS is in the region of $10^{-6}$, while maximum intensity in the generated images is 1, the stated value of $\Gamma$ is not indicative of any increase in aliasing distortions or unacceptably high arithmetic noise. Therefore, we conclude that the implementation of the Convolutional Gridding is valid for use as a reference in our subsequent experiments.


\section{Comparative analysis of the studied implementations.}
\label{sec:comparative:compalgos}
In this section, we will do the final analysis that shows if and how we reached the primary goal of this thesis. We also prove many of the claims we established theoretically in Chapter \ref{chap:theory}.  

This section is divided into two sub-sections. In the first sub-section, we do a comparative analysis on the Performance of the studied implementations to identify which modifications perform better than Convolutional Gridding. In the subsequent sub-section, we make sure that Pruned NN Interpolation and Hybrid Gridding do not decrease the level of aliasing suppression, which is a must mandated by the primary goal of this thesis.

\subsection{Performance comparison of the studied algorithms}
\label{sec:comparative:performancestudied}
Let us do a Performance comparative analysis of the three implemented algorithms. In previous chapters, we have already compared the Performance of the Gridders via the Gridder Advantage group of metrics, and we here complete the analysis by including the effects of finalisation in each algorithm. We will rely on the Minimum Record Density Function for this comparison which we will derive, explain and take measurements of in this section.

\subsubsection{Summary}

Let us summarise what we know so far. In the previous chapters, we showed that our implementations of the Hybrid and NN Gridders are faster than our implementation of the Convolutional Gridder. Nevertheless, in our test scenarios, the NN Gridder is faster than the Hybrid Gridder only for low values of $\beta$. The output grid of the Hybrid and NN Gridders are larger than that of the Convolutional Gridder by a factor of $\beta$ and $\beta^2$ respectively. Therefore, finalisation in Hybrid Gridding and Pruned NN Interpolation require more compute than in Convolutional Gridding, which risks cancelling the gains made by the respective Gridders over the Convolutional Gridder. In order to mitigate the risk, we prune the output grid using convolution, where we prove that it can reduce the execution time of the finalisation stage. However, in no way it can be as fast as the finalisation stage of the Convolutional Gridder. Therefore, we argue that finalisation plays a vital role in determining how the three implementations fare against each other and, that there should be a threshold number of input records at which two given implementations will fare equally from a Performance perspective. We will use such a number of records to provide a Performance comparison between the algorithms.

\subsubsection{The Minimum Record Density Function - $D(A_1,A_2)$}

Let $A_1$ and $A_2$ be any of the three implemented Algorithms \ref{algo:maths:convgriddingoversampling}, \ref{algo:maths:hybrid}, \ref{algo:maths:purenninterpolation}, and let $R(A_1,A_2)$ be the number of input records at which we evaluate that  $A_1$ and $A_2$ take the same time to execute, given $N_{\text{pol}}$, Precision,$N_{\text{nn}}$ and $\beta$ are kept equal. $R(A_1,A_2)$ is calculated using the relationship expressed in Equation \ref{equ:comparative:recs}. 

\begin{equation}
\label{equ:comparative:recs}
    \text{fin}(A_1)+\frac{R(A_1,A_2)}{\metric{Max Best Gridding Rate(A\textsubscript{1})}}=\text{fin}(A_2)+\\ \frac{R(A_1,A_2)}{\metric{Max Best Gridding Rate(A\textsubscript{2})}}
\end{equation}
where \metric{Max Best Gridding Rate(A)} is the highest \bestgridrate for the Gridder of Algorithm $A$ when considering all Ordering Modes. $\text{fin}(A)$ is the GPU execution time taken for finalisation in Algorithm A which includes the IFFT and correction together with pruning for Hybrid Gridding and Pruned NN Interpolation or a pre-FFT processor for Convolutional Gridding.

Since the total execution time of finalisation has some dependency on $N_{\text{nn}}$, we see it more meaningful to express $R(A_1,A_2)$ as a ratio with the size of the  NN Grid, and define such a ratio as the Minimum Record Density function $D(A_1,A_2)$ given by Equation \ref{equ:comparative:density}.

\begin{equation}
\label{equ:comparative:density}
    D(A_1,A_2)=\frac{R(A_1,A_2)\times100\%}{N_{\text{nn}} \times N_{\text{nn}} }
\end{equation}

We note that if $D(A_1,A_2)>100\%$, then $R(A_1,A_2)$ is larger than the size of the NN Grid, and with full compression enabled, it is impossible to have the input number of records equal or greater than $R(A_1,A_2)$. If $D(A_1,A_2)<0$, then $A_1$ is always faster or slower than $A_2$ irrespective to the number of input records.

We give arguments $A_1$ and $A_2$ in $D(A_1,A_2)$ the values of $\texttt{co}$, $\texttt{hy}$ and $\texttt{pn}$ to represent Convolutional Gridding (Algorithm \ref{algo:maths:convgriddingoversampling}), Hybrid Gridding (Algorithm \ref{algo:maths:hybrid}) and Pruned NN Interpolation (Algorithm \ref{algo:maths:purenninterpolation}) respectively. 

\subsubsection{Results and analyses}

\begin{table}[]
    \centering
    \begin{tabular}{@{}cc@{}ccccccccc}
\toprule
\multirow{2}{*}{$\beta$} & \rule{0.6cm}{0cm}&\multicolumn{4}{c}{Single-Precision} & \rule{0.6cm}{0cm}& \multicolumn{4}{c}{Double-Precision}\\
\cmidrule{3-6} \cmidrule{8-11}
& &\multicolumn{1}{@{}c}{\begin{tabular}[c]{@{}c@{}}$N_{\text{pol}}$\\ 1\end{tabular}} & \multicolumn{1}{c}{\begin{tabular}[c]{@{}c@{}}$N_{\text{pol}}$\\ 2\end{tabular}} & \multicolumn{1}{c}{\begin{tabular}[c]{@{}c@{}}$N_{\text{pol}}$\\ 3\end{tabular}} & \multicolumn{1}{c}{\begin{tabular}[c]{@{}c@{}}$N_{\text{pol}}$\\ 4\end{tabular}} & & \multicolumn{1}{c}{\begin{tabular}[c]{@{}c@{}}$N_{\text{pol}}$\\ 1\end{tabular}} & \multicolumn{1}{c}{\begin{tabular}[c]{@{}c@{}}$N_{\text{pol}}$\\ 2\end{tabular}} & \multicolumn{1}{c}{\begin{tabular}[c]{@{}c@{}}$N_{\text{pol}}$\\ 3\end{tabular}} & \multicolumn{1}{c}{\begin{tabular}[c]{@{}c@{}}$N_{\text{pol}}$\\ 4\end{tabular}} \\
\midrule
 \multicolumn{11}{c}{\cellcolor{mygrey}$D(\text{co},\text{pn})$, $N_{\text{nn}} =16384$} \\
 \midrule
4 &    &  30.8 &     36.9 &     43.2 &     38.4 & &    60.2 &     69.2 &      n/a &      n/a \\
8 & &    25.7 &     38.6 &     63.0 &     58.1 &   &  50.4 &     81.5 &      n/a &      n/a \\
16 & &    20.7 &     43.3 &     86.6 &    114.3 &   &  35.8 &     95.0 &    259.7 &      n/a \\
32 &  &   18.2 &     43.4 &     95.9 &    140.9 &   &  36.0 &     98.3 &  -3361.7 &      n/a \\
64 &   &  16.4 &     42.2 &    102.8 &    146.6 &    & 37.3 &    114.7 &   -586.4 &      n/a \\
128 &   & 16.7 &     40.2 &     85.8 &    134.6 &    & 25.1 &     56.3 &    117.9 &      n/a \\

\midrule
\multicolumn{11}{c}{\cellcolor{mygrey}$D(\text{hy},\text{pn})$, $N_{\text{nn}} =16384$} \\
\midrule
4 &  &  136.3 &    145.9 &    236.8 &    194.8 &   & 489.7 &    442.8 &      n/a &      n/a \\
8 &   & 285.9 &    579.3 &  -4127.7 &    635.7 &  & -559.4 &   -553.1 &      n/a &      n/a \\
16 & & -3758.2 &   -310.2 &   -155.1 &   -201.3 &  & -162.7 &   -152.3 &   -127.8 &      n/a \\
32 &  & -210.1 &   -126.4 &    -83.0 &    -96.8 &  & -105.0 &   -107.9 &    -91.4 &      n/a \\
64 &  & -154.4 &    -97.3 &    -65.9 &    -71.3 &   & -89.1 &    -90.6 &    -74.7 &      n/a \\
128 &  & -117.8 &    -86.7 &    -63.3 &    -61.4 &   & -86.0 &    -85.8 &    -71.8 &      n/a \\

\midrule
\multicolumn{11}{c}{\cellcolor{mygrey}$D(\text{co},\text{hy})$, $N_{\text{nn}} =16384$} \\
 \midrule
4 & &      9.4 &     10.6 &      9.2 &      8.5 &    & 14.1 &     15.5 &     14.2 &     13.4 \\
8 &  &    4.2 &      5.8 &      7.0 &      7.7 &    &  6.7 &      9.8 &      9.5 &      9.9 \\
16 &  &    1.6 &      2.8 &      3.7 &      5.2 &    &  2.3 &      4.4 &      6.3 &      6.5 \\
32 &   &   0.8 &      1.2 &      1.7 &      1.9 &    &  1.2 &      2.0 &      3.6 &      3.7 \\
64 &    &  0.4 &      0.6 &      0.9 &      0.8 &    &  0.6 &      0.9 &      1.4 &      1.6 \\
128 &  &    0.3 &      0.3 &      0.4 &      0.4 &    &  0.3 &      0.3 &      0.3 &      0.4 \\
\midrule
 \multicolumn{11}{c}{\cellcolor{mygrey}$D(\text{co},\text{hy})$, $N_{\text{nn}} =65536$} \\
 \midrule
 4 & &      9.4 &      n/a &      n/a &      n/a &    &  n/a &      n/a &      n/a &      n/a \\
 8 &  &    4.0 &      6.4 &      n/a &      n/a &     & 7.7 &      n/a &      n/a &      n/a \\
16 &   &   1.6 &      2.8 &      3.9 &      4.7 &     & 2.5 &      4.3 &      n/a &      n/a \\
32 &  &      0.7 &      1.2 &      1.8 &      2.5 &    &  0.9 &      2.0 &      2.9 &      3.9 \\
64 &   &   0.3 &      0.6 &      0.8 &      1.0 &     & 0.5 &      1.0 &      1.4 &      1.7 \\
128 &   &   0.1 &      0.2 &      0.4 &      0.4 &    &  0.2 &      0.3 &      0.4 &      0.4 \\
256 &    &  0.0 &      0.1 &      0.1 &      0.1 &    &  0.0 &      0.1 &      0.1 &      0.1 \\
512 &  &    0.0 &      0.0 &      0.0 &      0.0 &    &  0.0 &      0.0 &      0.0 &      0.1 \\

\bottomrule
\end{tabular}
\caption[Measured Minimum Density Functions of the studied algorithms]{Table giving the measured Minimum Density Functions $D(\text{co},\text{pn})$, $D(\text{hy},\text{pn})$ and $D(\text{co},\text{hy})$ for various values of$N_{\text{nn}}$ and Precision. Values given as "n/a" imply the Hybrid Gridding or the Pure NN Interpolation implementations could not be executed due to memory limitations on the P100. The GCF for the Gridding Step is with S=6, and the GCF of the Pruning Step is with $S_{\text{z}}=14$.}
\label{tab:comparative:mdf}
\end{table}
Table \ref{tab:comparative:mdf} tabulates the Minimum Record Density Function for various test scenarios on which we base the comparative analyses made in the next paragraphs. In calculating the Minimum Density Function, we extracted  the \maxbestgridrate from the relevant \bestgridrate results given in Figures \ref{fig:2dgridding:convbrute_single}, \ref{fig:2dgridding:convbrute_double}, \ref{fig:hybrid:brute-single}, \ref{fig:hybrid:brute-double} and \ref{fig:purenn:brute_nn}. We measured the time taken in finalisation for a given implementation via new experiments that we did for this purpose.

It is clear from Table \ref{tab:comparative:mdf}, that Pruned NN Interpolation requires very High Record Densities to beat Convolutional Gridding, because of the large output grid, that requires a lot of execution time to be finalised. When comparing Pruned NN Interpolation with Hybrid Gridding, $D(\text{hy},\text{pn})$ is either below 0 or larger than 100\%, implying that within our test scenarios with full compression enabled, Pruned NN Interpolation is always slower than Hybrid Gridding.

$D(\text{co},\text{hy})$ shows that Hybrid Gridding can  be faster than Convolutional Gridder provided a certain threshold in the number of input records is reached. We know from results in Chapter \ref{chap:hybrid} that best gains are for $\beta>64$, where for such values of $\beta$, $D(\text{co},\text{hy})$ is below 0.5\%. We remind the reader that the Hybrid Gridder requires an output grid which is larger than that of the Convolutional Gridder by a factor of $\beta$ and the Hybrid Gridder delivered good Performance when using the SS Ordering Mode. At the same time, the Convolutional Gridder can deliver good Performance in all Ordering Modes.

\subsubsection{Conclusion}

In conclusion, based on our test scenarios, we state that when comparing the studied implementations from a Performance perspective and as implemented on the P100, the following results:

\begin{enumerate}
\item Convolutional Gridding is fastest for Small Record Densities or when the SS Ordering Mode is not viable. One has to remember that Convolutional Gridding is the most memory-efficient from all the studied implementations. Therefore, in case of memory limitations, it might be the only viable implementation to use among the three studied implementations.

\item Hybrid Gridding is fastest when there is a high Record Density, and the SS Ordering Mode is viable.
\item Pruned NN Interpolation is not viable for any tested scenario, since Hybrid Gridding always provides less memory consumption and more Performance than Pruned NN Interpolation.
\end{enumerate}
The first and second results agree with the first and second claims we made at the end of the \textit{Time Complexity} sub-section in Section \ref{sec:maths:timecomplexitygrid}. Nevertheless, the third result does not entirely tally with the third claim made in said sub-section.  Sometimes the NN Gridder was less performant than the other Gridders, pushing  $D(\text{co},\text{pn})$ or $D(\text{hy},\text{pn})$ to be below zero. Alternatively, the records required for Pruned NN Interpolation to be more performant than the other implementations were higher than the size of the NN Grid, therefore pushing $D(\text{hy},\text{pn})$ or $D(\text{co},\text{pn})$ over 100\%.

\subsection{Aliasing comparison}
\label{sec:comparative:aliasing}

Let us now discuss experiments designed to ensure that Hybrid Gridding and Pruned NN Interpolation suppress aliasing at a level equal or higher than that of Convolutional Gridding. In this way, we prove the claims made in the \textit{Aliasing} sub-section of Section \ref{sec:maths:comparisons} and ensure that Hybrid Gridding is valid from an aliasing perspective and meets the aliasing requirement set by the primary goal of this thesis.

\subsubsection{Experimental setup and definition of $R(X)$}

We equipped the \textit{mt-imager} with an internal simulator capable of simulating a sky over the input $uv$-profile having one point source of unity intensity, positioned precisely on a pixel in the output image. The simulator computes using Double-Precision and  applies a DFT. In our experiments we shall simulate a point source at position $(X,X)$, where position $(0,0)$ is the origin on the central pixel of the image and $X\in \mathbb{Z}$, $0\le X<64$. We define $K(X)$ to be the intensity value of the pixel at position $(X,X)$ of the simulated output image generated by the studied implementations. We also define $R(X)$ to be equal to the absolute value of $[1 \text{--} K(X)]$. $R(X)$ measures aliasing effects, and arithmetic noise at position $(X,X)$ in such a way that if the two quantities are zero on the stated pixel, $R(X)$ will have a value of zero since the point source is simulated with unity intensity.

We used the same observations from GMRT, VLA and LOFAR used in Section  \ref{sec:comparative:convgridderaliasing}, and generate 64 images of size $128\times128$ for each observation and studied implementation in Single and Double-Precision. In each set of 64 images the point source of unit intensity moves diagonally for all possible positions $(X,X)$, beginning from $(0,0)$ and ending at $(63,63)$. 
Pixel interval is set as per Table \ref{tab:comparative:pixelinterval}, while the oversampling factor ($\beta$) is set to $128$. The GCF used in the Gridders is the Prolate Spheroidal of order 1, and the Pruners use the least-misfit function with $S_{\text{z}}=14$ and the parameter $x_0=0.25$. We first evaluate the effects of aliasing through a visual inspection of the generated images and then measure aliasing distortion and arithmetic noise through $R(X)$.

\subsubsection{Results and analyses}
\begin{figure}[h]

\includegraphics[page=1]{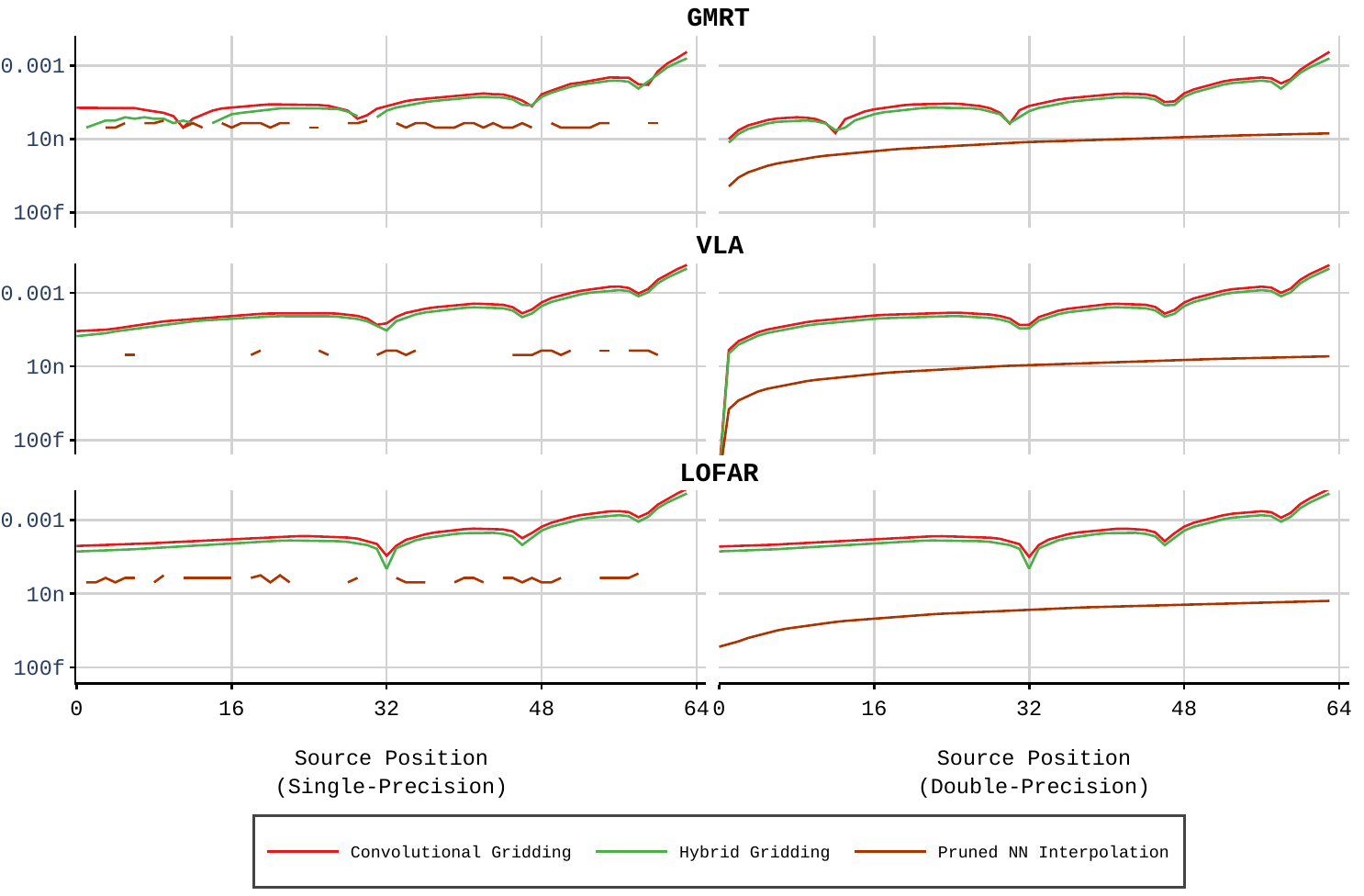}

\caption[Aliasing and arithmetic noise results for experiments made on the studied algorithms]{Various graphs plotting $R(X)$ against the point source position X for the aliasing experiments discussed in Section \ref{sec:comparative:aliasing}.}
\label{fig:comparative:aliascompare}
\end{figure}

\begin{figure}
\centering
\begin{tabular}{@{}c@{}c@{}c@{}}

\multicolumn{3}{c}{\small{\texttt{\bfseries X=0}}} \\ 
\includegraphics[height=135pt]{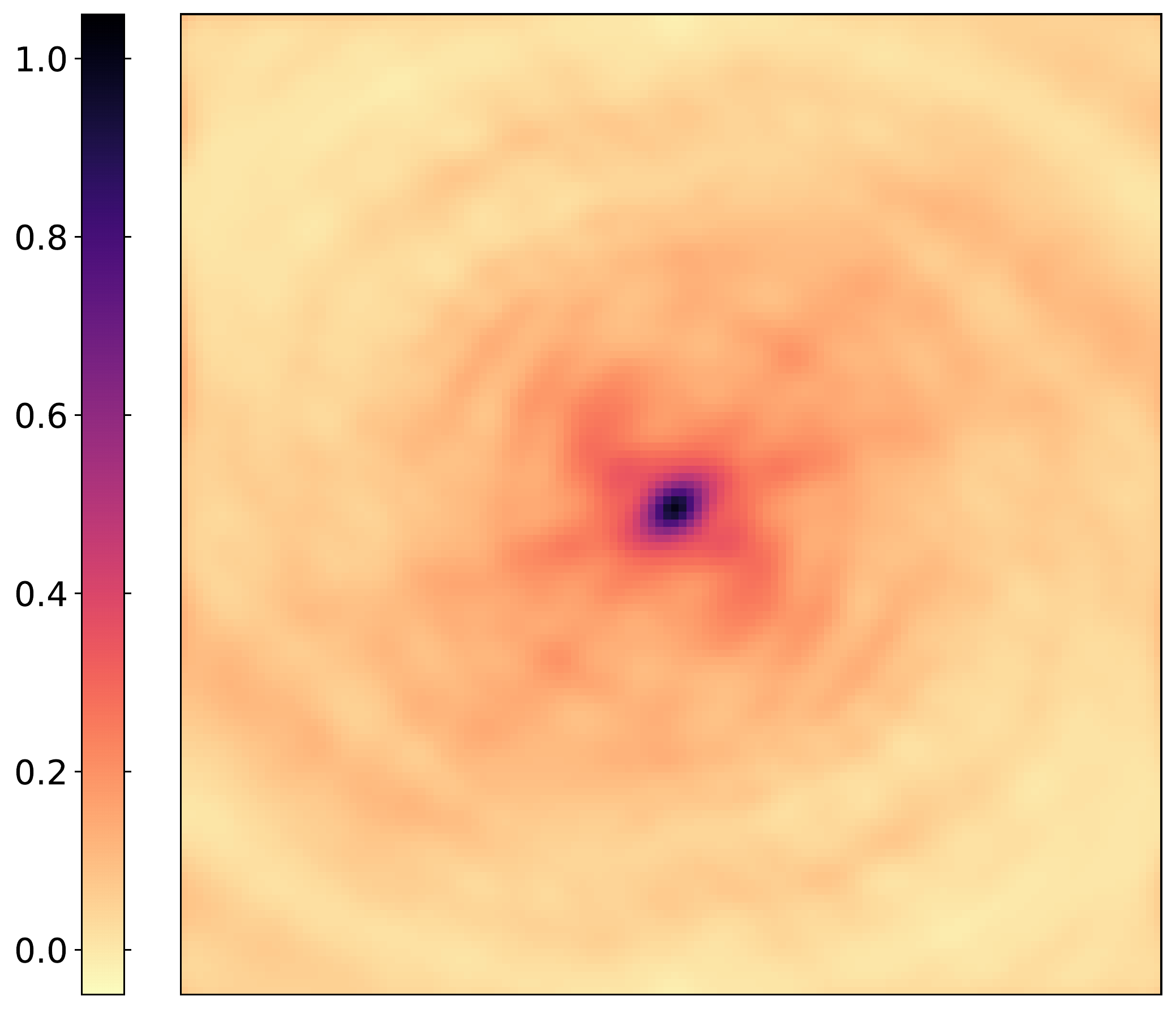} &
\includegraphics[height=135pt]{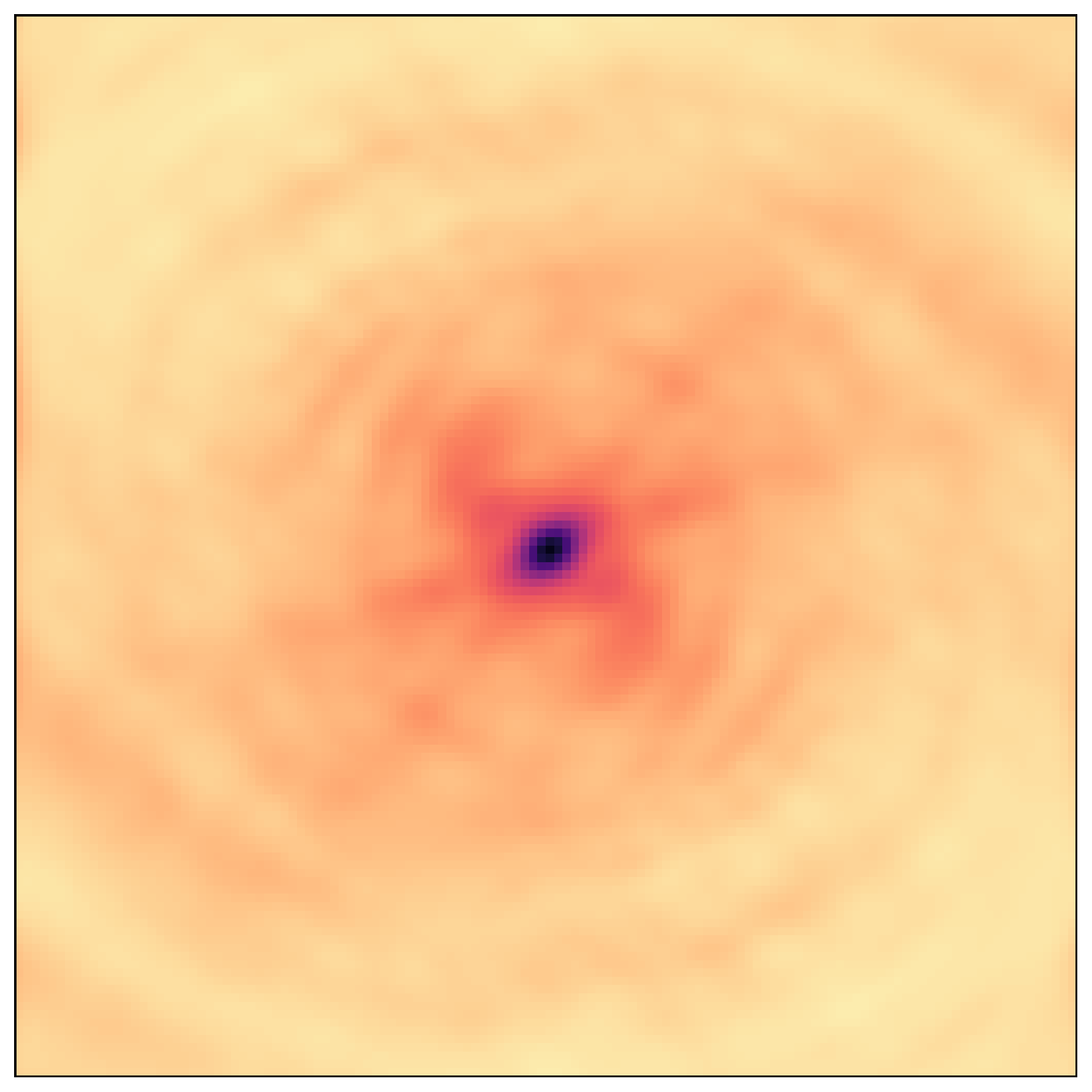} &
\includegraphics[height=135pt]{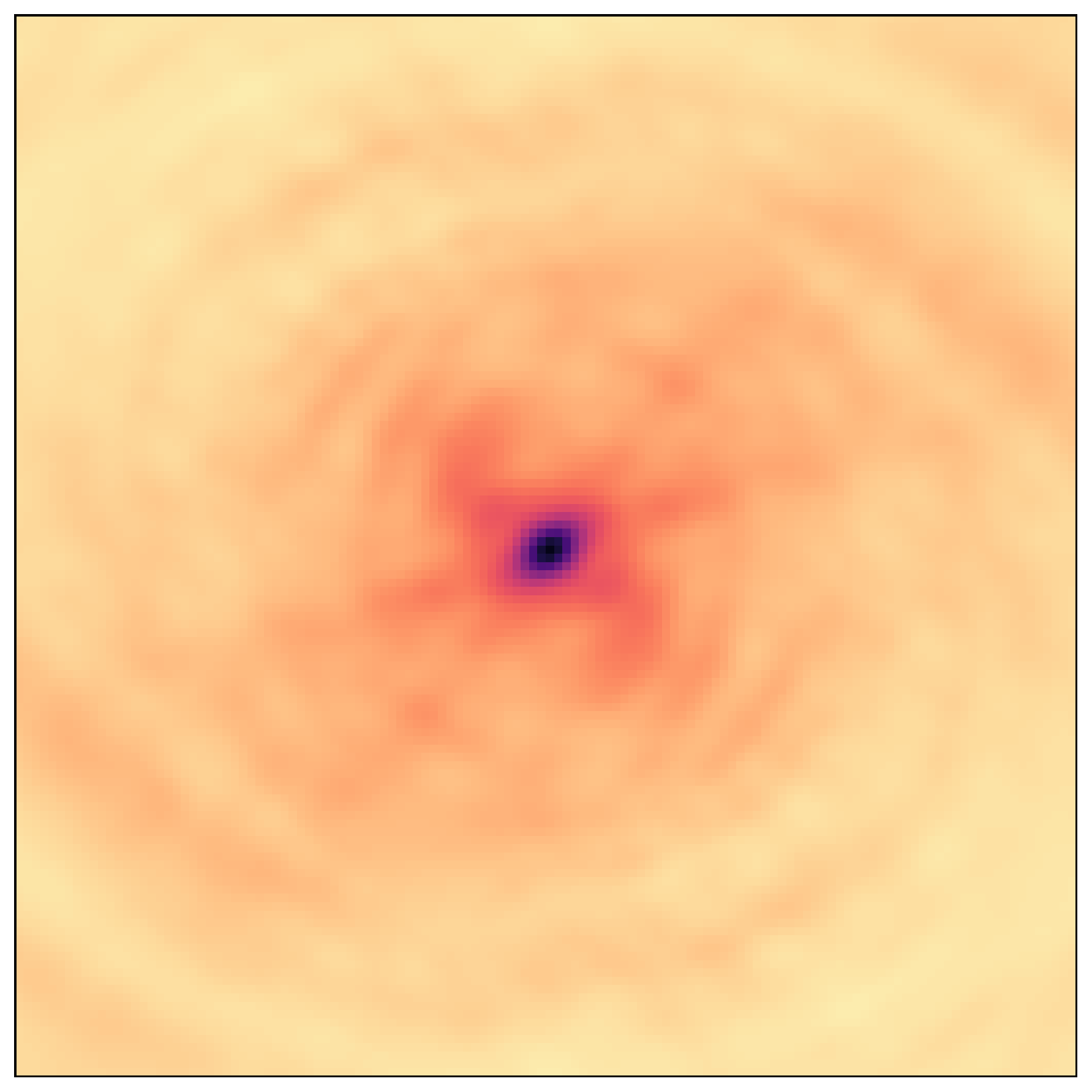} \\

\multicolumn{3}{c}{\small{\texttt{\bfseries X=30}}} \\
\includegraphics[height=135pt]{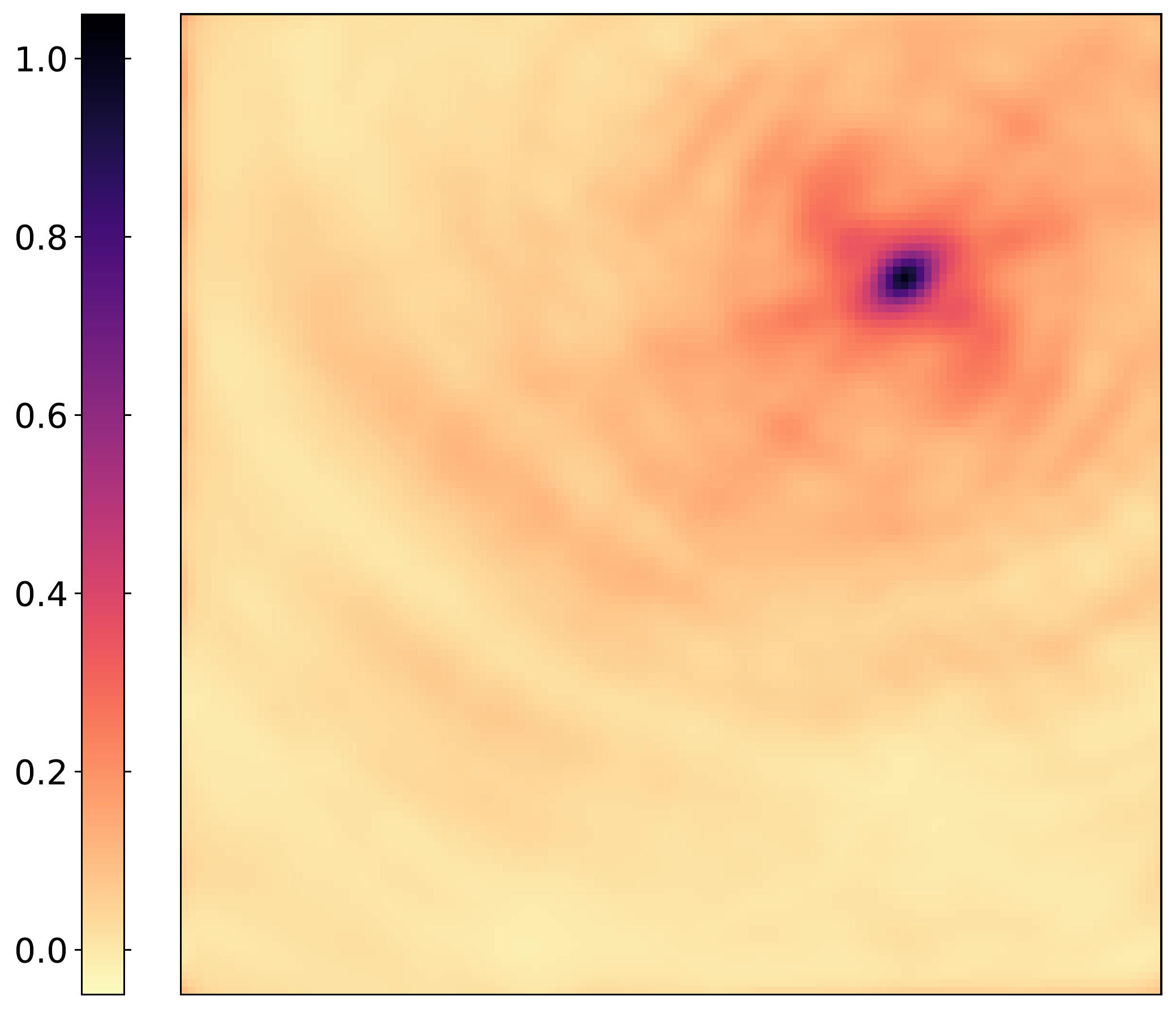} &
\includegraphics[height=135pt]{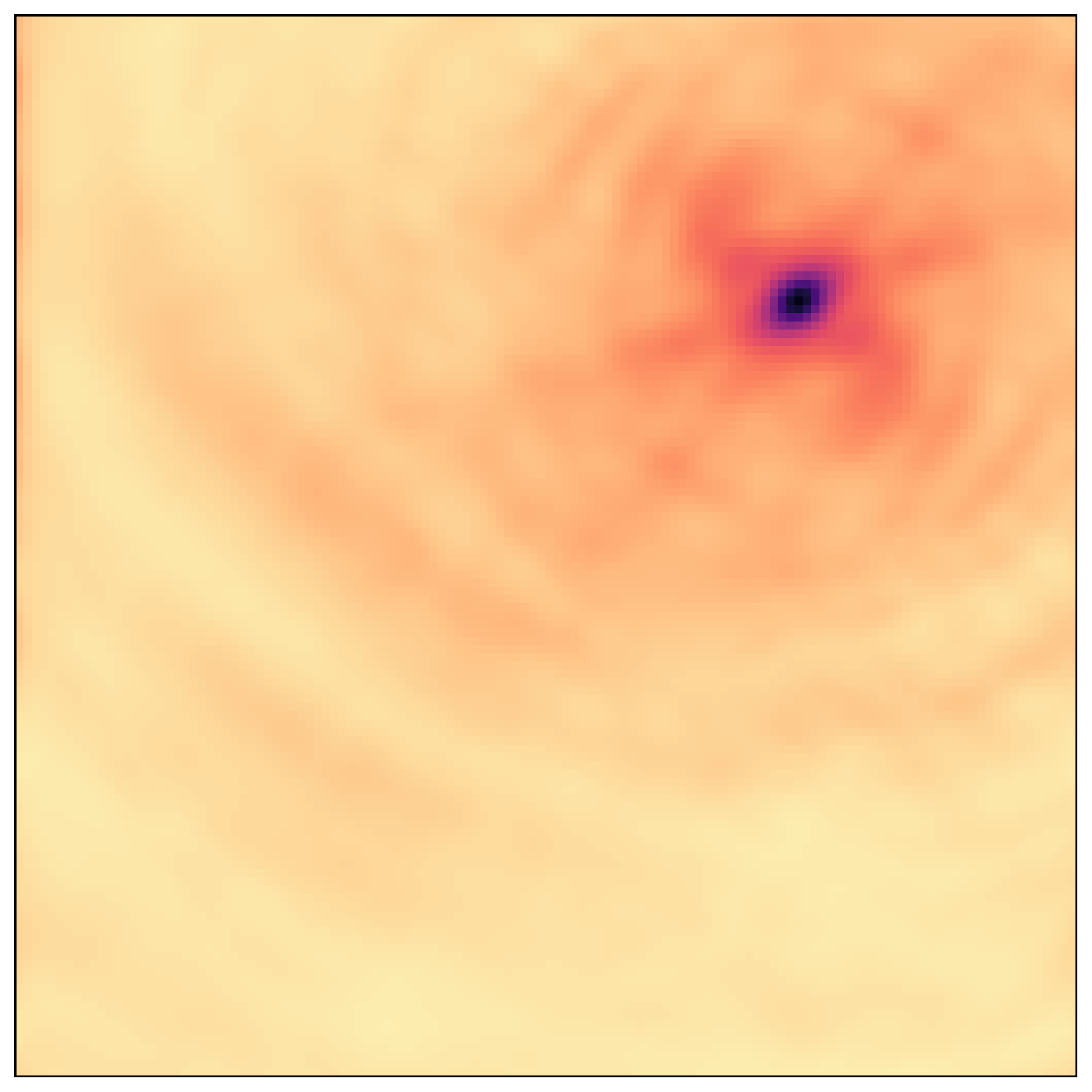} &
\includegraphics[height=135pt]{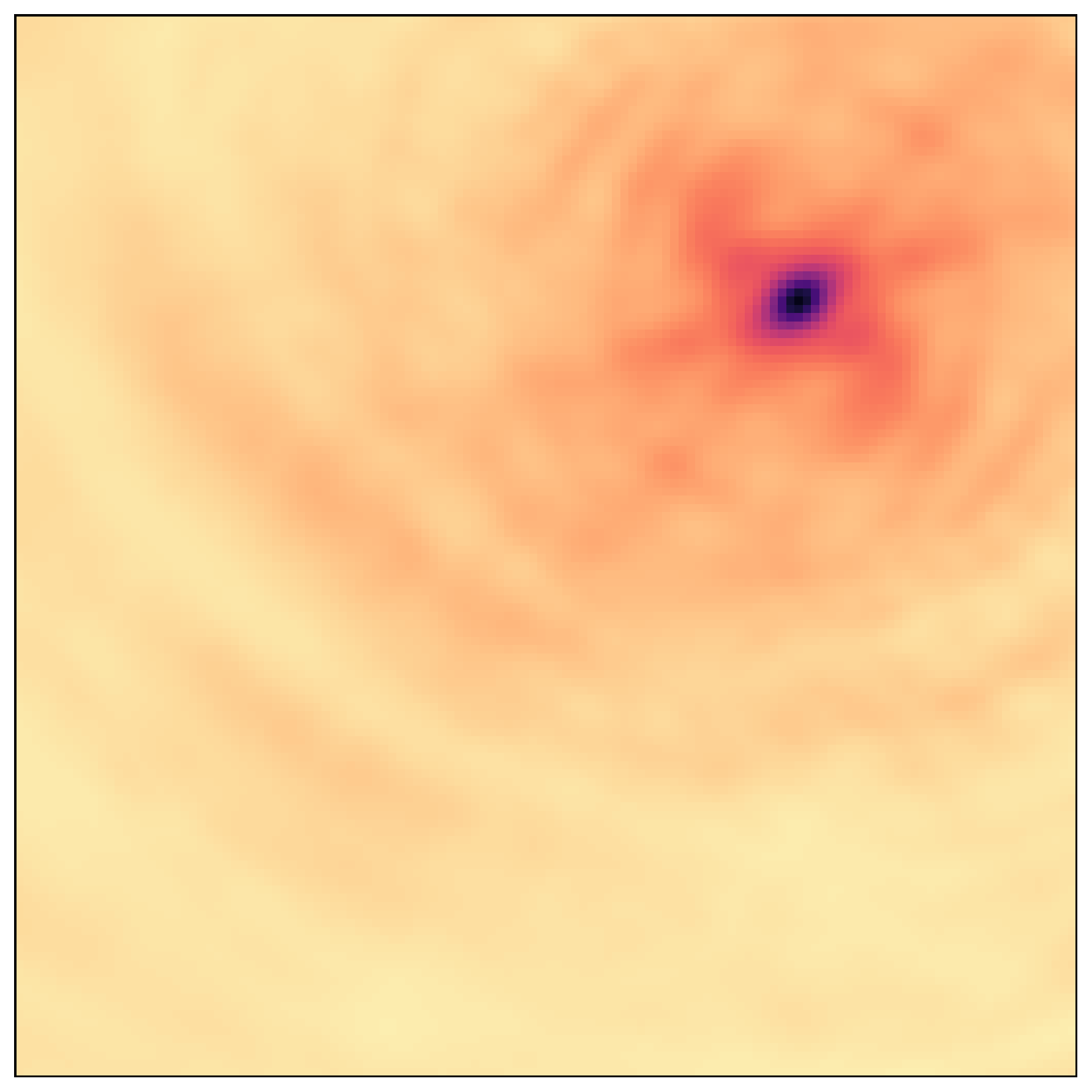} \\

\multicolumn{3}{c}{\small{\texttt{\bfseries X=62}}} \\ 
\includegraphics[height=135pt]{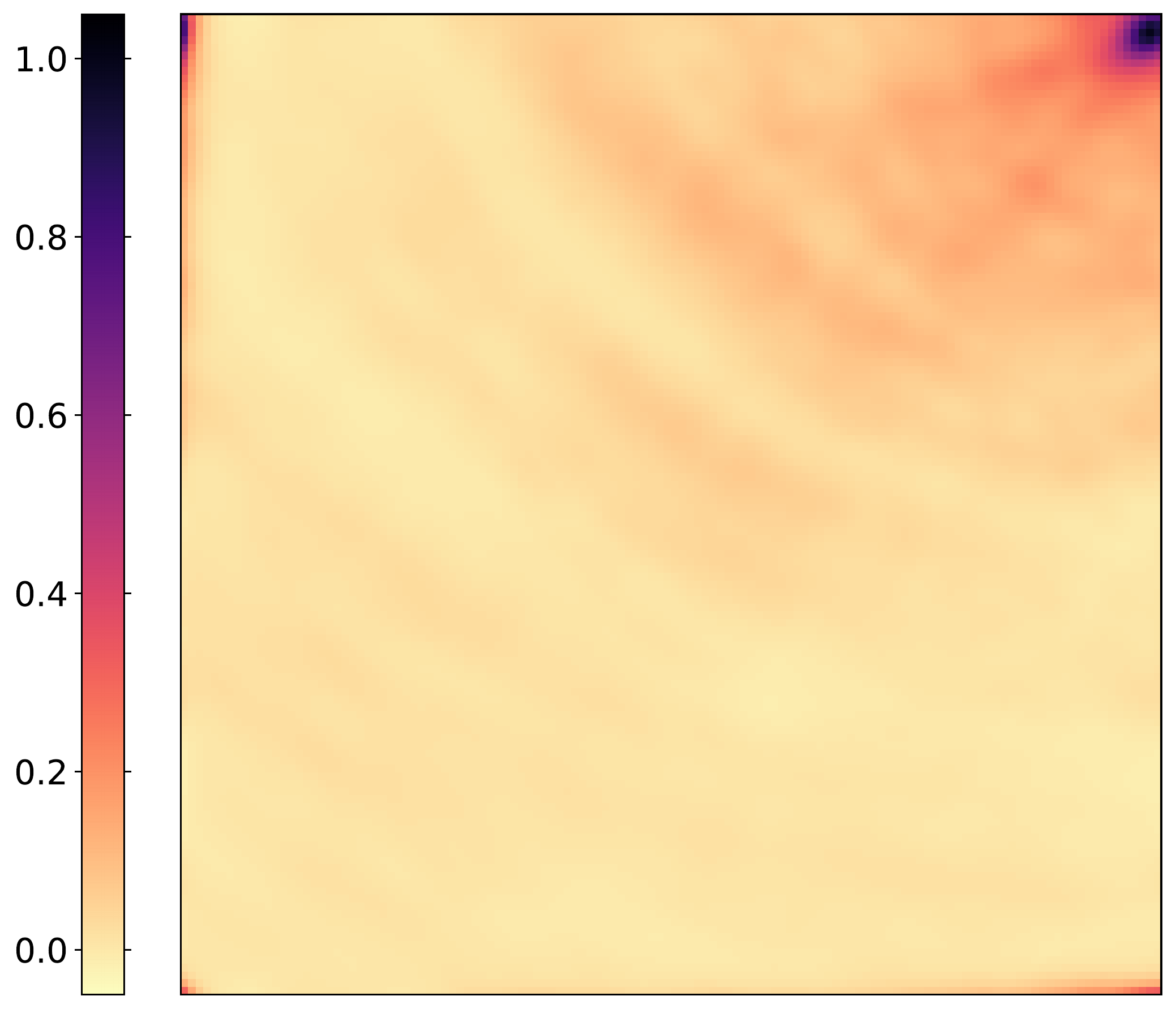} &
\includegraphics[height=135pt]{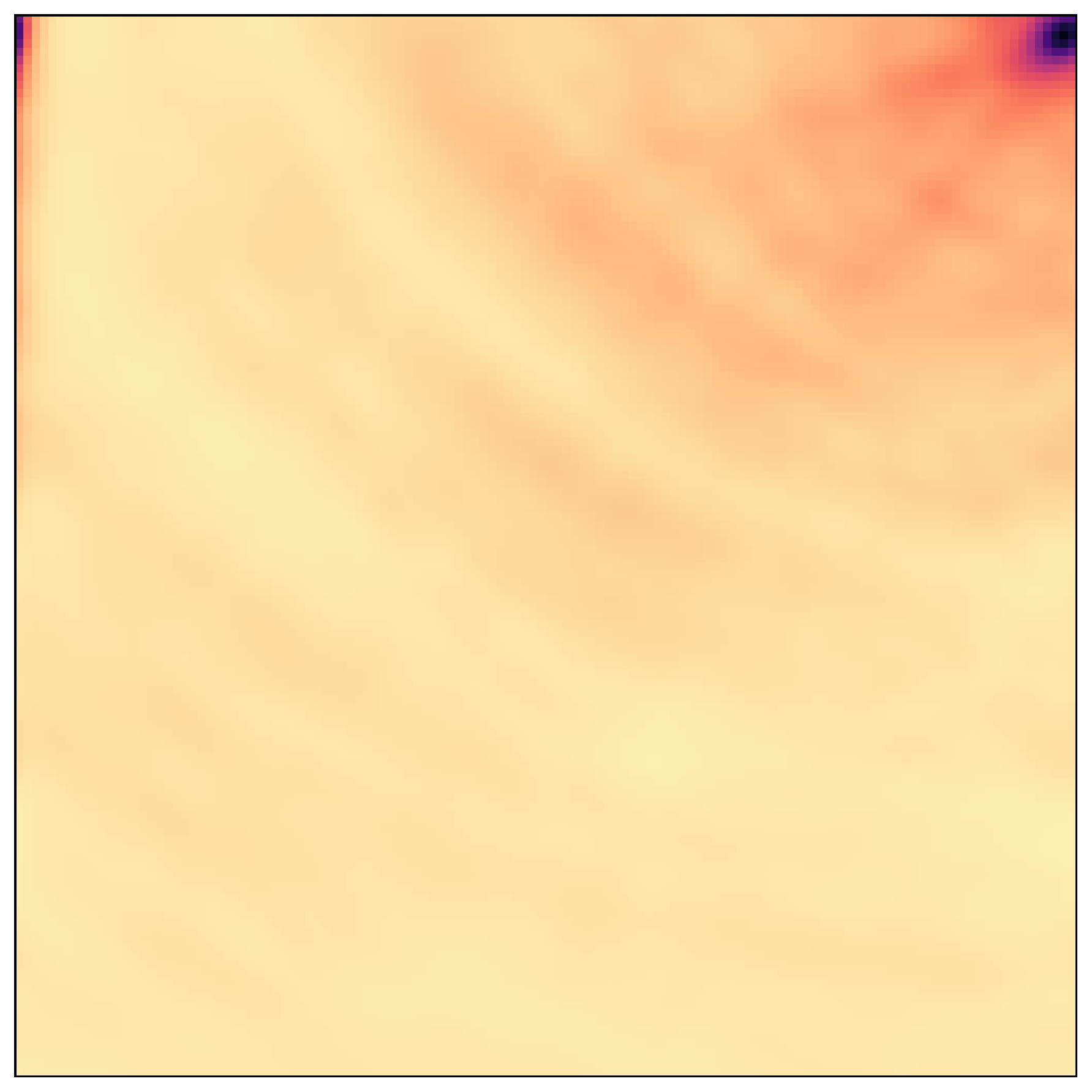} &
\includegraphics[height=135pt]{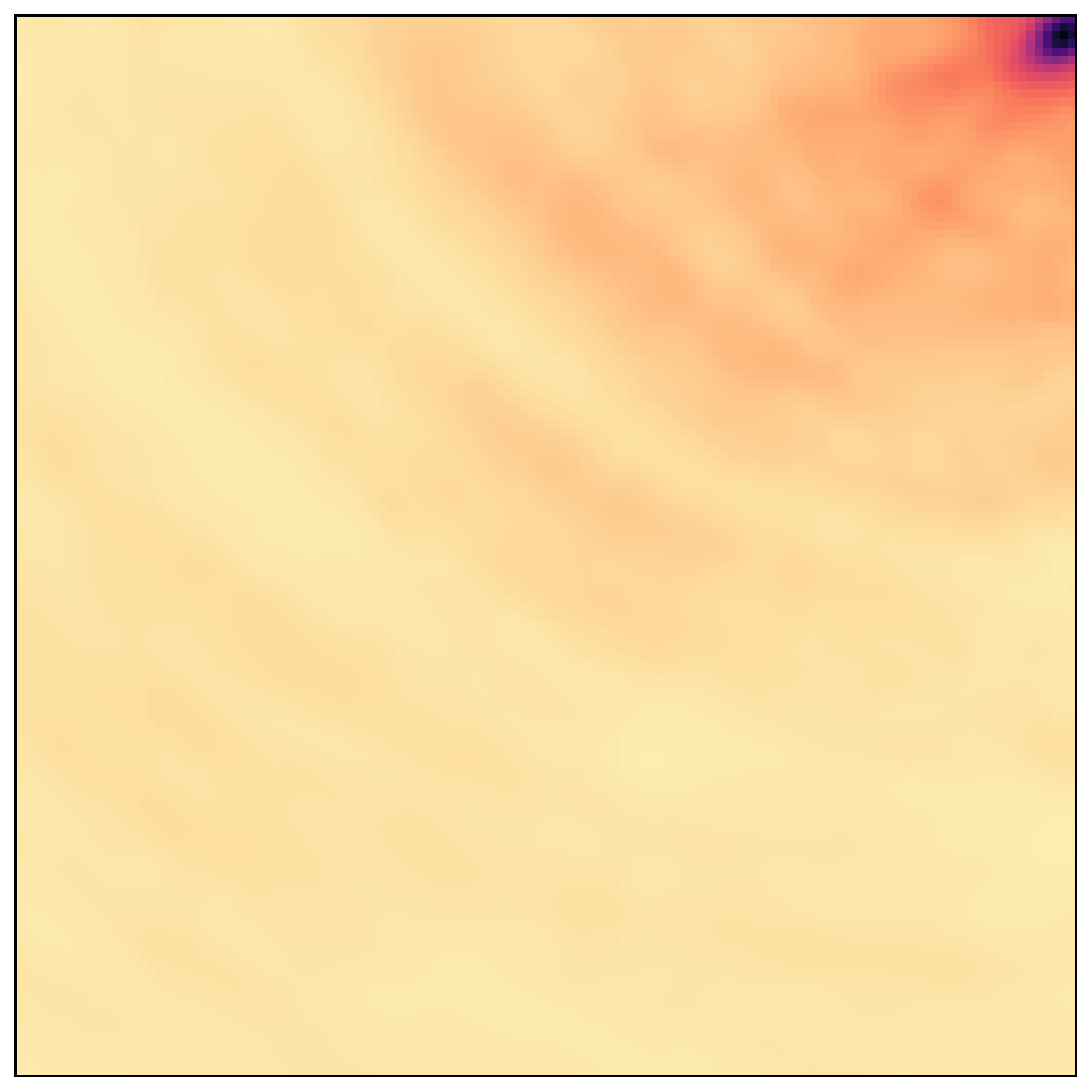} \\
\hspace{0.8cm}\footnotesize\texttt{Convolutional Gridding} & \footnotesize\texttt{Hybrid Gridding} & \footnotesize\texttt{Pruned NN Interpolation} \\
\\
\end{tabular}
\caption[Images generated by experiments of Section \ref{sec:comparative:aliasing}]{A selection of images computed in Single-Precision by the experiments of Section \ref{sec:comparative:aliasing} using the GMRT observation.} 
\label{fig:comparative:aliasinggmrtsingle}

\end{figure}
\begin{figure}
\centering
\begin{tabular}{@{}c@{}c@{}c@{}}

\multicolumn{3}{c}{\small{\texttt{\bfseries X=0}}} \\
\includegraphics[height=135pt]{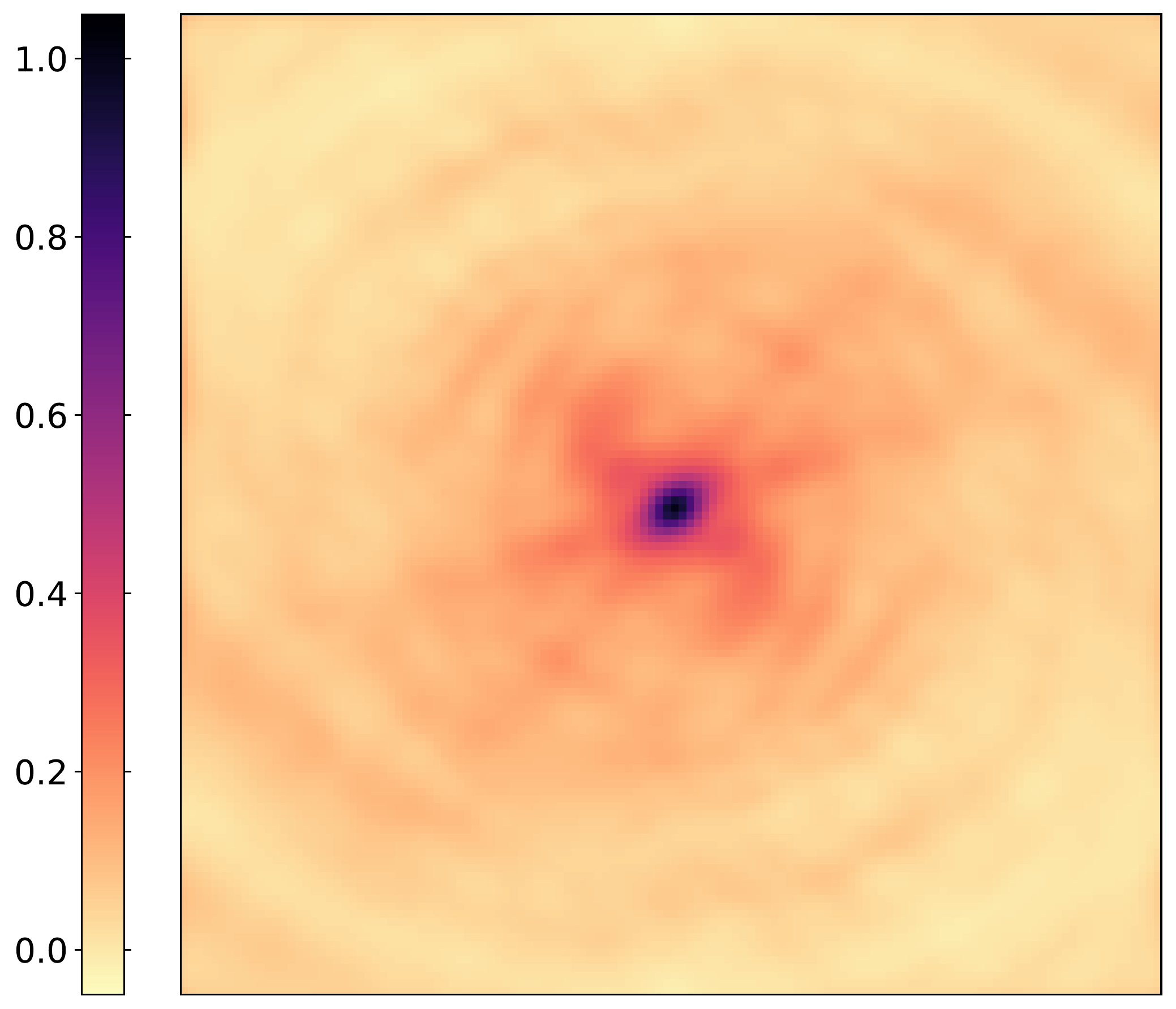} &
\includegraphics[height=135pt]{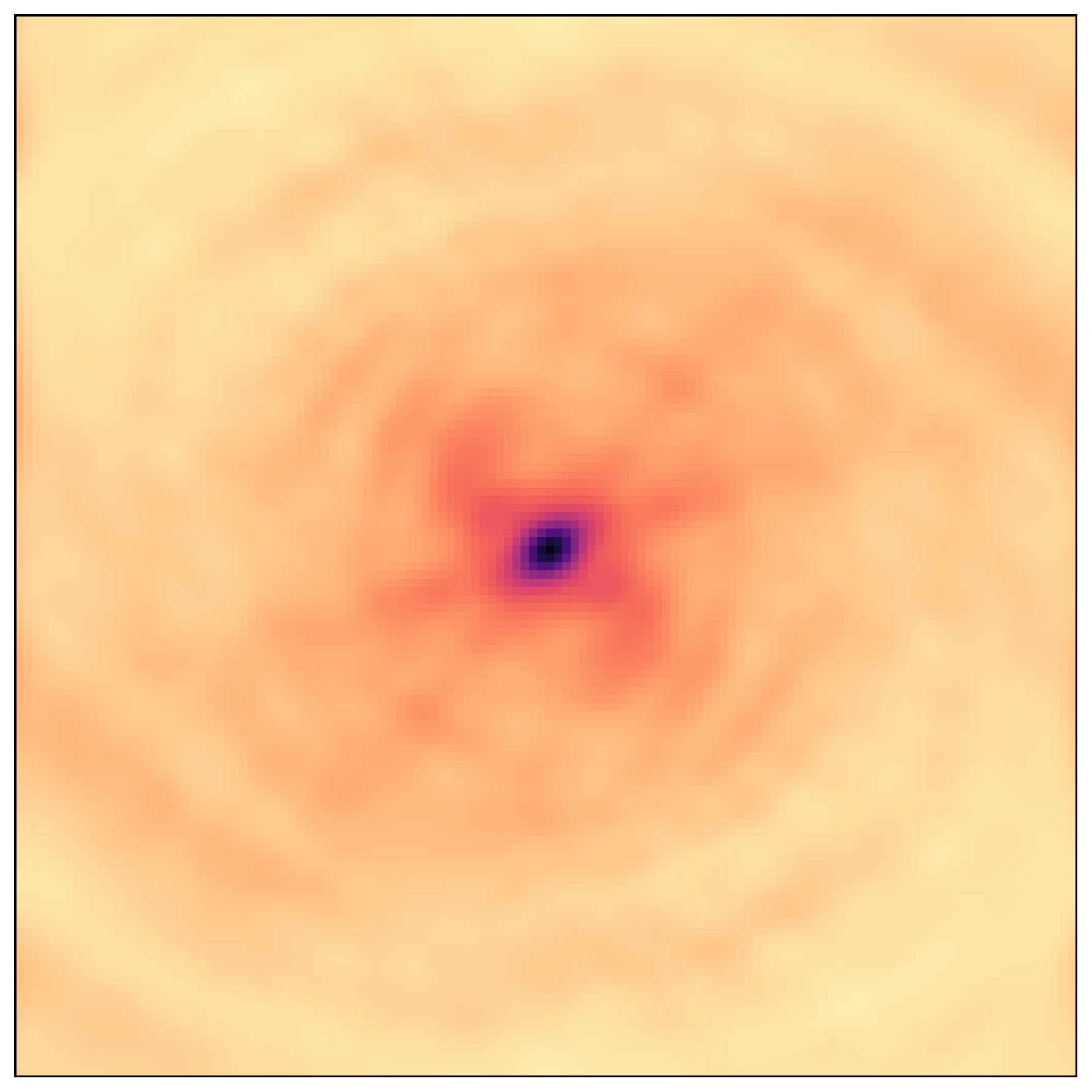} &
\includegraphics[height=135pt]{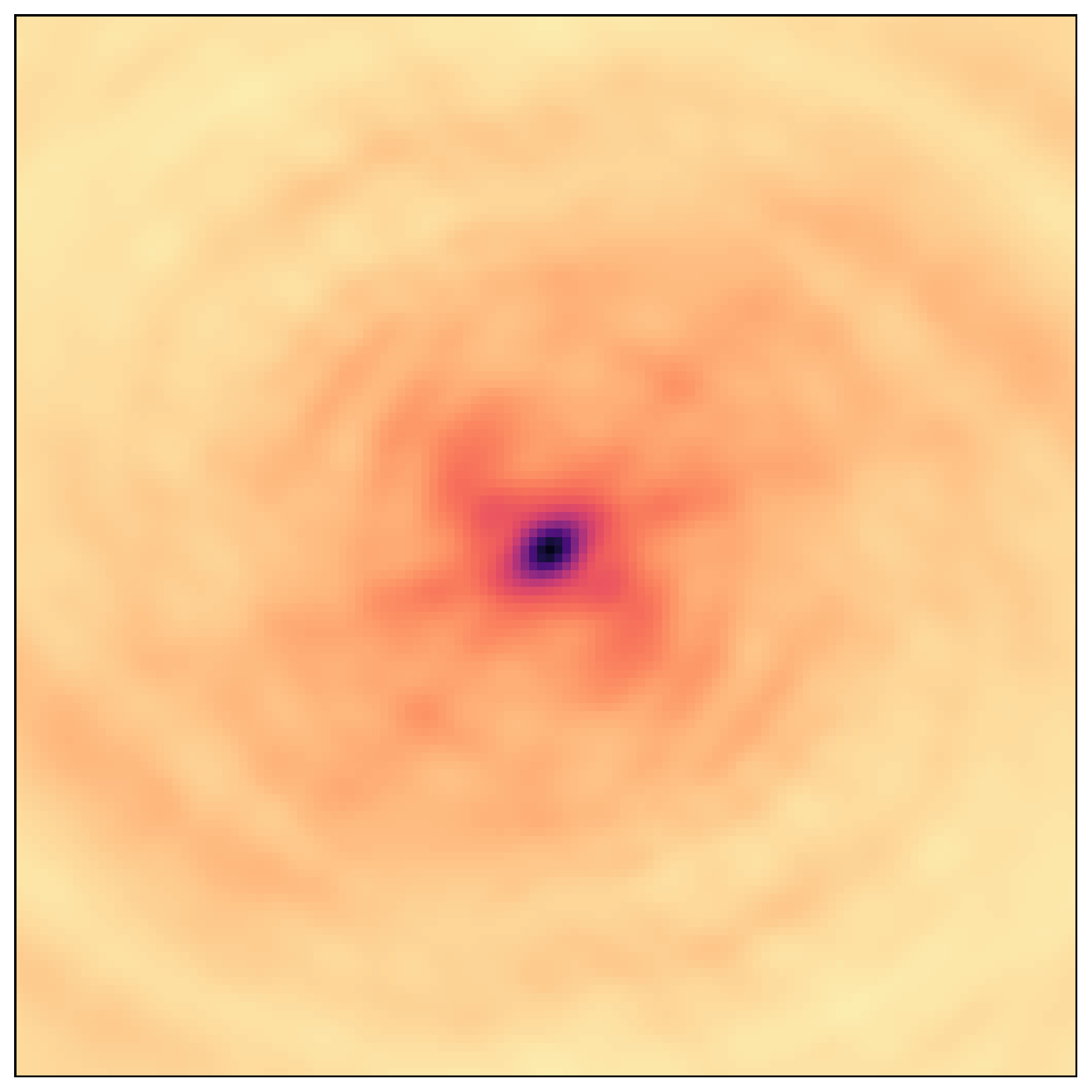} \\

\multicolumn{3}{c}{\small{\texttt{\bfseries X=30}}} \\
\includegraphics[height=135pt]{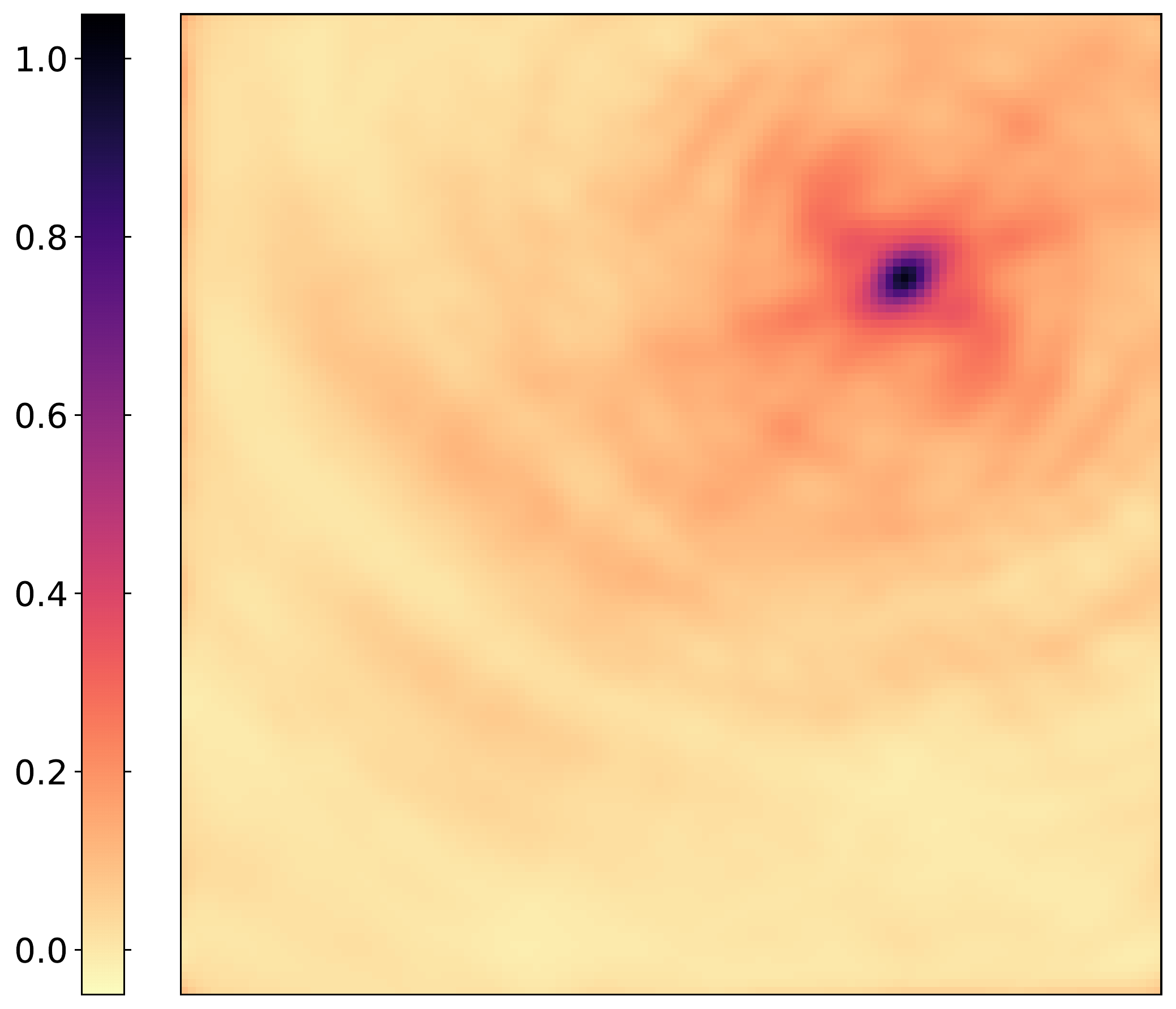} &
\includegraphics[height=135pt]{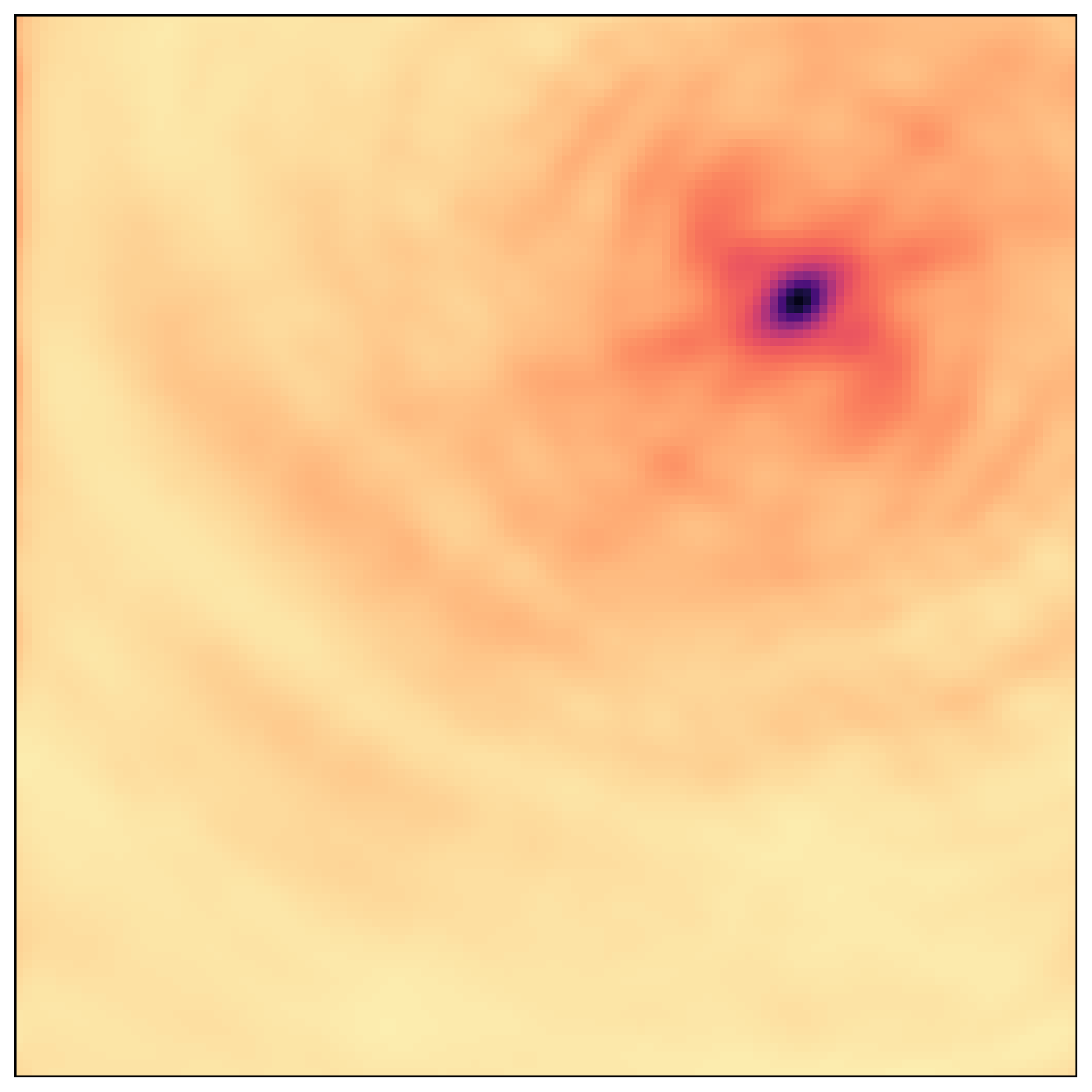} &
\includegraphics[height=135pt]{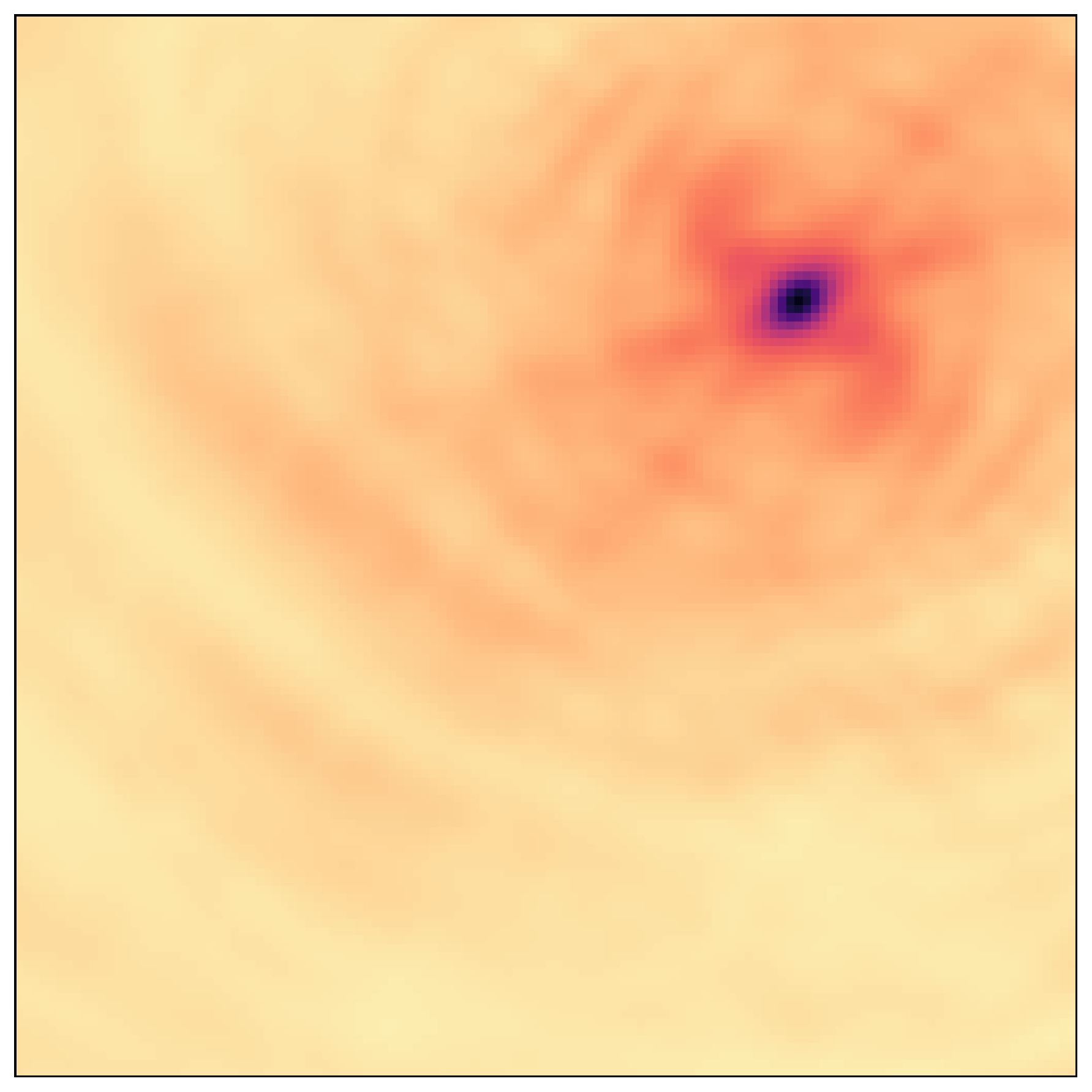} \\

\multicolumn{3}{c}{\small{\texttt{\bfseries X=62}}} \\
\includegraphics[height=135pt]{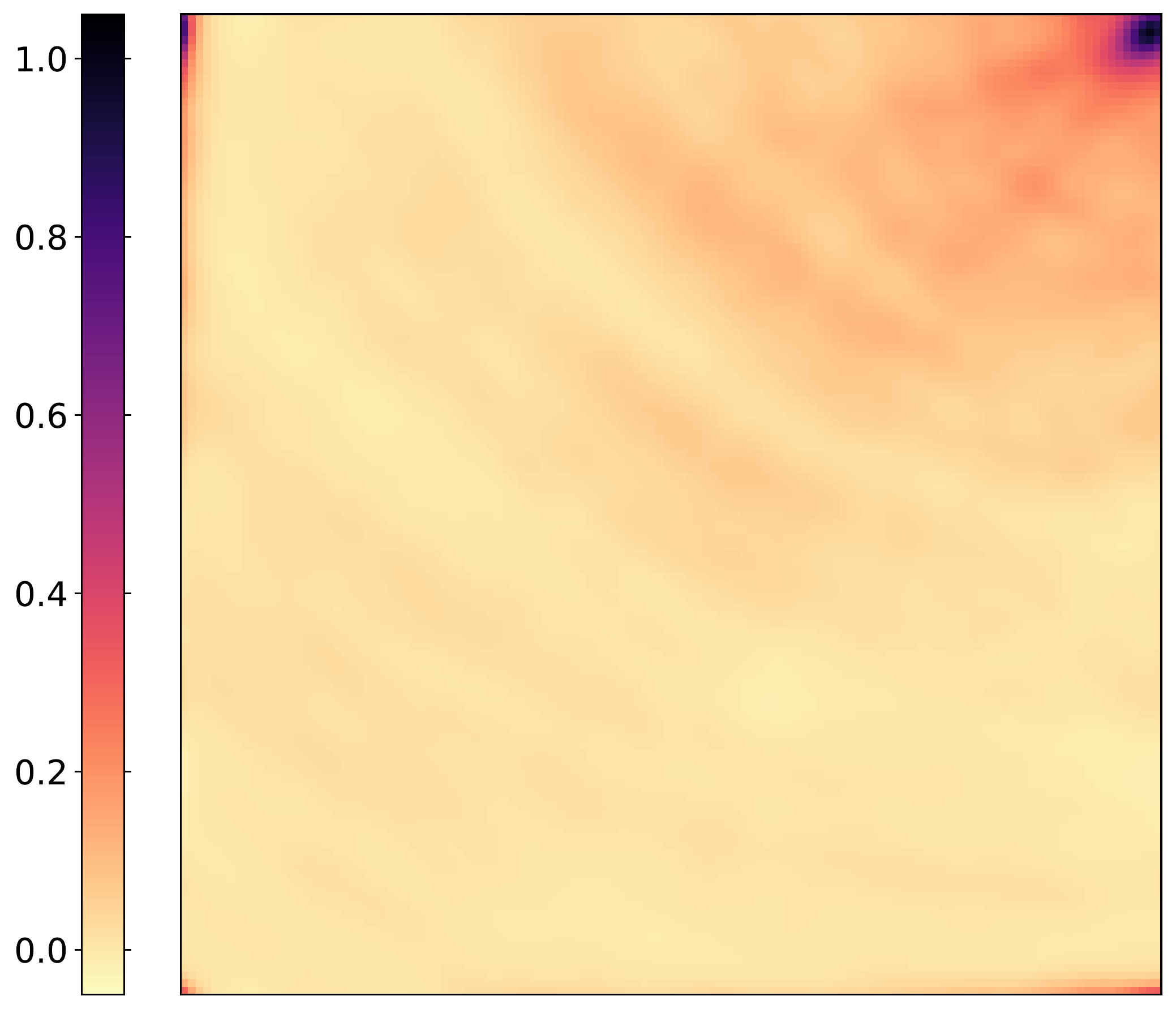} &
\includegraphics[height=135pt]{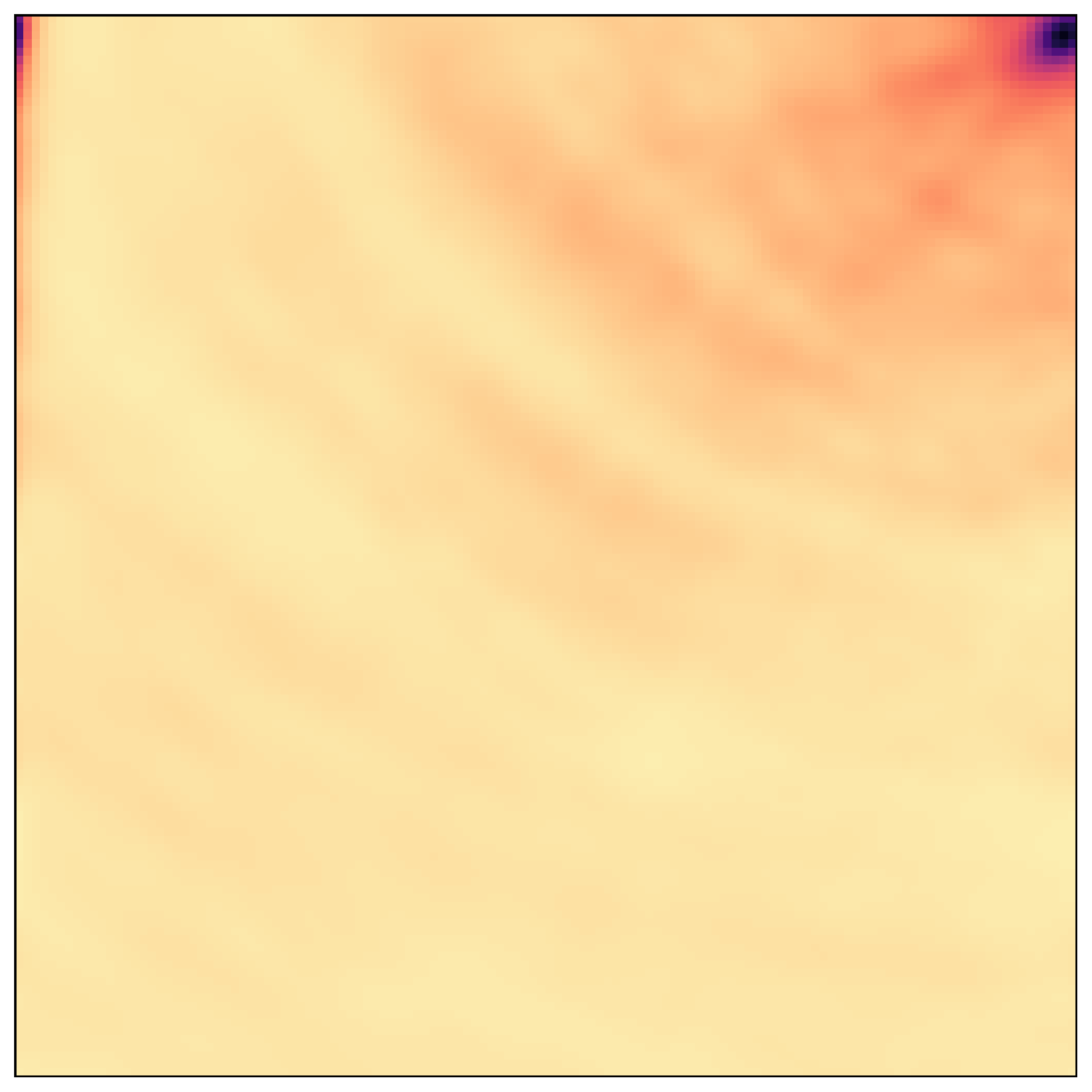} &
\includegraphics[height=135pt]{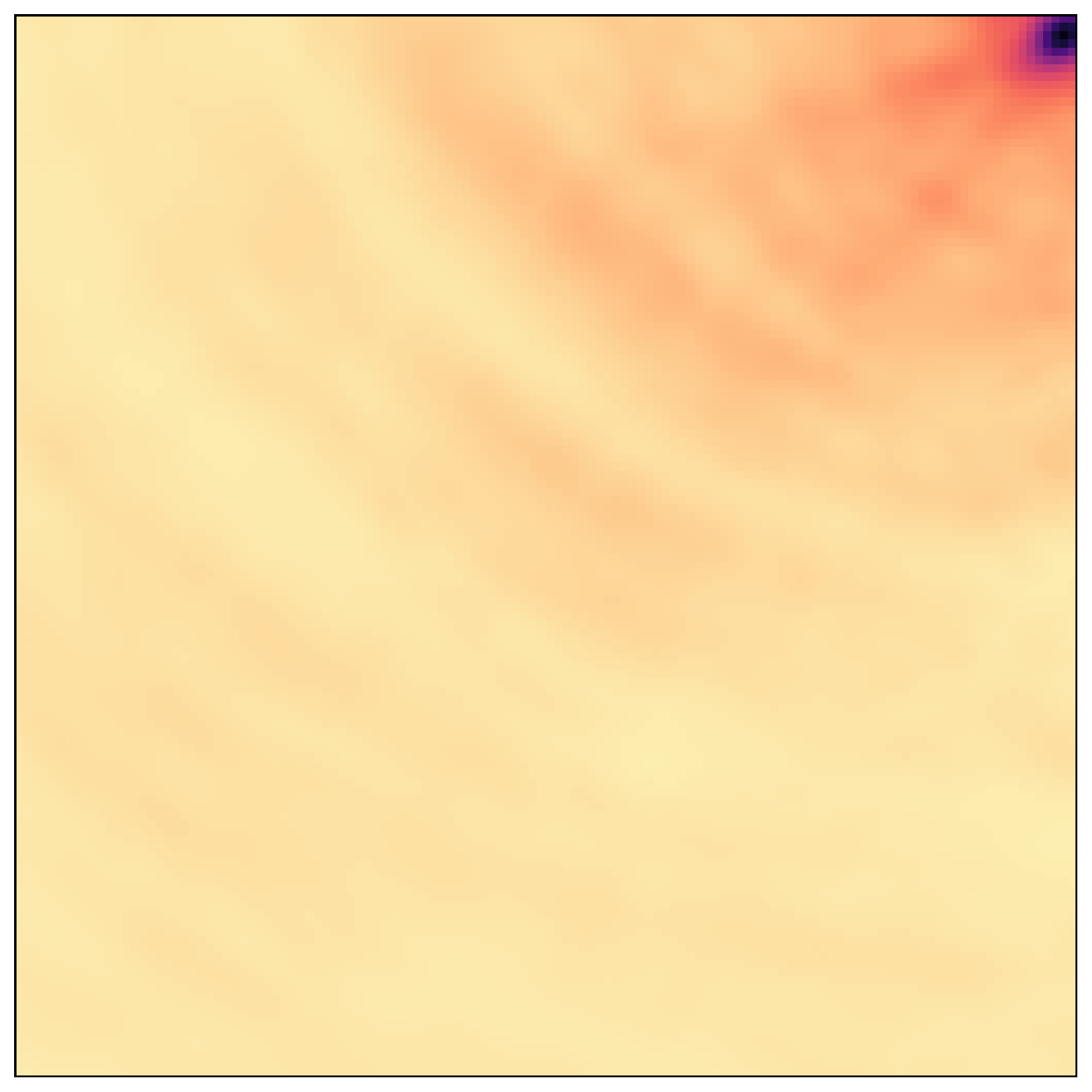} \\
\hspace{0.8cm}\footnotesize\texttt{Convolutional Gridding} & \footnotesize\texttt{Hybrid Gridding} & \footnotesize\texttt{Pruned NN Interpolation}\\
\\
\end{tabular}
\caption[Images generated by experiments in Section \ref{sec:comparative:aliasing}]{A selection of images computed in Double-Precision by the experiments in Section \ref{sec:comparative:aliasing} using the GMRT observation.} 
\label{fig:comparative:aliasinggmrtdouble}
\end{figure}
\begin{figure}
\centering
\begin{tabular}{@{}c@{}c@{}c@{}}

\multicolumn{3}{c}{\small{\texttt{\bfseries X=0}}} \\
\includegraphics[height=135pt]{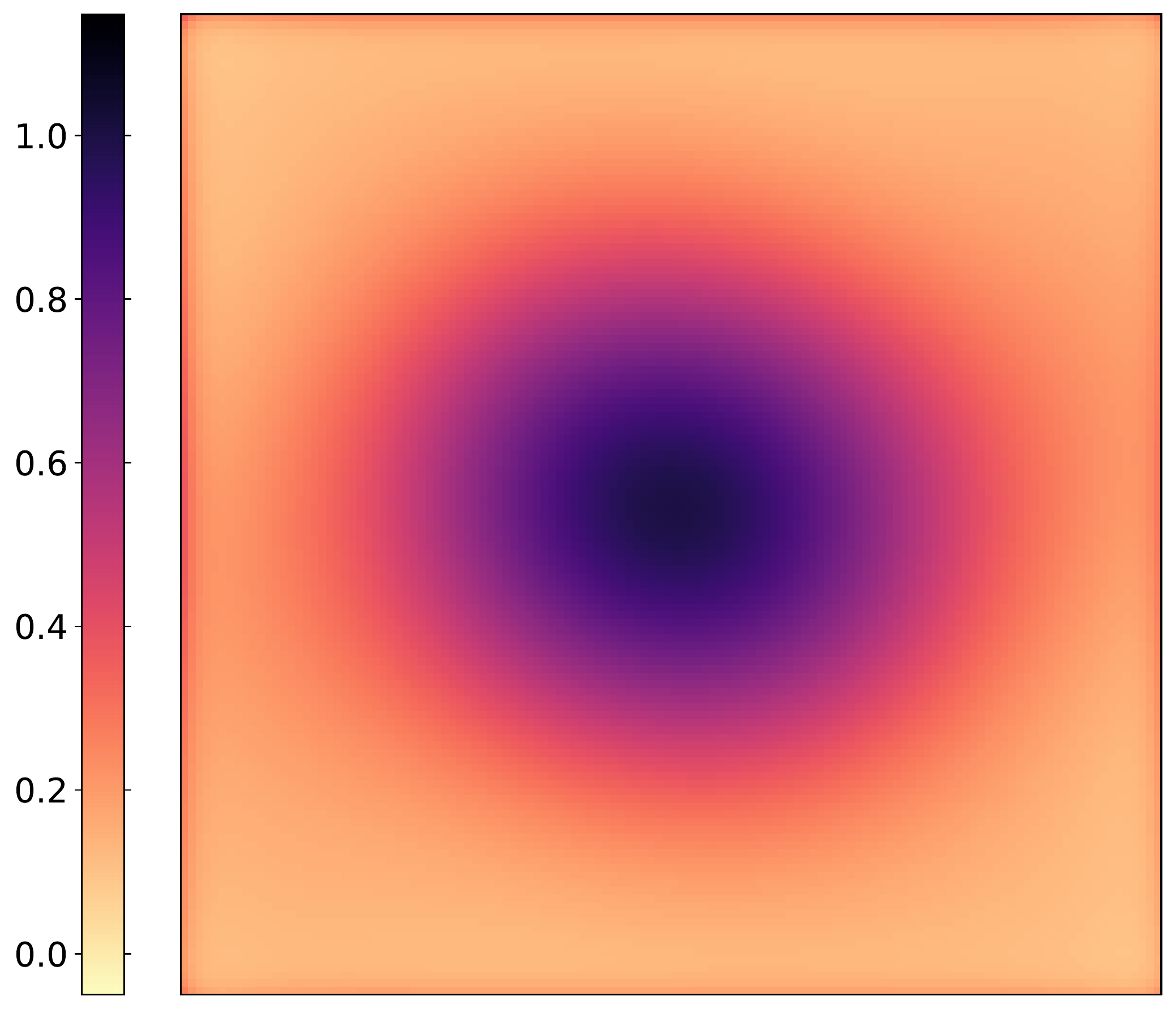} &
\includegraphics[height=135pt]{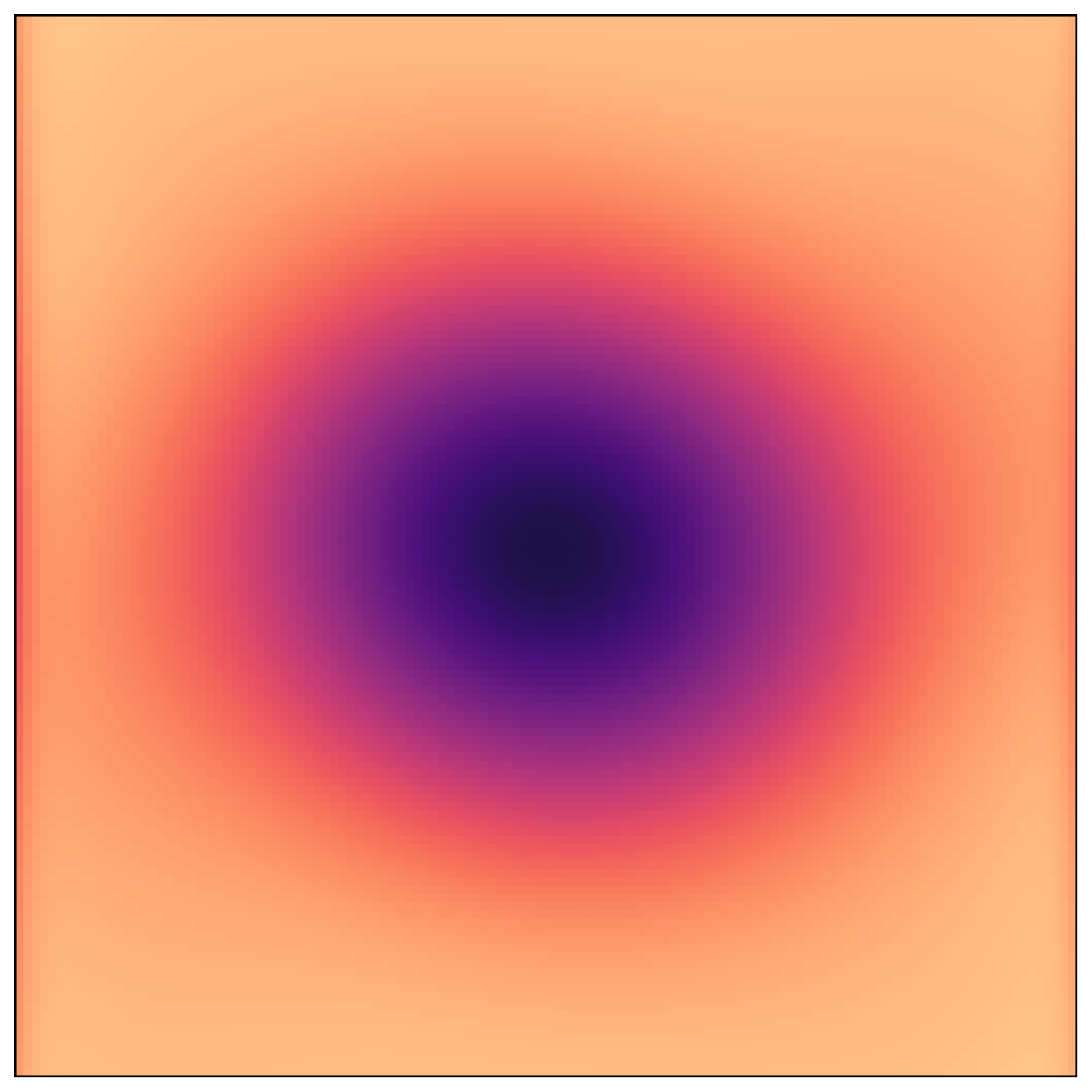} &
\includegraphics[height=135pt]{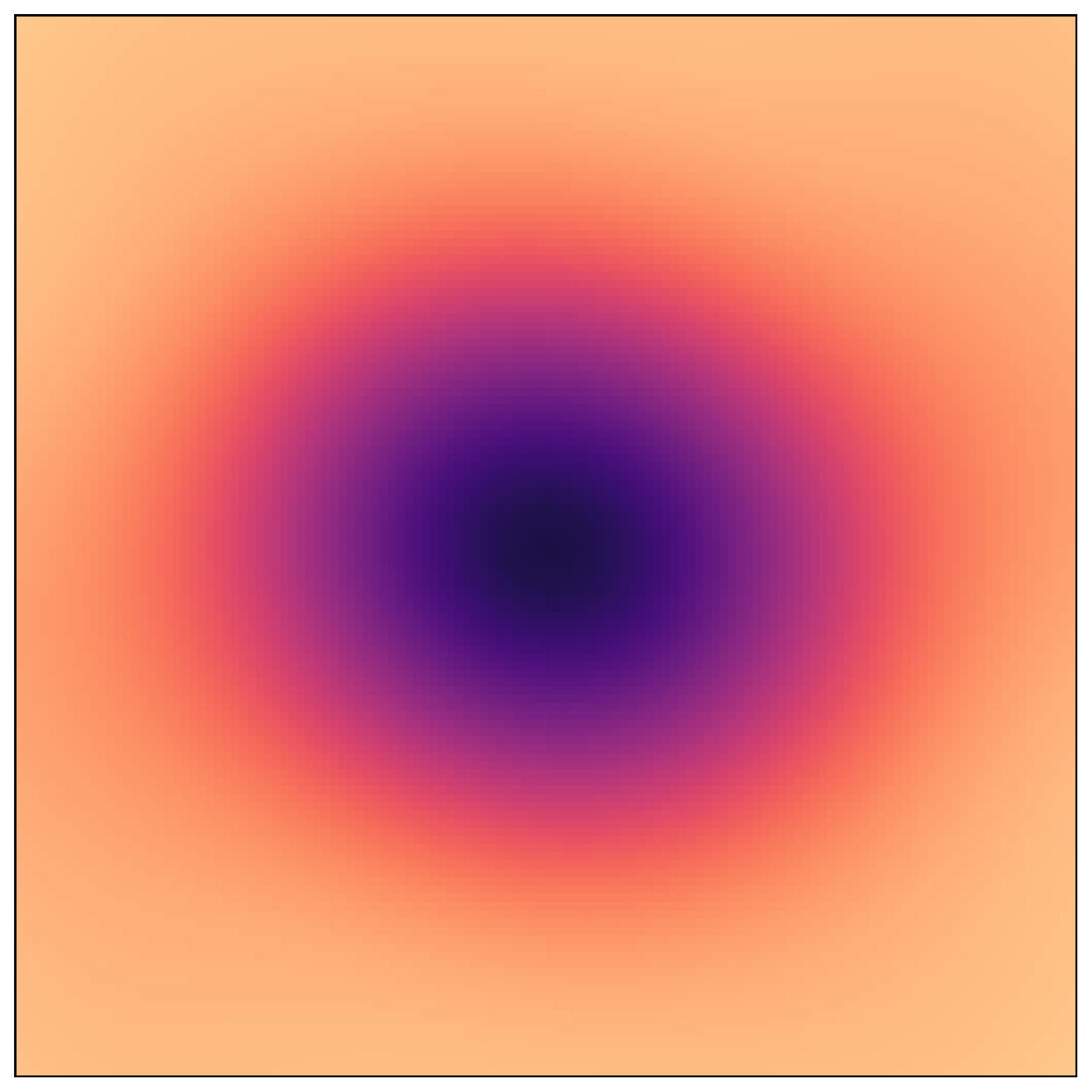} \\

\multicolumn{3}{c}{\small{\texttt{\bfseries X=30}}} \\
\includegraphics[height=135pt]{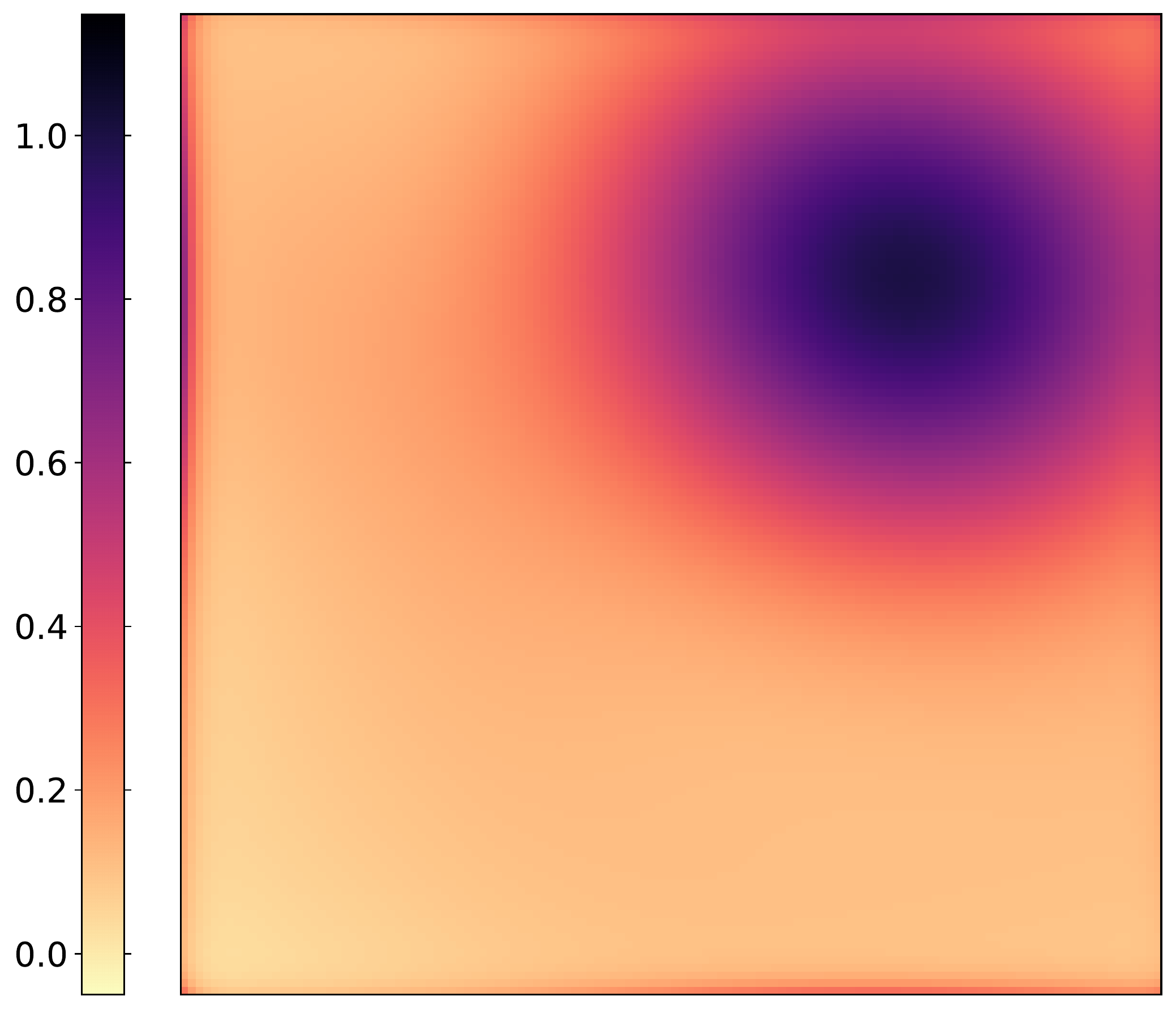} &
\includegraphics[height=135pt]{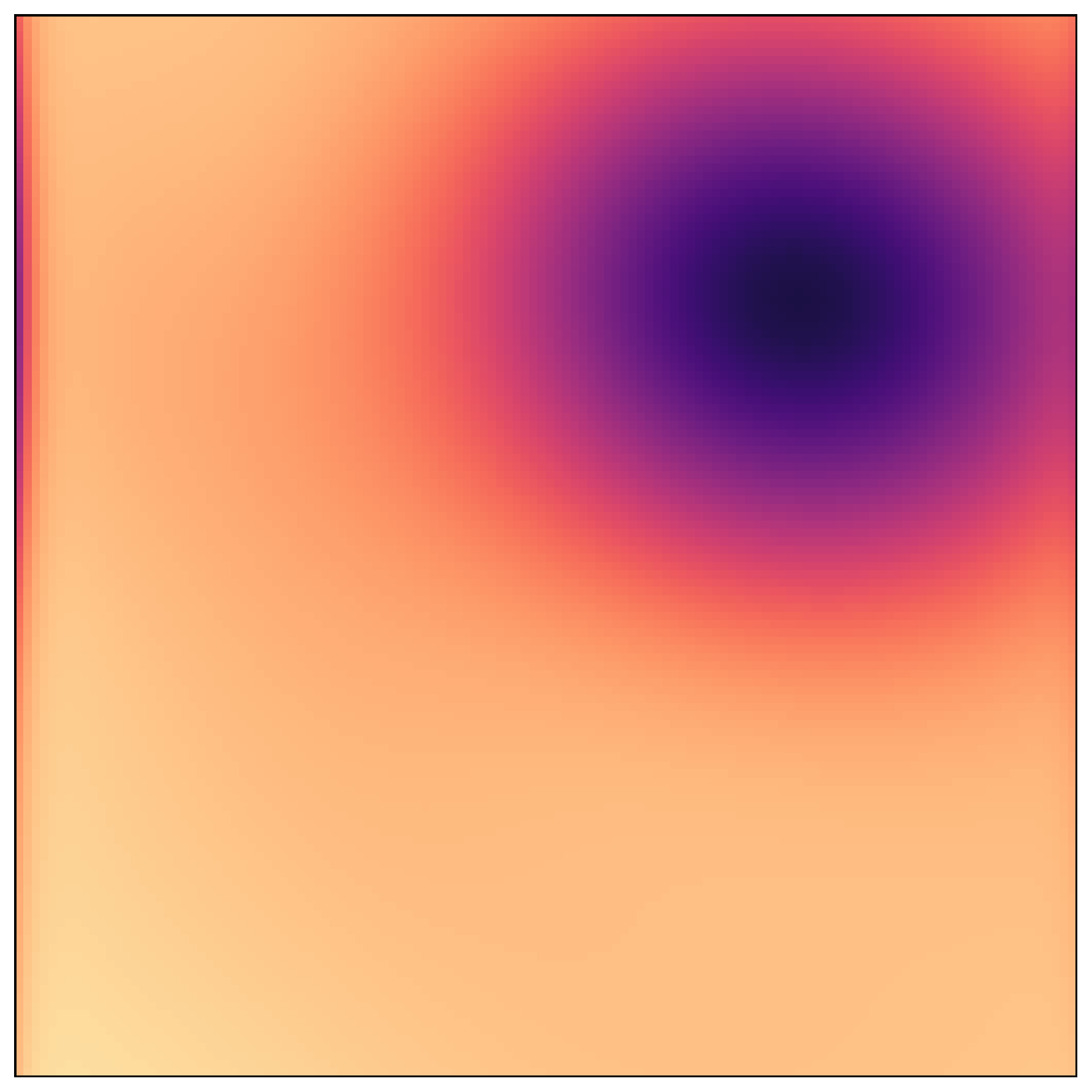} &
\includegraphics[height=135pt]{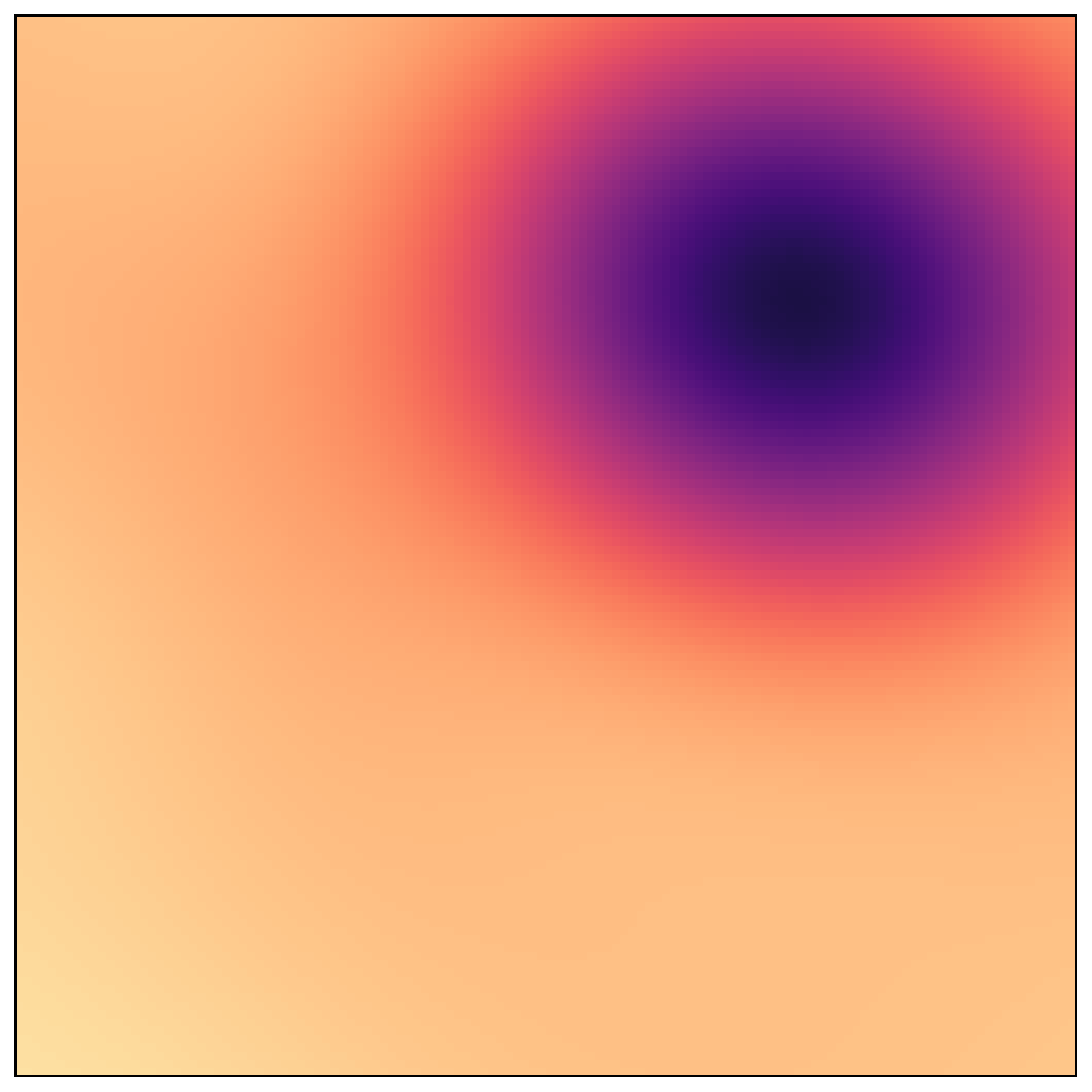} \\

\multicolumn{3}{c}{\small{\texttt{\bfseries X=62}}} \\
\includegraphics[height=135pt]{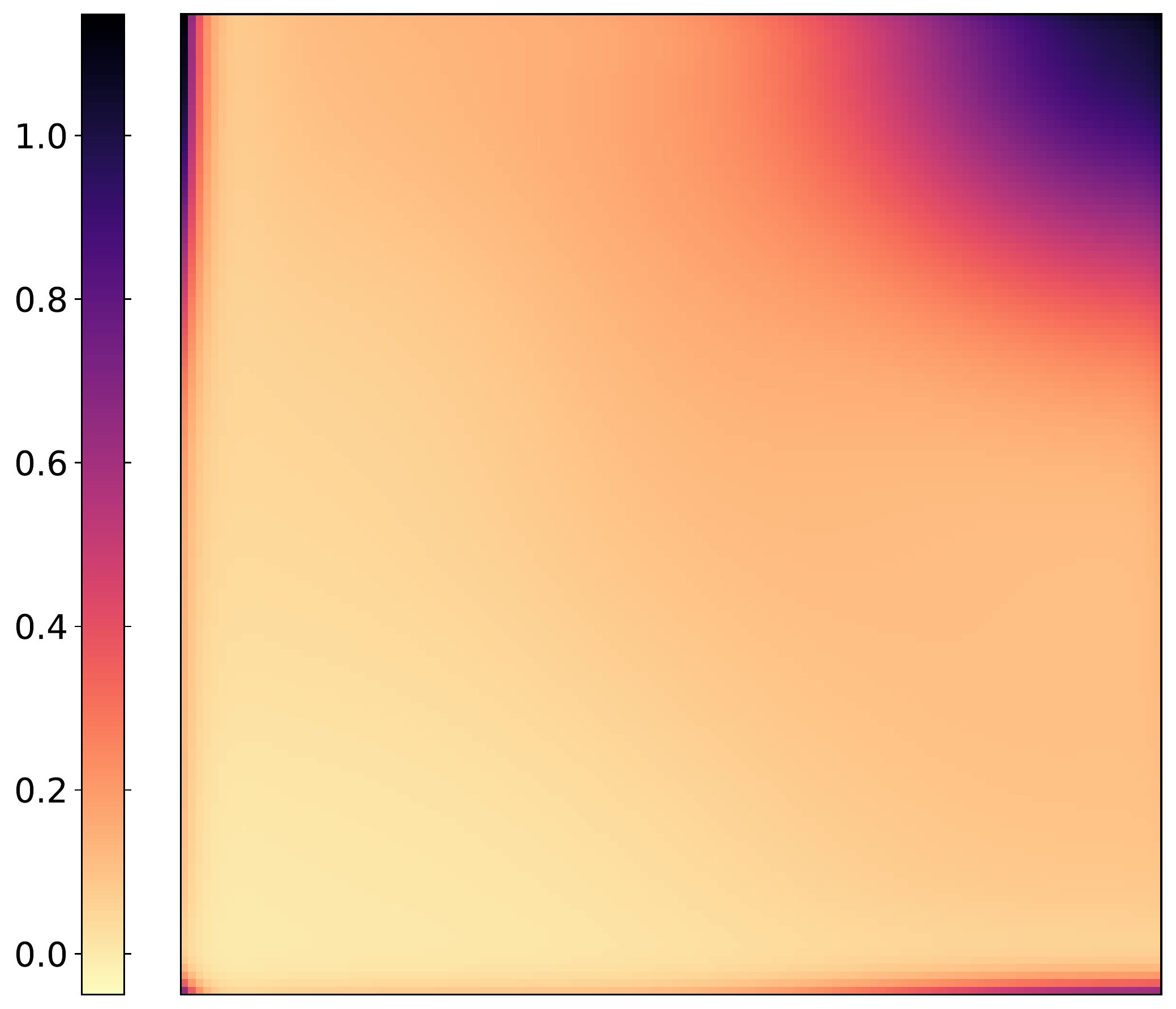} &
\includegraphics[height=135pt]{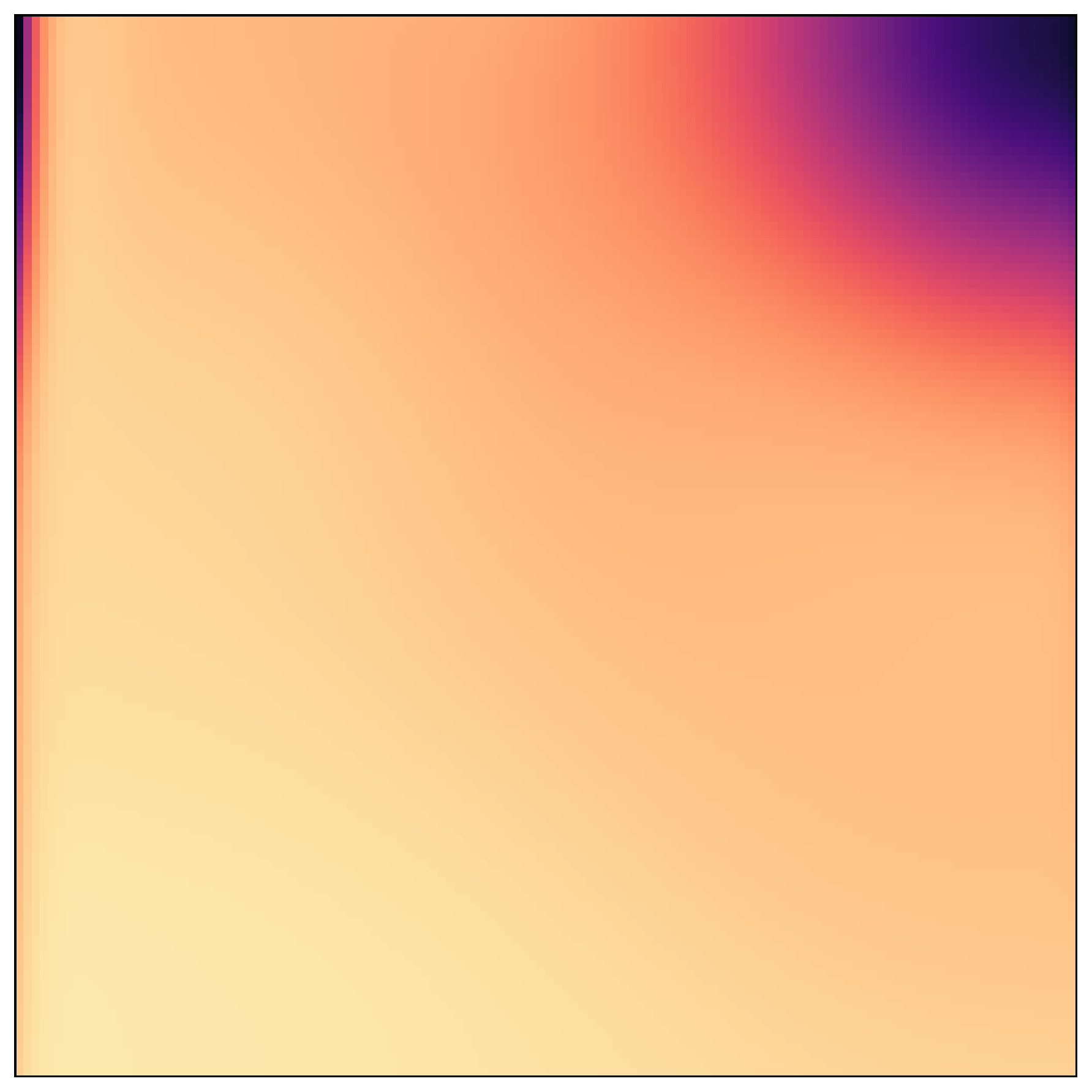} &
\includegraphics[height=135pt]{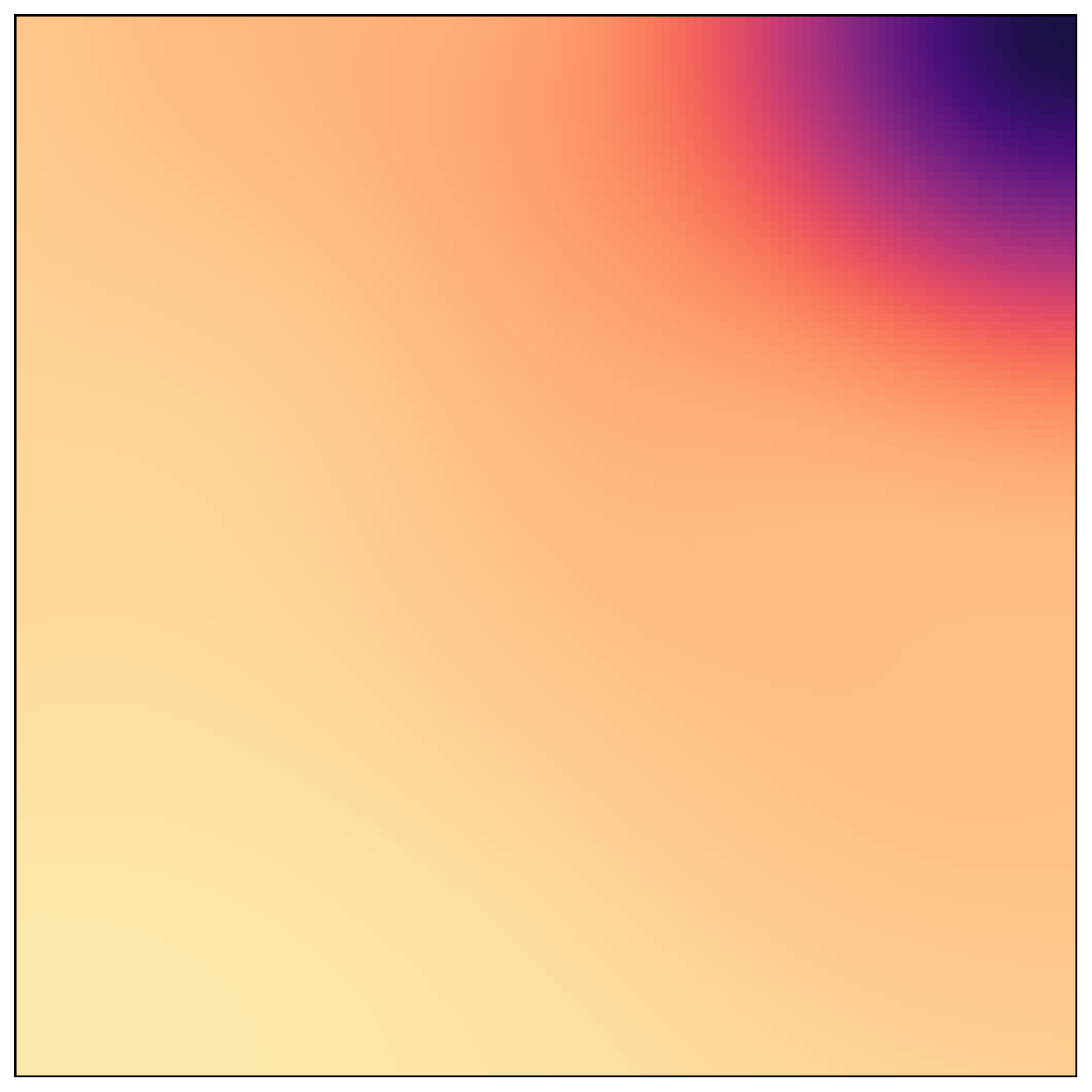} \\
\hspace{0.8cm}\footnotesize\texttt{Convolutional Gridding} & \footnotesize\texttt{Hybrid Gridding} & \footnotesize\texttt{Pruned NN Interpolation}\\
\\
\end{tabular}
\caption[Images generated by experiments of Section \ref{sec:comparative:aliasing}]{A selection of images computed in Single-Precision by the experiments of Section \ref{sec:comparative:aliasing} using the VLA observation.} 
\label{fig:comparative:aliasingvlasingle}
\end{figure}
\begin{figure}
\centering
\begin{tabular}{@{}c@{}c@{}c@{}}

\multicolumn{3}{c}{\small{\texttt{\bfseries X=0}}} \\
\includegraphics[height=135pt]{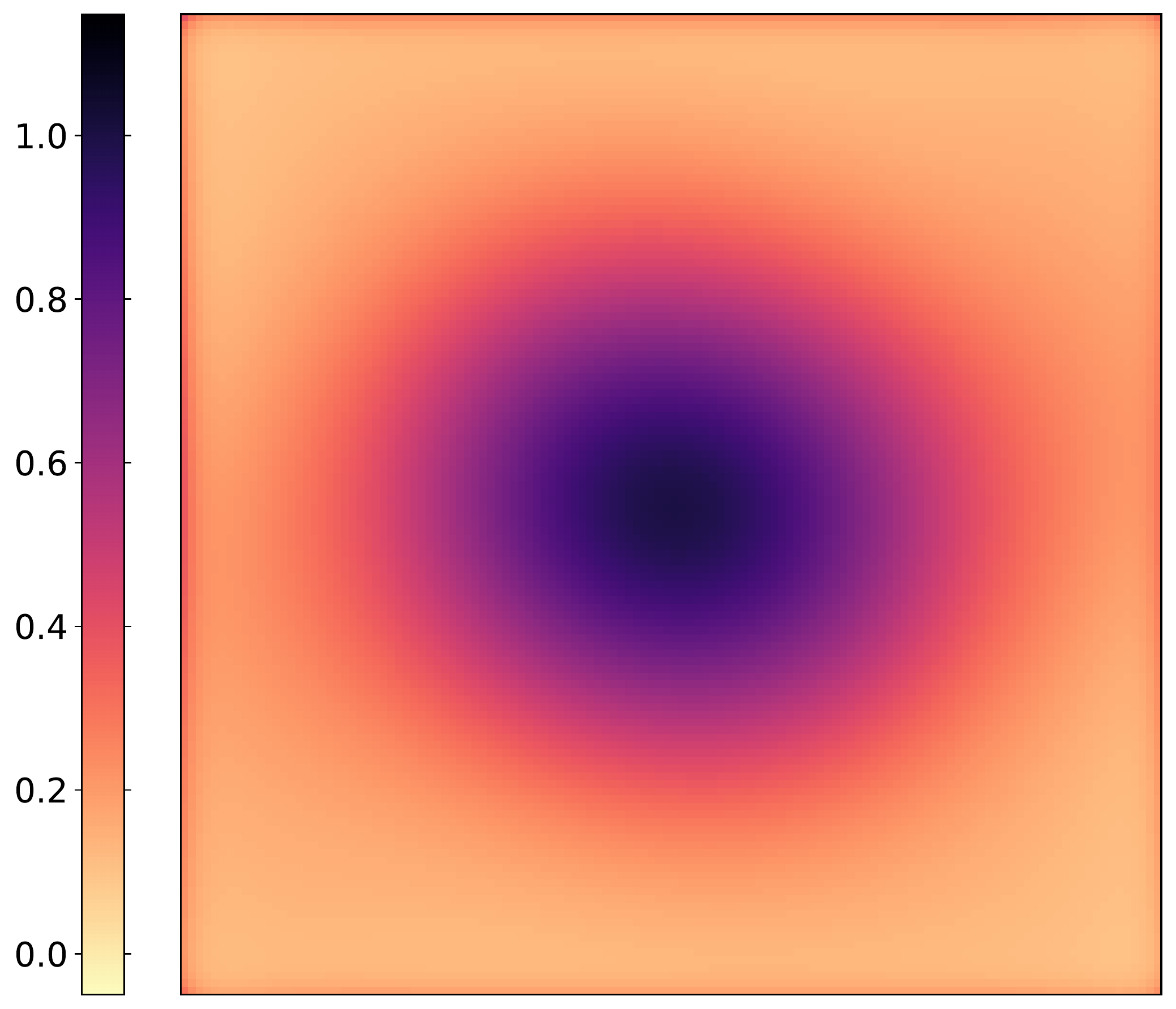} &
\includegraphics[height=135pt]{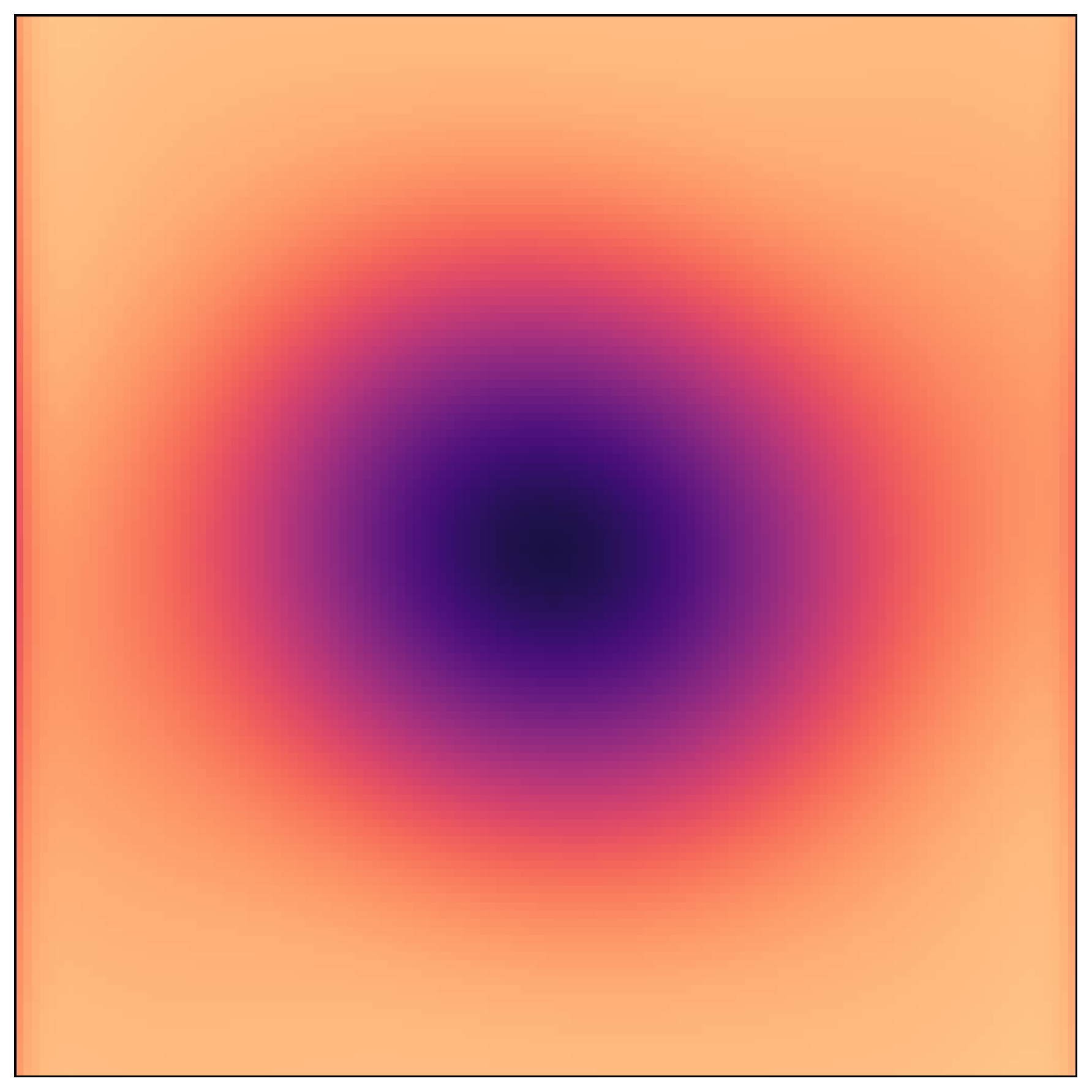} &
\includegraphics[height=135pt]{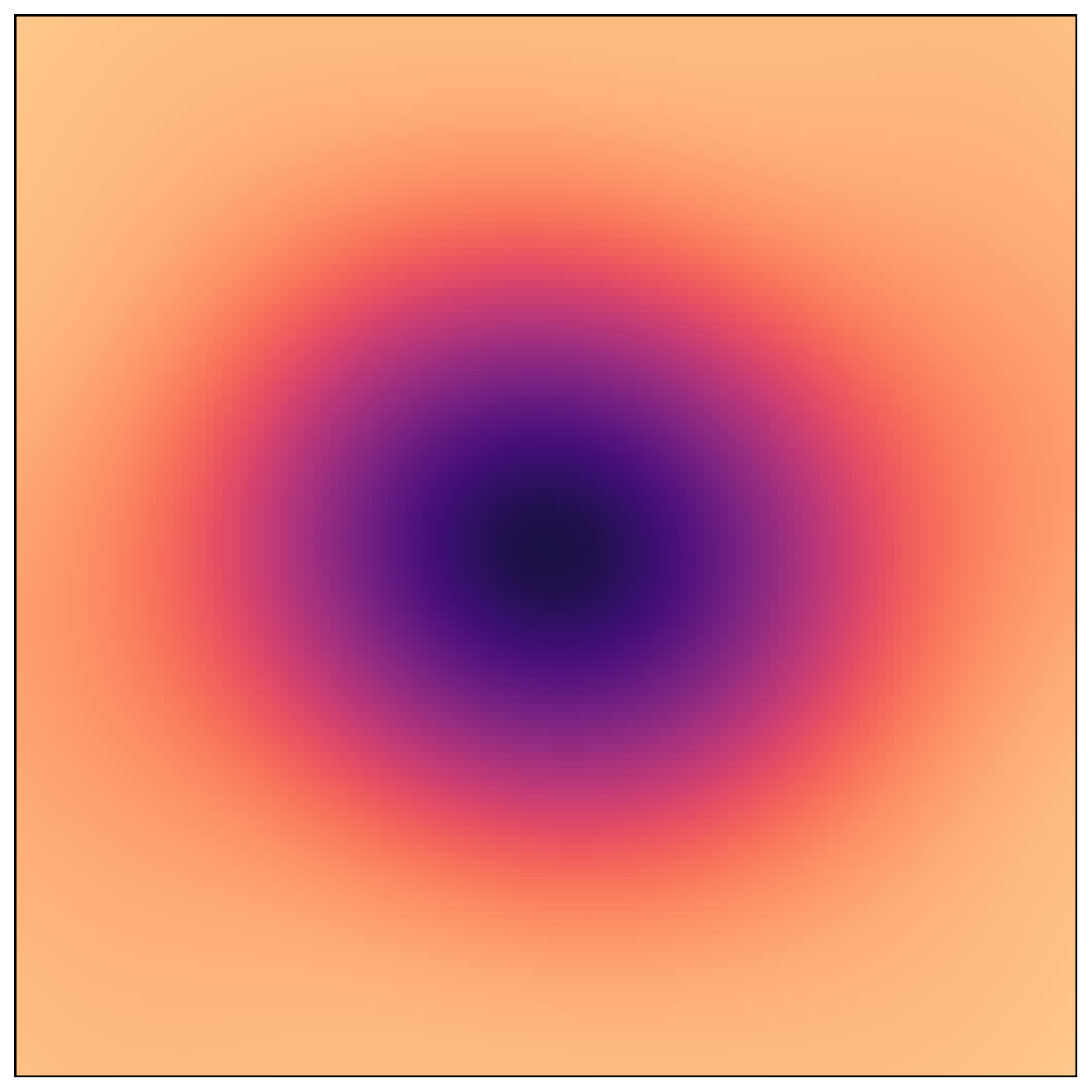} \\

\multicolumn{3}{c}{\small{\texttt{\bfseries X=30}}} \\
\includegraphics[height=135pt]{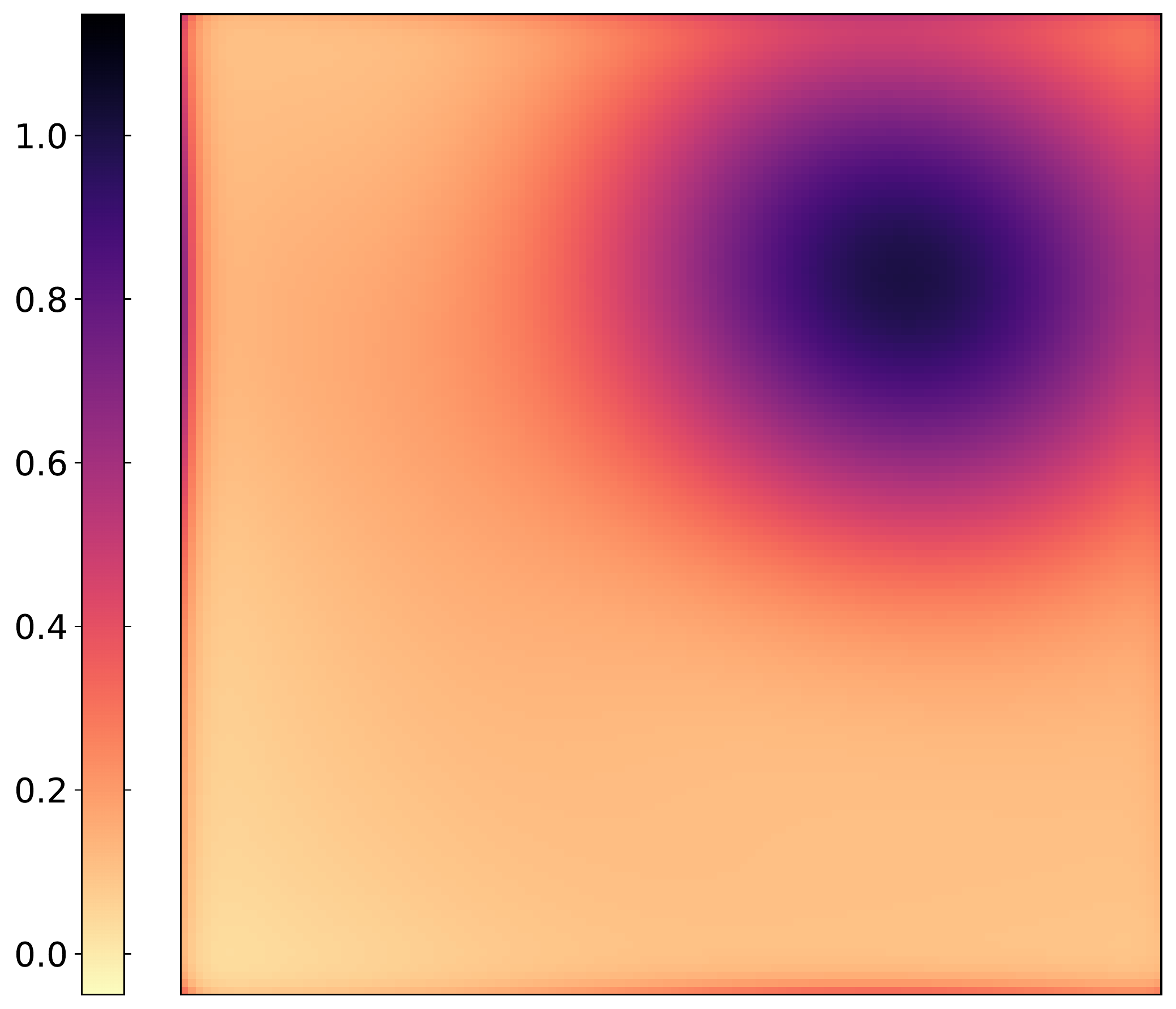} &
\includegraphics[height=135pt]{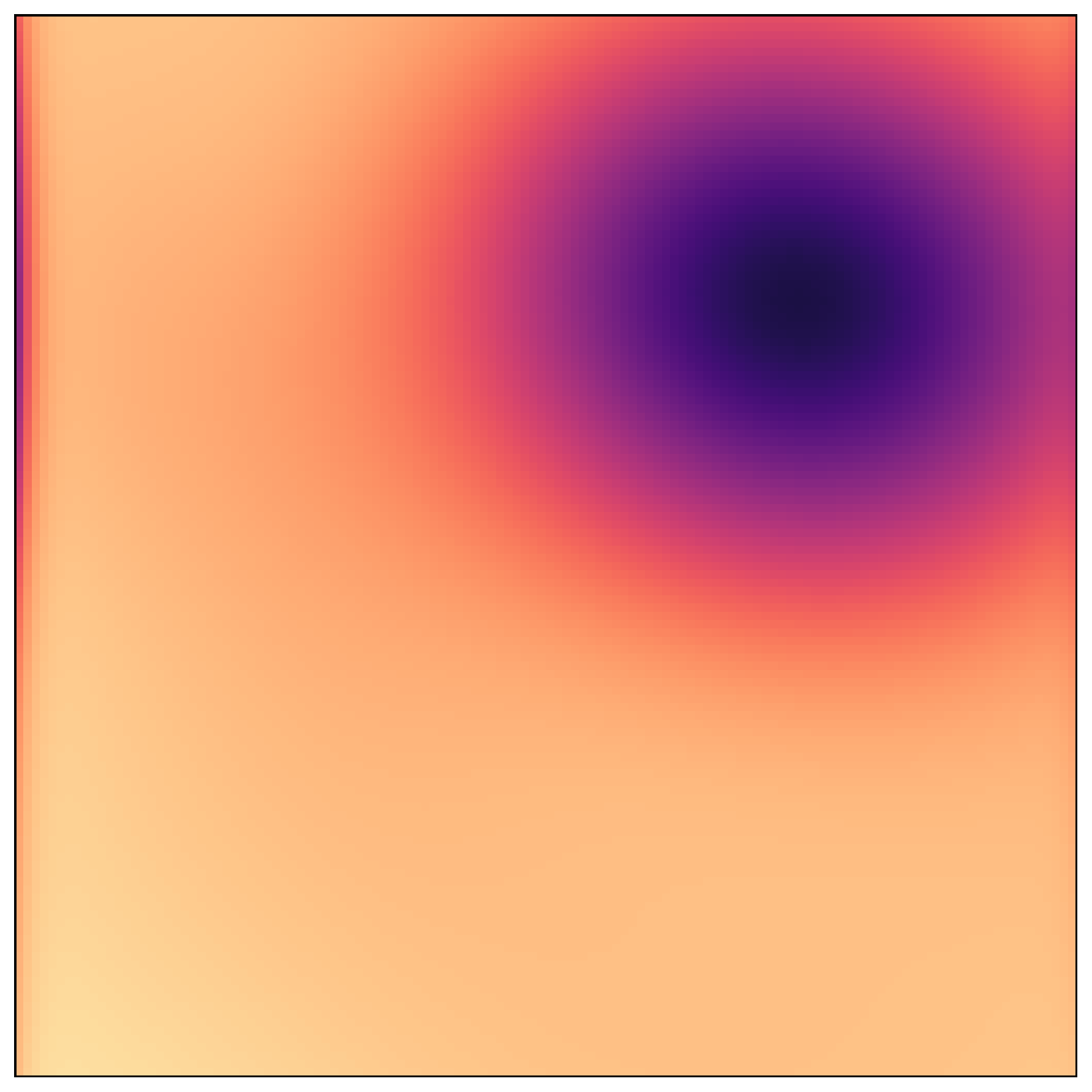} &
\includegraphics[height=135pt]{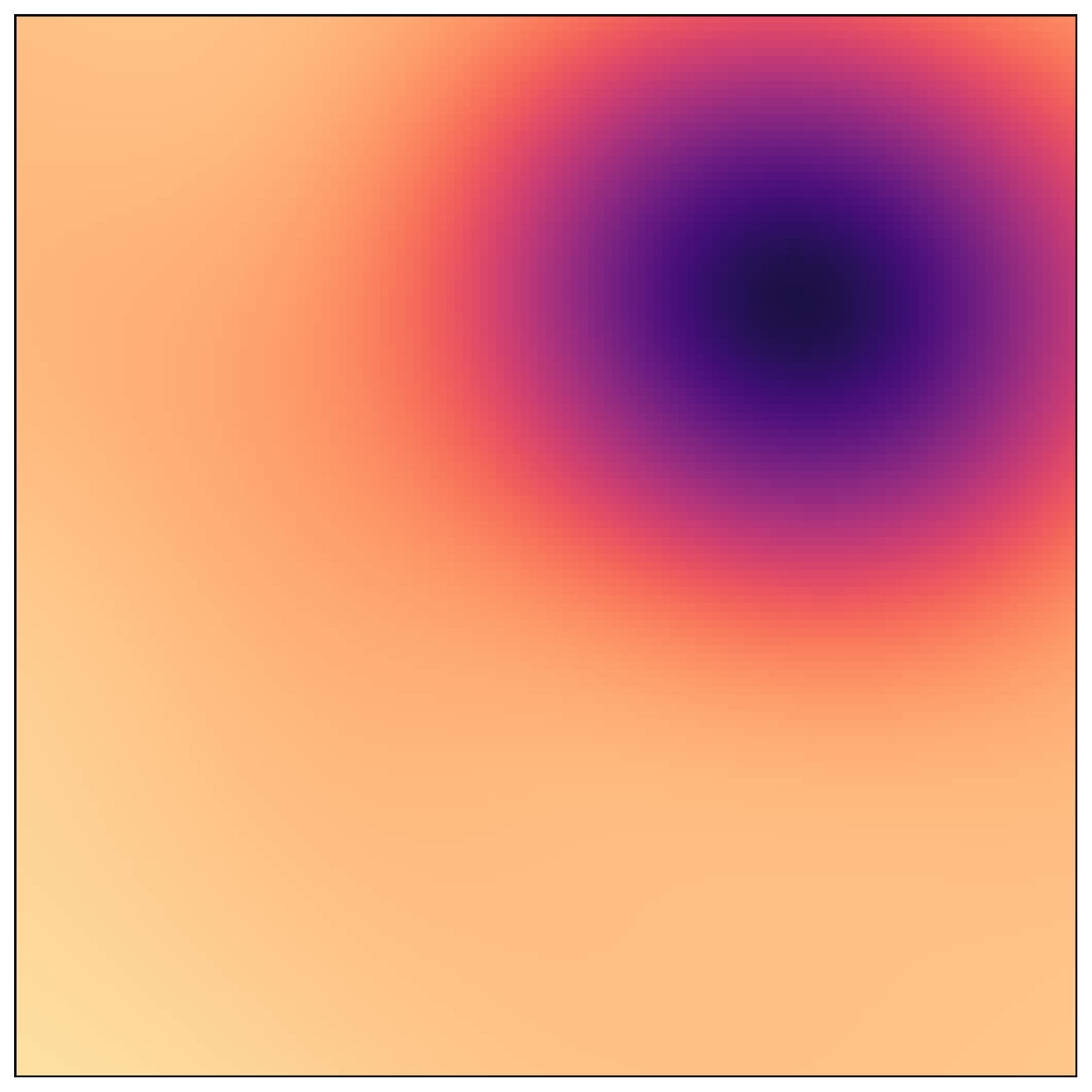} \\

\multicolumn{3}{c}{\small{\texttt{\bfseries X=62}}} \\
\includegraphics[height=135pt]{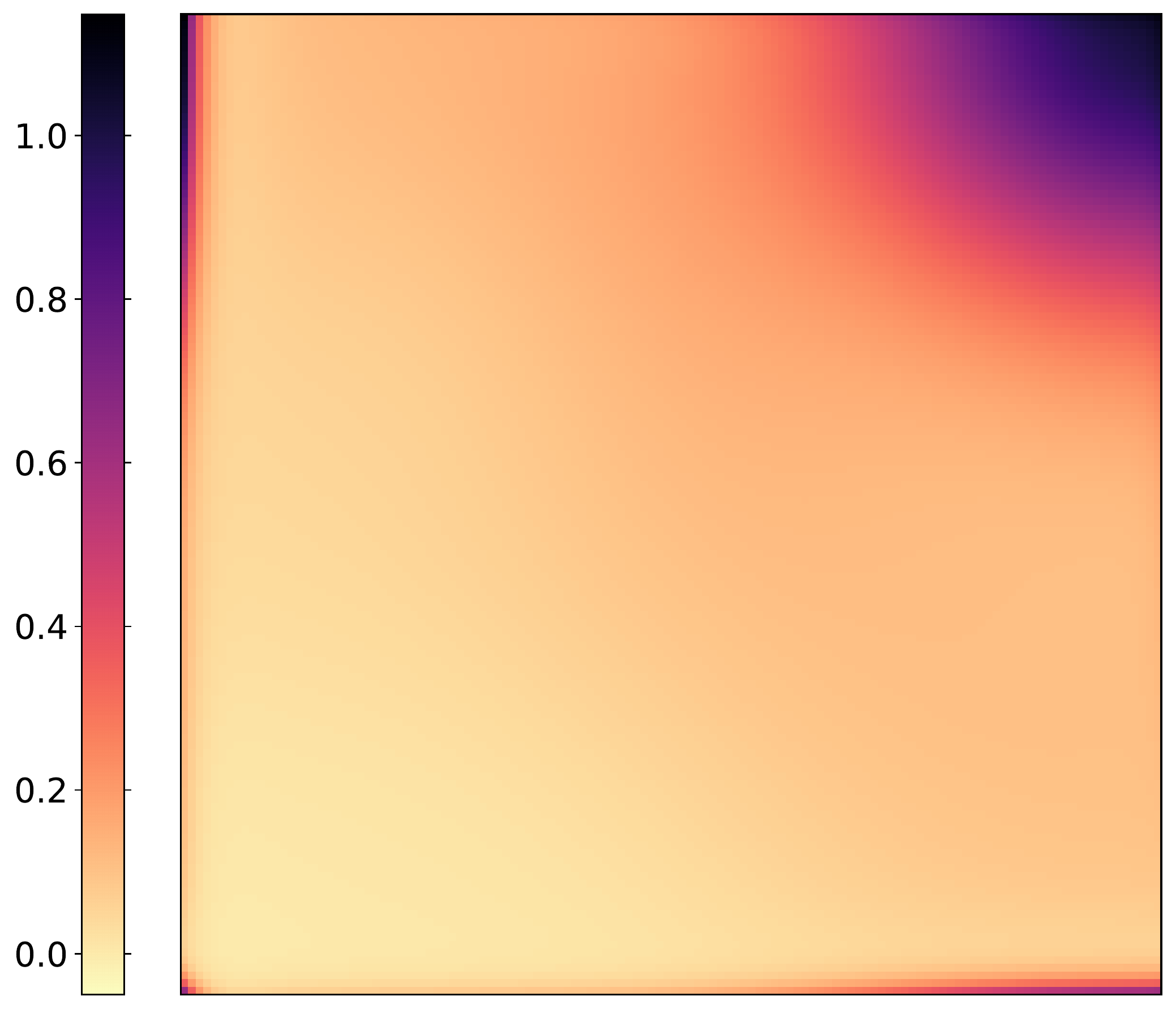} &
\includegraphics[height=135pt]{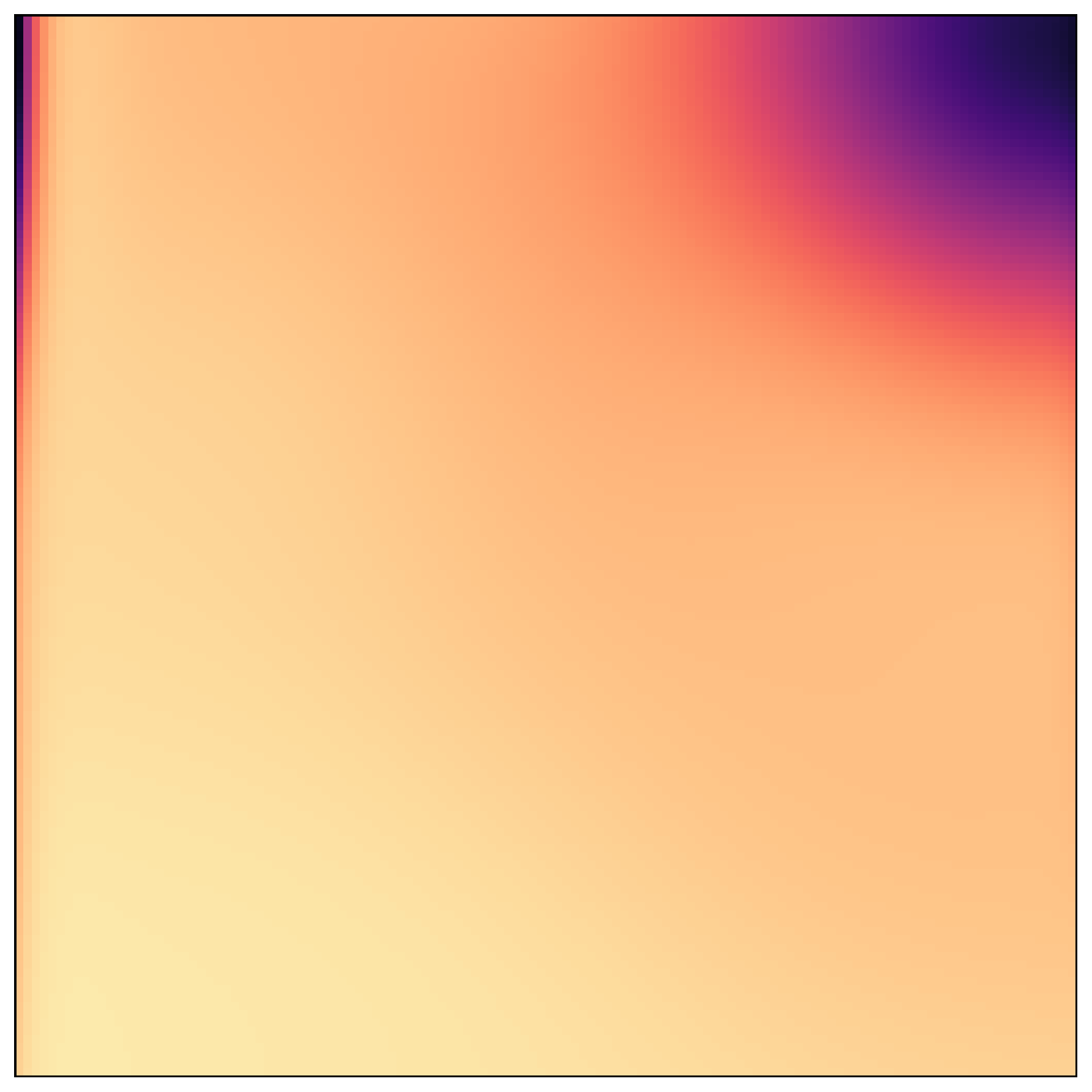} &
\includegraphics[height=135pt]{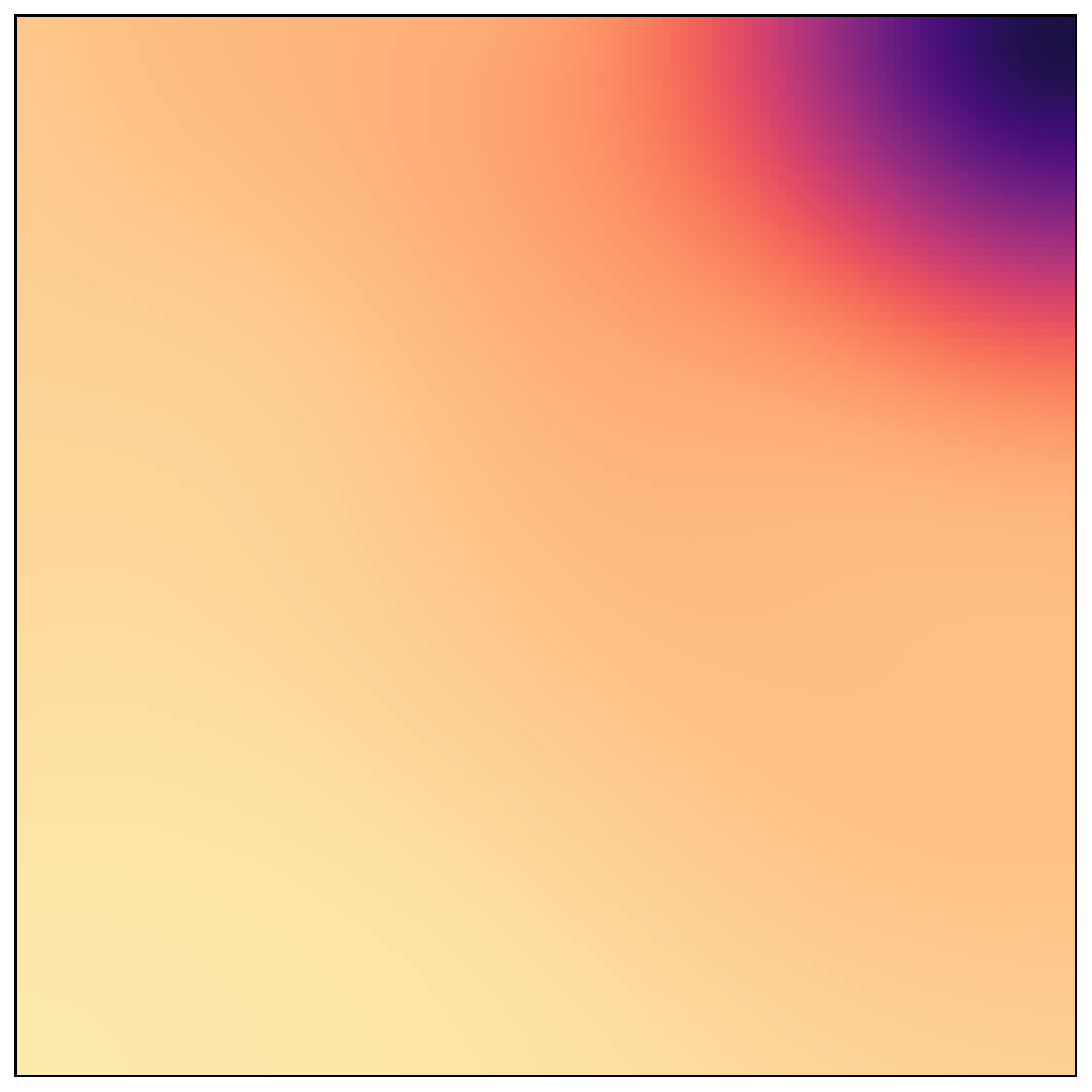} \\
\hspace{0.8cm}\footnotesize\texttt{Convolutional Gridding} & \footnotesize\texttt{Hybrid Gridding} & \footnotesize\texttt{Pruned NN Interpolation}\\
\\
\end{tabular}
\caption[Images generated by experiments of Section \ref{sec:comparative:aliasing}]{A selection of images computed in Double-Precision by the experiments of Section \ref{sec:comparative:aliasing} using the VLA observation.} 
\label{fig:comparative:aliasingvladouble}
\end{figure}
\begin{figure}
\centering
\begin{tabular}{@{}c@{}c@{}c@{}}

\multicolumn{3}{c}{\small{\texttt{\bfseries X=0}}} \\
\includegraphics[height=135pt]{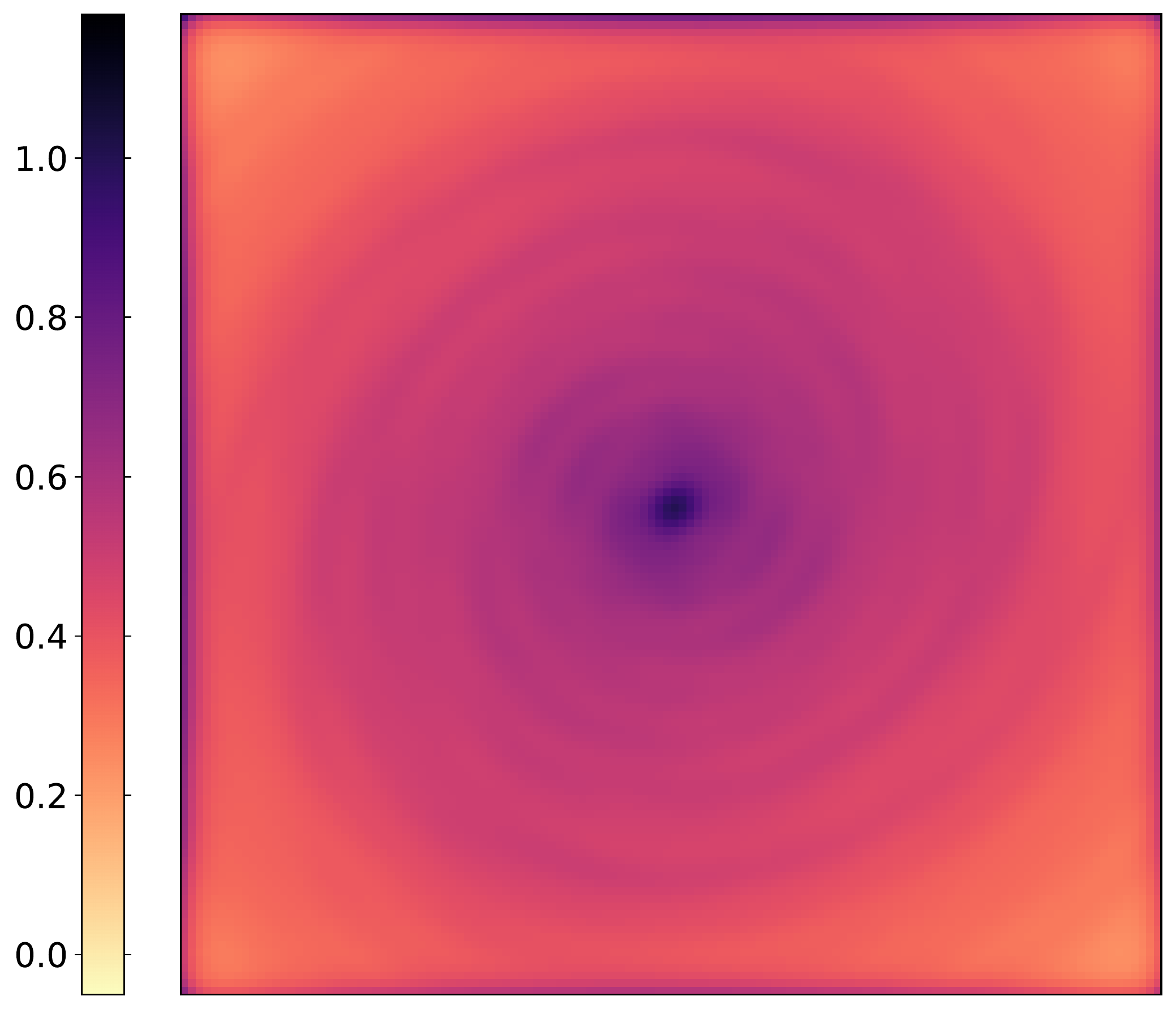} &
\includegraphics[height=135pt]{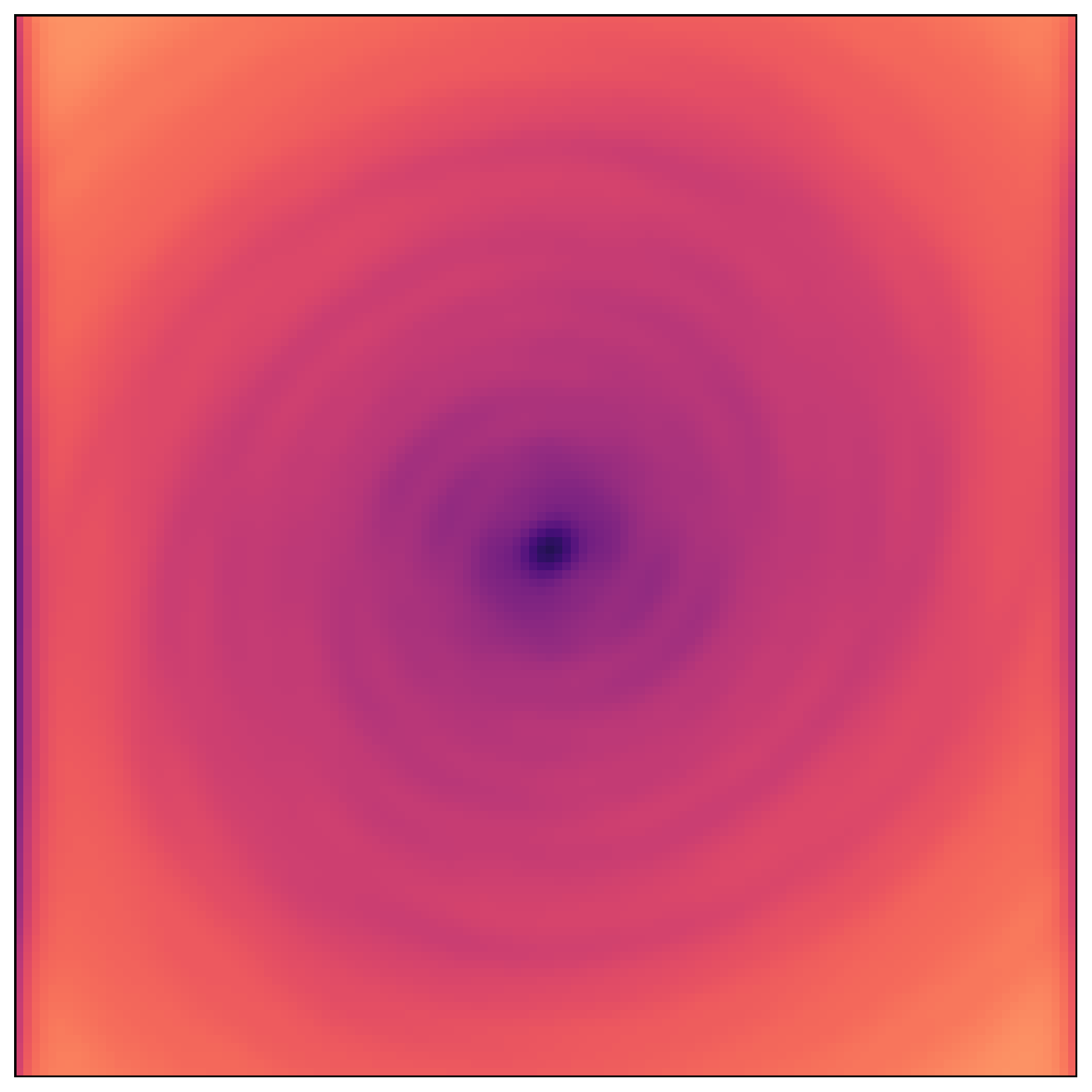} &
\includegraphics[height=135pt]{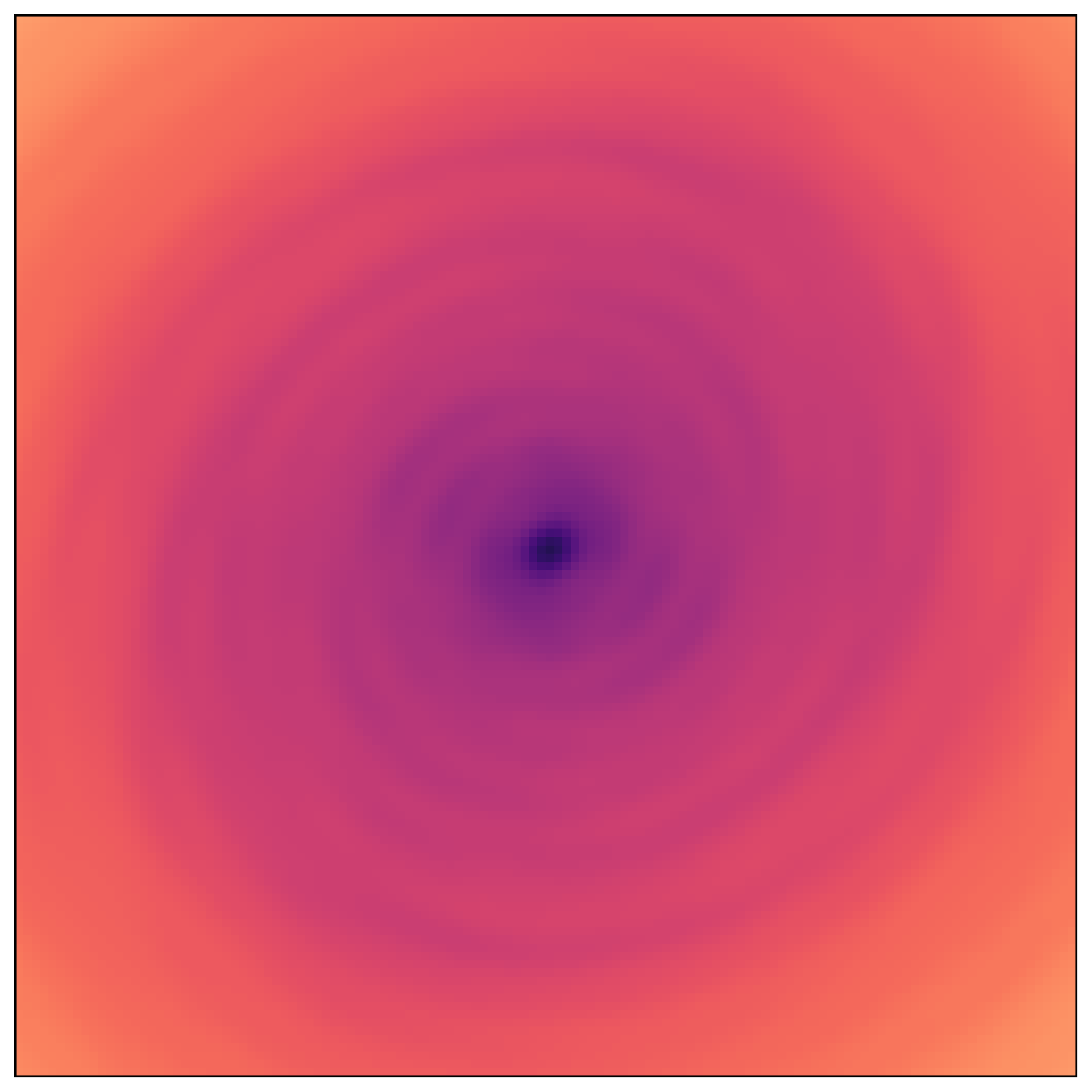} \\

\multicolumn{3}{c}{\small{\texttt{\bfseries X=30}}} \\
\includegraphics[height=135pt]{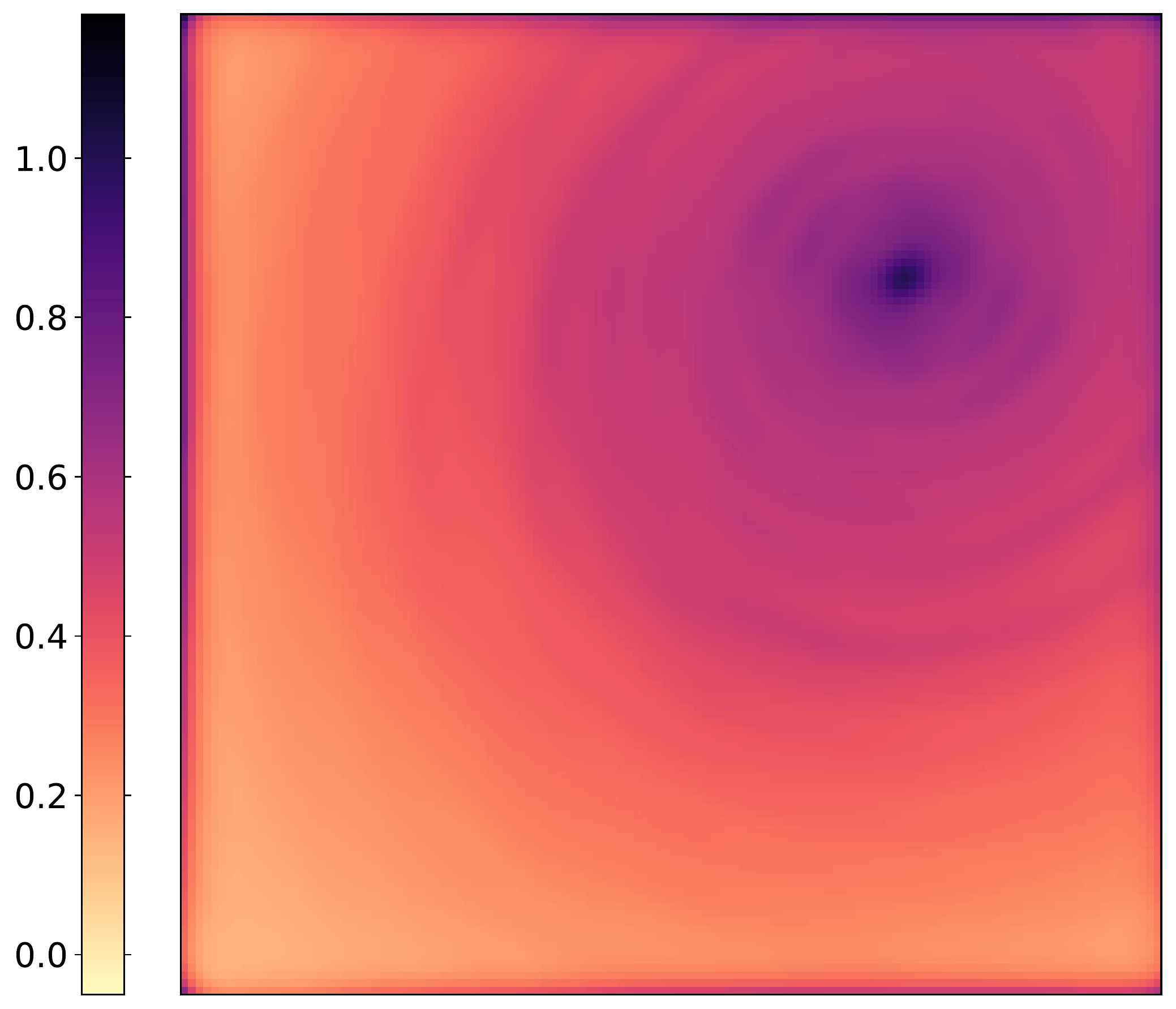} &
\includegraphics[height=135pt]{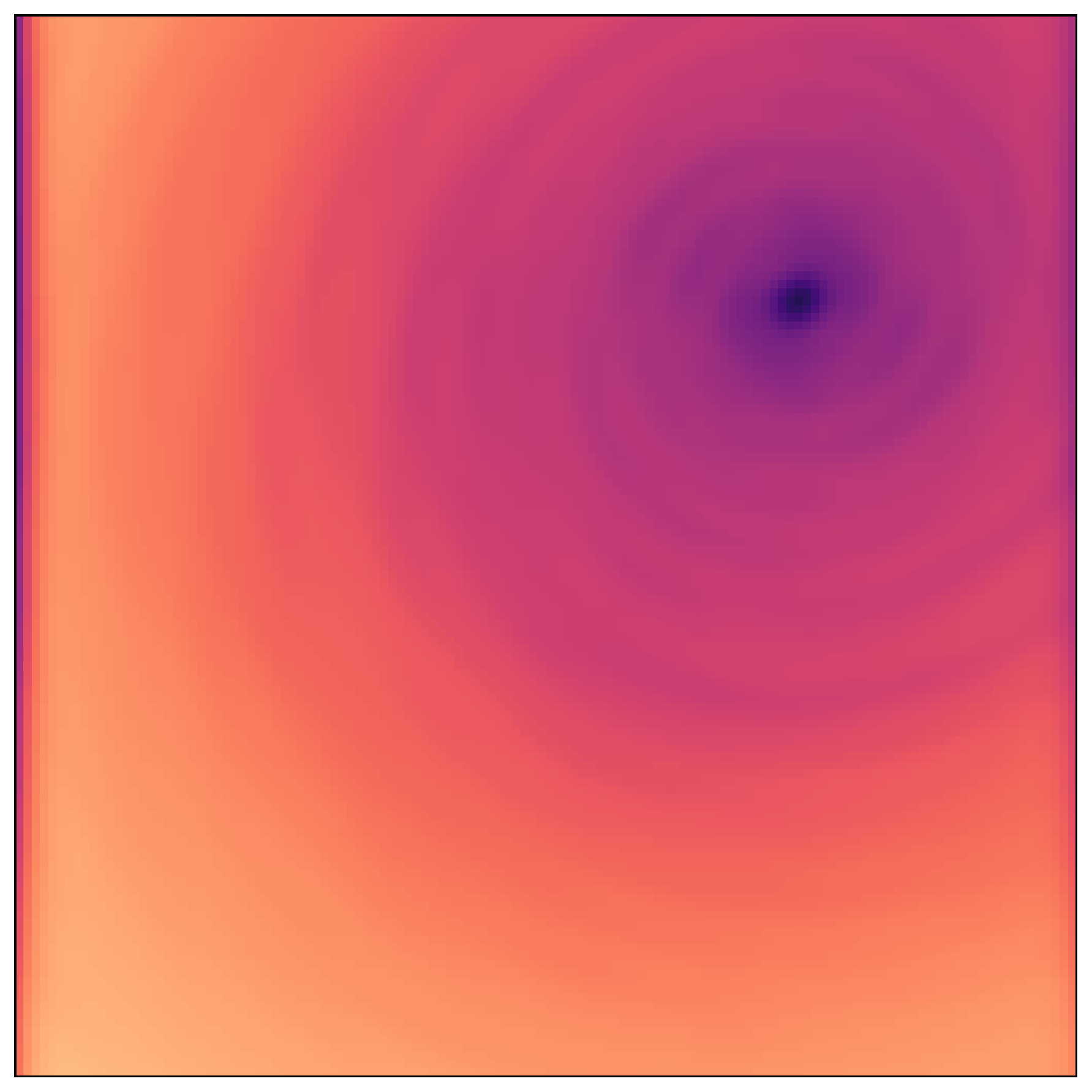} &
\includegraphics[height=135pt]{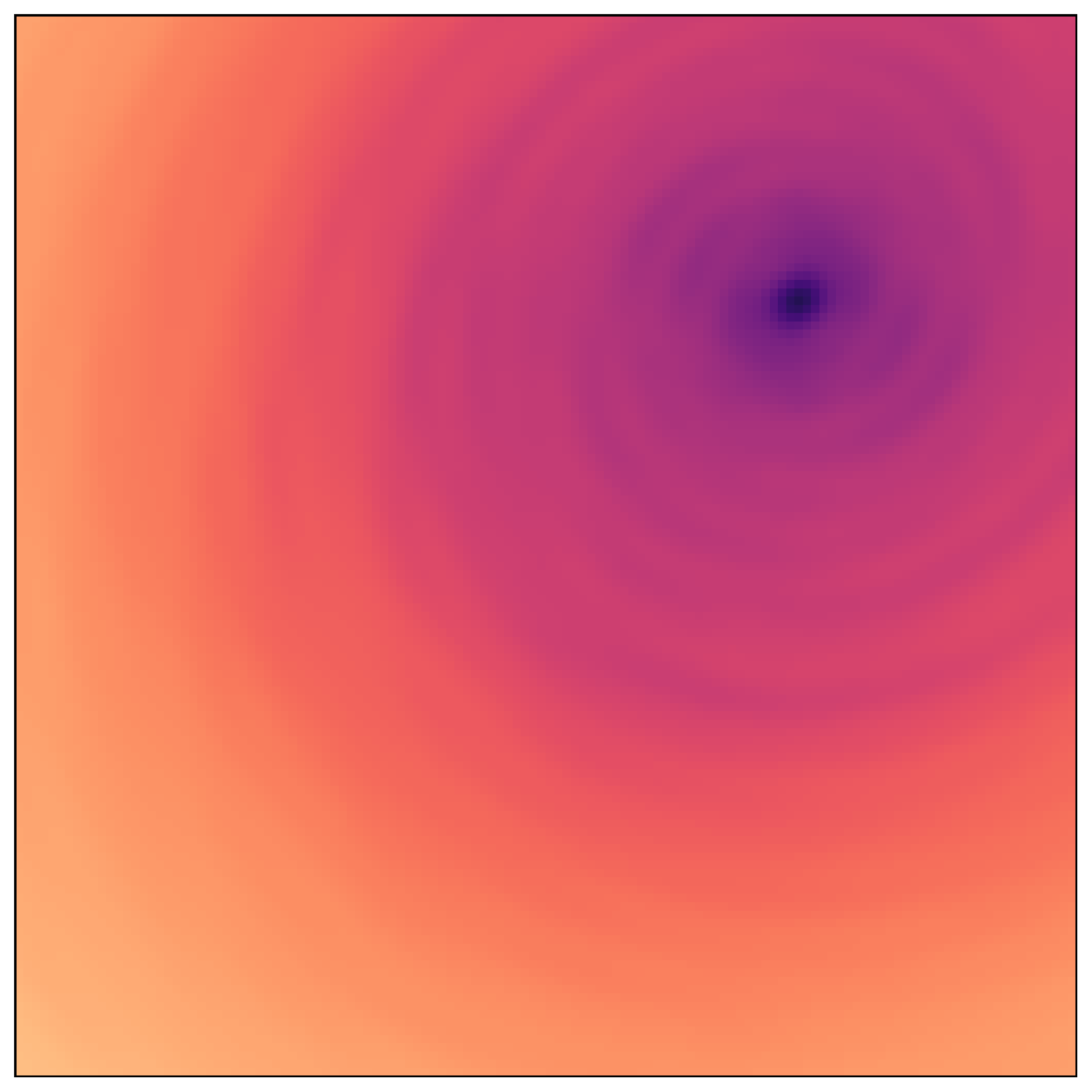} \\

\multicolumn{3}{c}{\small\texttt{\bfseries{X=62}}} \\
\includegraphics[height=135pt]{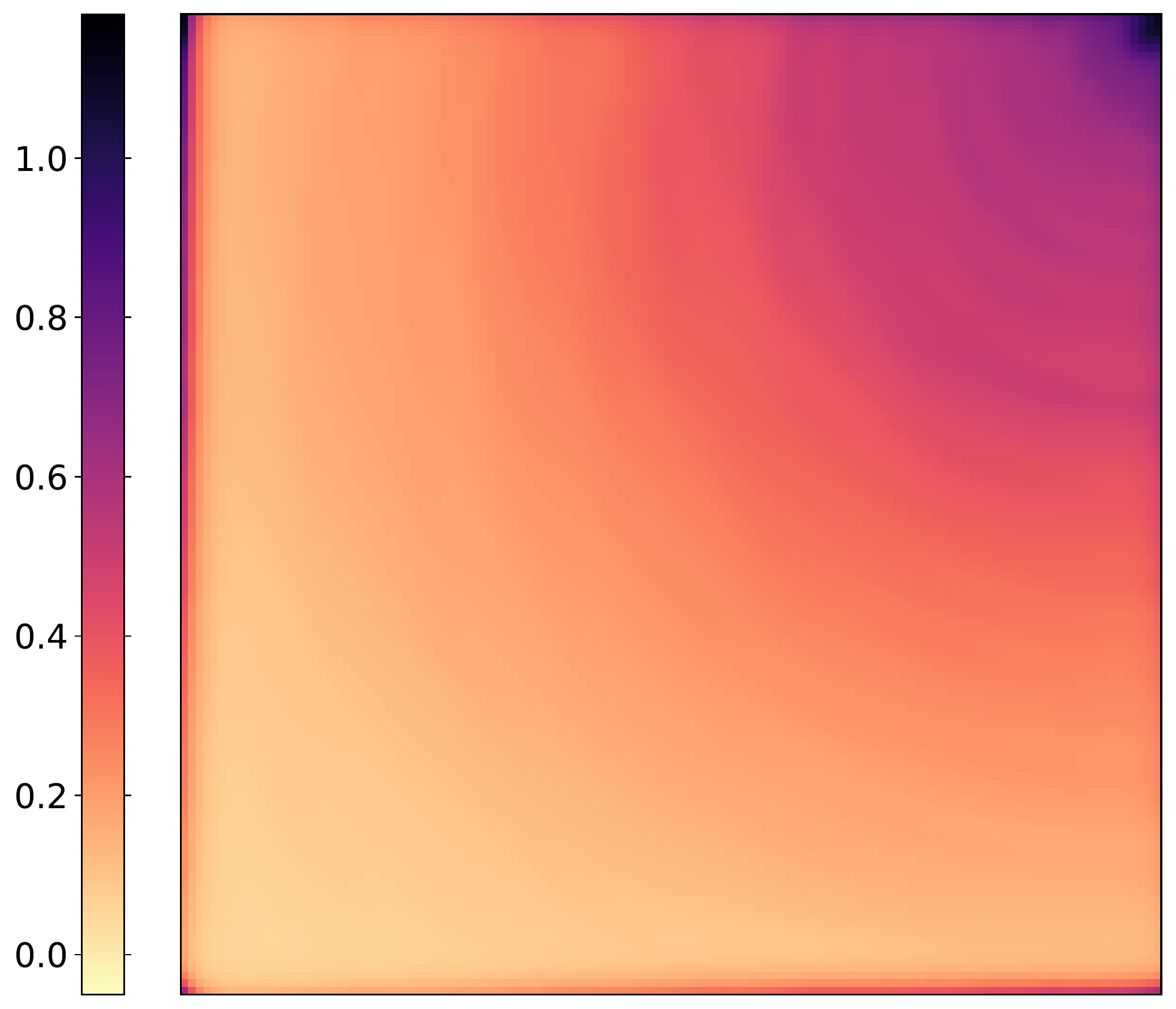} &
\includegraphics[height=135pt]{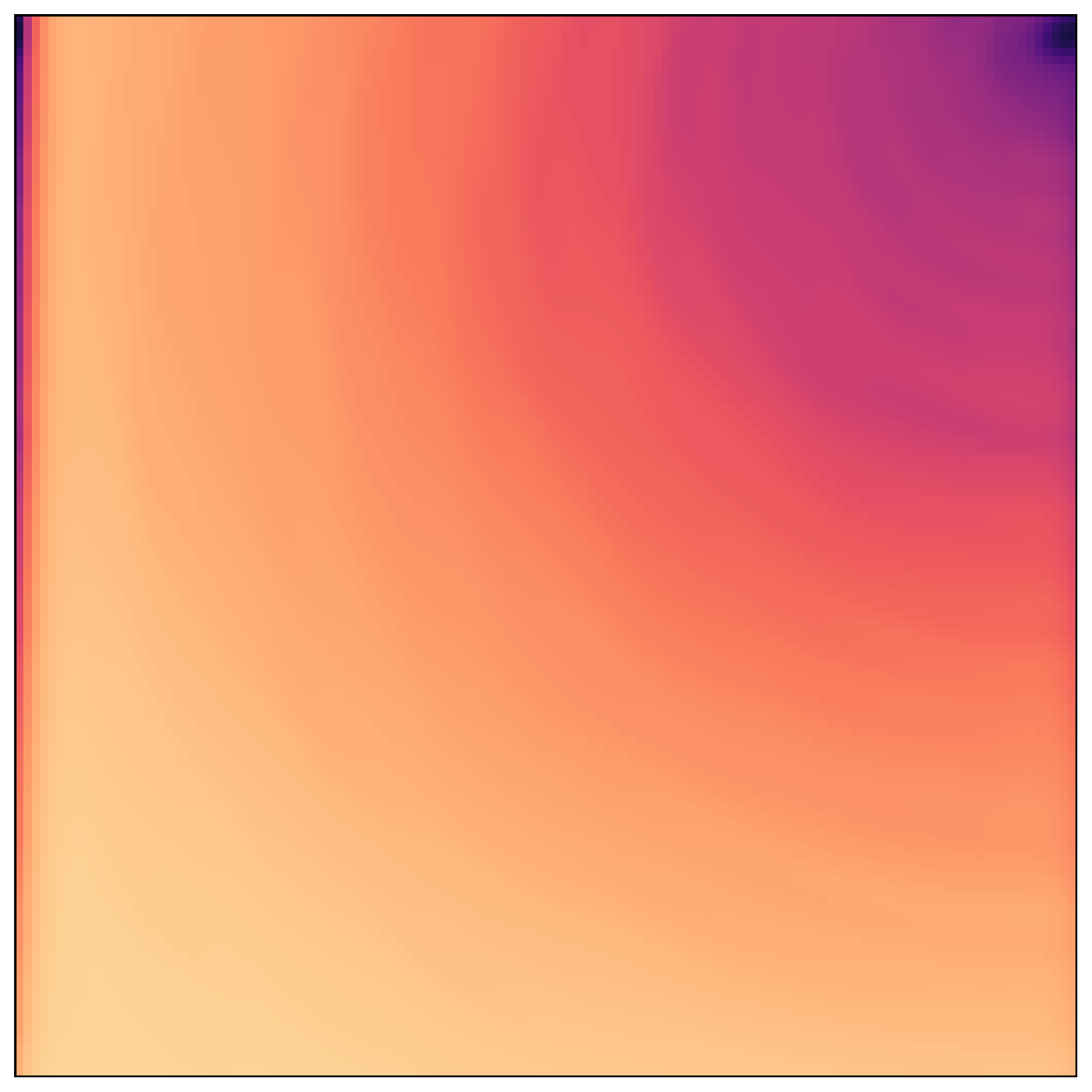} &
\includegraphics[height=135pt]{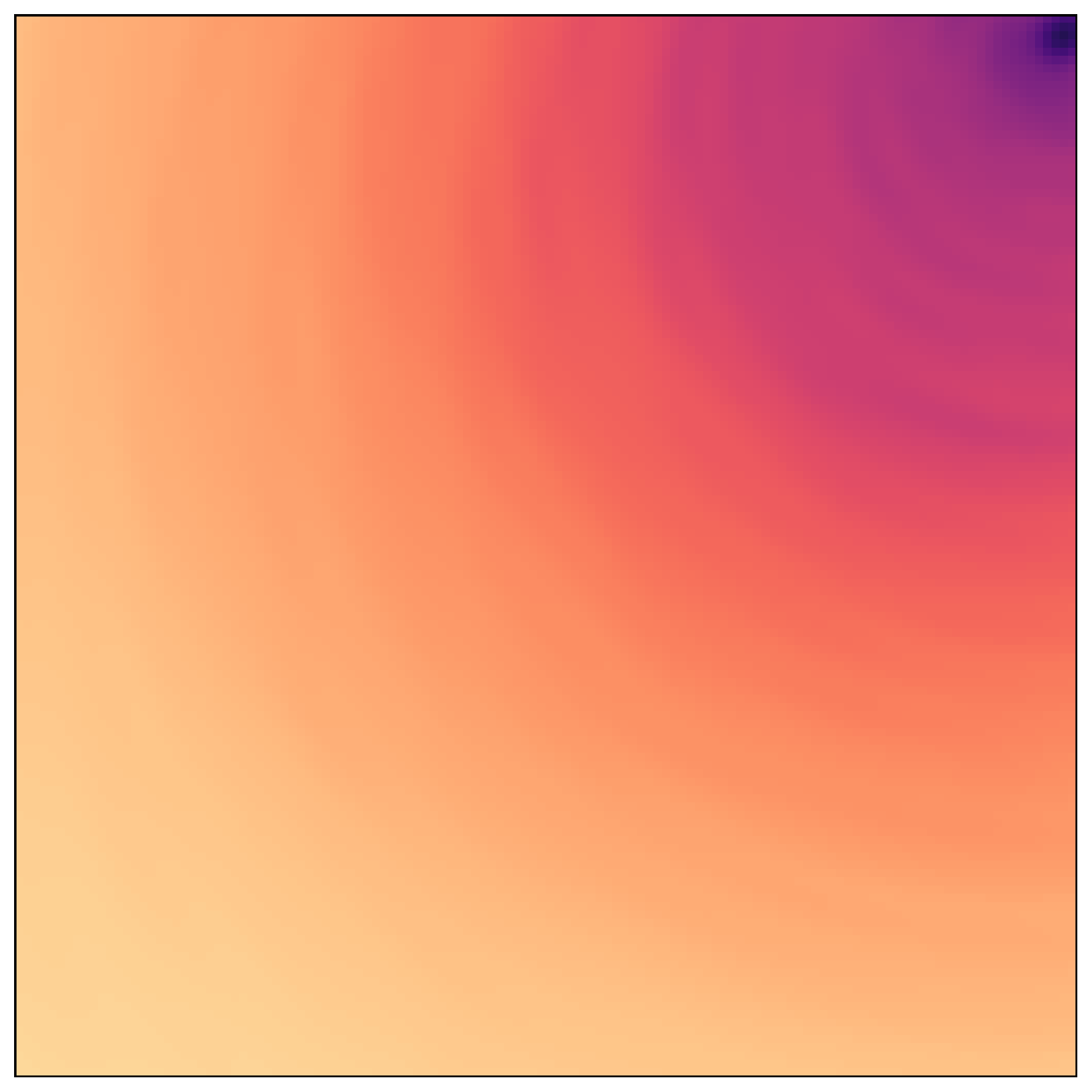} \\
\hspace{0.8cm}\footnotesize\texttt{Convolutional Gridding} & \footnotesize\texttt{Hybrid Gridding} & \footnotesize\texttt{Pruned NN Interpolation}\\
\\
\end{tabular}
\caption[Images generated by experiments of Section \ref{sec:comparative:aliasing}]{A selection of images computed in Single-Precision by the experiments of Section \ref{sec:comparative:aliasing} using the LOFAR observation.} 
\label{fig:comparative:aliasinglofarsingle}
\end{figure}
\begin{figure}
\centering
\begin{tabular}{@{}c@{}c@{}c@{}}

\multicolumn{3}{c}{\small{\texttt{\bfseries X=0}}} \\
\includegraphics[height=135pt]{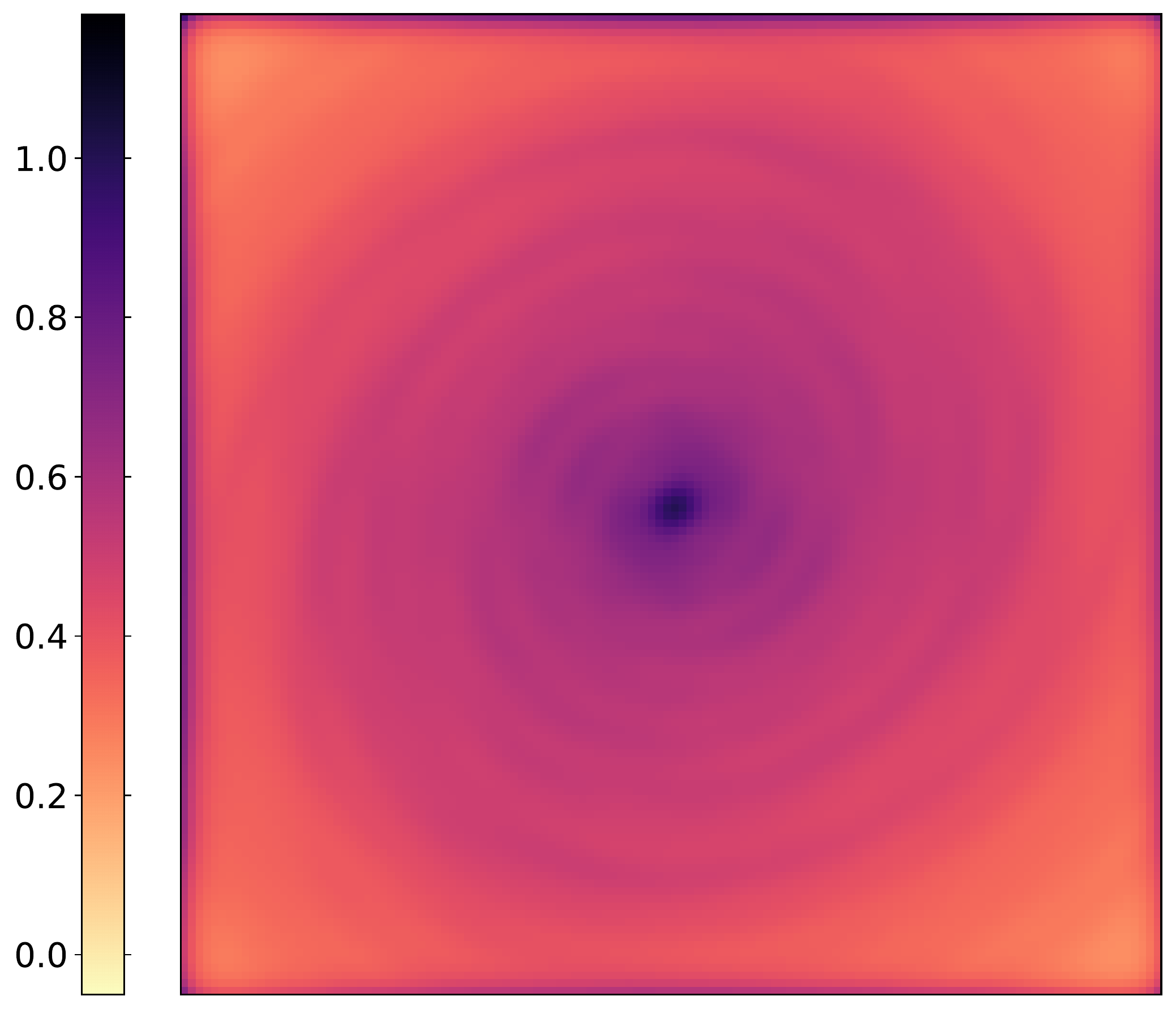} &
\includegraphics[height=135pt]{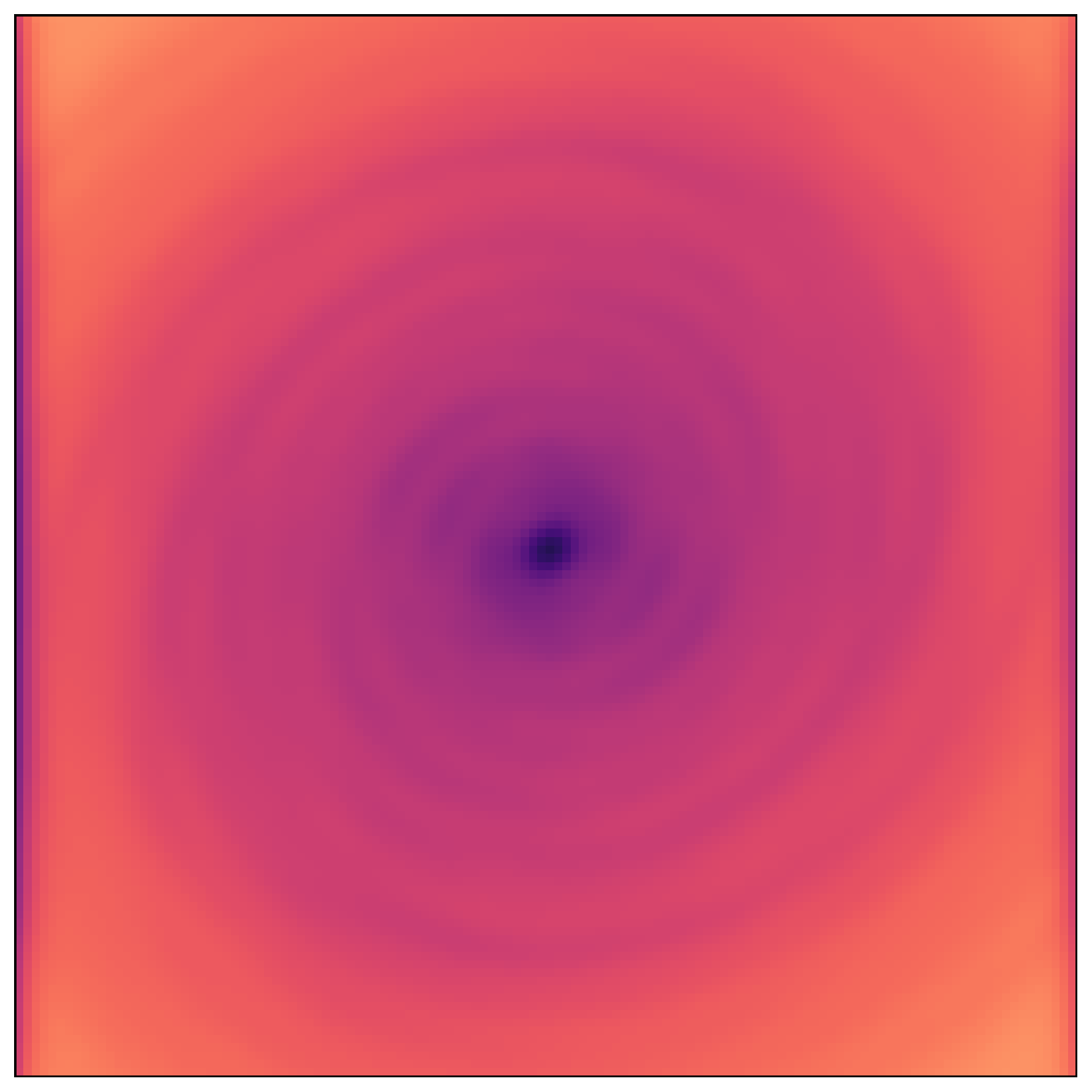} &
\includegraphics[height=135pt]{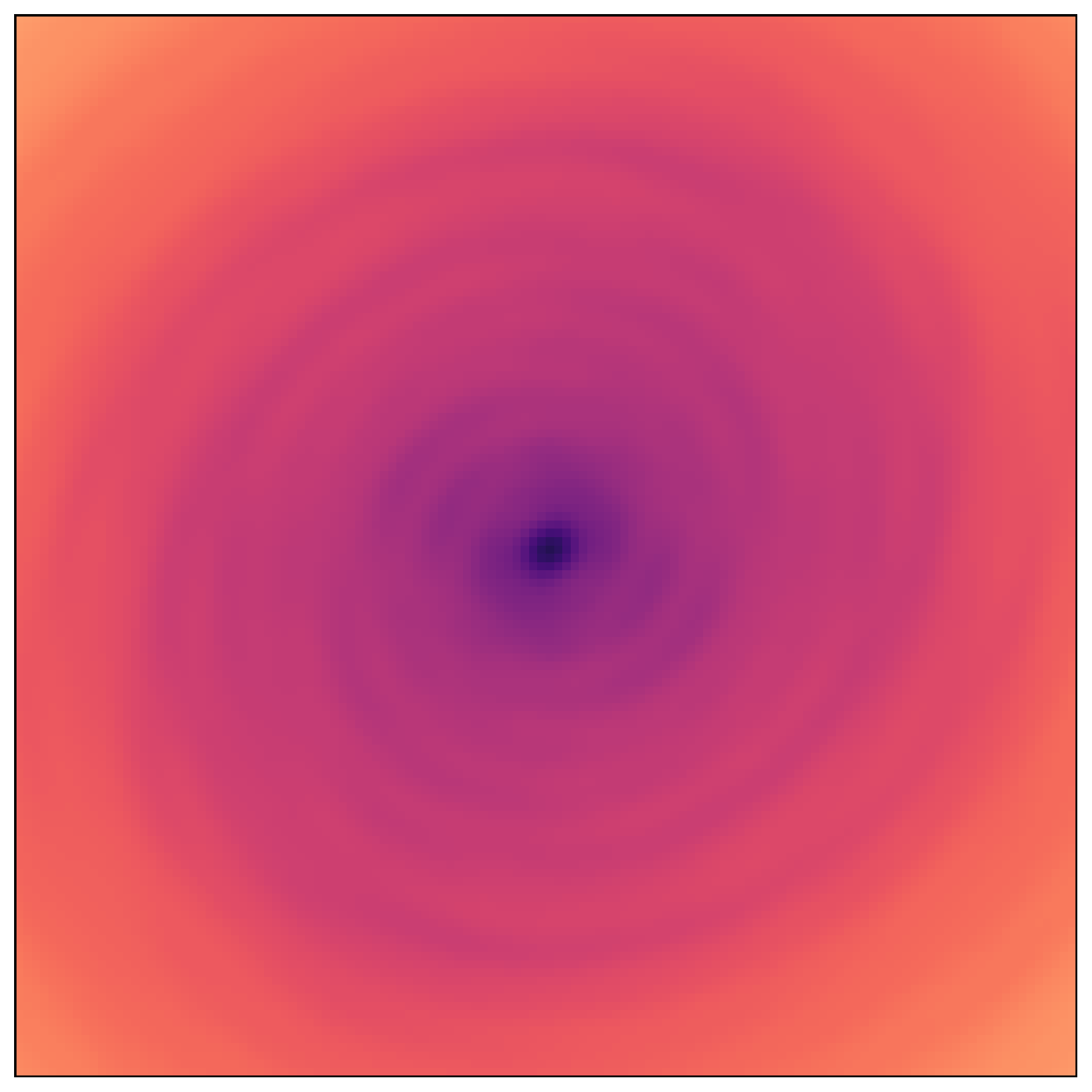} \\

\multicolumn{3}{c}{\small{\texttt{\bfseries X=30}}} \\
\includegraphics[height=135pt]{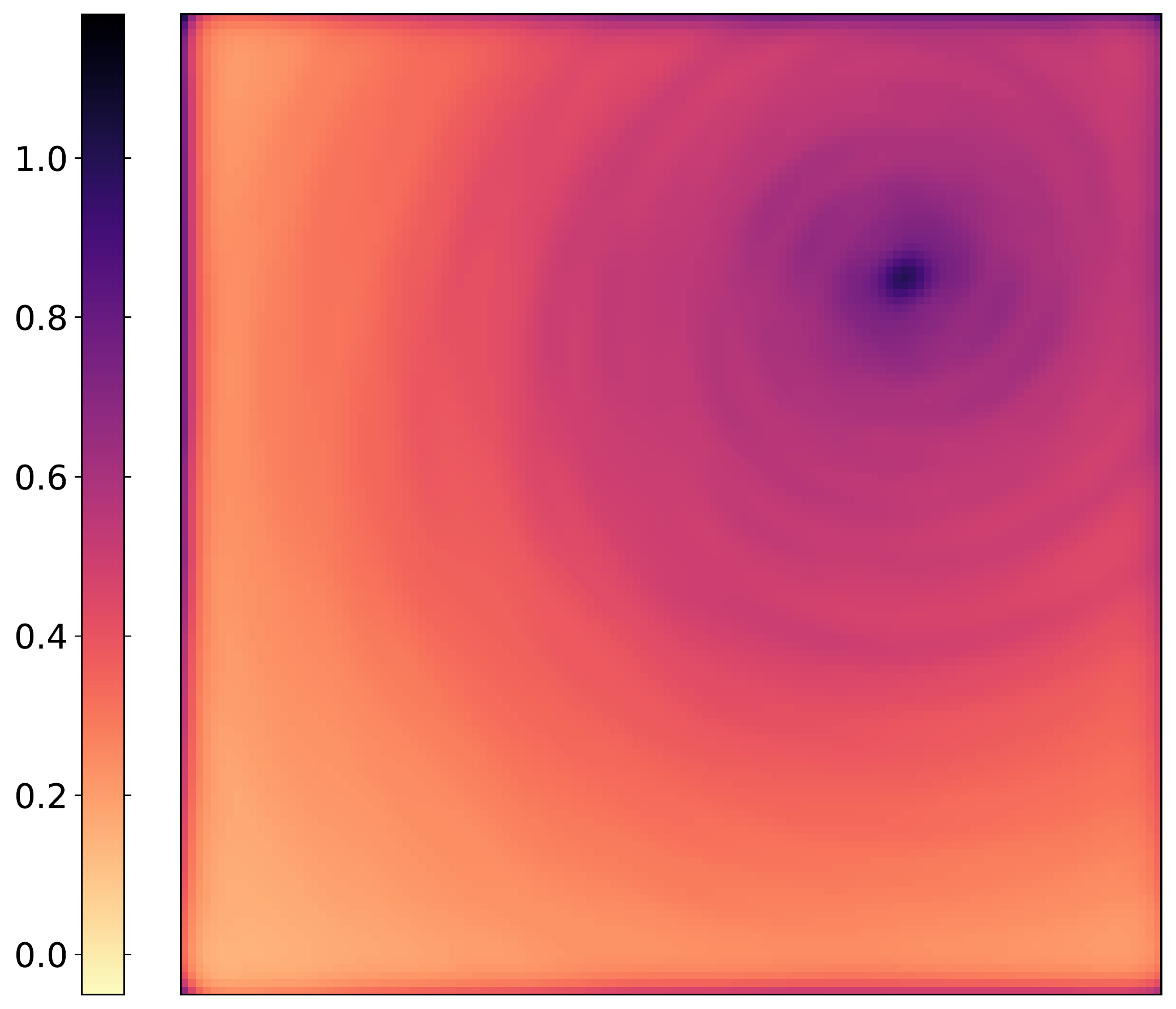} &
\includegraphics[height=135pt]{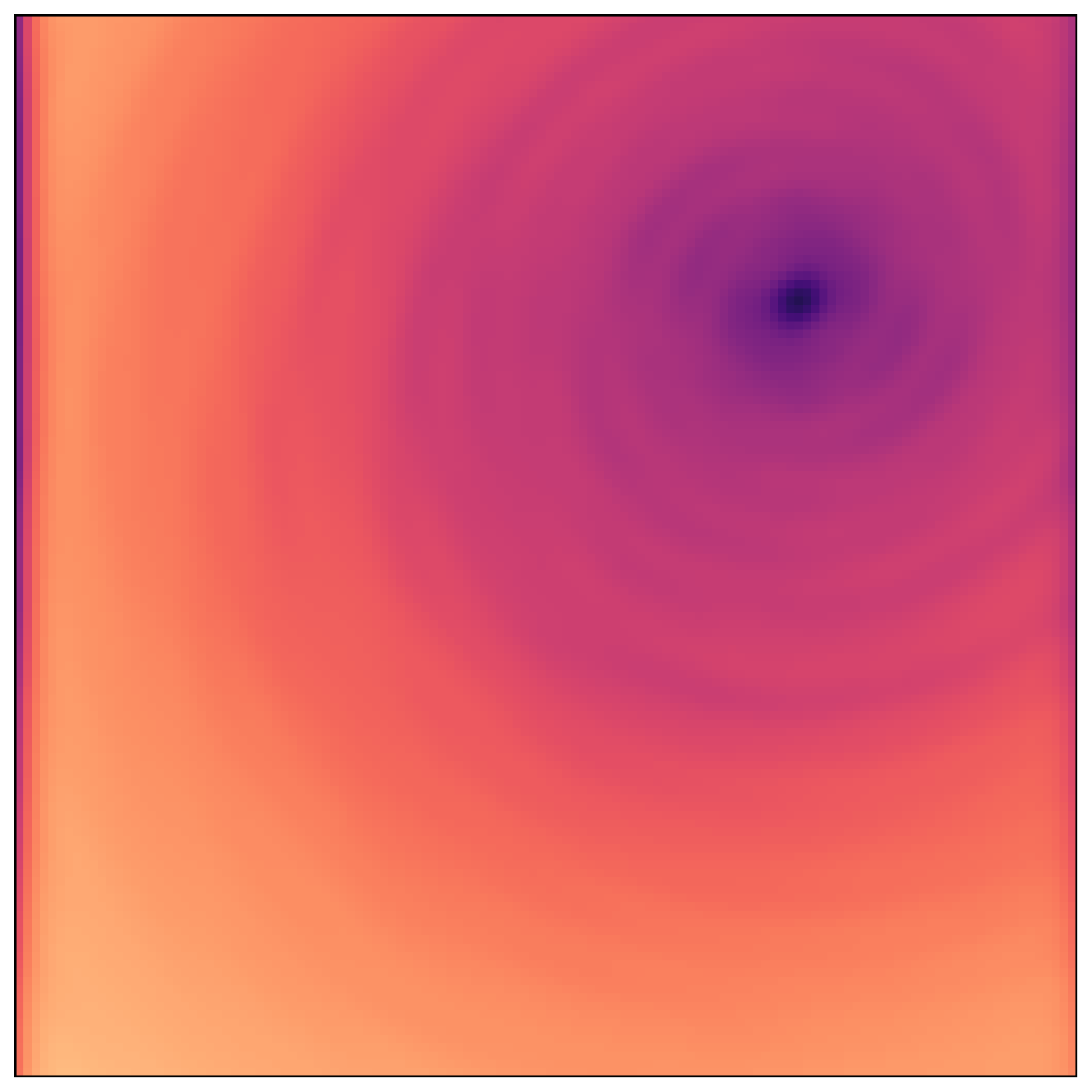} &
\includegraphics[height=135pt]{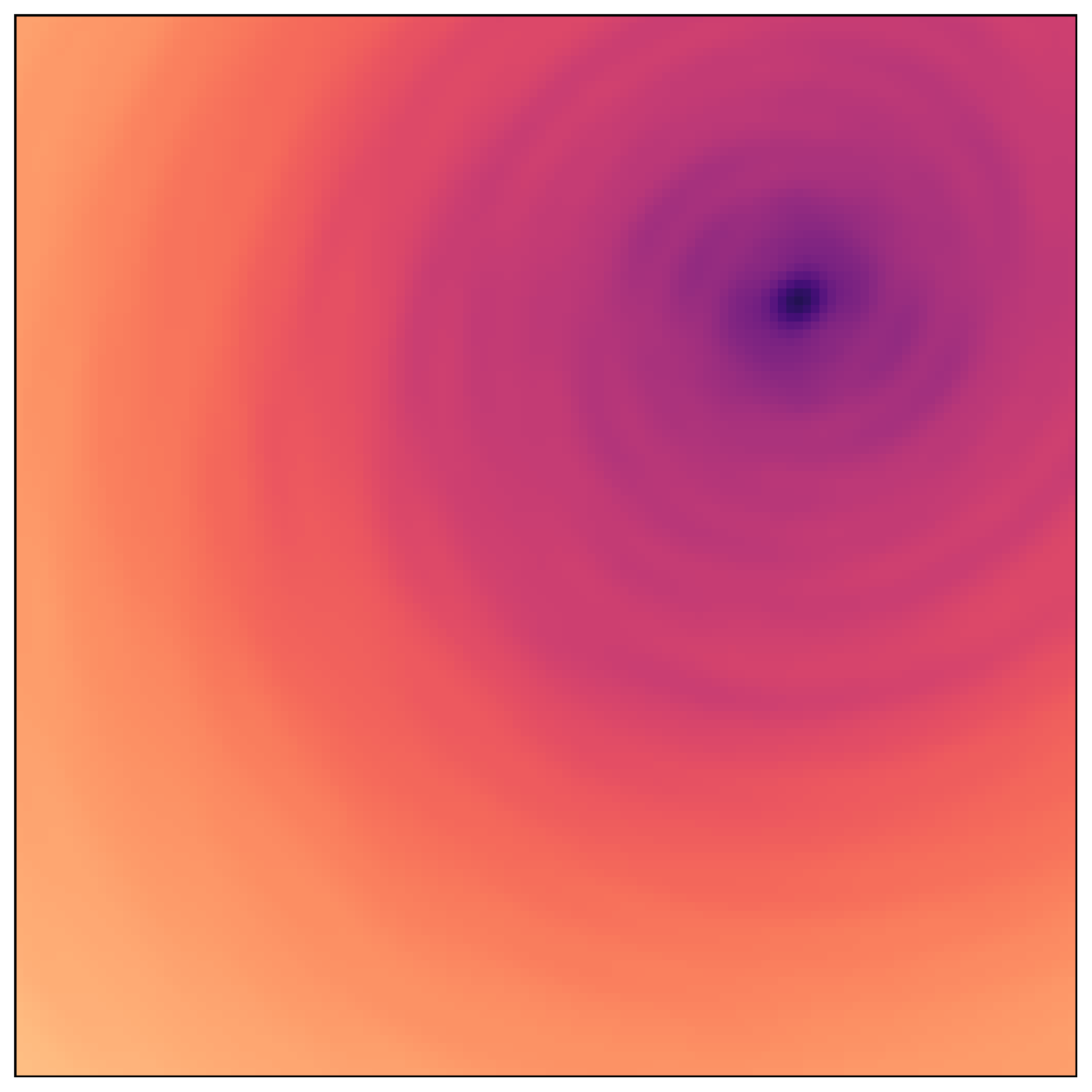} \\

\multicolumn{3}{c}{\small{\texttt{\bfseries X=62}}} \\
\includegraphics[height=135pt]{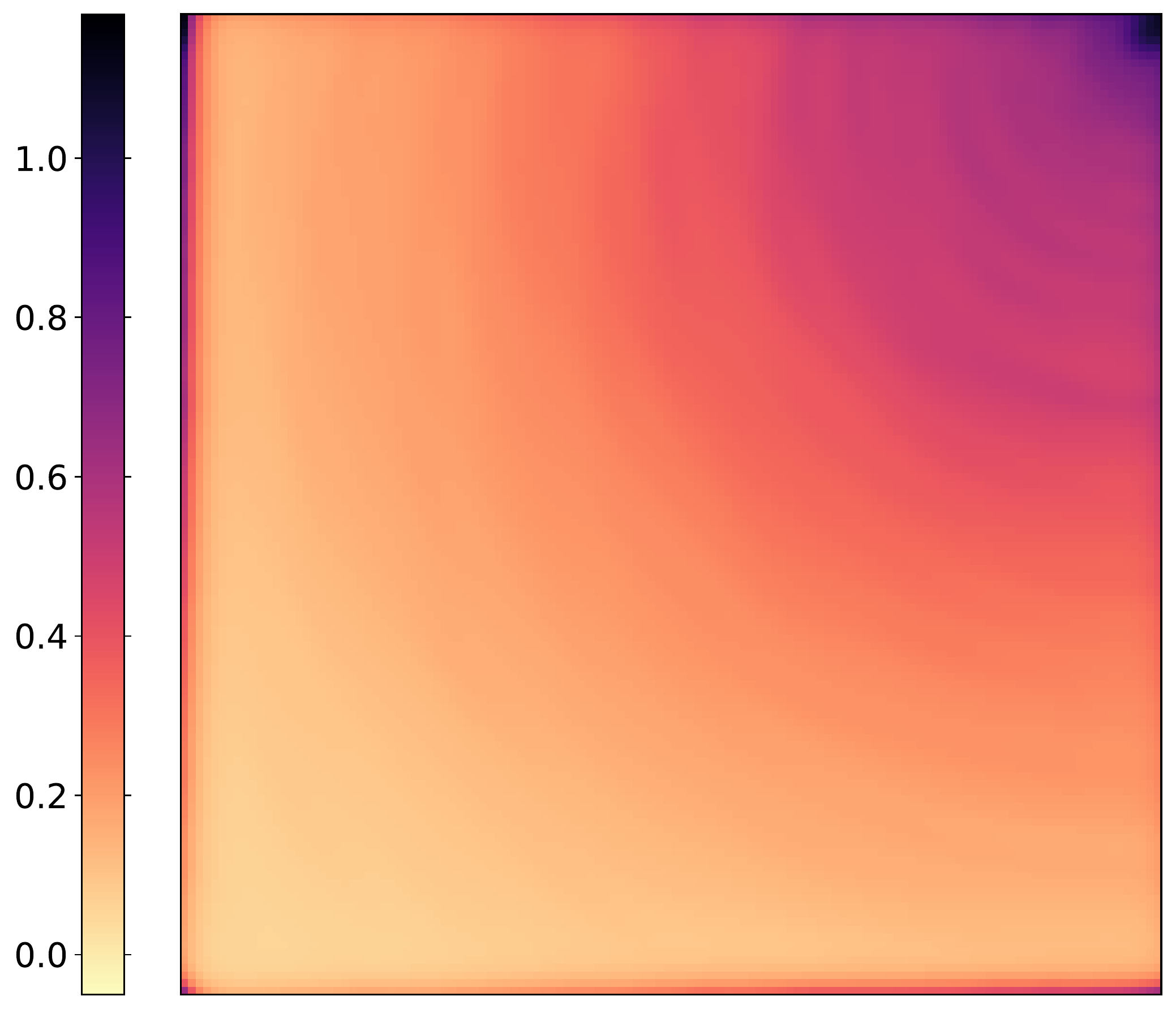} &
\includegraphics[height=135pt]{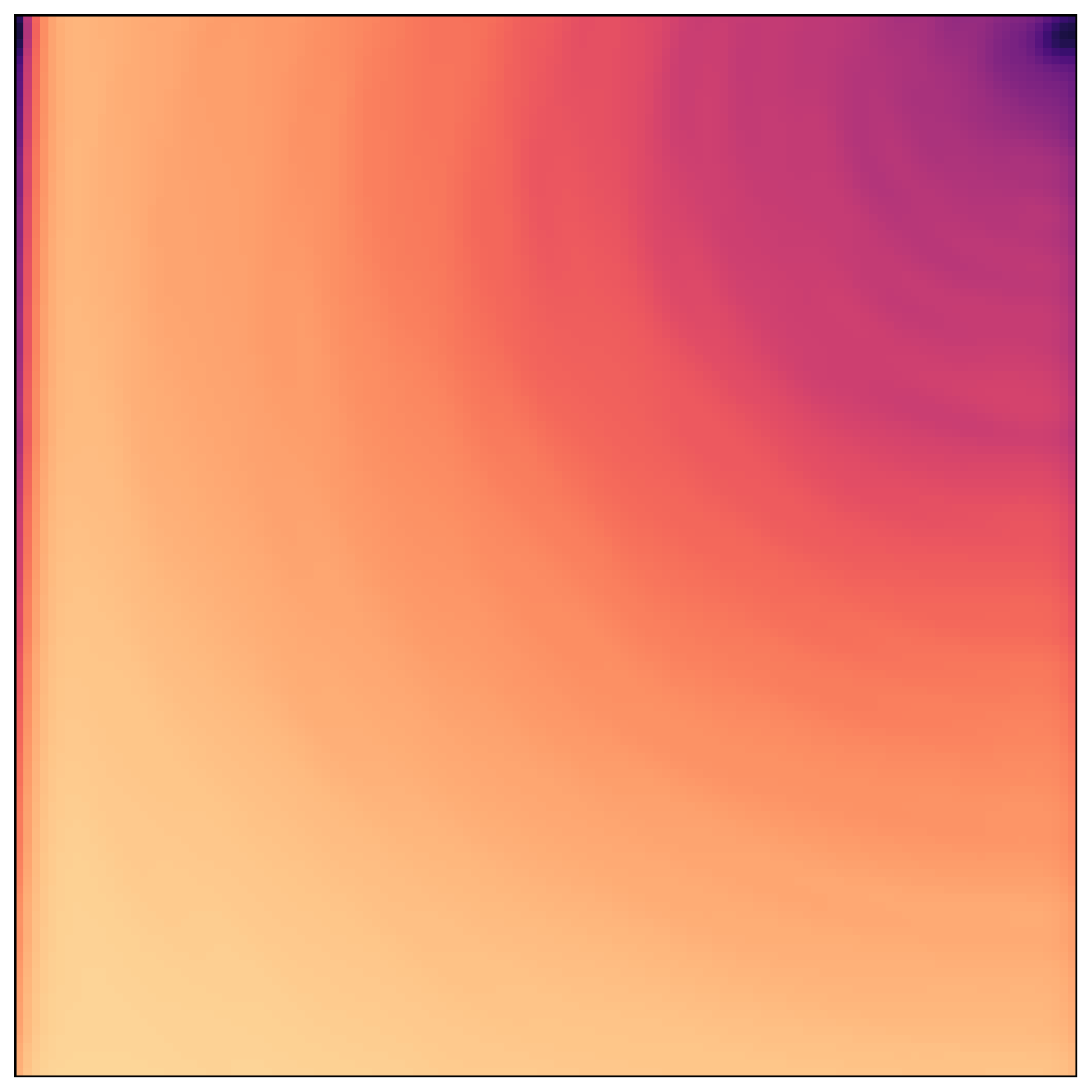} &
\includegraphics[height=135pt]{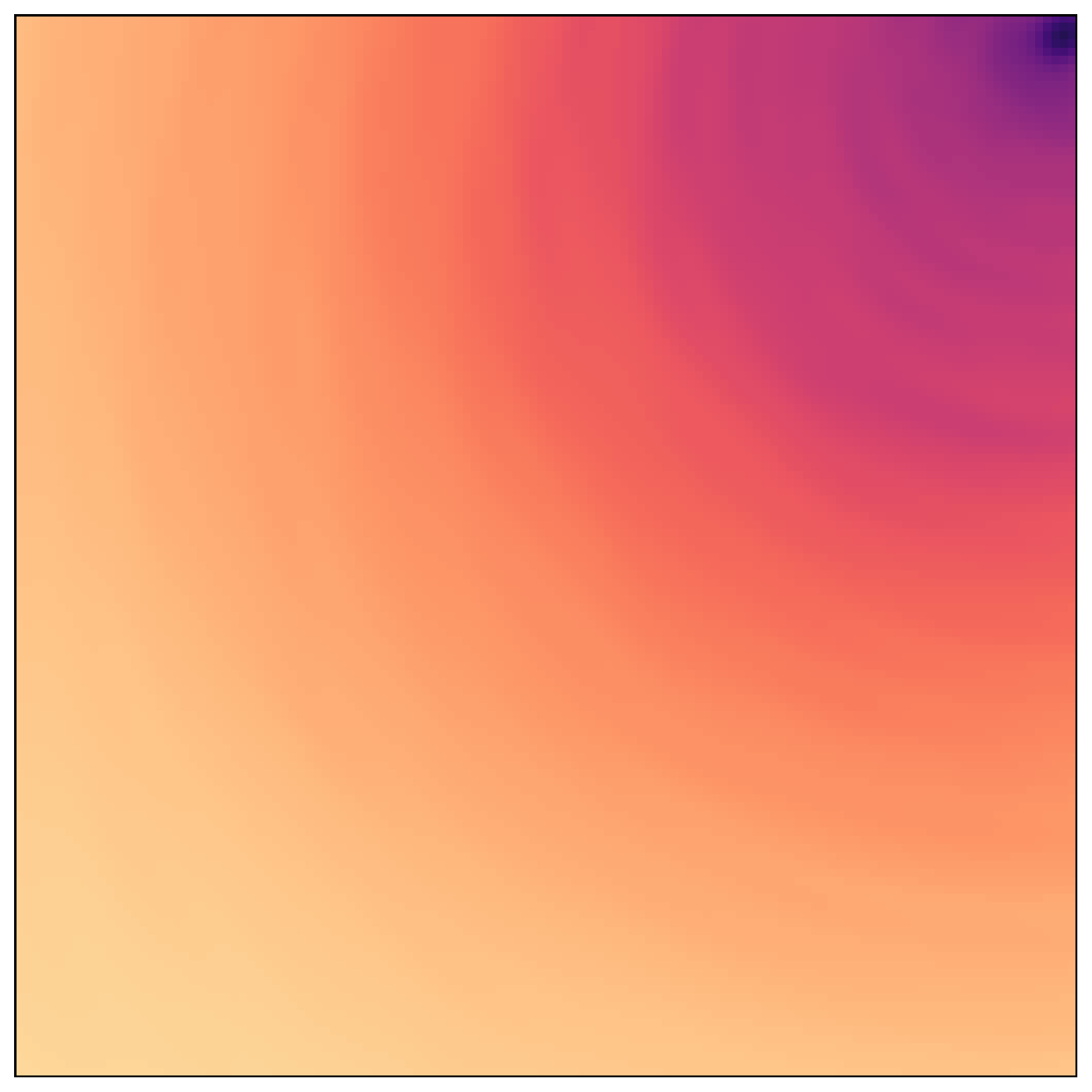} \\
\hspace{0.8cm}\footnotesize\texttt{Convolutional Gridding} & \footnotesize\texttt{Hybrid Gridding} & \footnotesize\texttt{Pruned NN Interpolation}\\
\\
\end{tabular}
\caption[Images generated by experiments of Section \ref{sec:comparative:aliasing}]{A selection of images computed in Double-Precision by the experiments of Section \ref{sec:comparative:aliasing} using the LOFAR observation.} 
\label{fig:comparative:aliasinglofardouble}
\end{figure}

We reproduce the output images for $X=0$, $X=30$ and $X=62$ in Figures \ref{fig:comparative:aliasinggmrtsingle} to \ref{fig:comparative:aliasinglofardouble} while plotting all measurements of $R(X)$ in Figure \ref{fig:comparative:aliascompare}.

Let us first do a visual inspection of the output images. It is easy to see that all images for $X=62$ and $X=30$ generated through Convolutional Gridding feature conspicuous aliasing distortion near the left and bottom edges, identified by a reddish glow near the stated edges.  The $X=0$ images generated by the VLA observation also features such visible distortions. Aliasing distortions at the bottom edge vanishes\footnote{In our discourse, we need to clarify that "vanishing" alias distortions do not mean that there is no more aliasing but that such distortions do not cause enough change in intensity to be visibly detectable.} in the respective images generated by Hybrid Gridder, while all visible aliasing distortions vanish in the respective images generated by Pruned NN Interpolation. Such reduction in aliasing is as expected and gives the first proof that Hybrid Gridding and Pruned NN Interpolation suppress aliasing at a level higher than that of Convolutional Gridding. 

 Let's now analyse the measured $R(X)$ plotted in Figure \ref{fig:comparative:aliascompare}. We first need to point out that aliasing distortions measured through $R(X)$ are independent of Precision. On the other hand, arithmetic noise depends on Precision, where Double-Precision should have less noise, as the results indicate. Therefore, it is best to look at experiments conducted in Double-Precision to analyse aliasing distortions. 
 
 In general, $R(X)$, in the  Double-Precision and Single-Precision experiments, was highest for  Convolutional Gridding, decreased for Hybrid Gridding, and was measured to be lowest for Pruned NN Interpolation. Therefore, we proved our arguments made in Section \ref{sec:maths:comparisons}, whereby we claimed that Pruned NN Interpolation should suppress aliasing more than Hybrid Gridding, which in turn suppresses aliasing more than Convolutional Gridding. Consequently,  these results show that  Hybrid Gridding suppresses aliasing at least at the same level as Convolutional Gridding.
 
The Single-Precision Experiments show an understandably relatively high level of arithmetic noise. Interestingly, aliasing effects in Pruned NN Interpolation were below arithmetic noise, and this gave rise to a situation whereby arithmetic noise cancelled aliasing in a way to have some experiments reporting an $R(X)$ of zero. Due to this phenomenon, we shied away from presenting averaging results over an image as we did for the validation of the Convolutional Gridding implementation against CASA. Such averaging can lead to results that falsely suggest that Single-Precision is more accurate than Double-Precision.

\section{Discussion}

Following the Performance results of Section \ref{sec:comparative:performancestudied}, it is essential to discuss how Hybrid Gridding can be applied in an actual image synthesis scenario as part of a Visibility-to-image transform in a major cycle in Deconvolution. We noted that Hybrid Gridding has two drawbacks compared to Convolutional Gridding: the need to sort the set of Visibility records in the SS Ordering Mode and the need for more memory than the Convolutional Gridder. 

Sorting might take a substantial amount of execution time. However, given that sorting is executed once in an initialisation phase and not repeated in every major cycle of which there may be thousands of such cycles, we expect the impact of sorting to be irrelevant to the imaging pipeline's total execution time. 

Memory needs can make some imaging scenarios impossible with Hybrid Gridding.  The major problems arise when $\beta$ is large, say 128. Small images of size, say $256 \times 256$ should be fine to image, but less likely for large wide-band images of say $8192 \times 8192$, imaged using  $w$-stacking with Hybrid Gridding at $\beta=128$. In this case, each $w$-plane will need an amount of memory far larger than what the P100 provides. One might think to use facets with $w$-stacking to work with the small facets. In such a case, we expect each facet to need less memory to be imaged than one full $w$-plane, potentially making it possible to image the facet on the P100. It is essential to keep in mind that $\beta$ in each facet will likely need to be increased (therefore requiring more memory) to keep under control the level of aliasing distortions. 

If we are required to image with a small value of $\beta$ say $\beta=8$, then large images of size, say $8192\times 8192$, ($N_{\text{nn}}=65536$), should be imageable with $w$-stacking using Hybrid Gridding as well revealed by the many results we gave in this chapter and Chapter \ref{chap:hybrid}.

\section{Conclusion}
In this chapter, we concluded our analyses for this thesis. We showed that as implemented on the P100, Hybrid Gridding can be more performant than Convolutional Gridding, with no reduction in the suppression of aliasing. Therefore, we reached the primary goal of this thesis.

\chapter{Conclusion}
\label{chap:conclusion}

In this thesis, we proposed modifications to  Convolutional Gridding for better Performance (faster execution). We took advantage of the oversampling of the GCF, to design Hybrid Gridding (Algorithm \ref{algo:maths:hybrid}) and Pruned NN Interpolation (Algorithm \ref{algo:maths:purenninterpolation}). Through novel implementations on the P100 of the three mentioned algorithms, we successfully proved that Hybrid Gridding could indeed perform faster than Convolutional Gridding. In some test scenarios, the Hybrid Gridder was more than $6\times$ faster than the Convolutional Gridder. In other test scenarios, the NN Gridder was also faster than the Convolutional Gridder, but many times slower than the Hybrid Gridder. To top it up finalisation of Pruned NN Interpolation made the implementation always slower in execution than Hybrid Gridding, with full compression enabled.

Based on the many results we gave in this thesis, we recommend the use of Hybrid Gridding in favour of Convolutional Gridding provided that sorting of records is viable and the number of records to grid is substantial. If any of the stated two conditions are not met, we recommend the use of the Convolutional Gridding. We do not recommend the use of the Pruned NN Interpolation in any scenario since Hybrid Gridding always performed better than Pruned NN Interpolation provided there is full compression enabled.

The two new algorithms feature FFT pruning through downsampling of an oversampled grid using a convolution. We proposed the use of the recently discovered least-misfit functions (Ye \etal \cite{Ye2019}) to potentially reduce the effects of aliasing caused by the downsampling below arithmetic noise. Our experiments showed that such pruning could reduce the inversion time of an oversampled grid by $8\times$. Results related to reducing aliasing effects below arithmetic noise were promising, but we were not always able to reproduce fully the works of Ye \etal \cite{Ye2019}. We recommend the use of downsampling of a grid using a convolution with the least-misfit functions as a possible Convolution-Based FFT Pruning technique. Nevertheless, we tag as future work the need for discovery of more least-misfit functions and more in-depth analyses of how they work out in the context Convolution-Based FFT Pruning.

\section{Future work}
As future work, we point out our intention to publish most of the contents of this thesis, in well respected peer-reviewed journals. Following are various ideas that we wish to propose going forward.

We heavily suggest the implementation of full faceting or w-stacking using Hybrid Gridding to have a comprehensive picture of how Hybrid Gridding plays out in a complete setup. We also suggest revisiting the new algorithms for possible implementation on a CPU and other computing devices such as FPGAs.

We did implement the Hybrid and Convolutional Gridders to handle GCFs of support eight. However, in this thesis, we provided no analyses of their Performance and we, therefore, suggest such analyses. It is also worth to modify the Convolutional Gridder to support complex-valued GCFs and review its Performance. A gridder handling complex GCFs is very useful in w-projection. 

In our analyses of the Convolutional Gridder, we raised our suspicion that the GCF data retrieval is choking grid committing and the loading of record data. It would be interesting to consider the possibility of calculating the GCF on the fly instead of loading any GCF data from memory, using the ideas proposed in Image Domain Gridding. We have conflicting thoughts on the Performance results of such an implementation. We expect the level of  compute to substantially increase, which might back-fire on Performance since the Convolutional Gridder is already high on the utilisation of compute. At the same time, such an implementation would reduce memory access which can help increase the Performance of the Gridder. We warn the reader that there is a national patent in the Netherlands (van der Tol \cite{VanderTol2017}) in regards to IDG. Therefore, any research considerations that includes IDG should include possible legal limitations that ASTRON might enforce through the patent.

In this thesis, we also worked on another algorithm whose analysis was not finalised due to time limitations. It is similar to Hybrid Gridding, but instead of a convolution, we considered a DFT. Finalisation is done using one-dimensional FFTs operated on the columns. A preliminary analysis of this algorithm gave promising performance results up to image sizes of dimensions $256\times 256$, and we think it is worth to finalise all the analyses of the algorithm and publish results.

In Section \ref{sec:introduction:residualcalculation} we briefly mentioned the process of degridding which calculates Visibility values from an intensity model. It forms part of the major cycle in most Deconvolution algorithms and dominates the execution time of a major cycle similar to gridding. Therefore, we recommend the re-adaption of the three studied implementations for degridding.  

Finally, we would like to conclude this thesis with a personal note. Through the tenure of our doctoral studies, we have been able to gather extensive expertise in GPU development. We have indications from correspondence with collaborators that the Gridders we developed will be quite useful to the Radio Interferometric community. We also believe that the ideas for future work should give results which are publishable and give further contribution to Radio Interferometry. We are grateful to have had the opportunity to do such work and eager and hopeful to be able to continue to evolve this work further.

\bibliomatter
\bibliographystyle{packages/IEEEtran}

\bibliography{manual.bib}
\end{document}